\documentclass[fleqn,twoside]{article}

\usepackage{espcrc2}

\usepackage[dvipsone]{graphicx}
\graphicspath{{lanl/}}

\usepackage{relsize}
\def\babar{\mbox{\slshape B\kern-0.1em{\smaller A}\kern-0.1em
    B\kern-0.1em{\smaller A\kern-0.2em R}}}

\def\en         {\ensuremath{e^-}}      
\def\ep         {\ensuremath{e^+}}
\def\epm        {\ensuremath{e^\pm}}
\def\epem       {\ensuremath{e^+e^-}}

\def\mumu       {\ensuremath{\mu^+\mu^-}}
\def\mtau       {\ensuremath{\tau}}

\def\taum       {\ensuremath{\tau^-}}
\def\tautau     {\ensuremath{\tau^+\tau^-}}

\def\qqbar {\ensuremath{q\overline q}}

\def\s  {\ensuremath{s}}

\def\ccbar {\ensuremath{c\overline c}}

\def\bbbar {\ensuremath{b\overline b}}
\def\t  {\ensuremath{t}}

\def\piz   {\ensuremath{\pi^0}}

\def\pip   {\ensuremath{\pi^+}}

\def\pipi  {\ensuremath{\pi^+\pi^-}}

\def\Kbar  {\kern 0.2em\overline{\kern -0.2em K}{}}

\def\KS    {\ensuremath{K^0_{\scriptscriptstyle S}}}
\def\KL    {\ensuremath{K^0_{\scriptscriptstyle L}}}

\def\Kzb   {\ensuremath{\Kbar^0}}
\def\KzKzb {\ensuremath{K^0 \kern -0.16em \Kzb}}

\def\Dbar  {\kern 0.2em\overline{\kern -0.2em D}{}}

\def\Dzb   {\ensuremath{\Dbar^0}}
\def\DzDzb {\ensuremath{D^0 {\kern -0.16em \Dzb}}}
\def\Dstar   {\ensuremath{D^*}}

\def\Bz    {\ensuremath{B^0}}
\def\B     {\ensuremath{B}}
\def\Bbar  {\kern 0.18em\overline{\kern -0.18em B}{}}

\def\Bzb   {\ensuremath{\Bbar^0}}

\def\Bub   {\ensuremath{B^-}}

\def\BB    {\ensuremath{B\Bbar}}
\def\BzBzb {\ensuremath{B^0 {\kern -0.16em \Bzb}}}

\def\jpsi  {\ensuremath{{J\mskip -3mu/\mskip -2mu\psi\mskip 2mu}}}
\mathchardef\Upsilon="7107
\def\Y#1S{\ensuremath{\Upsilon{(#1S)}}}

\def\FourS {\Y4S}

\mathchardef\Deltares="7101
\mathchardef\Xi="7104
\mathchardef\Lambda="7103
\mathchardef\Sigma="7106
\mathchardef\Omega="710A
\def\Deltabar   {\kern 0.25em\overline{\kern -0.25em \Deltares}{}}
\def\Lbar {\kern 0.2em\overline{\kern -0.2em\Lambda\kern 0.05em}\kern-0.05em{}}
\def\Sigbar{\kern 0.2em\overline{\kern -0.2em \Sigma}{}}
\def\Xibar{\kern 0.2em\overline{\kern -0.2em \Xi}{}}
\def\Obar{\kern 0.2em\overline{\kern -0.2em \Omega}{}}
\def\Nbar{\kern 0.2em\overline{\kern -0.2em N}{}}
\def\Xbar{\kern 0.2em\overline{\kern -0.2em X}{}}

\def\bpsiks{\ensuremath{B^0 \to \jpsi \KS}}

\def\Bztopipi   {\ensuremath{B^0 \to \pipi}}
\def\Bztokpi    {\ensuremath{B^0 \to K^{\pm}\pi^{\mp}}}

\def\Bztodd     {\ensuremath{B^0 \to D^+D^-}}

\def\pxy        {\mbox{$p_T$}}
\def\pt         {\mbox{$p_T$}}
\def\mes        {\mbox{$m_{\rm ES}$}}

\def\mphi       {\mbox{$\phi$}}

\def\ev   {\ensuremath{\rm \,e\kern -0.08em V}}
\def\kev  {\ensuremath{\rm \,ke\kern -0.08em V}}
\def\mev  {\ensuremath{\rm \,Me\kern -0.08em V}}
\def\gev  {\ensuremath{\rm \,Ge\kern -0.08em V}}
\def\gevc {\ensuremath{{\rm \,Ge\kern -0.08em V\!/}c}}
\def\tev  {\ensuremath{\rm \,Te\kern -0.08em V}}
\def\mevc {\ensuremath{{\rm \,Me\kern -0.08em V\!/}c}}
\def\gevcc{\ensuremath{{\rm \,Ge\kern -0.08em V\!/}c^2}}
\def\mevcc{\ensuremath{{\rm \,Me\kern -0.08em V\!/}c^2}}

\def\km   {\ensuremath{\rm \,km}}
\def\m    {\ensuremath{\rm \,m}}
\def\cm   {\ensuremath{\rm \,cm}}
\def\cma  {\ensuremath{{\rm \,cm}^2}}
\def\mm   {\ensuremath{\rm \,mm}}

\def\mum  {\ensuremath{\,\mu\rm m}} 

\def\nm   {\ensuremath{\rm \,nm}}

\def\invpb {\ensuremath{\mbox{\,pb}^{-1}}}

\def\invfb   {\ensuremath{\mbox{\,fb}^{-1}}}
\def\mus  {\ensuremath{\rm \,\mus}}
\def\ns   {\ensuremath{\rm \,ns}}

\def\gm   {\ensuremath{\rm \,g}}
\def\Xrad {\ensuremath{X_0}}

\def\cms         {\ensuremath{\,\mathrm{cm}^{-2} \mathrm{s}^{-1}}}
\def\nm         {\ensuremath{\,\mathrm{nm}}}    
\def\nb         {\ensuremath{\,\mathrm{nb}}}
\def\ms         {\ensuremath{\,\mathrm{ms}}}      
\def\mus        {\ensuremath{\,\mu\mathrm{s}}}    
\def\ns         {\ensuremath{\,\mathrm{ns}}}      
\def\kbytes     {\ensuremath{\,\mathrm{kbytes}}}

\def\mbytes     {\ensuremath{\,\mathrm{Mbytes}}}

\def\mbps       {\ensuremath{\,\mathrm{Mbytes/s}}}
\def\mbsps      {\ensuremath{\,\mathrm{Mbits/s}}}

\def\gbytes     {\ensuremath{\,\mathrm{Gbytes}}}

\def\gbps       {\ensuremath{\,\mathrm{Gbytes/s}}}
\def\gbsps      {\ensuremath{\,\mathrm{Gbits/s}}}

\def\tbytes     {\ensuremath{\,\mathrm{Tbytes}}}

\def\krad {\ensuremath{\,\mathrm{krad}}}

\def\degc {\ensuremath{^\circ}{C}}
\def\degk {\ensuremath{\mathrm{K}}}
\def\degrees{\ensuremath{^{\circ}}}
\def\mrad{\ensuremath{\,\mathrm{mr}}}                

\def\gsim{{~\raise.15em\hbox{$>$}\kern-.85em 
          \lower.35em\hbox{$\sim$}~}}        
\def\lsim{{~\raise.15em\hbox{$<$}\kern-.85em
          \lower.35em\hbox{$\sim$}~}}

\def\CP                 {\ensuremath{C\!P}}
\def\ra                 {\ensuremath{\rightarrow}}

\def\pep2{PEP-II}
\def\BF{$B$ Factory}

\newcommand{\dedx}{\ensuremath{\mathrm{d}\hspace{-0.1em}E/\mathrm{d}x}}

\providecommand{\eqref}[1]{Eq.~(\ref{eq:#1})}

\newcommand{\itns}      [1]  {{IEEE Trans.\ Nucl.\ Sci.\ {#1}}}

\newcommand{\nim}       [1]  {{Nucl.\ Instr.\ and Methods~{#1}}}
\newcommand{\nima}      [1]  {{Nucl.\ Instr.\ Methods {A{#1}}}}

\newcommand{\npb}       [1]  {{Nucl.\ Phys.\ {B{#1}}}}

\newcommand{\ncim}      [1]  {{Nuo.\ Cim.\ {#1}}}

\providecommand{\pr}        [1]  {{Phys.\ Rev.\ {#1}}}

\def\jetset74   {\mbox{\tt Jetset \hspace{-0.5em}7.\hspace{-0.2em}4}}

\newcommand{\comment}[1]{}

\newcommand{\sectiondir}{}
\newcommand{\secname}{}

\def\etal{{\em et al.,}}

\def\ie{{\em i.e.,}}
\def\eg{{\em e.g.,}}

\def\to{\rightarrow}
\def\ra{\rightarrow}

\def\hz	  {\ensuremath{\mbox{\,Hz}}}
\def\khz  {\ensuremath{\mbox{\,kHz}}}
\def\mhz  {\ensuremath{\mbox{\,MHz}}}

\def\mv  	{\ensuremath{\mbox{\,mV}}}
\def\volt  	{\ensuremath{\mbox{\,V}}}
\def\kv  	{\ensuremath{\mbox{\,kV}}}
\def\mw  	{\ensuremath{\mbox{\,mW}}}
\def\kw  	{\ensuremath{\mbox{\,kW}}}
\def\watt  	{\ensuremath{\mbox{\,W}}}
\def\ohm  	{\ensuremath{\mbox{\,$\Omega$}}}
\def\Mohm  	{\ensuremath{\mbox{\,M$\Omega$}}}

\def\kohm       {\ensuremath{\mbox{\,k$\Omega$}}}

\def\Amp        {\ensuremath{\mbox{\,A}}}
\def\mA         {\ensuremath{\mbox{\,mA}}}
\def\muA        {\ensuremath{\mbox{\,$\mu$A}}}
\def\nA         {\ensuremath{\mbox{\,nA}}}

\def\fC         {\ensuremath{\mbox{\,fC}}}
\def\pf         {\ensuremath{\mbox{\,pF}}}

\def\mbar       {\ensuremath{\mbox{\,mbar}}}

\def\kN         {\ensuremath{\mbox{\,kN}}}
\def\kg   {\ensuremath{\,\mbox{kg}}}
\def\liter{\ensuremath{\,\ell}}

\def\tesla{\ensuremath{\mbox{\,T}}}
\def\mtesla{\ensuremath{\mbox{\,mT}}}
\def\mrad       {\ensuremath{\mbox{\,mrad}}}
\def\pxy        {\mbox{$p_{\rm t}$}}
\def\pt         {\mbox{$p_{\rm t}$}}

\renewcommand{\dedx}{\ensuremath{dE/dx}}

\renewcommand{\epsilon}{\varepsilon}
\begin{document}

\title{The \babar\ Detector}

\author{The \babar\ Collaboration
\vspace{2.5\baselineskip}}

\begin{abstract}
 \babar, the detector for the SLAC \pep2\ asymmetric \epem\ \BF\
 operating at the \FourS\ resonance, was designed to
 allow comprehensive studies of \CP-violation in $B$-meson decays.
 Charged particle tracks are measured in a multi-layer
 silicon vertex tracker surrounded by a cylindrical wire drift
 chamber.  Electromagnetic showers from electrons and photons are
 detected in an array of CsI crystals located just inside the
 solenoidal coil of a superconducting magnet. Muons and neutral
 hadrons are identified by arrays of resistive plate chambers inserted
 into gaps in the steel flux return of the magnet. Charged hadrons are
 identified by \dedx\ measurements in the tracking detectors and in a
 ring-imaging Cherenkov detector surrounding the drift chamber.
 The trigger, data acquisition and data-monitoring systems , VME-
 and network-based, are controlled by custom-designed
 online software.  Details of the layout and performance of the detector
 components and their associated electronics and software are
 presented.
\end{abstract}

\maketitle

\renewcommand{\secname}{intro_}
\renewcommand{\sectiondir}{sec01_intro}
\section{Introduction}

The primary physics goal of the \babar\ experiment is the
systematic study of \CP -violating asymmetries in the decay of
neutral $B$ mesons to \CP\ eigenstates. Secondary goals are
precision measurements of decays of bottom and charm mesons and of
$\tau$ leptons, and searches for rare processes that become
accessible with the high luminosity of the \pep2\ \BF\
\cite{R_PEP}.  The design of the detector is optimized for \CP\
violation studies, but it is also well suited for these other
physics topics.  The scientific goals of the \babar\ experiment
were first presented in the Letter of Intent \cite{R_LOI} and the
Technical Design Report \cite{ref:R_TDR}; detailed physics studies
have been documented in the \babar\ Physics Book \cite{R_PBook}
and earlier workshops \cite{R_PWS}.

The \pep2\ \BF\ is an asymmetric \epem\ collider designed to
operate at a luminosity of $3 \times 10^{33}$ \cms\ and above, at
a center-of-mass energy of 10.58\gev, the mass of the \FourS\
resonance. This resonance decays exclusively to \Bz \Bzb\ and
$B^+B^-$ pairs and thus provides an ideal laboratory for the study
of $B$ mesons.  In \pep2, the electron beam of 9.0\gev\ collides
head-on with the positron beam of 3.1\gev\ resulting in a Lorentz
boost to the \FourS\ resonance of $\beta\gamma =0.56$. This boost
makes it possible to reconstruct the decay vertices of the two $B$
mesons, to determine their relative decay times, and thus to
measure the time dependence of their decay rates.  The crucial
test of \CP\ invariance is a comparison of the time-dependent
decay rates for \Bz\ and \Bzb\ to a self-conjugate state.  For the
cleanest experimental test, this requires events in which one $B$
meson decays to a \CP\ eigenstate that is fully reconstructed and
the other $B$ meson is tagged as a \Bz\ or a \Bzb\ by its decay
products: a charged lepton, a charged kaon, or other flavor
sensitive features such as a low momentum charged pion from a
$D^*$ decay.

The very small branching ratios of $B$ mesons to \CP\ eigenstates,
typically $10^{-4}$, the need for full reconstruction of final
states with two or more charged particles and several \piz s, plus
the need to tag the second neutral $B$, place stringent
requirements on the detector, which should have

\begin{figure*}
\centering
\includegraphics*[width=.95\textwidth]{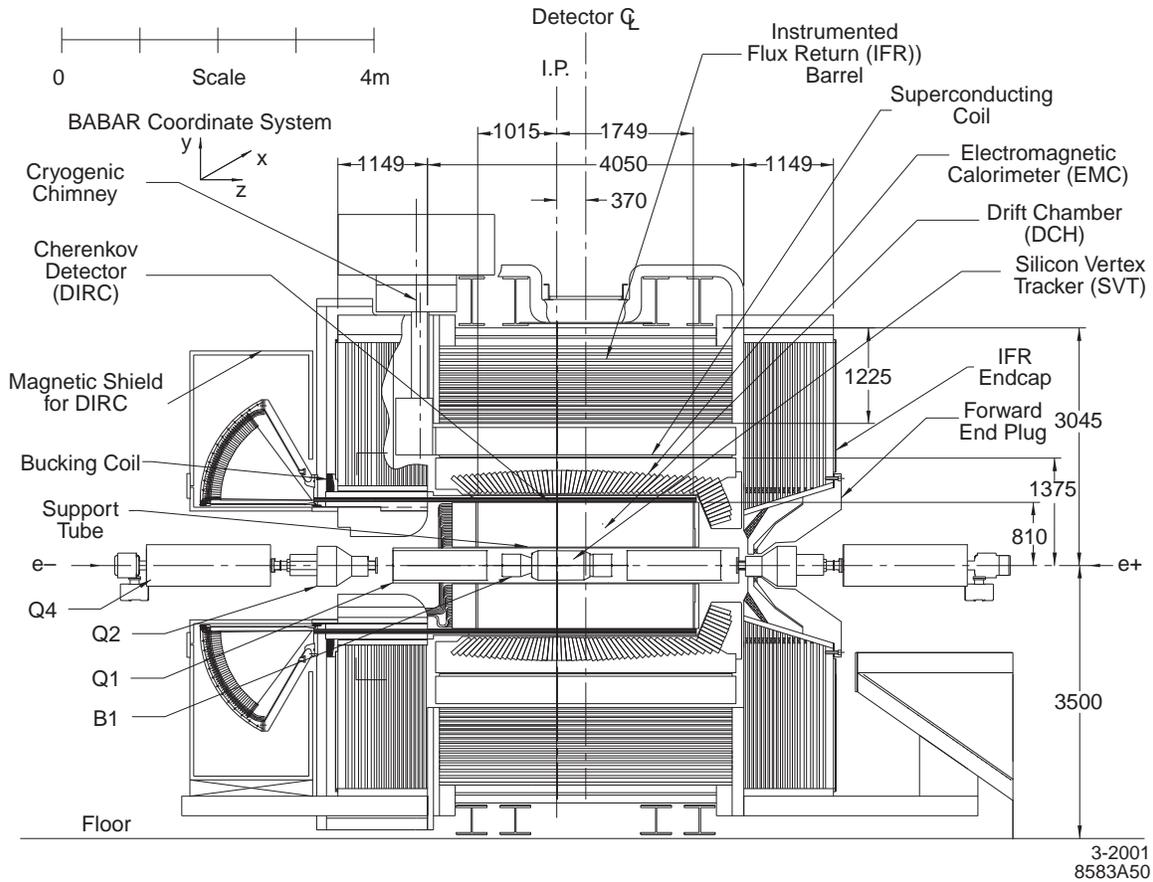}
\vspace{-1.5pc}
\caption{\babar\ detector longitudinal section.}
\label{layout_fig:detectorElevation}
\end{figure*}

\begin{figure*}[t]
\centering
\includegraphics*[width=.95\textwidth]{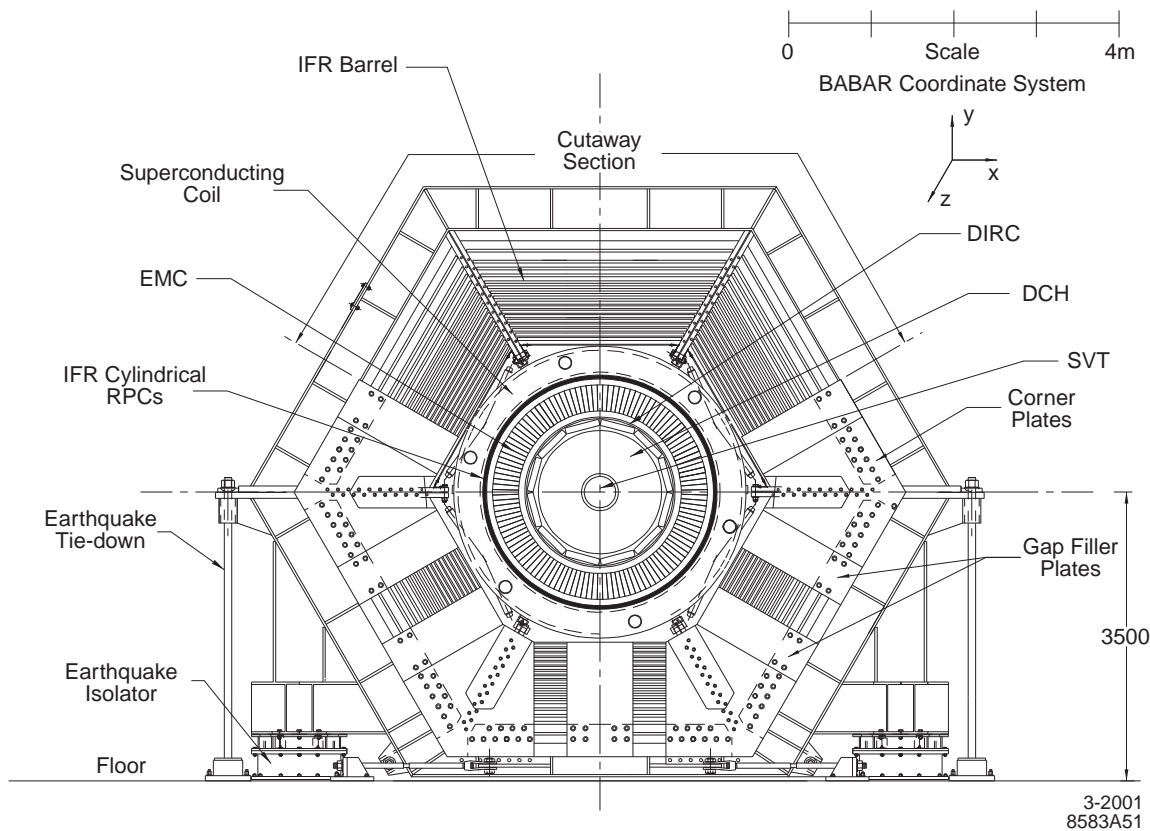}
\vspace{-1.5pc}
\caption{\babar\ detector end view.}
\label{layout_fig:detectorEnd}
\end{figure*}

\begin{itemize}

 \item a large and uniform acceptance down to small polar angles
 relative to the boost direction;

 \item excellent reconstruction efficiency for charged particles down
 to 60\mevc\ and for photons to 20\mev;

 \item very good momentum resolution to separate small signals from
 background;

\item excellent energy and angular resolution for the detection of
photons from \piz and $\eta^0$ decays, and from radiative decays in
the range from 20\mev\ to 4\gev;

 \item very good vertex resolution, both transverse and parallel to
 the beam direction;

 \item efficient electron and muon identification, with low
 misidentification probablities for hadrons. This feature is crucial
 for tagging the $B$ flavor, for the reconstruction of charmonium
 states, and is also important for the study of decays involving leptons;

 \item efficient and accurate identification of hadrons over a wide
 range of momenta for $B$ flavor-tagging, and for the reconstruction
 of exclusive states; modes such as \Bztokpi\ or \Bztopipi, as well as
 in charm meson and $\tau$ decays;

 \item a flexible, redundant, and selective trigger system;

 \item low-noise electronics and a reliable, high bandwidth
 data-acquisition and control system;

 \item detailed monitoring and automated calibration;

 \item an online computing and network system that can control,
 process, and store the expected high volume of data; and

 \item detector components that can tolerate significant radiation
 doses and operate reliably under high-background conditions.
\end{itemize}

To reach the desired sensitivity for the most interesting
measurements, data sets of order $10^8$ $B$ mesons will be needed.
For the peak cross section at the \FourS\ of about 1.1\nb, this
will require an integrated luminosity of order 100\invfb\ or three
years of reliable and highly efficient operation of a detector
with state-of-the art capabilities.

In the following, a brief overview of the principal components of
the detector, the trigger, the data-acquisition, and the online
computing and control system is given.  This overview is followed
by brief descriptions of the \pep2\ interaction region, the beam
characteristics, and of the impact of the beam generated
background on the design and operation of the detector. Finally, a
detailed presentation of the design, construction, and performance
of all \babar\ detector systems is provided.

\renewcommand{\secname}{layout_}
\renewcommand{\sectiondir}{sec02_layout}

\section{Detector Overview}
\label{sec:layout}

The \babar\ detector was designed and built by a large international
team of scientists and engineers.  Details of its original design are
documented in the Technical Design Report \cite{ref:R_TDR}, issued in
1995.

Figure~\ref{\secname fig:detectorElevation} shows a longitudinal
section through the detector center, and Figure~\ref{\secname
fig:detectorEnd} shows an end view with the principal dimensions.
The detector surrounds the \pep2 interaction region. To maximize
the geometric acceptance for the boosted \FourS\ decays, the whole
detector is offset relative to the beam-beam interaction point
(IP) by 0.37\m\ in the direction of the lower energy beam.

The inner detector consists of a silicon vertex tracker, a drift
chamber, a ring-imaging Cherenkov detector, and a CsI calorimeter.
These detector systems are surrounded by a superconducting
solenoid that is designed for a field of 1.5\tesla. The steel flux
return is instrumented for muon and neutral hadron detection.  The
polar angle coverage extends to 350\mrad\ in the forward direction
and 400\mrad\ in the backward direction, defined relative to the
high energy beam.  As indicated in the two drawings, the right
handed coordinate system is anchored on the main tracking system,
the drift chamber, with the $z$-axis coinciding with its principal
axis. This axis is offset relative to the beam axis by about
20\mrad\ in the horizontal plane.  The positive $y$-axis points
upward and the positive $x$-axis points away from the center of
the \pep2\ storage rings.

The detector is of compact design, its transverse dimension being
constrained by the 3.5\m\ elevation of the beam above the floor.
The solenoid radius was chosen by balancing the physics
requirements and performance of the drift chamber and calorimeter
against the total detector cost. As in many similar systems, the
calorimeter was the most expensive single system and thus
considerable effort was made to minimize its total volume without
undue impact on the performance of either the tracking system or
the calorimeter itself.  The forward and backward acceptance of
the tracking system are constrained by components of \pep2, a pair
of dipole magnets (B1) followed by a pair of quadrupole magnets
(Q1). The vertex detector and these magnets are placed inside a
support tube (4.5\m\ long and 0.217\m\ inner diameter) that is
cantilevered from beamline supports.  The central section of this
tube is fabricated from a carbon-fiber composite.

\begin{figure}
\centering
\includegraphics[width=.9\columnwidth]{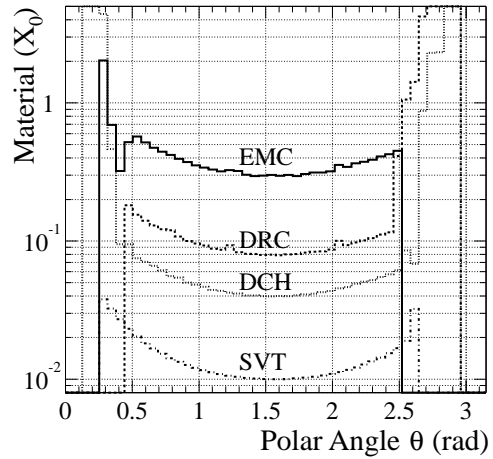}
\vspace{-2pc} \caption{Amount of material (in units of radiation
lengths) which a high energy particle, originating from the center
of the coordinate system at a polar angle $\theta$, traverses
before it reaches the first active element of a specific detector
system.} \label{\secname fig:F-mat}
\end{figure}

Since the average momentum of charged particles produced in
$B$-meson decay is less than 1\gevc, the precision of the measured
track parameters is heavily influenced by multiple Coulomb
scattering. Similarly, the detection efficiency and energy
resolution of low energy photons are severely impacted by material
in front of the calorimeter.  Thus, special care has been taken to
keep material in the active volume of the detector to a minimum.
Figure~\ref{\secname fig:F-mat} shows the distribution of material
in the various detector systems in units of radiation lengths.
Each curve indicates the material that a high energy particle
traverses before it reaches the first active element of a specific
detector system.

\subsection{Detector Components}

An overview of the coverage, the segmentation, and performance of the
\babar\ detector systems is presented in Table~\ref{\secname
tb:electronics}.

The charged particle tracking system is made of two components, the
silicon vertex tracker (SVT) and the drift chamber (DCH).

The SVT has been designed to measure angles and positions of
charged particles just outside the beam pipe.  The SVT is composed
of five layers of double-sided silicon strip detectors that are
assembled from modules with readout at each end, thus reducing the
inactive material in the acceptance volume.  The inner three
layers primarily provide position and angle information for the
measurement of the vertex position. They are mounted as close to
the water-cooled beryllium beam pipe as practical, thus minimizing
the impact of multiple scattering in the beam pipe on the
extrapolation to the vertex.  The outer two layers are at much
larger radii, providing the coordinate and angle measurements
needed for linking SVT and DCH tracks.

The principal purpose of the DCH is the momentum measurement for
charged particles. It also supplies information for the charged
particle trigger and a measurement of \dedx\ for particle
identification. The DCH is of compact design, with 40 layers of
small, approximately hexagonal cells. Longitudinal information is
derived from wires placed at small angles to the principal axis.
By choosing low-mass wires, and a helium-based gas mixture the
multiple scattering inside the DCH is minimized. The readout
electronics are mounted on the backward endplate of the chamber,
minimizing the amount of material in front of the calorimeter
endcap.

The DIRC, the detector of internally reflected Cherenkov light, is
a novel device providing separation of pions and kaons from about
500\mevc\ to the kinematic limit of 4.5\gevc.  Cherenkov light is
produced in 4.9\m\ long bars of synthetic fused silica of
rectangular cross section, $1.7\cm \times 3.5\cm$, and transported
by total internal reflection, preserving the angle of emission, to
an array of photomultiplier tubes.  This array forms the backward
wall of a toroidal water tank that is located beyond the backward
end of the magnet. Images of the Cherenkov rings are reconstructed
from the position and time of arrival of the signals in the
photomultiplier tubes.

The electromagnetic calorimeter (EMC) is designed to detect
electromagnetic showers with excellent energy and angular resolution
over the energy range from 20\mev\ to 4\gev. This coverage allows the
detection of low energy $\pi^0$s and $\eta^0$s from $B$ decays and
higher energy photons and electrons from electromagnetic, weak, and
radiative processes. The EMC is a finely segmented array of projective
geometry made of thallium doped cesium iodide (CsI(Tl)) crystals.  The
crystals are arranged in modules that are supported individually from
an external support structure.  This structure is built in two
sections, a barrel and a forward endcap.  To obtain the desired
resolution, the amount of material in front of and in-between the
crystals is held to a minimum. The individual crystals are read out by
pairs of silicon PIN diodes.  Low noise analog circuits and frequent,
precise calibration of the electronics and energy response over the
full dynamic range are crucial for maintaining the desired
performance.

The instrumented flux return (IFR) is designed to identify muons
and to detect neutral hadrons.  For this purpose, the magnet flux
return steel in the barrel and the two end doors is segmented into
layers, increasing in thickness from 2\cm\ on the inside to 10\cm\
at the outside. Between these steel absorbers, single gap
resistive plate chambers (RPCs) are inserted which detect
streamers from ionizing particles via external capacitive readout
strips. There are 19 layers of RPCs in the barrel sectors and 18
layers in the end doors. Two additional cylindrical layers of RPCs
with four readout planes are placed at a radius just inside the
magnet cryostat to detect particles exiting the EMC.

\begin{table*}
\caption{Overview of the coverage, segmentation, and performance
of the \babar\ detector systems.  The notation (C), (F), and (B)
refers to the central barrel, forward and backward components of
the system, respectively.  The detector coverage in the laboratory
frame is specified in terms of the polar angles $\theta_1$
(forward) and $\theta_2$ (backward).  The number of readout
channels is listed. The dynamic range (resolution) of the FEE
circuits is specified for pulse height (time) measurements by an
ADC (TDC) in terms of the number of bits (nsec).  Performance
numbers are quoted for 1\gevc\ particles, except where noted. The
performances for the SVT and DCH are quoted for a combined Kalman
fit (for the definition of the track parameters, see Section
\ref{sec:trk}.)} \vspace{.2cm} \label{\secname tb:electronics}

\begin{tabular}{lcccccccc} \hline\hline \rule{0pt}{12pt}&$\theta_1$
& & No.  & ADC & TDC&No. &  &  \\
System\rule[-5pt]{0pt}{0pt}&($\theta_2)$& & Channels &
(bits)&(ns)&Layers &Segmentation &Performance \\ \hline\hline
SVT\rule{0pt}{12pt}  & 20.1$^{\circ}$ & & 150K& 4&-&5 & 50-100
$\mu m~r-\phi$&
 $\sigma_{d_0}=55\mum$ \\
\rule[-5pt]{0pt}{0pt} & (-29.8$^{\circ}$) & & & & & & 100-200
$\mum~z$ & $\sigma_{z_0}=65\mum$ \\ \cline{1-8}
DCH\rule{0pt}{12pt} & 17.2$^{\circ}$ & & 7,104 &8&2& 40 & 6-8\mm &
$\sigma_{\phi}=1 $
 \mrad\\
 & (-27.4$^{\circ}$) & & & & & & drift distance & $\sigma_{tan\lambda}=0.001$\\
 & & & & & & & & $\sigma_{\pt}/\pt = 0.47\%$ \\
\rule[-5pt]{0pt}{0pt} & & & & & & & & $\sigma(dE/dx)=7.5$\%\\
\hline DIRC\rule{0pt}{12pt}   & 25.5$^{\circ}$ & & 10,752&-&0.5&1
&35 $\times$ 17\mm$^2$&
  $\sigma_{\theta_C} =2.5\mrad$\\
  & (-38.6$^{\circ}$) & & & & & &  ($r\Delta\phi\times\Delta r$)& per track \\
          & & & & & & & 144 bars & \\ \hline
          EMC(C)\rule{0pt}{12pt} & 27.1$^{\circ}$ &
          & $2 \times 5760$ &17--18&---&1 & 47 $\times$ 47\mm$^2$ &
          $\sigma_E/E=3.0\%$ \\
          & (-39.2$^{\circ}$) & & & & & & 5760
          cystals &$\sigma_{\phi}= 3.9$\mrad \\
          EMC(F) &15.8$^{\circ}$ & &$2 \times 820$ & & &1 & 820 crystals
          &$\sigma_{\theta}=3.9$\mrad \\
         & \rule[-5pt]{0pt}{0pt} (27.1$^{\circ}$) & & & & & & & \\
\hline
IFR(C)\rule{0pt}{12pt}  & 47$^{\circ}$ & & 22K+2K &1& 0.5 & 19+2 & 20-38\mm & 90\%
 $\mu^{\pm}$ eff.  \\
 & (-57$^{\circ}$) & & & & & & & 6-8\%
 $\pi^{\pm}$ mis-id \\ IFR(F) & 20$^{\circ}$ & & 14.5K & & & 18 &
 28-38\mm & (loose selection, \\
 & (47$^{\circ}$) & & & & && &
 $1.5$--$3.0$\gevc) \\
 IFR(B) & -57$^{\circ}$ & & 14.5K & & & 18 & 28-38 mm \\
 &\rule[-5pt]{0pt}{0pt} (-26$^{\circ}$) & & & & & & \\
\hline\hline
\end{tabular}
\end{table*}

\subsection {Electronics, Trigger, Data Acquisition\\* and Online Computing}

The electronics, trigger, data acquisition, and online computing
represent a collection of tightly coupled hardware and software
systems.  These systems were designed to maximize the physics data
acceptance, maintainability, and reliability while managing
complexity, and minimizing deadtime, and cost.

\begin{itemize}

\item
\emph{Front-End Electronics} (FEE) assemblies,
located on the detector, consist of signal processing and digitization
electronics along with the data transfer via optical fiber to the data
acquisition system.

\item
A robust and flexible two-level trigger copes with the full beam-beam
interaction rate.  The first level, \emph{Level 1} (L1), is
implemented in hardware, the other, \emph{Level 3} (L3), in software.
Provision is made for an intermediate trigger (Level 2) should severe
conditions require additional sophistication.

\item
The \emph{Online Dataflow} (ODF), handles digitized data from the FEE
through the event building.  ODF includes the \emph{fast control and
timing system} (FCTS).

\item
A farm of commercial Unix processors and associated software,
\emph{Online Event Processing} (OEP), provides the realtime
environment within which complete events are processed by the L3
trigger algorithms, partial event reconstruction is performed for
monitoring, and event data are logged to an intermediate storage.

\item
Software running on a second farm of processors, \emph{Online
Prompt Reconstruction} (OPR), completely reconstructs all physics
events, and performs monitoring and constants generation in near
realtime. Physics event data are transferred to an object database
\cite{ref:objectivity}  and are made available for further
analyses.

\item
An \emph{Online Run Control} (ORC) system implements the logic for
managing the state of the detector systems, starting and stopping
runs, and performing calibrations as well as providing a user
control interface.

\item
A system to control and monitor the detector and its support systems,
\emph{Online Detector Control} (ODC), is based upon the Experimental
Physics Industrial Control System (EPICS) toolbox~\cite{ref:EPICS}.
This system includes communication links with PEP-II.

\end{itemize}

\subsubsection{Electronics}

All \babar\ detector systems share a common electronics architecture.
Event data from the detector flows through the FEE, while monitoring
and control signals are handled by a separate, parallel system.  All
FEE systems are mounted directly on the detector to optimize
performance and to minimize the cable plant, thereby avoiding noise
pickup and ground loops in long signal cables.  All detector systems
utilize standard \babar\ interfaces to the data acquisition
electronics and software.

Each FEE chain consists of an amplifier, a digitizer, a trigger
latency buffer for storing data during the L1 trigger processing,
and an event buffer for storing data between the \emph{L1 Accept}
and subsequent transfer to the data acquisition system (see
Figure~\ref{\secname fig:elect01}). Custom integrated circuits
(ICs) have been developed to perform the signal processing.  The
digital L1 latency buffers function as fixed length data pipelines
managed by common protocol signals generated by the FCTS.  All
de-randomizing event buffers function as FIFOs
(first-in-first-out) capable of storing a fixed number of events.
During normal operation, analog signal processing, digitization,
and data readout occur continuously and simultaneously.

\begin{figure}[htb]
\centering
\includegraphics[width=.9\columnwidth]{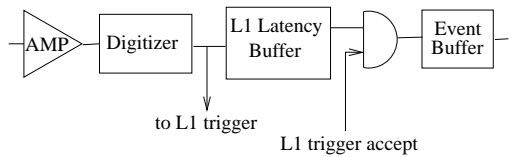}
\vspace{-2pc} \caption{Schematic diagram of the Front-End
Electronics (FEE). Analog signals arrive from the left, proceed
conditionally through the indicated steps and are injected into
the remainder of the data acquisition system.} \label{\secname
fig:elect01}
\end{figure}

Since many of the front-end circuits are inaccessible or require
significant downtime for access, stringent requirements were placed on
reliability.  Most components underwent comprehensive mean time
between failure (MTBF) studies.  All circuits underwent a burn-in
procedure prior to installation with the goal of minimizing initial
failure rates.

\subsubsection{Trigger}

The trigger system operates as a sequence of two independent stages,
the second conditional upon the first.  The L1 trigger is responsible
for interpreting incoming detector signals, recognizing and removing
beam-induced background to a level acceptable for the L3 software
trigger which runs on a farm of commercial processors.

L1 consists of pipelined hardware processors designed to provide
an output trigger rate of \lsim2\khz.  The L1 trigger selection is
based on data from DCH, EMC, and IFR.  The maximum L1 response
latency for a given collision is 12\mus.  Based on both the
complete event and L1 trigger information, the L3 software
algorithms select events of interest which are then stored for
processing. The L3 output rate is administratively limited to
120\hz\ so as not to overload the downstream storage and
processing capacity.

\babar\ has no fast counters for triggering purposes, and bunch
crossings are nearly continuous at a 4.2\ns\ spacing.  Dedicated
L1 trigger processors receive data continuously clocked in from
the DCH, EMC, and IFR detector systems.  These processors produce
clocked outputs to the fast control system at 30\mhz, the time
granularity of resultant \emph{L1 Accept} signals.  The arrival of
an \emph{L1 Accept} signal by the data acquisition system causes a
portion of each system's L1 latency buffer to be read out, ranging
from about 500\ns\ for the SVT to 4--16\mus\ for the EMC. Absolute
timing information for the event, \ie\ associating an event with a
particular beam crossing, is determined offline, using DCH track
segment timing, waveforms from the EMC, and accelerator timing
fiducials.

\subsubsection{Data Acquisition\\  and Online Computing}

The data acquisition and computing systems, responsible for the
transport of event data from the detector FEE to mass storage with a
minimum of dead time are shown schematically in Figure~\ref{\secname
fig:daq}.  These systems also interface with the trigger to enable
calibrations and testing.  Other parts of these systems provide for
the control and monitoring of the detector and supporting facilities.

\begin{figure*}
\centering
\includegraphics[width=\textwidth]{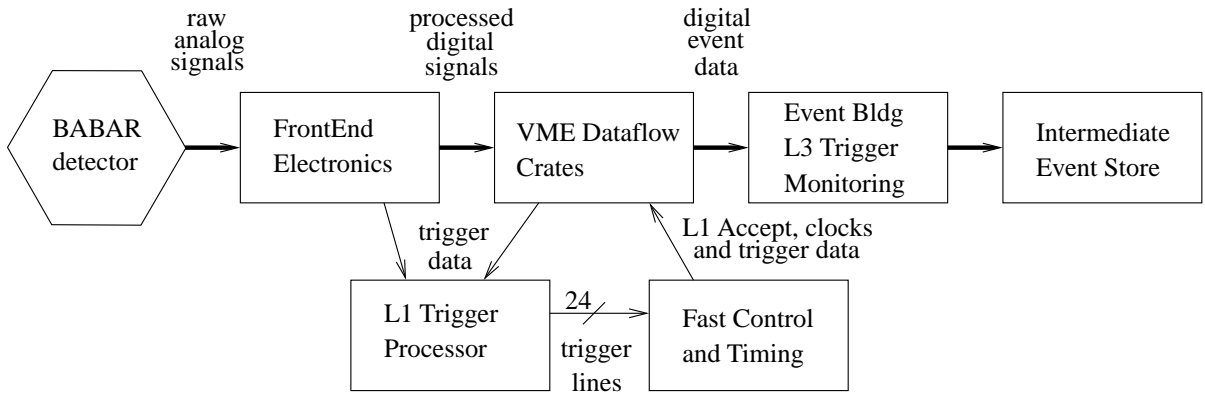}
\vspace{-2pc} \caption{Schematic diagram of the data acquisition.}
\label{\secname fig:daq}
\end{figure*}

\paragraph{Hardware}

The data acquisition system hardware consists of VME crates,
specialized VME-based processors called \emph{readout modules} (ROMs), the
FCTS, a Unix processor farm, various server machines and an Ethernet
network.  A ROM consists of a Motorola MVME2306 PowerPC single board
computer, event buffers, an interface to the FCTS, and a custom
\emph{Personality Card} that connects with the FEE circuits via
1.2\gbsps\ fiber optic cables.  The ROM provides the standard
interface between the detector specific FEE, the FCTS, and the
\emph{event builder}.  There are 157 ROMs in the system located in 19
physical VME crates divided into 24 logical crates by virtue of
segmented backplanes.  The FCTS system consists of a VME crate plus
individual Fast Control Distribution Modules in each of the data
acquisition VME crates.  The Unix processor farm consists of 32 Sun
workstations.

The detector monitoring and control system consists of a standard
set of components, including Motorola MVME177 single-board
computers, and other VME modules.  With the exception of the
solenoid magnet, which has its own control and monitoring, all
\babar\ detector components use this system.

The online computing system relies on a complex of workstation
consoles and servers with ~0.8\tbytes\ of attached storage, all
interconnected with switched 100\mbsps\ and 1\gbsps\ Ethernet
networks.  Multiple 1\gbsps\ Ethernet links connect the experimental
hall with the SLAC computing center.

\paragraph{Online Dataflow (ODF)}

The ODF software connects, controls and manages the flow of data
in the acquisition hardware with little dead time. This code is
divided between embedded processors in the ROMs running the
VxWorks \cite{ref:vxworks} realtime operating system and Unix
processors running the Solaris operating system.  ODF provides
configuration and readout of the FEE over fiber links to the ROMs;
data transport, buffering, and event building from the ROMs to the
Unix farm over a switched 100\mbsps\ Ethernet network; masking and
prescaling of L1 triggers; and logical partitioning of DAQ
hardware into multiple, independent data acquisition systems for
parallel calibrations and diagnostics.  Additional \emph{feature
extraction} (FEX) code in the ROMs extracts physical signals from
the raw data, performs gain and pedestal correction, data
sparsification, and data formatting. Data from electronics
calibrations are accumulated in the ROMs, channel response
functions are evaluated, results are compared to reference data
and subsequently applied in feature extraction.  Calibration data
are stored in a \emph{conditions database}.

\paragraph{Online Event Processing (OEP)}

OEP receives and processes data from the ODF event builder on each of
the Unix processors.  OEP orchestrates the following tasks: L3 trigger
algorithms; \emph{fast monitoring} to assure data quality; and logging
the selected events to disk while merging the multiple data output
streams to a single file.

\paragraph{Online Prompt Reconstruction (OPR)}

OPR bridges the online and offline systems~\cite{ref:opr}.  This
system reads raw data recorded to disk by OEP and, operating on a
farm of ~150 Unix processors, selects physics events, performs
complete reconstruction, performs  \emph{rolling calibrations},
collects extensive monitoring data for quality assurance, and
writes the result into an \emph{event store}.  A \emph{rolling
calibration} is the generation of reconstruction constants during
normal event processing, which are then applied to the processing
of subsequent data.

\paragraph{Online Detector Control (ODC)\\* and Run Control (ORC)}

The ODC system controls and extensively monitors the electronics,
the environment, and assures the safety of the detector.  Its
implementation is based on EPICS, providing detector-wide
standardization for control and monitoring, diagnostics and alarm
handling.  ODC also provides communication with \pep2 and the
magnet control system.  Monitoring data are archived in an
\emph{ambient database}.

The ORC system ties together the various components of the online
system and provides the operator with a single graphical interface
to control detector operation.  Complex configurations are stored
in a \emph{configuration database}; keys to the configuration used
for any run are stored along with the data.  The event store,
conditions, ambient, and configuration databases are implemented
in an object database~\cite{ref:objectivity}, while other data are
stored in a relational database.

\renewcommand{\secname}{ir_}
\renewcommand{\sectiondir}{sec03_ir}
\section{The PEP II Storage Rings and Their Impact on the \babar\ Detector}
\label{sec:ir}

\subsection{\pep2\ Storage Rings }
\pep2\ is an \epem\ storage ring system designed to operate at a
center of mass (c.m.) energy of 10.58\gev, corresponding to the
mass of the \FourS\ resonance.  The parameters of these energy
asymmetric storage rings are presented in Table~\ref{\secname
tab:tabpep2}. \pep2\ has surpassed its design goals, both in terms
of the instantaneous and the integrated daily luminosity, with
significantly fewer bunches than anticipated. A detailed
description of the design and operational experience of \pep2\ can
be found in references \cite{pepnim} and \cite{epac2000}.

\begin{table}[htb]
\caption{\pep2\ beam parameters. Values are given both for the design and
for typical colliding beam operation in the first year. HER and LER
refer to the high energy \en\ and low energy \ep\ ring, respectively.
$\sigma_{Lx}$, $\sigma_{Ly}$, and $\sigma_{Lz}$ refer to the
horizontal, vertical, and longitudinal rms size of the luminous
region.}

\vspace{\baselineskip}

\label{\secname tab:tabpep2}

\small
\begin{tabular}{lcc}
\hline Parameters\rule[-5pt]{0pt}{17pt} &Design &Typical\\ \hline
Energy HER/LER (GeV)\rule{0pt}{12pt}&9.0/3.1&9.0/3.1 \\ Current
HER/LER (A) &0.75/2.15 &0.7/1.3\\ \# of bunches    &1658
&553-829\\ Bunch spacing~(ns) &4.2 &6.3-10.5\\ $\sigma_{Lx}$
 (\mum) &110 &120\\ $\sigma_{Ly}$ (\mum) &3.3 &5.6\\ $\sigma_{Lz}$
(mm) &9 &9\\ Luminosity ($10^{33}$\cms) &3 &2.5\\ Luminosity
(\invpb/d) &135 \rule[-5pt]{0pt}{0pt}&120\\ \hline
\end{tabular}
\normalsize
\end{table}

\pep2\ typically operates on a 40--50 minute fill cycle. At the
end of each fill, it takes about three minutes to replenish the
beams. After a loss of the stored beams, the beams are refilled in
approximately 10--15 minutes. \babar\ divides the data into runs,
defined as periods of three hour duration or less during which
beam and detector conditions are judged to be stable.  While most
of the data are recorded at the peak of the \FourS\ resonance,
about 12\% are taken at a c.m. energy 40\mev\ lower to allow for
studies of non-resonant background.

\subsection{Impact of \pep2\ on \babar\ Layout}

The high beam currents and the large number of closely-spaced
bunches required to produce the high luminosity of \pep2\ tightly
couple the issues of detector design, interaction region layout,
and remediation of machine-induced background.  The bunches
collide head-on and are separated magnetically in the horizontal
plane by a pair of dipole magnets (B1), followed by a series of
offset quadrupoles. The tapered B1 dipoles, located at $\pm$
21\cm\ on either side of the IP, and the Q1 quadrupoles are
permanent magnets made of samarium-cobalt placed inside the field
of the \babar\ solenoid, while the Q2, Q4, and Q5 quadrupoles,
located outside or in the fringe field of the solenoid, are
standard iron magnets. The collision axis is off-set from the
$z$-axis of the \babar\ detector by about 20\mrad\ in the
horizontal plane \cite{sullivan97} to minimize the perturbation of
the beams by the solenoidal field.

The interaction region is enclosed by a water-cooled beam pipe of
27.9\mm\ outer radius, composed of two layers of beryllium (0.83~mm
and 0.53\mm\ thick) with a 1.48\mm\ water channel between them.  To
attenuate synchrotron radiation, the inner surface of the pipe is
coated with a 4\mum\ thin layer of gold. In addition, the beam pipe is
wrapped with 150\mum\ of tantalum foil on either side of the IP,
beyond $z=+10.1$\cm\ and $z=-7.9$\cm.  The total thickness of the
central beam pipe section at normal incidence corresponds to 1.06\% of
a radiation length.

The beam pipe, the permanent magnets, and the SVT were assembled and
aligned, and then enclosed in a 4.5\m-long support tube which spans
the IP.  The central section of this tube was fabricated from a
carbon-fiber epoxy composite with a thickness of 0.79\% of a radiation
length.

\subsection{Monitoring of Beam Parameters}

The beam parameters most critical for \babar\ performance are the
luminosity, the energies of the two beams, and the position,
angles, and size of the luminous region.

\subsubsection{Luminosity}

While \pep2\ measures radiative Bhabha scattering to provide a
fast monitor of the relative luminosity for operations, \babar\
derives the absolute luminosity offline from other QED processes,
primarily \epem, and $\mu^+\mu^-$ pairs. The measured rates are
consistent and stable as a function of time. For a data sample of
1\invfb, the statistical error is less than 1\%. The systematic
uncertainty on the relative changes of the luminosity is less than
0.5\%, while the systematic error on the absolute value of the
luminosity is estimated to be about 1.5\%.  This error is
currently dominated by uncertainties in the Monte Carlo generator
and the simulation of the detector. It is expected that with a
better understanding of the efficiency, the overall systematic
error on the absolute value of the luminosity will be
significantly reduced.

\subsubsection{Beam Energies}

During operation, the mean energies of the two beams are calculated
from the total magnetic bending strength (including the effects of
off-axis quadrupole fields, steering magnets, and wigglers) and the
average deviations of the accelerating frequencies from their central
values.  While the systematic uncertainty in the PEP-II calculation of
the absolute beam energies is estimated to be 5--10\mev, the relative
energy setting for each beam is accurate and stable to about
1\mev. The rms energy spreads of the LER and HER beams are 2.3\mev\
and 5.5\mev, respectively.

To ensure that data are recorded close to the peak of the \FourS\
resonance, the observed ratio of \BB\ enriched hadronic events to
lepton pair production is monitored online. Near the peak of the
resonance, a 2.5\% change in the \BB\ production rate corresponds
to a 2\mev\ change in the c.m. energy, a value that is close to
the tolerance to which the energy of \pep2\ can be held. However,
a drop in the \BB\ rate does not distinguish between energy
settings below or above the \FourS\ peak. The sign of the energy
change must be determined from other indicators.  The best monitor
and absolute calibration of the c.m. energy is derived from the
measured c.m. momentum of fully reconstructed \B\ mesons combined
with the known $B$-meson mass. An absolute error of 1.1\mev\ is
obtained for an integrated luminosity of 1\invfb. This error is
presently limited by the uncertainty in the $B$-meson
mass~\cite{cleo-mass} and by the detector resolution.

The beam energies are necessary input for the calculation of two
kinematic variables that are commonly used to separate signal from
background in the analysis of exclusive $B$-meson decays. These
variables, which make optimum use of the measured quantities and
are largely uncorrelated, are Lorentz-invariants which can be
evaluated both in the laboratory and c.m. frames.

The first variable, $\Delta E$, can be expressed in Lorentz invariant
form as
\begin{equation}
    \Delta E = (2 q_B q_0 - s) / 2 \sqrt{s} ,
\end{equation}
where $\sqrt{s}=2 E^*_{beam}$ is the total energy of the \epem\
system in the c.m. frame, and $q_B$ and $q_0=(E_0, \vec{p}_0)$ are
the Lorentz vectors representing the momentum of the $B$ candidate
and of the \epem\ system, $q_0 = q_{e+} + q_{e-}$.  In the
c.m.~frame, $\Delta E$ takes the familiar form
\begin{equation}
 \Delta E = E^*_B - E^*_{beam} ,
\end{equation}
here $E^*_B$ is the reconstructed energy of the $\B$~meson.  The
$\Delta E$ distribution receives a sizable contribution from the beam
energy spread, but is generally dominated by detector resolution.

\begin{figure}
\centering
\includegraphics[width=6.1cm]{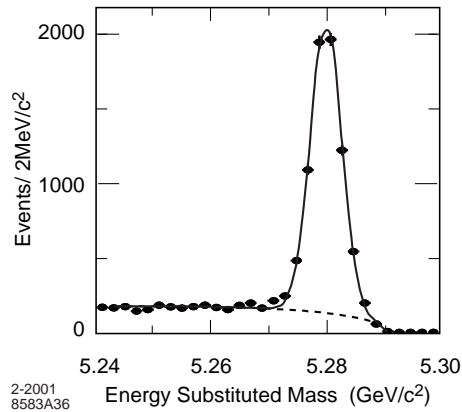}

\vspace{-2pc}

\caption{The energy-substituted mass for a sample of 6,700 neutral
$B$ mesons reconstructed in the final states $D^{(*)-} \pip, D^{(*)-}
\rho^+,  D^{(*)-} a_1^+$, and $\jpsi K^{*0}$. The background is extrapolated
from events outside the signal region.
}
\label{\secname fig:mes}
\end{figure}

The second variable is the energy-substituted mass, $\mes$,
defined as $\mes^2 = q_B^2$. In the laboratory frame, \mes\ can be
determined from the measured three-momentum $\vec p_B$ of the \B\
candidate without explicit knowledge of the masses of the decay
products:
\begin{equation}
   \mes = \sqrt { ( s/2 + \vec p_B \vec \cdot p_0)^2/ E_0^2 -p_B^2} .
\end{equation}
In the c.m.~frame ($\vec{p}_0 = 0$), this variable takes the familiar
form
\begin{equation}
   \mes = \sqrt { E_{beam}^{*2} - p_B^{*2}} ,
\end{equation}
where $p_B^*$ is the c.m.~momentum of the $\B$~meson, derived from the
momenta of its decay products, and the $B$-meson energy is substituted
by $E^*_{beam}$.  Figure~\ref{\secname fig:mes} shows the $\mes$
distribution for a sample of fully reconstructed $B$~mesons.  The
resolution in $\mes$ is dominated by the spread in $E^*_{beam}$,
$\sigma_{E^*_{beam}}= 2.6$\mev.

\subsubsection{Beam Direction}

The direction of the beams relative to \babar\ is measured
iteratively run-by-run using $e^+e^- \ra e^+e^-$ and $e^+e^- \ra
\mu^+\mu^-$ events. The resultant uncertainty in the direction of
the boost from the laboratory to the c.m. frame,~$\vec{\beta}$, is
about 1\mrad, dominated by alignment errors. This translates into
an uncertainty of about 0.3\mev\ in $m_{ES}$.  $\vec{\beta}$ is
consistent to within 1\mrad\ with the orientation of the elongated
beam spot (see below). It is stable to better than 1\mrad\ from
one run to the next.

\subsubsection{Beam Size and Position}

The size and position of the luminous region are critical
parameters for the decay-time-dependent analyses and their values
are monitored continuously online and offline. The design values
for the size of the luminous region are presented in
Table~\ref{\secname tab:tabpep2}. The vertical size is too small
to be measured directly. It is inferred from the measured
luminosity, the horizontal size, and the beam currents; it varies
typically by 1--2\mum.

The transverse position, size, and angles of the luminous region
relative to the \babar\ coordinate system are determined by
analyzing the distribution of the distance of closest approach to
the $z$-axis of the tracks in well measured two-track events as a
function of the azimuth $\phi$. The longitudinal parameters are
derived from the longitudinal vertex distribution of the two
tracks.  A combined fit to nine parameters (three average
coordinates, three widths, and three small angles) converges
readily, even after significant changes in the beam position.  The
uncertainties in the average beam position are of the order of a
few $\mu$m in the transverse plane and 100\mum\ along the
collision axis. Run-by-run variations in the beam position are
comparable to these measurement uncertainties, indicating that the
beams are stable over the period of a typical run.  The fit
parameters are stored run-by-run in the \emph{conditions
database}. These measurements are also checked offline by
measuring the primary vertices in multi-hadron events. The
measured horizontal and longitudinal beam sizes, corrected for
tracking resolution, are consistent with those measured by \pep2.

\subsection{Beam Background Sources }

The primary sources of steady-state accelerator backgrounds are, in
order of increasing importance: synchrotron radiation in the vicinity
of the interaction region; interactions between the beam particles and
the residual gas in either ring; and electromagnetic showers generated
by beam-beam collisions \cite{mattison,B99,hlbtf}. In addition, there
are other background sources that fluctuate widely and can lead to
very large instantaneous rates, thereby disrupting stable operation.

\subsubsection{Synchrotron Radiation}
Synchrotron radiation in nearby dipoles, the interaction-region
quadrupole doublets and the B1 separation dipoles generates many kW of
power and is potentially a severe background. The beam orbits,
vacuum-pipe apertures and synchrotron-radiation masks have been
designed such that most of these photons are channeled to a distant
dump; the remainder are forced to undergo multiple scatters before
they can enter the \babar\ acceptance.  The remaining synchrotron
radiation background is dominated by x-rays (scattered from tungsten
tips of a mask) generated by beam tails in the high-field region of
the HER low-$\beta$ quadrupoles.  This residual background is
relatively low and has not presented significant problems.

\subsubsection{Beam-Gas Scattering}
Beam-gas bremsstrahlung and Coulomb scattering off residual gas
molecules cause beam particles to escape the acceptance of the ring if
their energy loss or scattering angle are sufficiently large. The
intrinsic rate of these processes is proportional to the product of
the beam current and the residual pressure (which itself increases
with current). Their relative importance, as well as the resulting
spatial distribution and absolute rate of lost particles impinging the
vacuum pipe in the vicinity of the detector, depend on the beam
optical functions, the limiting apertures, and the entire
residual-pressure profile around the rings.  The separation dipoles
bend the energy-degraded particles from the two beams in opposite
directions and consequently most \babar\ detector systems exhibit
occupancy peaks in the horizontal plane, \ie\ the LER background near
$\phi =0^{\degrees}$ and HER background near $\phi =180^{\degrees}$.

During the first few months of operation and during the first
month after a local venting of the machine, the higher pressures
lead to significantly enhanced background from beam-gas
scattering.  The situation has improved significantly with time
due to \emph{scrubbing} of the vacuum pipe by synchrotron
radiation. Towards the end of the first year of data-taking, the
dynamic pressure in both rings had dropped below the design goal,
and the corresponding background contributions were much reduced.
Nevertheless, beam-gas scattering remains the primary source of
radiation damage in the SVT and the dominant source of background
in all detectors systems, except for the DIRC.

\subsubsection{Luminosity Background}
Radiative Bhabha scattering results in energy-degraded electrons
or positrons hitting aperture limitations within a few meters of
the IP and spraying \babar\ with electromagnetic shower debris.
This background is directly proportional to the instantaneous
luminosity and thus is expected to contribute an increasing
fraction of the total background in the future. Already this is
the dominant background in the DIRC.

\subsubsection{Background Fluctuations}
In addition to these steady-state background sources, there are
instantaneous sources of radiation that fluctuate on diverse time
scales:
\begin{itemize}
\item
beam losses during injection,
\item
intense bursts of radiation, varying in duration from a few ms to
several minutes, currently attributed to very small dust particles,
which become trapped in the beam,
and
\item
non-Gaussian tails from beam-beam interactions (especially of the
e$^+$ beam) that are highly sensitive to adjustments in collimator
settings and ring tunes.
\end{itemize}
These effects typically lead to short periods of high background and
have resulted in a large number of \babar -initiated beam aborts (see
below).

\subsection{Radiation Protection\\ and Monitoring}

A system has been developed to monitor the instantaneous and
integrated radiation doses, and to either abort the beams or to
halt or limit the rate of injection, if conditions become
critical.  In addition, DCH and IFR currents, as well as DIRC and
IFR counting rates, are monitored;  abnormally high rates signal
critical conditions.

Radiation monitoring and protection systems are installed for the
SVT, the DCH electronics, and the EMC. The radiation doses are
measured with silicon photodiodes.  For the SVT, 12 diodes are
arranged in three horizontal planes, at, above, and below the beam
level, with four diodes in each plane, placed at $z= +12.1$\cm\
and $z= -8.5$\cm\ and at a radial distance of 3\cm\ from the beam
line~\cite{tim}. The diode leakage current, after correction for
temperature and radiation damage effects, is proportional to the
dose rate.  The four diodes in the middle are exposed to about ten
times more radiation than the others. These mid-plane diodes are
connected to the beam abort system, while the remaining eight
diodes at the top and bottom are used to monitor the radiation
dose delivered to the SVT. The accuracy of the measured average
dose rate is better than 0.5 mRad/s.  The integrated dose, as
measured by the SVT diodes, is presented in Figure~\ref{\secname
fig:svtdose}.

\begin{figure}
\centering
\includegraphics [width=6.2cm] {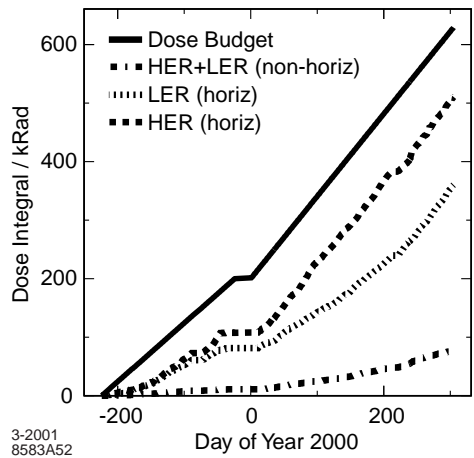}
\vspace{-1.5pc}

\caption {The integrated radiation dose as measured by PIN diodes
located at three different positions, showing contributions from the
 HER ($\phi =180^{\circ}$),
and the LER ($\phi =0^{\circ}$) in the horizontal plane, and from both
beams combined elsewhere.  Also shown is the SVT radiation budget for
the first 500 days of operation.}
\label{\secname fig:svtdose}
\end{figure}

The radiation level at the DCH and the EMC is more than two orders
of magnitude lower than at the SVT. To amplify the signal, the PIN
diodes for the DCH and EMC are mounted on small CsI(Tl) crystals
(with a volume of about 10\cm$^3$). These silicon diodes are
installed in sets of four.  Three sets are mounted on the front
face of the endcap calorimeter and one set on the backward
endplate of the DCH, close to the readout electronics. The signals
of the four diodes in each set are summed, amplified, and fed into
the radiation protection electronics.  Only one of the three diode
sets of the EMC is used at any given time.  The DCH and the EMC
use identical hardware and decision algorithms.  They limit
injection rates whenever an instantaneous dose equivalent to about
1~Rad/day is exceeded.

The SVT employs a different strategy and circuitry to assess
whether the measured radiation levels merit a beam abort or a
reduction in single-beam injection rate. Every beam dump initiated
by \babar\ is followed by a 10--15 minute period of injection with
significant radiation exposure. Thus, to minimize the ratio of the
integrated radiation dose to the integrated luminosity, it has
been beneficial to tolerate transient high-dose events as long as
the integrated dose remains less than the typical dose accumulated
during injection.  To differentiate between very high
instantaneous radiation and sustained high dose rates, trip
thresholds are enforced on two different time scales: an
instantaneous dose of 1~Rad accumulated over 1\ms, and an average
of 50~mRad/s measured over a 5-minute period.  During injection,
higher thresholds are imposed, since an aborted injection will
delay the return to taking data.

\begin{figure}
\centering
\includegraphics [width=6.5cm] {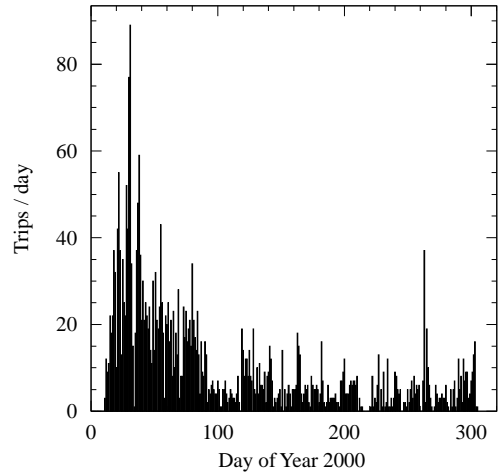}

\vspace{-1.5pc}

\caption{Daily rates of beam aborts initiated by the SVT radiation
protection diodes, summed over regular data-taking and \pep2\ injection.}
\label{\secname fig:trip_rate}
\end{figure}

Figure \ref{\secname fig:trip_rate} shows the daily rate of beam
aborts initiated by the SVT protection diodes during the year
2000. Initially, as many as 80 beam aborts were triggered per day,
while the average for stable operation was significantly below ten
at the end of the run. The measures described above, combined with
a significant reduction in large background fluctuations, have
been very effective in protecting the detector against radiation
damage, as well increasing the combined live time of the machine
and detector to greater than 75\%.

\subsection{Impact of Beam-Generated\\ Background on \babar\ }

Beam-generated backgrounds affect the detector in multiple ways.
They cause radiation damage to the detector components and the
electronics and thus may limit the lifetime of the experiment.
They may also cause electrical breakdown and damage or generate
large numbers of extraneous signals leading to problems with
bandwidth limitations of the data acquisition system and with
event reconstruction.  Backgrounds can degrade resolution and
decrease efficiency.

The impact of the beam-generated background on the lifetime and on
the operation of the different detector systems varies
significantly. Table~\ref{\secname tab:bbrlims} lists the limits
on the instantaneous and integrated background levels in terms of
the total dose and instantaneous observables. These limits are
estimates derived from beam tests and experience of earlier
experiments.  For each detector system, an annual radiation
allowance has been established taking into account the total
estimated lifetime of the components and the expected annual
operating conditions.  The typical values accumulated for the
first year of operation are also presented in the table.

\begin{table*}[htb]
\caption{\babar\ background tolerance. Operational limits are
expressed either as lifetime limits (radiation-damage and
aging-related quantities), or in terms of instantaneous observables
(DCH current, DIRC and L1-trigger rates).}

\vspace{\baselineskip}

\centering
\begin{tabular}{lccc}
\hline \hline
\rule{0pt}{12pt} & Limiting factor & Operational &First-year \\
Detector system& \rule[-5pt]{0pt}{0pt}and impact & limit & typical
\\ \hline \hline
SVT sensors\rule{0pt}{12pt}  & Integrated dose: & 2 MRad & 0.33
MRad \\
and electronics & radiation damage & & (hor.-plane modules) \\
& & & 0.06 MRad \\
& & & (other modules)\\
SVT sensors & Instantaneous dose: &1~Rad/ms & N/A \\
& diode shorts & \rule[-5pt]{0pt}{0pt}& \\ \hline
DCH: electronics &Integrated dose:\rule{0pt}{12pt} & 20 kRad & $\le$ 100 Rad \\
&radiation damage && \\
DCH: wire current & Accumulated charge: &100 mC/cm & 8 mC/cm\\
& wire aging & & \\
DCH: total current & HV system limitations & 1000 $\mu$A & 250 $\mu$A \\
& \rule[-5pt]{0pt}{0pt}& & (steady-state) \\ \hline
DIRC PMTs \rule{0pt}{12pt}& Counting rate: & 200 kHz & 110 kHz (steady-state,\\
& TDC deadtime &\rule[-5pt]{0pt}{0pt} & well-shielded sector) \\
\hline
EMC crystals & Integrated dose:\rule{0pt}{12pt} & 10 kRad & 0.25 kRad \\
& radiation damage & \rule[-5pt]{0pt}{0pt}& (worst case) \\ \hline
L1 trigger & Counting rate:\rule{0pt}{12pt} & 2 kHz & 0.7 kHz \\ & DAQ dead time & &
(steady-state)\\
\hline\hline
\end{tabular}
\label{\secname tab:bbrlims}
\end{table*}

\begin{figure}
\centering
\includegraphics [width=6.5cm] {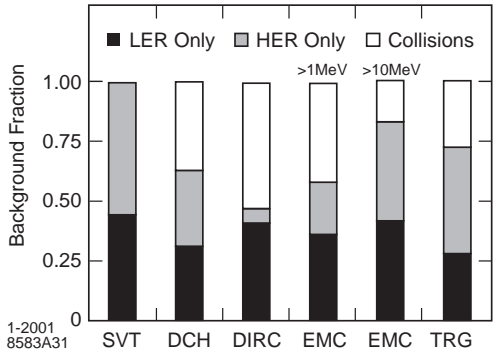}

 \vspace{-1.5pc}

\caption{Fractional steady-state background contributions in
\babar\ detector systems, measured for single beams and colliding
beams under stable conditions ($I^+=1.25~A$, $I^-=0.75~A$,
$L=2.3\times10^{33}\,$cm$^{-2}\,$s$^{-1}$) in July 2000.  The
contributions are derived from the measured doses in the
horizontal plane for the SVT, the total currents in the DCH, the
rates in the DIRC photomultipliers, the occupancy and number of
photons above 10~\mev\ in the EMC, and the L1 trigger rates.
}
\label{\secname fig:towersnow}
\end{figure}

Systematic studies of background rates were performed with stable
stored beams.  Measurements of the current-dependence of the
backgrounds were carried out for single beams, two beams not
colliding, and colliding beams with the goal to identify the principal
background sources, to develop schemes of reducing these sources, and
to extrapolate to operation at higher luminosity \cite{hlbtf}.  These
experimental studies were complemented by Monte Carlo simulations of
beam-gas scattering and of the propagation of showers in the
detector. The studies show that the relative importance of the
single-beam and luminosity background contributions varies, as
illustrated in Figure~\ref{\secname fig:towersnow}.  Data for the IFR
are not shown because this system is largely insensitive to
beam-generated backgrounds, except for the outer layer of the forward
endcap, due to insufficient shielding of the external beam line
components.

The experience of the first year of operation and the concern for
future operation for each of the detectors are summarized as follows.

\textbf{SVT}: ~~The most significant concern for the SVT with
regard to machine background is the integrated radiation dose. The
instantaneous and integrated dose rates in the radiation
protection diodes are representative, to within about a factor of
two, of the radiation doses absorbed by the SVT modules.  The
exposure in the horizontal planes is an order of magnitude larger
than elsewhere, averaging 15--25~mRad/s during stable beam
operation.  The highest integrated dose is 450~kRad, roughly
1~kRad/day. This dose is about 30\% below the allowance, giving
some confidence that the SVT can sustain operation for several
more years (see Figure \ref{\secname fig:svtdose}).

\textbf{DCH}: ~~For the DCH, the currents on the wires are the main
concern, both because of the limited capacity of the HV power supplies
and the effect of wire aging.  The currents drawn are approximately
uniformly distributed among the 44 HV supplies, one for each quadrant
of superlayers 2--10, and two per quadrant for superlayer 1.
Consequently, the total current limit is close to the sum of the
limits of the individual supplies.  During stable operation the total
chamber current is 200--300\muA.  However, radiation spikes can lead
to currents that occasionally exceed the limit of 1000\muA, causing HV
supplies to trip.  Other background effects are measured to be well
below the estimated lifetime limits and thus are not a serious issue
at this time.  The average wire occupancy has not exceeded 1--2\%
during stable operation, but the extrapolation to future operation at
higher luminosity and currents remains a major concern.

\textbf{DIRC}: ~~The DIRC radiators, made of synthetic fused silica, were
tested up to doses of 100~kRad without showing any measurable effects
and thus radiation damage is not a concern.  The present operational
limit of the DIRC is set by the TDC electronics which induce
significant dead time at rates above 250\khz, well above the stable
beam rate of 110\khz\ in well shielded areas.  Roughly half of the
present rate is luminosity-related and can be attributed to radiative
Bhabha scattering. The counting rate is due to debris from
electromagnetic showers entering the water-filled \emph{stand-off box}.
Efforts are underway to improve the shielding of the beam pipe nearby.

\textbf{EMC}: ~~The lifetime of the EMC is set by the reduction in light
collection in the CsI crystals due to radiation damage.  The
cumulative dose absorbed by the EMC is measured by a large set of
RadFETs placed in front of the barrel and endcap crystals.
RadFETs~\cite{radfets} are realtime integrating dosimeters based on
solid-state Metal Oxide Semiconductor (MOS) technology. The absorbed
dose increases approximately linearly with the integrated luminosity.
The highest dose to date was observed in the innermost ring of the
endcap, close to 250~Rad, while the barrel crystals accumulated about
80~Rad.  The observed reduction in light collection of 10--15\% in the
worst place, and 4--7\% in the barrel, is consistent with expectation
(see Section~\ref{sec:emc}).

\begin{figure}
\centering
\includegraphics [width=6.5cm] {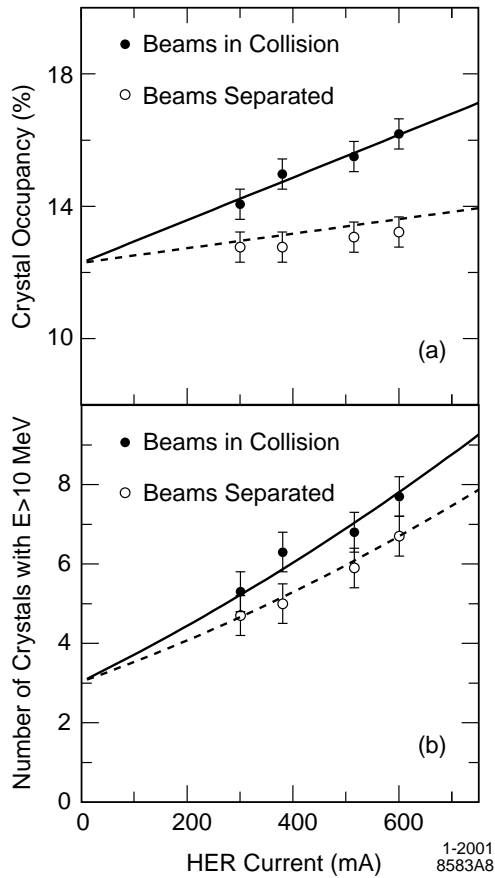}
\vspace{-2pc}

\caption{Average rates in the EMC for random triggers as a function of the
HER current for a fixed LER current of 1.1A, both for separated and colliding
beams; a) the single crystal occupancy for thresholds of 1\mev\ and b) the
number of crystals with a deposited energy greater than 10\mev.  The solid curves
represent a fit to the colliding beam data, the dashed curves indicate the sum of
rates recorded for single beams.}
\label{\secname fig:EMC_bgrate}
\end{figure}

The energy resolution is dependent on the single crystal readout
threshold, currently set at 1\mev.  During stable beam conditions
the average crystal occupancy for random triggers is 16\%, with
10\% originating from electronics noise in the absence of any
energy deposition.  The spectrum of photons observed in the EMC
from the LER and HER is presented in Figure \ref{\secname
fig:EMC_bg}. The HER produces a somewhat harder spectrum.  The
average occupancy for a threshold of 1\mev\ and the average number
of crystals with a deposited energy of more than 10\mev\ are shown
in Figure \ref{\secname fig:EMC_bgrate} as a function of beam
currents for both single and colliding beams.  The occupancy
increases significantly at smaller polar angles, in the forward
endcap and the backward barrel sections, and in the horizontal
plane.  The rate increase is approximately linear with the single
beam currents. Background rates recorded with separated beams are
consistent with those produced by single beams.  For colliding
beams, there is an additional flux of photons originating from
small angle radiative Bhabha scattering.  This effect is larger
for low energy photons and thus it is expected that at higher
luminosities the low energy background will raise the occupancy
and thereby limit the EMC energy resolution.

\begin{figure}
\centering
\includegraphics[width=6.5cm]{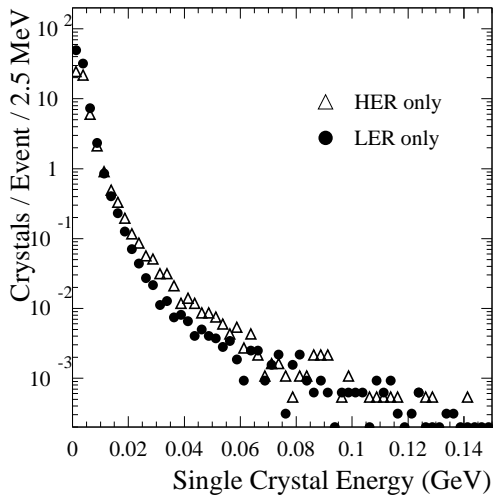}
\vspace{-1.5pc}
\caption{The energy spectrum of photons recorded in the EMC by random
triggers with single beams at typical operating currents, LER at 1.1A and
HER at 0.7A.  The electronic noise has been subtracted. }
\label{\secname fig:EMC_bg}
\end{figure}

\textbf{L1 Trigger}: ~~During stable beam operation, the typical
L1 trigger rate is below 1\khz, more than a factor of two below
the data acquisition bandwidth limit of about 2.0--2.5\khz.
Experience shows that background bursts and other rate spikes can
raise the data volume by as much as a factor of two and thus it is
necessary to aim for steady state rates significantly below the
stated limit.  For the L1 trigger, the dominant sources of DCH
triggers are particles generated by interactions in vacuum flanges
and the B1 magnets (see Figure~\ref{trg_fig:trg-l3trkz0} in
Section~\ref{trg_sec:trg}). This effect is most pronounced in the
horizontal plane. At present, the HER background is twice as high
as that of the LER, and the colliding beams contribute less than
half of the combined LER and HER single beam triggers.

\begin{figure}
\centering
\includegraphics [width=6.2cm] {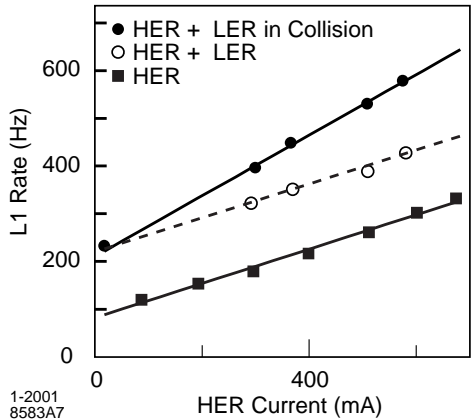}
\vspace{-2pc}
\caption{The L1 trigger rate as a function of the HER current for single beam
only, for both beams, separated and colliding (with a LER current of 1.1A).}
\label{\secname fig:TRG_bg}
\end{figure}

\subsection{Summary and Outlook}

Towards the end of the first year of data-taking, \pep2\ routinely
delivered beams close to design luminosity. Due to the very close
cooperation with the \pep2\ operations team, the machine-induced
backgrounds have not been a major problem once stable conditions
were established.  The background monitoring and protection system
has become a reliable and useful tool to safeguard the detector
operation.

Currently planned upgrades are expected to raise the luminosity to
$1.5 \times 10^{34}\,$cm$^{-2}\,$s$^{-1}$ within a few years.  The
single beam backgrounds will increase with beam currents and the
luminosity background is projected to exceed, or at best remain
comparable to, the beam-gas contribution.  Measures are being prepared
to reduce the sources and the impact of machine-related background on
\babar, among them upgrades to the DCH power supply system and to the
DIRC TDC electronics, the addition of localized shielding against
shower debris (especially for the DIRC \emph{stand-off box}), new
vacuum chambers, adjustable collimators, and additional pumping
capacity in critical regions upbeam of the interaction point.

With the expected increase in LER current and in luminosity, both the
single-beam and the luminosity-generated L1 trigger rates will
increase and are projected to exceed 2\khz\ (see Figure \ref{\secname
fig:TRG_bg}).  Therefore, the DCH trigger is being upgraded to improve
the rejection of background tracks originating from outside the
luminous region. In addition, the data acquisition and data processing
capacity will need to be expanded to meet the demands of higher
luminosity.

Overall, the occupancy in all systems, except the IFR, will
probably reach levels that are likely to impact the resolution and
reconstruction efficiency.  For instance, the occupancy in the EMC
is expected to more than double. Thus, beyond the relatively
straight forward measures currently planned for \babar\ system
upgrades, detailed studies of the impact of higher occupancy will
be necessary for all systems.

\renewcommand{\secname}{magnet_}
\renewcommand{\sectiondir}{sec08_magnet}
\section{The Solenoid Magnet and Flux Return}
\label{sec:magnet}

\subsection {Overview}

The \babar\ magnet system consists of a super-conducting
solenoid~\cite{IEEEmagnet}, a segmented flux return and a field
compensating or \emph{bucking coil}.  This system provides the
magnetic field which enables charged particle momentum measurement,
serves as the hadron absorber for hadron/muon separation, and provides
the overall structure and support for the detector components.
Figures~\ref{layout_fig:detectorElevation}
and~\ref{layout_fig:detectorEnd} show key components of the \babar\
magnet system and some of the nearby \pep2\ magnets.

The magnet coil cryostat is mounted inside the hexagonal barrel flux
return by four brackets on each end.  The flux return end doors are
each split vertically and mounted on skids that rest on the floor.  To
permit access to the inner detector, the doors can be raised and moved
on rollers.  At the interface between the barrel and the end doors,
approximately 60\% of the area is occupied by structural steel and
filler plates; the remaining space is reserved for cables and
utilities from the inner detectors.  A vertical, triangular chase cut
into the backward \emph{end doors} contains the cryostat chimney.
Table~\ref{\secname tab:MagProp} lists the principal parameters of the
magnet system.  The total weight of the flux return is approximately
870 metric tons.

To optimize the detector acceptance for unequal beam energies, the
center of the \babar\ detector is offset by 370\mm\ in the
electron beam direction. The principal component of the magnetic
field, $B_z$, lies along the $+z$ axis; this is also the
approximate direction of the electron beam. The backward end door
is tailored to accommodate the DIRC bar boxes and to allow access
to the drift chamber electronics.  Both ends allow space and
adequate shielding for the \pep2\ quadrupoles.

\begin{table}
\caption {Magnet Parameters}
\label {\secname tab:MagProp}

\vspace{\baselineskip}

\begin {tabular}{lrl}
\hline\hline
 \rule[-5pt]{0pt}{17pt} Field Parameters\\
\hline
\rule{0pt}{12pt}   Central Field             &     1.5 & \tesla\\
   Max. Radial Field         & $<$0.25 & \tesla\\
\quad at Q1 and $r=200$\mm\\
   Leakage into \pep2\      & $<$0.01 & \tesla\\
   Stored Energy             &    27 & MJ\\ 
\hline
 \rule[-5pt]{0pt}{17pt} Steel Parameters\\
\hline
   \rule{0pt}{12pt}Overall Barrel Length     &  4050 & \mm\\
   Overall Door Thickness    & 1149 & \mm\\\
   \quad (incl. gaps for RPCs)\\
   Overall Height            &  6545 & \mm\\
   Plates in Barrel             &    18\\
   \quad   9                    &    20 & \mm\\
   \quad   4                    &    30 & \mm\\
   \quad   3                    &    50 & \mm\\
   \quad   2                    &   100 & \mm\\
   Plates in Each Door          &    18\\
   \quad   9                    &    20 & \mm\\
   \quad   4                    &    30 & \mm\\
   \quad   4                    &    50 & \mm\\
   \quad   1                    &   100 & \mm\\
\hline
 \rule[-5pt]{0pt}{17pt} Main Coil Parameters\\
\hline
  \rule{0pt}{12pt} Mean Diameter of          &  3060 & \mm\\
   \quad Current Sheet                          \\
   Current Sheet Length      &  3513 & \mm\\
   Number of layers          &     2    \\
   Operating Current         &  4596 & A\\
   Conductor Current         &   1.2 & kA/\mm$^2$\\
   \quad Density                        \\
   Inductance                &  2.57 & H\\
\hline
 \rule[-5pt]{0pt}{17pt} Bucking Coil Parameters\\
\hline
 \rule{0pt}{12pt}  Inner Diameter            &  1906 & \mm\\
   Operating Current         &   200 & A\\
   Number of Turns           &   140 &\\
\hline
 \rule[-5pt]{0pt}{17pt} Cryostat Parameters\\
\hline
 \rule{0pt}{12pt}  Inner Diameter            &1420   &\mm\\
   Radial Thickness          &350    &\mm\\
   Total Length              &3850   &\mm\\
   Total Material (Al)       &$\sim126$&\mm\\
\hline\hline
\end {tabular}
\normalsize
\end {table}

\subsection {Magnetic Field Requirements\\ and Design}

\subsubsection {Field Requirements}

A solenoid magnetic field of 1.5\tesla\ was specified in order to
achieve the desired momentum resolution for charged particles.  To
simplify track finding and fast and accurate track fitting, the
magnitude of the magnetic field within the tracking volume was
required to be constant within a few percent.

The magnet was designed to minimize disturbance of operation of the
\pep2\ beam elements. The samarium-cobalt B1 dipole and Q1 quadrupole
magnets are located inside the solenoid as shown in
Figure~\ref{layout_fig:detectorElevation}.  Although these magnets can
sustain the high longitudinal field of 1.5\tesla, they cannot tolerate a
large radial component. Specifically, the field cannot exceed 0.25\tesla\
at a radius $r=200$\mm\ (assuming a linear dependence of $B_r$ on
$r$) without degrading their field properties due to partial
demagnetization.  The conventional iron quadrupoles Q2, Q4, and Q5 are
exposed to the solenoid stray fields. To avoid excessive induced skew
octupole components, the stray field leaking into these beam elements
is required to be less than 0.01\tesla\, averaged over their apertures.

\subsubsection {Field Design Considerations}

Saturation of the steel near the coil and the gap between the coil and
the steel leads to field non-uniformities.  To control these
non-uniformities, the current density of the coil is increased at the
ends relative to the center by reducing the thickness of the aluminum
stabilizer.  While the requirements on the radial field component at
Q1 inside the solenoid can be satisfied easily at the forward end, the
shape of the backward plug had to be specifically designed to
simultaneously control field uniformity and unwanted radial
components.

Leakage of magnetic flux is a problem, in particular at the backward
end.  A bucking coil, mounted at the face of the backward door
and surrounding the DIRC strong support tube, is designed to reduce
the stray field to an acceptable level for the DIRC photomultipliers
and the \pep2\ quadrupoles.

\subsubsection {Magnetic Modeling}

Extensive calculations of the magnetic field were performed to develop
the detailed design of the flux return, the solenoid coil, and the
bucking coil.  To crosscheck the results of these calculations
the fields were modeled in detail in two and three dimensions using
commercial software~\cite{MagMapping}.  The shape of the hole in each
end door was designed by optimizing various parameters, such as the
minimum steel thickness in areas of saturation.  The design of the
hole in the forward door was particularly delicate because the highly
saturated steel is very close to the Q2 quadrupole.  The multiple
\emph{finger} design of the hole was chosen to control the saturation of
the steel.

Most of the design work was performed in two dimensions, but some
three dimensional calculations were necessary to assure the accuracy
of modeling the transitions between the end doors and the barrel, the
leakage of field into the \pep2\ magnets, and the impact of that
leakage on the multipole purity~\cite{BBR344,BBR370}.  The
computations of the leakage field were done for central field of 1.7\tesla\
instead of 1.5\tesla\ to provide some insurance against uncertainties in
the modeling of complex steel shapes and the possible variations of
the magnetic properties of the steel.

\subsection {Steel Flux Return}

\subsubsection {Mechanical\\ and Magnetic Forces}

The magnet flux return supports the detector components on the inside,
but this load is not a major issue. Far greater demands are placed on
the structural design by the magnetic forces and the mechanical forces
from a potential earthquake.

Magnetic forces are of three kinds.  First, there is a symmetric
magnetic force on the end doors which was taken into consideration in
their design and construction.  Second, there is an axial force on the
solenoid due to the forward-backward asymmetry of the steel.  Because
the steel is highly saturated in places, the magnitude of the field
asymmetry changes when the current is raised from zero, and there is
no position of the solenoid at which the force remains zero at all
currents.  Because it is important that this axial force should not
change sign, which could cause a quench, the superconducting solenoid
was deliberately offset by 30\mm\ towards the forward door.  This
offset was chosen to accommodate a worst case scenario, including
uncertainties in the calculation.  Third, during a quench of the
superconducting coil, eddy currents in conducting components inside
the magnetic volume could generate sizable forces.  These forces were
analyzed for components such as the endplates of the drift chamber and
the electromagnetic calorimeter and were found not to be a problem.

\subsubsection {Earthquake Considerations}

Because SLAC is located in an earthquake zone, considerable attention
has been given to protecting the detector against severe damage from
such an event.  The entire detector is supported on four earthquake
isolators, one at each corner, which limit the component acceleration
in the horizontal plane to 0.4\gm.  However, these isolators offer no
protection in the vertical direction.  Vertical ground accelerations
of 0.6\gm\ are considered credible and actual component accelerations
may be considerably larger due to resonances.  By taking into account
resonant frequencies and the expected frequency spectra of
earthquakes, the magnet and all detector components have been designed
to survive these accelerations without serious damage.  Because the
magnet is isolated from the ground moving beneath it, worst case
clearances to external components, \eg \pep2\ components, are
provided.  It is expected that even during a major earthquake, damage
would be modest.

\subsubsection {Fabrication}

The flux return was fabricated~\cite{Steel} from drawings prepared by
the \babar\ engineering team.  A primary concern was the magnetic
properties of the steel.  The need for a high saturation field
dictated the choice of a low carbon steel, specified by its chemical
content (close to AISI 1006).  The manufacturer supplied sample steel
for critical magnetic measurements and approval. The availability of
very large tools at the factory made it possible to machine the entire
face of each end of the assembled barrel, thus assuring a good fit of
the end doors.  The entire flux return was assembled at the factory,
measured mechanically, and inspected before disassembly for shipment.

\subsection {Magnet Coils}

The design of the superconducting solenoid is conservative and follows
previous experience. The superconducting material is composed of
niobium-titanium (46.5\% by weight Nb) filaments, each less than
40\mum\ in diameter.  These filaments are then wound into 0.8\mm\
strands, 16 of which are then formed into \emph{Rutherford cable}
measuring 1.4\,x\,6.4\mm.  The final conductor~\cite{conductor}
consists of Rutherford cable co-extruded with pure aluminum stabilizer
measuring 4.93\,x\,20.0\mm\ for use on the outer, high current density
portion of the solenoid, and 8.49\,x\,20.0\mm\ for the central, lower
current density portion.  The conductor is covered in an insulating
dry wrap fiberglass cloth which is vacuum impregnated with epoxy.  The
conductor has a total length of 10.3\km.

The solenoid is indirectly cooled to an operating point of 4.5\degk\
using a thermo-syphon technique.  Liquid helium~\cite{LHe} is
circulated in channels welded to the solenoid support cylinder.
Liquid helium and cold gas circulate between the solenoid, its
shields, the liquefier-refrigerator and a 4000\liter\ storage dewar
via 60\m\ of coaxial, gas-screened, flexible transfer line.  The
solenoid coil and its cryostat were fabricated~\cite{Ansaldo} to
drawings prepared by the \babar\ engineering team.  Before
shipment~\cite{airforce}, the fully assembled solenoid was cooled to
operating temperature and tested with currents of up to 1000\Amp,
limited by coil forces in the absence of the iron flux return.

A portion of the cryostat assembly, containing the solenoid coil, its
support cylinder and heat shield, is shown in Figure~\ref{\secname
fig:cryostat}.

\begin{figure}
\centering
\includegraphics[width=.9\columnwidth]{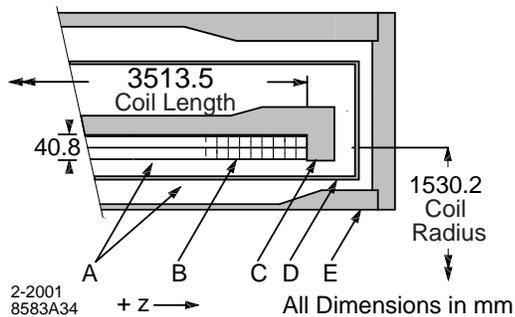}
\vspace{-2pc} \caption {A portion of cryostat assembly.  The
forward end is shown. Legend: (A) evacuated spaces filled with
IR-reflective insulator; (B) superconducting coil (2-layers); (C)
aluminum support cylinder; (D) aluminum heat shield; (E) aluminum
cryostat housing.} \label {\secname fig:cryostat}
\end{figure}

To reduce the leakage fields into the \pep2\ components and the
DIRC photomultipliers, an additional external bucking coil is
installed~\cite{bfield}. This is a conventional water cooled
copper coil consisting of ten layers.  Although the nominal
operating current is 200\Amp, a current of up to 575\Amp\ is
attainable, if needed, to demagnetize the DIRC shield.

To optimally control the stray fields and avoid a magnetization of the
DIRC magnetic shield, the currents in the solenoid and the
bucking coil are ramped together under computer control.  High
precision transducers are used to measure the currents and provide the
feedback signals to the power supplies.  The values of the currents
are recorded in the \babar\ database.

\subsection {Magnetic Field Map}

The goal of the magnetic field mapping and subsequent corrections was
to determine the magnetic field in the tracking volume to a precision
of 0.2\mtesla.

\subsubsection {Mapping Procedure}

A field mapping device was built specifically for the \babar\ magnet
based on a design concept developed at Fermilab~\cite{FNALMapper}. The
magnetic field sensors were mounted on a \emph{propeller} at the end of a
long cantilevered spindle which reached through the hole in the
forward end door.  The spindle in turn rode on a carriage which moved
on precision-aligned rails.  The propeller rotated to sample the
magnetic volume in $\phi$, and the carriage moved along its axis to
cover $z$.  Measurements were obtained from five sets of $B_r$ and
$B_z$ and two $B_\phi$ Hall probes, all of which were mounted on a
plate at different radial positions.  This plate was attached to the
propeller and its position could be changed to cover the desired range
in the radial distance $r$ from the axis.  Precision optical alignment
tools were used to determine the position of the sensors transverse to
the $z$-axis.

The $B_r$ and $B_z$ probes were two-element devices with a short-term
(few month) precision of 0.01\%, the $B_\phi$ probes were single
element devices with a precision of 0.1\%~\cite{HallProbes}.  In
addition to the Hall probes, an NMR probe~\cite{NMRProbe} was mounted
at a radius of 89\mm\ on the propeller to provide a very precise field
reference near the $z$-axis as a function of $z$ for $|z| < 1000$\mm,
where $z=0$ at the magnet center.  The NMR measurements set the
absolute scale of the magnetic field.

The magnetic field was mapped at the nominal central operating field
of 1.5\tesla, as well as at 1.0\tesla.  Measurements were recorded in
100\mm\ intervals from $-$1800 to $+$1800\mm\ in $z$, and in 24
azimuthal positions spaced by 15$^\circ$ for each of three different
radial positions of the Hall probe plate. Thus for each $z$ and $\phi$
position, the components $B_r$ and $B_z$ were measured at 13 distinct
radii from 130\mm\ to 1255\mm\ and $B_\phi$ at six radii between
505\mm\ and 1180\mm.

The field map was parameterized in terms of a polynomial of degree
up to 40~in $r$ and $z$ plus additional terms to account for
expected perturbations~\cite{BBR514}. The fit reproduced the
measurements to within an average deviation of 0.2\mtesla\
throughout the tracking volume. The fitting procedure also served
as a means of detecting and removing questionable measurements.

\subsubsection {Perturbations to the Field Map}

During the mapping process, the permanent magnet dipoles (B1) and
quadrupoles (Q1) were not yet installed.  Their presence inside the
solenoid results in field perturbations of two kinds.  The first is
due to the fringe fields of the B1 and Q1 permanent magnets, and of
the dipole and quadrupole trim coils mounted on Q1.  The B1 field
strength reaches a maximum of $\sim$20\mtesla\ close to the surface of
the B1 casing and decreases rapidly with increasing radius.  The
fields associated with the trim coils were measured and parameterized
prior to installation; they are essentially dipole in character.

The second field perturbation is due to the permeability of the
permanent magnet material.  Sintered samarium-cobalt has a
relative permeability of 1.11 to 1.13 in the $z$ direction, and as
a result the solenoid field is modified significantly.  Probes
between the B1 and Q1 magnets at a radius of about 190\mm\ measure
the effect of the permeability.  The field perturbation is
obtained from a two-dimensional, finite element analysis which
reproduces the $r$ and $z$ dependence of $B_r$ and $B_z$.  The
induced magnetization increases $B_z$ by about 9\mtesla\ at the
interaction point; the effect decreases slowly with increasing
radius.

\begin{figure}
\centering
\includegraphics [width=6cm] {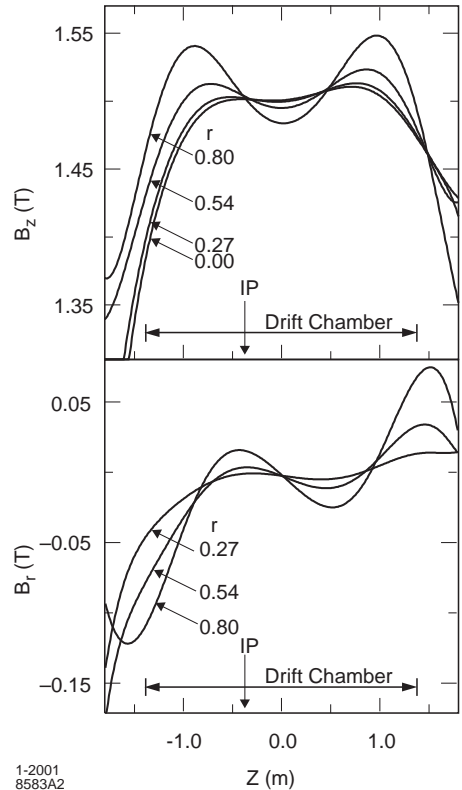}
\vspace{-2pc}

 \caption {The magnetic field components $B_z$ and $B_r$ as a function
           of $z$ for various radial distances $r$ (in \m).  The extent of
           the DCH and the location of the interaction point (IP) are
           indicated.}
 \label {\secname fig:BzBrField}
\end {figure}

\begin{figure}
\centering
\includegraphics [width=6cm] {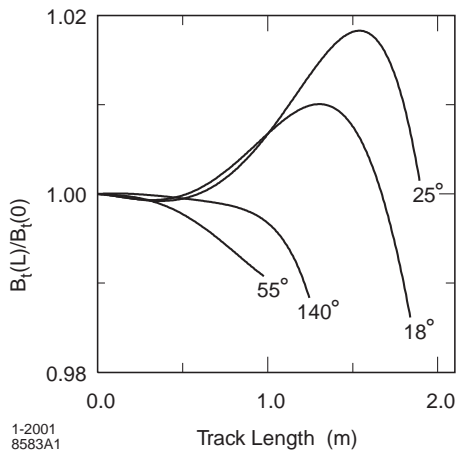}

\vspace{-2pc}

 \caption {Relative magnitude of magnetic field transverse to a high momentum
           track as a function
           of track length from the IP for various polar angles (in
           degrees). The data are normalized to the field at
           the origin.}
\label {\secname fig:RelBend}
\end{figure}

\subsubsection {Field Quality}

To illustrate the quality of magnetic field, Figure~\ref {\secname
fig:BzBrField} shows the field components $B_z$ and $B_r$ as a
function of $z$ for various radial distances $r$.  In the tracking
volume the field is very uniform, the $B_\phi$ component does not
exceed 1\mtesla.  The variation of the bend field, \ie\ the field
transverse to the trajectory, along the path of a high momentum track
is at most 2.5\% from maximum to minimum within the tracking region,
as shown in Figure~\ref{\secname fig:RelBend}.

\subsubsection {Field Computation}

In order to reduce the computation of the magnetic field for track
reconstruction and momentum determination, the field values averaged
over azimuth are stored in a grid of $r$--$z$ space points spanning
the volume interior to the cryostat.  Local values are obtained by
interpolation.  Within the volume of the SVT, a linear interpolation is
performed in a 20\mm\ grid; elsewhere the interpolation is quadratic
in a 50\mm\ grid.  Azimuthal dependence is parameterized by means of
a Fourier expansion at each $r$--$z$ point. The Fourier coefficients at
the point of interest are obtained by interpolation on the $r$--$z$ grid,
and the average field value is corrected using the resulting Fourier
series.

\subsection {Summary}

Since its successful commissioning, the magnet system has
performed without problems. There have been no spontaneous
quenches of the superconducting solenoid. In the tracking region,
the magnetic field meets specifications, both in magnitude and
uniformity.  The field compensation and magnetic shielding work
well for the DIRC photomultiplier array and the external
quadrupoles.  Measurements indicate that the bucking coil reduces
the field at the face of Q2 from $\sim$50\mtesla\ to
$\sim$1\mtesla~\cite{bfield}, in agreement with calculations.

\renewcommand{\secname}{svt_}
\renewcommand{\sectiondir}{sec04_svt}
\section{Silicon Vertex Tracker}
\label{sec:svt}
\subsection{ Charged Particle Tracking}
The principal purpose of the \babar\ charged particle tracking
systems, the SVT and the DCH, is the efficient detection of
charged particles and the measurement of their momentum and angles
with high precision.  Among many applications, these precision
measurements allow for the reconstruction of exclusive $B$- and
$D$-meson decays with high resolution and thus minimal background.
The reconstruction of multiple decay vertices of weakly decaying
$B$ and $D$ mesons is of prime importance to the physics goals.

Track measurements are also important for the extrapolation
to the DIRC, EMC, and IFR.
At lower momenta, the DCH measurements are more important,
while at higher momenta the SVT measurements dominate.
Most critical are the angles at the DIRC,
because the uncertainties in the charged
particle track parameters add to the uncertainty in the
measurement of the Cherenkov angle.
Thus, the track errors from the combined SVT and DCH measurements
should be small compared to the average DIRC Cherenkov angle measurements,
\ie\ of order of 1\mrad, particularly at the highest momenta.

\subsection{SVT Goals and Design Requirements}

The SVT has been designed to provide precise reconstruction of
charged particle trajectories and decay vertices near the
interaction region. The design choices were driven primarily by
direct requirements from physics measurements and constraints
imposed by the PEP-II interaction region and \babar\ experiment.
In this section  the mechanical and electronic design of the SVT
are discussed, with some discussion of the point resolution per
layer and \dedx\ performance. The tracking performance and
efficiency of the SVT alone and in combination with the DCH are
described in Section~\ref{sec:trk}.

\subsubsection{SVT Requirements and Constraints}
The SVT is critical for the measurement of the time-dependent \CP\
asymmetry. To avoid significant impact of the resolution on the
\CP\ asymmetry measurement the mean vertex resolution along the
$z$-axis for a fully reconstructed $B$ decay must be better than
80\mum~\cite{R_LOI}. The required resolution in the $x$--$y$ plane
arises from the need to reconstruct final states in $B$ decays as
well as in \mtau\ and charm decays. For example, in decays of the
type \Bztodd, separating the two $D$ vertices is important. The
distance between the two $D$'s in the $x$--$y$ plane for this
decay is typically $\sim 275 $\mum. Hence, the SVT needs to
provide  resolution of order $\sim$100\mum\ in the plane
perpendicular to the beam line.

Many of the decay products of $B$ mesons have low \pt. The SVT
must provide standalone tracking for particles with transverse
momentum less than 120\mevc, the minimum that can be measured
reliably in the DCH alone.  This feature is fundamental for the
identification of slow pions from \Dstar-meson decays: a tracking
efficiency of 70\% or more is desirable for tracks with a
transverse momentum in the range 50--120\mevc. The standalone
tracking capability and the need to link SVT tracks to the DCH
were crucial in choosing the number of layers.

Beyond the standalone tracking capability, the SVT provides the
best measurement of track angles, which is required to achieve
design resolution for the Cherenkov angle for high momentum
tracks.

Additional constraints are imposed by the storage ring components.
The SVT is located inside the $\sim$4.5\m-long support tube, that
extends all the way through the detector.  To maximize the angular
coverage, the SVT must extend down to 350\mrad\ (20\degrees) in
polar angle from the beam line in the forward direction.  The
region at smaller polar angles is occupied by the B1 permanent
magnets.  In the backward direction, it is sufficient to extend
the SVT sensitive area down to 30\degrees.

The SVT must withstand 2 MRad of ionizing radiation. A radiation
monitoring system capable of aborting the beams is required. The
expected radiation dose is 1~Rad/day in the horizontal plane
immediately outside the beam pipe (where the highest radiation is
concentrated), and 0.1~Rad/day on average otherwise.
\label{req:radhard}

The SVT is inaccessible during normal detector operations. Hence,
reliability and robustness are essential: all components of the
SVT inside the support tube should have long mean-time-to-failure,
because the time needed for any replacement is estimated to be
4--5 months. Redundancies are built in whenever possible and
practical.

The SVT is cooled to remove the heat generated by the electronics.
In addition, it operates in the 1.5\tesla\ magnetic field.

To achieve the position resolution necessary to carry out physics
analyses, the relative position of the individual silicon
\emph{sensors} should be stable over long time periods. The
assembly allows for relative motion of the support structures with
respect to the B1 magnets.

\begin{figure}
\includegraphics[width=7.5cm]{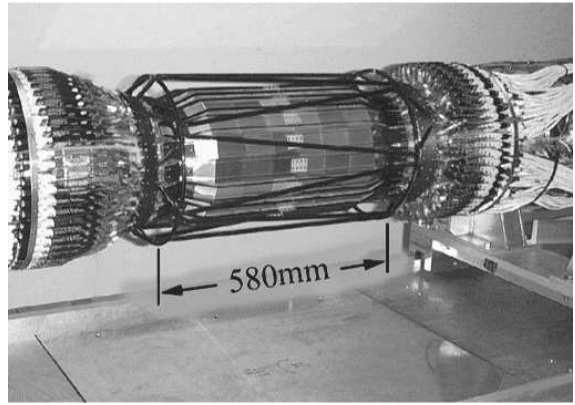}

\vspace{-2pc}
 \caption{Fully assembled SVT. The silicon sensors of
the outer layer are visible, as is the carbon-fiber space frame
(black structure) that surrounds the silicon.} \label{\secname
fig:svt_overall_pic}
\end{figure}

\begin{figure*}[htb]
\includegraphics[width=16cm]{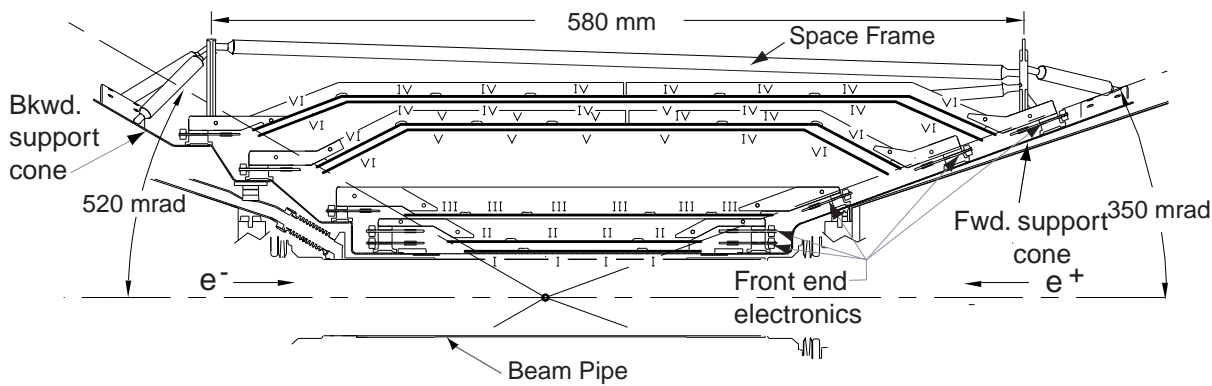}
\vspace{-2pc}

\caption{Schematic view of SVT: longitudinal section. The roman
numerals label the six different types of sensors.}
\label{\secname fig:sideview}
\end{figure*}

These requirements and constraints have led to the choice of a SVT
made of five layers of double-sided silicon strip sensors. To
fulfill the physics requirements, the spatial resolution, for
perpendicular tracks, must be 10--15\mum\ in the three inner
layers and about 40\mum\ in the two outer layers.  The inner three
layers perform the impact parameter measurements, while the outer
layers are necessary for pattern recognition and low \pt\
tracking.

\subsection{SVT Layout}

\begin{figure}[htb]
\includegraphics[width=7.5cm]{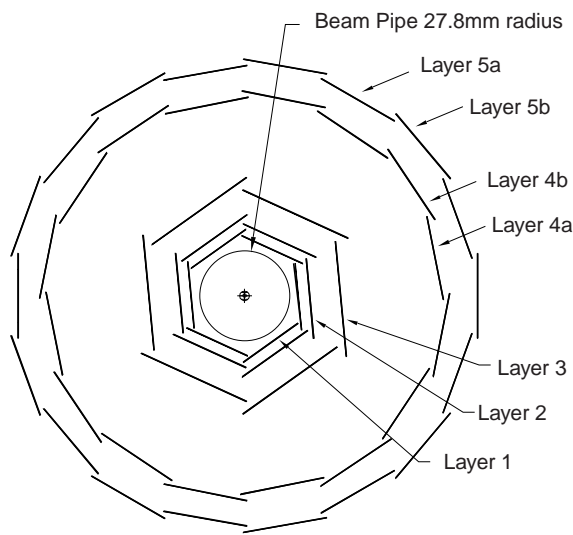}
\vspace{-2pc} \caption{Schematic view of SVT: tranverse section.}
\label{\secname fig:endview}
\end{figure}

The five layers of double-sided silicon strip sensors, which form
the SVT detector, are organized in 6, 6, 6, 16, and 18 modules,
respectively; a photograph is shown in Figure~\ref{\secname
fig:svt_overall_pic}. The strips on the opposite sides of each
sensor are oriented orthogonally to each other: the \mphi\
measuring strips (\emph{\mphi\ strips}) run parallel to the beam
and the $z$ measuring strips (\emph{$z$ strips}) are oriented
transversely to the beam axis.  The modules of the inner three
layers are straight, while the modules of layers 4 and 5 are
\emph{arch}-shaped (Figures~\ref{\secname fig:sideview} and
\ref{\secname fig:endview}).

\begin{figure}
\includegraphics[width=7.5cm]{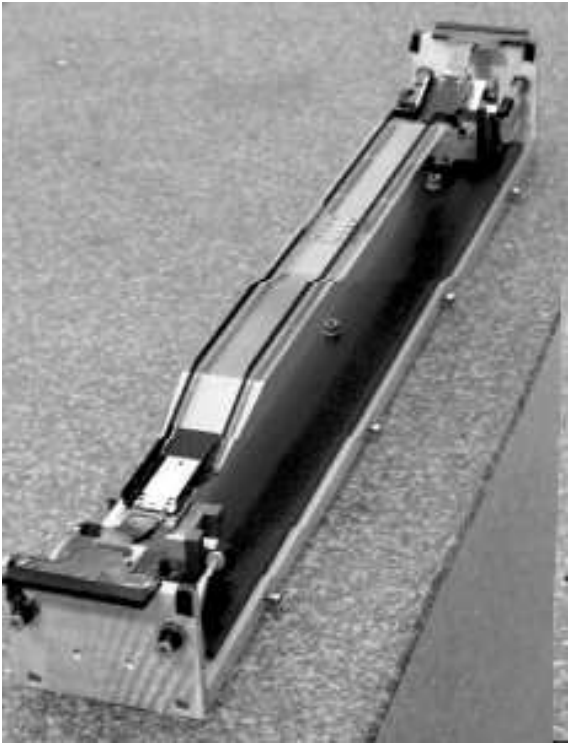}
\vspace{-2.1pc}

 \caption{Photograph of an SVT arch module in an
assembly jig.} \label{\secname fig:arch}
\end{figure}

This arch design was chosen to minimize the amount of silicon required
to cover the solid angle, while increasing the crossing angle for
particles near the edges of acceptance.  A photograph of an outer
layer arch module is shown in Figure~\ref{\secname fig:arch}.  The
modules are divided electrically into two half-modules, which are read
out at the  ends.

To satisfy the different geometrical requirements of the five SVT
layers, five different sensor shapes are required to assemble the
planar sections of the layers. The smallest detectors are 43
$\times$ 42\mm$^2$ ($z\ \times\ \phi$), and the largest are 68
$\times$ 53\mm$^2$. Two identical trapezoidal sensors are added
(one each at the forward and backward ends) to form the arch
modules.  The half-modules are given mechanical stiffness by means
of two carbon fiber/kevlar ribs, which are visible in
Figure~\ref{\secname fig:arch}. The \mphi\ strips of sensors in
the same half-module are electrically connected with wire bonds to
form a single readout strip.  This results in a total strip length
up to 140\mm\ (240\mm) in the inner (outer) layers.

The signals from the $z$ strips are brought to the readout
electronics using fanout circuits consisting of conducting traces
on a thin (50 \mum) insulating Upilex~\cite{upilex} substrate. For
the innermost three layers, each $z$
strip is connected to its own preamplifier channel, while in layers 4
and 5 the number of $z$ strips on a half-module exceeds the number of
electronics channels available, requiring that two $z$ strips on
different sensors be
electrically connected (ganged) to a single electronics channel.  The
length of a $z$ strip is about 50\mm\ (no ganging) or 100\mm\ (two strips
connected).  The ganging introduces an ambiguity on the $z$ coordinate
measurement, which must be resolved by the pattern recognition
algorithms.  The total number of readout channels is approximately
150,000.

The inner modules are tilted in \mphi\ by 5\degrees, allowing an
overlap region between adjacent modules, a feature that provides
full azimuthal coverage and is advantageous for alignment. The
outer modules cannot be tilted, because of the arch geometry. To
avoid gaps and to have a suitable overlap in the \mphi\
coordinate, layers 4 and 5 are divided into two sub-layers (4a,
4b, 5a, 5b) and placed at slightly different radii (see
Figure~\ref{\secname fig:endview}). The relevant geometrical
parameters of each layer are summarized in Table~\ref{\secname
tab:geometrical}.

\begin{table}[!htb]
\caption{Geometric parameters for each layer and readout plane of
the SVT. Floating strips refers to the number of strips between
readout (R-O) strips. Note: parts of the $\phi$ sides of layers 1
and 2 are bonded at 100 \mum\ and 110 \mum\ pitch, respectively,
with one floating strip. Strip length of $z$-strips for layers 4
and 5  includes ganging. The radial range for layers 4 and 5
includes the radial extent of the arched sections.}
\medskip
\begin{tabular}{lcccr} \hline
\rule{0pt}{12pt}  &  &R-O  & & Strip\\
Layer/&Radius &pitch &Floating   &length\\
view \rule[-5pt]{0pt}{0pt}    &(mm) &(\mum) &strips  &(mm)\\ \hline
\rule{0pt}{12pt}1 z & 32       & 100
& 1 & 40 \\ 1 $\phi$ & 32 & 50-100& 0-1 & 82\\ 2 z        & 40
& 100 & 1 & 48  \\ 2 $\phi$   & 40       & 55-110 & 0-1 & 88\\ 3 z
& 54 & 100 & 1 & 70  \\ 3 $\phi$ & 54       & 110  & 1 & 128 \\ 4
z & 91-127 & 210 & 1 & 104 \\ 4 $\phi$   & 91-127  & 100 & 1 & 224
\\ 5 z        & 114-144 & 210 & 1 & 104 \\
5\rule[-5pt]{0pt}{0pt}  $\phi$   & 114-144 &
100 & 1 & 265 \\ \hline
\end{tabular}
\label{\secname tab:geometrical}
\end{table}

In order to minimize the material in the acceptance region, the
readout electronics are mounted entirely outside the active
detector volume.  The forward electronics must be mounted in the
10\mm\ space between the 350\mrad\ stayclear space and B1 magnet.
This implies that the hybrids carrying the front-end chip must be
positioned at an angle of 350\mrad\ relative to the sensor for the
layers 3, 4, and 5 (Figure~\ref{\secname fig:sideview}).  In the
backward direction, the available space is larger and the inner
layer electronics can be placed in the sensor plane, allowing a
simplified assembly.

The module assembly and the mechanics are quite complicated,
especially for the arch modules, and are described in detail
elsewhere~\cite{module_assembly}. The SVT support structure
(Figure~\ref{\secname fig:svt_overall_pic}) is a rigid body made
from two carbon-fiber cones, connected by a \emph{space frame},
also made of carbon-fiber epoxy laminate.

An  optical survey of the SVT on its assembly jig indicated that
the global error in placement of the sensors with respect to
design was $\sim$200\mum, FWHM.  Subsequently, the detector was
disassembled and shipped to SLAC, where it was re-assembled on the
IR magnets. The SVT is attached to the B1 magnets by a set of
gimbal rings in such a way as to allow for relative motion of the
two B1 magnets while fixing the position of the SVT relative to
the forward B1 and the orientation relative to the axis of both B1
dipoles.  The support tube structure is mounted on the PEP-II
accelerator supports, independently of \babar, allowing for
movement between the SVT and the rest of \babar. Precise
monitoring of the beam interaction point is necessary, as is
described in Section \ref{sec:svtmonitoring}.

The total active silicon area is 0.96\m$^2$ and the material
traversed by particles is $\sim 4$\% of a radiation length (see
Section~\ref{sec:layout}). The geometrical acceptance of SVT is
90\% of the solid angle in the c.m.~system, typically 86\% are
used in charged particle tracking.

\subsection{SVT Components}

A block diagram of SVT components is shown in Figure~\ref{\secname
fig:block}. The basic components of the detector are the silicon
sensors, the \emph{fanout} circuits, the \emph{Front End
Electronics} (FEE) and the data transmission system. Each of these
components is discussed below.

\begin{figure}[htb]
\centering
\includegraphics[width=7.5cm]{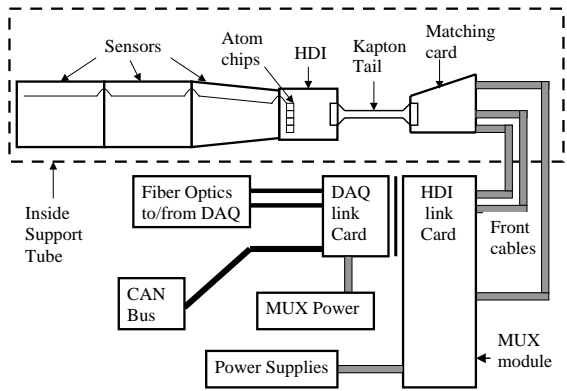}
\vspace{-2.5pc}
 \caption{Schematic block diagram showing the
different components of the SVT.} \label{\secname fig:block}
\end{figure}

\subsubsection{Silicon Sensors}

The SVT sensors~\cite{silicon_detectors} are 300\mum\ thick
double-sided silicon strip devices.  They were designed at INFN
Pisa and Trieste (Italy) and fabricated
commercially~\cite{micron}. They are built on high-resistivity
(6--15\kohm-cm) n-type substrates with p$^+$ strips and n$^+$
strips on the two opposite sides.  The insulation of the n$^+$
strips is provided by individual p-stops, so as to achieve an
inter-strip resistance greater than 100\Mohm\ at operating bias
voltage, normally about 10 V above the depletion voltage.

Typical depletion voltages are in the range 25--35\volt. The
strips are biased on both sides with polysilicon resistors
(4--20\Mohm) to ensure the required radiation hardness, keeping
the voltage drop across resistors and the parallel noise as low as
possible. Strips are AC-coupled to the electronics via integrated
decoupling capacitors, the capacitance of which depends on the
sensor shape, but is always greater than 14\pf/cm.  The sensors
were designed to maximize the active area, which extends to within
0.7\mm\ of the physical edges.  Another design goal was to control
the inter-strip capacitance: values between 0.7\pf/cm and
1.1\pf/cm were obtained for the various sensor shapes.  To achieve
the required spatial resolution, while keeping the number of
readout channels as low as possible, most of the modules have a
\emph{floating strip} (\ie\ not read out) between two readout
strips.

\begin{table}
\centering
 \caption{Electrical parameters of the SVT, shown for
the different layers and views. $C_{input}$ refers to the total
input capacitance, $R_{series}$ is the series resistance. The
amplifier peaking time is 200\ns\ for layers 1--3 and 400\ns\ for
layers 4--5.}

\vspace{\baselineskip}

\begin{tabular}{lrrrrrr} \hline
\rule{0pt}{12pt} &  &  &
\multicolumn{2}{c}{Noise,}\\
Layer/  &$C_\mathrm{input}$   &$R_\mathrm{series}$ &calc.  &meas. \\
\rule[-5pt]{0pt}{0pt}view &(pF)  &($\Omega$)  &(elec)& (elec)\\ \hline
\rule{0pt}{12pt}1 z        &  6.0 &  40.  &  550 & 880  \\ 1
$\phi$ & 17.2 & 164.  & 990  & 1200 \\
2 z   &  7.2 & 48.   &600  & 970
\\ 2 $\phi$   & 18.4 & 158.  & 1030 & 1240 \\ 3 z        & 10.5 &
70. & 700  & 1180  \\ 3 $\phi$   & 26.8 & 230.  & 1470 & 1440 \\ 4
z & 16.6 & 104.  & 870  & 1210 \\ 4 $\phi$   & 33.6 & 224.  & 1380
& 1350 \\ 5 z        & 16.6 & 104.  & 870  & 1200 \\
\rule[-5pt]{0pt}{0pt}5 $\phi$   & 39.7 & 265.  & 1580 & 1600 \\
\hline
\end{tabular}
\label{\secname tab:electrical}
\end{table}

The leakage currents, because of the excellent performance of the
manufacturing process, were as low as 50\nA/cm$^2$ on average,
measured at 10\volt\ above depletion voltage.  The silicon sensor
parameters have been measured after irradiation with $^{60}$Co
sources. Apart from an increase in the inter-strip capacitance of
about 12\% during the first 100\krad, the main effect was an
increase of the leakage current by 0.7\muA/cm$^2$/MRad.  However,
in a radiation test performed in a 1\gevc\ electron beam, an
increase in leakage current of about 2\muA/cm$^2$/MRad and a
significant shift in the depletion voltage, dependent on the
initial dopant concentration, were observed.  A shift of about
8--10\volt\ was seen for irradiation corresponding to a dose of
approximately 1~MRad. These observations indicate significant bulk
damage caused by energetic electrons. As indicated by the change
in depletion voltage, the SVT sensors could undergo type inversion
after about 1--3~MRad. Preliminary tests show that the sensors
continue to function after type inversion~\cite{rad_damage}.
Studies of the behavior of SVT modules as a function of radiation
dose continue.

\subsubsection{Fanout Circuits}

The fanout circuits, which route the signals from the strips to the
electronics, have been designed to minimize the series resistance and
the inter-strip capacitance.  As described in ref.~\cite{fanouts},
a trace on the fanout has a series resistance about 1.6~\ohm/cm,
an inter-strip resistance $>20$\Mohm, and an inter-strip
capacitance $<0.5$\pf/cm.  The electrical parameters of the final
assembly of sensors and fanouts (referred to as  \emph{Detector Fanout
Assemblies} or DFAs) are summarized in Table~\ref{\secname tab:electrical}. Due
to the different strip lengths, there are large differences between
the inner and the outer layers.  Smaller differences are also present
between the forward and backward halves of the module, that are of
different lengths.

\subsubsection{Front End Electronics}

The electrical parameters of a DFA and the general \babar\
requirements are the basic inputs that drove the design of the SVT
front-end custom IC; the ATOM (\emph{A Time-Over-Threshold
Machine}).  In particular, the front-end IC had to satisfy the
following requirements:

\begin{itemize}

\item signal to noise ratio greater than 15 for \emph{minimum ionizing
particle} (MIP) signals for all modules;

\item signals from all strips must be retained, in
order to improve the spatial resolution through interpolation,
while keeping the number of transmitted hits as low as possible. A
\emph{hit} refers to a deposited charge greater than $ 0.95$\fC,
corresponding to 0.25 MIP;

\item the amplifier must be sensitive to both negative and positive
charge;

\item the peaking time must be programmable, with a minimum of 100\ns\
  (in layers 1 and 2, because of the high occupancy), up to 400\ns\
  (outer layers, with high capacitance);

\item capability to accept random triggers with a latency up to 11.5\mus\ and
a programmable jitter up to $\pm 1$\mus, without dead
  time;

\item radiation hardness greater than 2.5~MRad;

\item small dimensions: 128 channels in a 6.2\mm-wide chip.

\end{itemize}

\begin{figure}[htb]
\centering
\includegraphics[width=7.5cm]{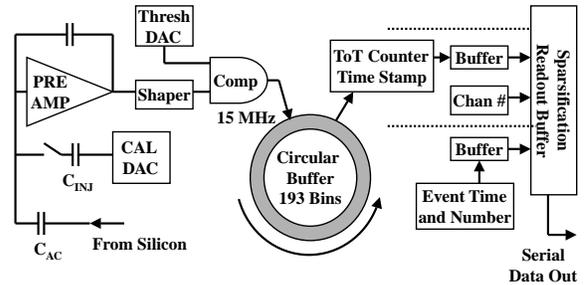}
\vspace{-2pc}
\caption{Schematic diagram of the ATOM front end
IC.} \label{\secname fig:atom}
\end{figure}

These requirements are fully satisfied by the ATOM IC~\cite{atom},
which is depicted schematically in Figure~\ref{\secname fig:atom}.

The linear analog section consists of a charge-sensitive
preamplifier followed by a shaper. Gains of 200\mv/fC (low) or
300\mv/fC (high) may be selected. The channel gains on a IC are
uniform to 5\mv/fC.  Signals are presented to a
programmable-threshold comparator, designed so that the output
width of the pulse (\emph{Time over Threshold} or ToT) is a
quasi-logarithmic function of the collected charge. This output is
sampled at 30\mhz\ and stored in a 193 location buffer.  Upon
receipt of a \emph{Level 1} (L1) trigger, the time and ToT is
retrieved from this latency buffer, sparcified, and stored in a
four event buffer. Upon the receipt of an \emph{L1 Accept} command
from the data acquisition system, the output data (the 4 bits for
the ToT, 5 bits for the time stamp, and 7 bits for the strip
address) are formatted, serialized, and delivered to the ROM.  The
IC also contains a test charge injection circuit.  The typical
noise behavior of the ATOM, as described by the \emph{Equivalent
Noise Charge} (ENC) of the linear analog section is given in
Table~\ref{\secname tab:atom}.

\begin{table}
\caption{ATOM chip ENC parameters at different
 peaking times}
\vspace{\baselineskip}
\begin{center}
\begin{tabular}{ccc} \hline
Peaking\rule{0pt}{12pt}    &ENC   &Noise  \\
time\rule[-5pt]{0pt}{0pt}  &(0\pf)  & slope \\   \hline
\rule{0pt}{12pt}100\ns     &380 \en &40.9 \en /pF\\
   200\ns  &280 \en   &33.9 \en /pF \\
400\ns\rule[-5pt]{0pt}{0pt}     &220 \en   &25.4\en /pF\\ \hline
\end{tabular}
\end{center}
\label{\secname tab:atom}
\end{table}

The average noise for the various module types is shown in
Table~\ref{\secname tab:electrical}. Given that shot noise due to
sensor leakage current is negligible, the expected noise may be
calculated from the parameters of Tables~\ref{\secname
tab:electrical} and \ref{\secname tab:atom}. The results of such a
calculation are also shown in Table~\ref{\secname tab:electrical}.
The maximum average noise is 1,600 electrons, leading to a
signal-to-noise ratio greater than 15.

The power consumption of the IC is about 4.5\mw/channel. Radiation
hardness was studied up to 2.4~MRad with a $^{60}$Co source. At
that dose, the gain decreased 20\%, and the noise increased less
than 15\%.

The ATOM ICs are mounted on thick-film double-sided hybrid
circuits (known as \emph{High Density Interconnects} or HDIs)
based on an  aluminum-nitride substrate with high thermal
conductivity. The electronics are powered through a floating power
supply system, in such a way as to guarantee a small voltage drop
($<1$\volt) across the detector decoupling capacitors.

\subsubsection{Data Transmission}

The digitized signals are transmitted from the ATOM ICs through a
thin \emph{kapton tail} or cable to the \emph{matching cards},
from where they are routed to more conventional cables.  Just
outside the detector, signals are \emph{multiplexed} by the MUX
modules, converted into optical signals and transmitted to the
\emph{Readout Modules} (ROMs). The MUX modules also receive
digital signals from the DAQ via a fiber optical connection.  The
SVT is connected to the \babar\ online detector control and
monitoring system via the industry standard CAN bus.  Details on
SVT data transmission system and DAQ can be found in
references~\cite{dat,daq}. Power to SVT modules (silicon sensor
bias voltage and ATOM low voltages) is provided by a CAEN A522
power supply system~\cite{power_supply}.

\subsection{Monitoring and Calibration}
\label{sec:svtmonitoring}

To identify immediately any operational problems, the SVT is
integrated in the  control and monitoring system
(see Section~\ref{sec:online}).  Major
concerns for SVT monitoring are temperature and humidity, mechanical
position, and radiation dose.

\subsubsection{Temperature\\ and Humidity Monitors}
The total power dissipation of the SVT modules is about 350\watt,
mainly dissipated in the ATOM ICs. External cooling is provided by
chilled water at 8\degc. In addition, humidity is reduced by a
stream of dry air in the support tube.

Since condensation or excessive temperature can permanently
damage the FEE, temperature and humidity monitoring are very
important to the safe operation of the SVT.  Thermistors are
located on the HDIs (for the measurements of FEE temperature),
around the SVT, along the cooling systems, and in the electronics
(MUX) crates.  The absolute temperatures are monitored to
0.2\degc\ and relative changes of 0.1\degc. Additionally, a series
of humidity sensors are employed to monitor the performance of the
dry air system.  The temperature and humidity monitors also serve
as an interlock to the HDI power supplies.

\subsubsection{Position Monitors}
\begin{figure}
\centering
\includegraphics[width=.9\columnwidth]{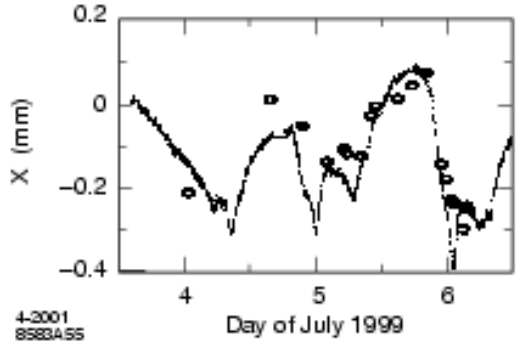}
\vspace{-2pc}
\caption{Horizontal motion between the DCH and the support tube
measured with the capacitive sensors (curve) compared to the mean
$x$ coordinate of the interaction point (circles) measured with
$e^+e^-$ and $\mu^+\mu^-$ events over a three-day period in July
1999. An arbitrary offset and scale has been applied to the beam
position data.} \label{\secname fig:pos}
\end{figure}

A system of capacitive sensors was installed to identify and track
changes in the position of the SVT with respect to the PEP-II B1
magnets and the position of the support tube with respect to the
DCH. An example of the understanding that can be achieved by this
system is given in Figure~\ref{\secname fig:pos}, where the
measured changes in the horizontal position of the SVT relative to
the DCH are shown for a period of six day in the summer of 1999.
These position changes can be attributed to local temperature
variations.  The sensor data are compared to measurements of the
mean position of the interaction point (in the horizontal plane)
determined with $e^+e^-$ and $\mu^+\mu^-$ events recorded over
this period.  While the amplitude of motion at the time was
uncharacteristically large, the strong correlation between these
independent measurements is quite evident.  Alignment with charged
particle tracks is now performed routinely to correct for relative
motion of the tracking systems, as described in Section 5.6.2.

\subsubsection{Radiation Monitors}

Radiation monitoring is extremely important to ensure the SVT does not
exceed its radiation budget, which could cause permanent damage to the
device. To date, the measured radiation absorbed by the SVT is well
within the allowed budget.

The monitoring of radiation dose to the SVT is discussed in detail in
Section~\ref{sec:ir}.

\subsubsection{Calibrations}

Once a day, and each time the SVT configuration has changed,
calibrations are performed in absence of circulating beams. All
electronic channels are tested with pulses through test
capacitors, for different values of the injected charge.  Gains,
thresholds, and electronic noise are measured, and defective
channels are identified. The calibration results have proven very
stable and repeatable. The main variation in time is the
occasional discovery of a new defective channel. The calibration
procedures have also been very useful for monitoring noise sources
external to the SVT.

\subsubsection{Defects}

Due to a series of minor mishaps incurred during the installation of
the SVT, nine out of 208 readout sections (each corresponding
to one side of a half-module) were damaged and are currently not
functioning. There is no single failure mode, but several
causes: defective connectors, mishandling during installation, and
not-fully-understood problems on the FEE
hybrid. There has been no module failure due to radiation damage.
It should be noted that
due to the redundancy afforded by the five layers of the SVT, the
presence of the defective modules has minimal impact on physics analyses.

In addition, there are individual channel
defects, of various types, at a level of about 1\%. Calibrations
have revealed an increase in the number of defective channels at a rate of
less than 0.2\%/year.

\subsection{Data Analysis and Performance}

This section describes the reconstruction of space points from
signals in adjacent strips on both sides of the sensors, the SVT
internal and global alignment, single hit efficiency, and
resolution  and \dedx\ performance of the SVT.

\subsubsection{Cluster and Hit Reconstruction}
\begin{figure}[htb]
\centering
\includegraphics[width=6.5cm]{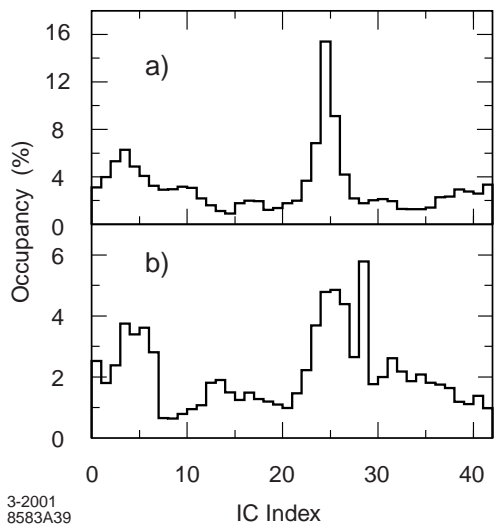}
\vspace{-1.5pc}

\caption{Typical occupancy in percent as a function of IC index in layer
1, \mphi\ side for a) forward half-modules and b) backward
half-modules. The IC index increases with azimuthal angle and
the higher occupancy in the horizontal plane is visible near
chip indices 3 and 25.}
\label{\secname fig:occupancy}
\end{figure}

Under normal running conditions, the average occupancy of the SVT
in a time window of 1\mus\ is about 3\% for the inner layers, with a
significant azimuthal variation due to beam-induced backgrounds, and
less than 1\% for the outer layers, where noise hits dominate.
Figure~\ref{\secname fig:occupancy} shows the typical occupancy as a
function of IC index (equivalent to azimuthal angle, in this case)
for layer 1, \mphi\ side. In the inner layers, the occupancy is
dominated by machine backgrounds, which are significantly higher in
the horizontal plane, seen in the plot as the peaks near IC indices
3 and 25.

The first step of the reconstruction program consists in
discarding out-of-time channels. A time correction, \ie\ the time
between the passage of the particle and the time the shaper
exceeds threshold, is performed, after which hits with times more
than 200\ns\ from the event time (determined by the DCH) are
discarded. The loss of real hits from this procedure is
negligible.  The resulting in-time hits are then passed to the
cluster finding algorithm.  First, the charge pulse height (Q) of
a single pulse is calculated from the ToT value, and clusters are
formed grouping adjacent strips with consistent times. In a second
pass, clusters separated by just one strip are merged into one
cluster.  The two original clusters plus the merged cluster are
made available to the pattern recognition algorithm, which chooses
among the three.

The position $x$ of a cluster formed by $n$ strips is determined,
with the ``head-to-tail'' algorithm:
\begin{displaymath}
   x = \frac{(x_1+x_n)}{2} + \frac{p}{2}
   \frac{(Q_n-Q_1)}{(Q_n+Q_1)},
\end{displaymath}
where $x_i$ and $Q_i$ are the position and collected charge of $i$-th
strip, respectively, and $p$ is the readout pitch.  This formula
results in a cluster position that is always within $p/2$ of the
geometrical center of the cluster. The cluster pulse height is simply
the sum of the strip charges, while the cluster time is the average of
the signal times.

\subsubsection{Alignment}

The alignment of the SVT is performed in two steps. The first step
consists of determining the relative positions of the 340 silicon
sensors. Once this is accomplished, the next step is to align the
SVT as a whole within the global coordinate system defined by the
DCH. The primary reason for breaking the alignment procedure into
these two steps is that the local positions are relatively stable
in time compared to the global position. Also, the local alignment
procedure is considerably more complex than the global alignment
procedure. Thus, the global alignment can be updated on a
run-by-run basis, while the local alignment constants are changed
as needed, typically after magnet quenches or detector access.

The local alignment procedure is performed with tracks from
$e^+e^- \rightarrow \mu^+\mu^-$ events and cosmic rays. Well
isolated, high momentum tracks from hadronic events are also used
to supplement di-muon and cosmic data. Data samples sufficient to
perform the local alignment are collected in one to two days of
typical running conditions.

In $\mu^+\mu^-$  events, the two tracks are simultaneously fit
using a Kalman filter technique and the known beam momentum. The
use of tracks from cosmic rays reduces any systematic distortion
that may be introduced  due to imprecise knowledge of the beam
momenta. No information from the DCH is used, effectively
decoupling the SVT and DCH alignment.

In addition to the information from tracks, data from an optical
survey performed during the assembly of the SVT are included in
the alignment procedure. The typical precision of these optical
measurements is 4\mum. This survey information is only used to
constrain sensors relative to other sensors in the same module,
but not one module to another or one layer to another.
Furthermore, only degrees of freedom in the plane of the sensor
are constrained as they are expected to be the most stable, given
the assembly procedure.

Using the hit residuals from the aforementioned set of tracks and
the optical survey information, a $\chi^2$ is formed for each
sensor and minimized with respect to the sensor's six local
parameters.  The constraints coming from the overlapping regions
of the silicon sensors, the di-muon fit, the cosmic rays, and the
optical survey result in internally consistent local alignment
constants.

\begin{figure}[htb]
\centering
\includegraphics[width=7.5cm]{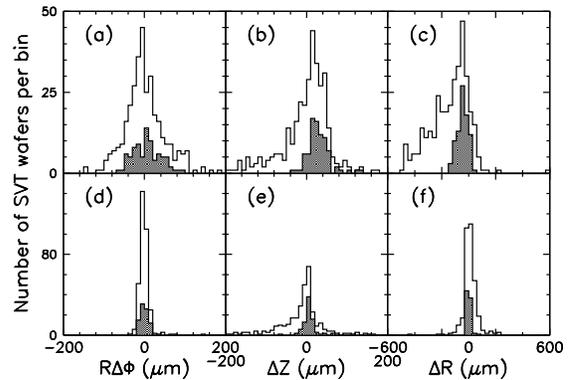}
\vspace{-2pc}
 \caption{Comparison of a local alignment of all the
sensors in the SVT using data from January 2000 with the optical
survey of the SVT made during assembly in February 1999 in the (a)
$r\Delta\phi$, (b) $\Delta z$ and (c) $\Delta r$ coordinates.
Plots (d), (e), and (f) show the difference between two local
alignments using data from January 15-19 and March 6-7, 2000 for
the  $r\Delta\phi$, $\Delta z$, and $\Delta r$ coordinates,
respectively. In all the plots, the shaded regions correspond to
the sensors in the first three layers. In comparing the different
alignments and optical survey, a six parameter fit (three global
translations and three global rotations) has been applied between
the data sets.} \label{\secname fig:SvtAlign1}
\end{figure}

Figure \ref{\secname fig:SvtAlign1} shows a comparison between the
optical alignment made during the SVT assembly in February 1999
and a local alignment using data taken during January 2000.  The
alignment from tracking data was made without using cosmics or
constraints from the optical survey. The width of the
distributions in these  plots has four contributions: 1)
displacement during the transfer of the SVT from the assembly jig
to the IR magnets,  2) time dependent motion of the SVT after
mounting, 3) statistical errors, and 4) systematic errors. The
second set of plots shows the difference in two alignment sets for
data taken in January 2000 as compared to March 2000.  In general,
the stability of the inner three layers is excellent, with
slightly larger tails in the outer two layers. The radial
coordinate is less tightly constrained in all measurements because
the radial location of the charge deposition is not well known,
and most of the information about the radial locations comes only
from constraints in the overlap region of the sensors.

After the internal alignment, the SVT is considered as a rigid
body. The second alignment step consists in determining the
position of the SVT with respect to the DCH. Tracks with
sufficient numbers of SVT and DCH hits are fit two times: once
using only the DCH information and again using only the SVT hits.
The six global alignment parameters, three translations and three
rotations, are determined by minimizing the difference between
track parameters obtained with the SVT-only and the DCH-only fits.
As reported above, because of the diurnal movement of the SVT with
respect to the DCH, this global alignment needs to be performed
once per run ($\sim$ every 2--3 hours). The alignment constants
obtained in a given run are then used to reconstruct the data in
the subsequent run.  This procedure, known as \emph{rolling
calibration}, ensures that track reconstruction is always performed
with up-to-date global alignment constants.

\begin{figure}
\centering
\includegraphics[scale=.8]{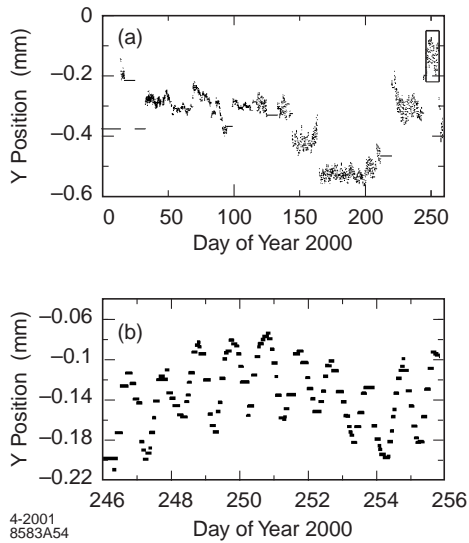}

\caption{ Global alignment of the SVT relative to the DCH based on
$e^+e^-$ and $\mu^+\mu^-$ events: changes in the relative vertical
placement measured a) over the entire ten-month run in the year
2000, and b) a ten-day period, illustrating diurnal variations.  }

\label{\secname fig:8583A53}
\end{figure}

A record of the changes in the relative position of the SVT as
determined by {\emph rolling calibrations} is shown in
Figure~\ref{\secname fig:8583A53}.  The position is stable to
better than $\pm 100 \mum$ over several weeks, but changes
abruptly from time to time, for instance, during access to the
detector. The calibrations track diurnal variations of typically
$\pm 50 \mum$ that have been correlated with local changes in
temperature of about $\pm 2 \degc$. Movements within a single run
are small compared to the size of the beam.

\subsubsection{Performance}
The SVT  efficiency can be calculated for each half-module by
comparing the number of associated hits to the number of tracks
crossing the active area of the  module. As can be seen in
Figure~\ref{\secname fig:efficiency}, a combined hardware and
software efficiencies of 97\% is measured, excluding defective
readout sections (9 out of 208), but employing no special
treatment for other defects, such as broken AC coupling capacitors
or dead channels on front-end chips.
\begin{figure}[htb]
\centering
\includegraphics[width=7.5cm]{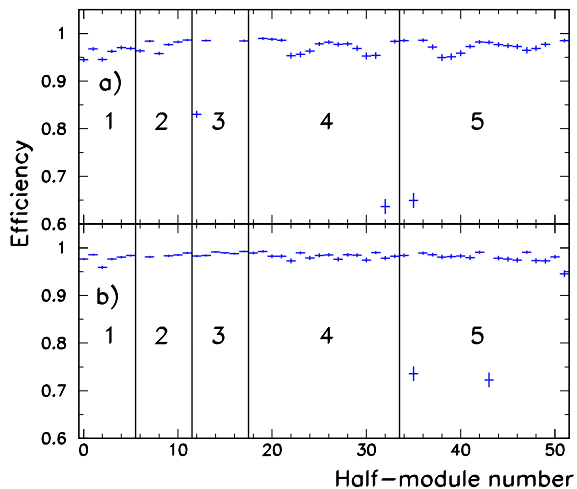}
\vspace{-2pc} \caption{SVT hit reconstruction efficiency, as
measured on $\mu^+\mu^-$ events for a) forward half-modules and b)
backward half-modules.  The plots show the probability of
associating both a $\phi$ and $z$ hit to a track passing through
the active part of the detector. The horizontal axis corresponds
to the different modules, with the vertical lines separating the
different layers as numbered.  Missing values correspond to
non-functioning half-modules.} \label{\secname fig:efficiency}
\end{figure}
Actually, since most of the defects affect a single channel, they
do not contribute to the inefficiency, because most tracks deposit
charge in two or more strips due to track crossing angles, and
charge diffusion.

\begin{figure}[htb]
\includegraphics[width=7.5cm]{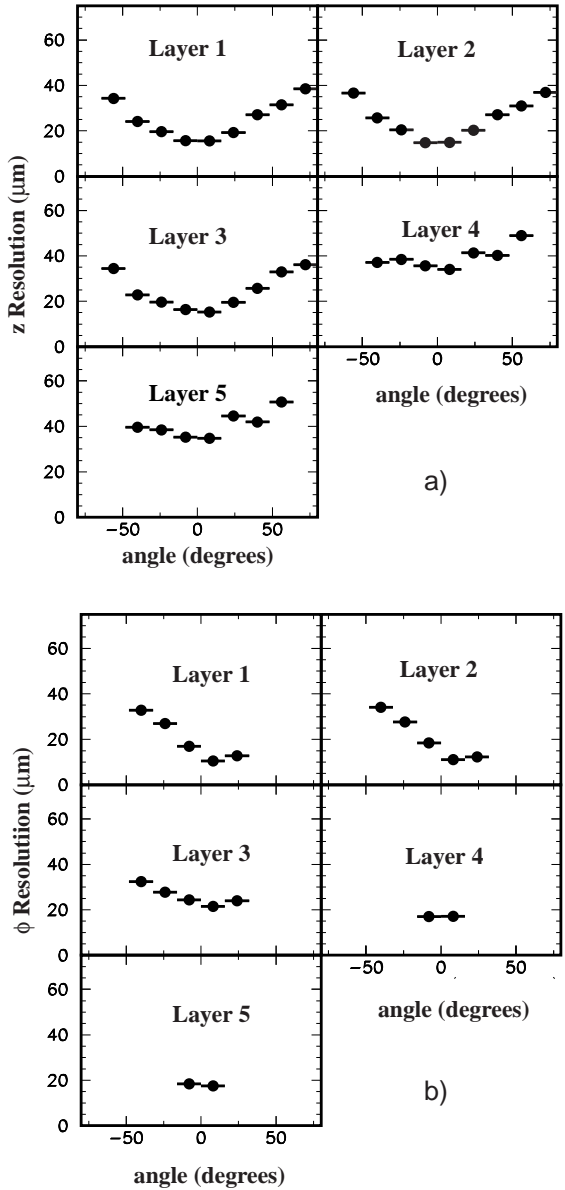}

\vspace{-2pc} \caption{SVT hit resolution in the a) $z$ and b)
$\phi$ coordinate in microns, plotted as a function of track
incident angle in degrees. Each plot shows a different layer of
the SVT. The plots in the $\phi$ coordinate for layers 1-3 are
asymmetric around $\phi=0$ because of the ``pinwheel'' design of
the inner layers. There are fewer points in the $\phi$ resolution
plots for the outer layers as they subtend smaller angles than the
inner layers. } \label{\secname fig:resolution}
\end{figure}

The spatial resolution of SVT hits is determined by measuring the
distance (in the plane of the sensor) between the track trajectory and the
hit, using high-momentum tracks in two prong events.  The uncertainty
due to the track trajectory is subtracted from the width of the
residual distribution to obtain the hit resolution.
Figure~\ref{\secname fig:resolution} shows the SVT hit resolution for
$z$  and $\phi$ side hits as a function of track incident angle, for each of the
five layers. The measured resolutions are in excellent agreement
with expectations from Monte Carlo simulations.

Initial studies have shown that hit reconstruction efficiency and
spatial resolution are effectively independent of occupancy for the
occupancy levels observed so far.

Measurement of the ToT value by the ATOM ICs enables one to obtain
the pulse height, and hence the ionization \dedx\ in the SVT
sensor. The values of ToT are converted to pulse height using a
lookup table computed from the pulse shapes obtained in the bench
measurements. The pulse height is corrected for track length
variation. The double-sided sensors provide up to ten measurements
of \dedx\ per track. For every track with signals from at least
four sensors in the SVT, a 60\% truncated mean \dedx\ is
calculated. The cluster with the smallest \dedx\ energy is also
removed to reduce sensitivity to electronics noise. For MIPs, the
resolution  on the truncated mean \dedx\ is approximately 14\%. A
$2\sigma$ separation between the kaons and pions can be achieved
up to momentum of 500\mevc, and between kaons and protons beyond
1\gevc.

\subsection{Summary and Outlook}
The SVT has been operating efficiently since its installation in
the \babar\ experiment in May 1999. The five layer device, based
on double-sided silicon sensors, has satisfied the original design
goals, in particular the targets specified for efficiency, hit
resolution, and low transverse momentum track reconstruction. The
radiation dose during the first 25\invfb\ of integrated luminosity
is within the planned budget, and  no modules have failed due to
radiation damage. The performance of the SVT modules at high
radiation dose is currently being studied. Early results indicate
that the sensors will continue to function after type inversion
(at 1--3~MRad), but further tests with irradiated sensors and ATOM
ICs need to be performed. A program of spare module production has
commenced, with the goal of replacing modules that are expected to
fail due to radiation damage. Beam-generated backgrounds are
expected to rise with increasing luminosity.  Physics studies at
five times the current backgrounds levels indicate no change in
mass or vertex resolution for the mode \bpsiks\ and a $\sim 20$\%
loss of resolution in the $D^{*+}-D^0$ mass difference. In this
study the detector efficiency for both decay modes was lower by
15--20\%.

\renewcommand{\secname}{dch_}
\renewcommand{\sectiondir}{sec05_dch}
\section{Drift Chamber}
\label{sec:dch}

\subsection {Purpose and Design Requirements}

The principal purpose of the drift chamber (DCH) is the efficient
detection of charged particles and the measurement of their
momenta and angles with high precision.  These high precision
measurements enable the reconstruction of exclusive $B$- and
$D$-meson decays with minimal background.  The DCH complements the
measurements of the impact parameter and the directions of charged
tracks provided by the SVT near the IP.  At lower momenta, the DCH
measurements dominate the errors on the extrapolation of charged
tracks to the DIRC, EMC, and IFR.

The reconstruction of decay and interaction vertices outside of
the SVT volume, for instance the \KS\ decays, relies solely on the
DCH. For this purpose, the chamber should be able to measure not
only the transverse momenta and positions, but also the
longitudinal position of tracks, with a resolution of $\sim$1\mm.

The DCH also needs to supply information for the charged particle
trigger with a maximum time jitter of 0.5\mus\
(Section~\ref{trg_sec:trg}).

For low momentum particles, the DCH is required to provide particle
identification by measurement of ionization loss (\dedx). A resolution
of about 7\% will allow $\pi/K$ separation up to 700\mevc. This
capability is complementary to that of the DIRC in the barrel region,
while in the extreme backward and forward directions, the DCH is the
only device providing some discrimination of particles of different
mass.

Since the average momentum of charged particles produced in $B$- and
$D$-meson decays is less than 1\gevc, multiple scattering is a
significant, if not the dominant limitation on the track parameter
resolution. In order to reduce this contribution, material in front of
and inside the chamber volume has to be minimized.

Finally, the DCH must be operational in the presence of large
beam-generated backgrounds, which were predicted to generate rates of
$\sim$5\khz/cell in the innermost layers.

\subsection{Mechanical Design and Assembly}

\subsubsection{Overview}

\begin{figure*}
\centering
\includegraphics[width=12cm]{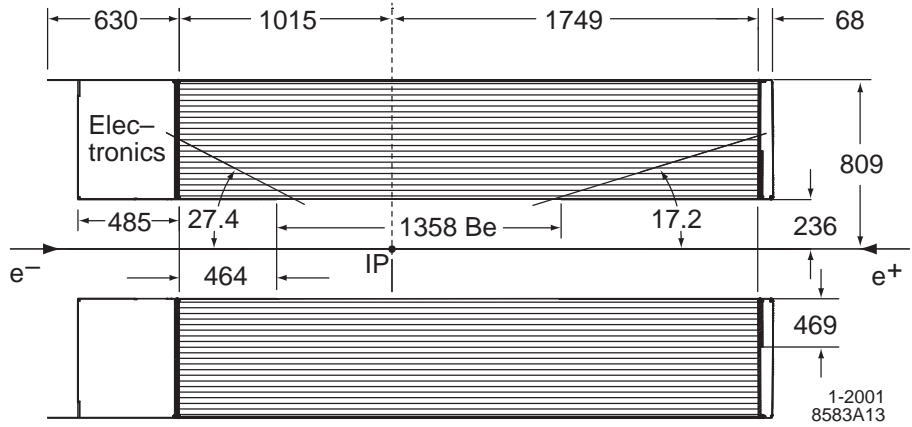}

\vspace{-2pc}
\caption{Longitudinal section of the DCH with
principal dimensions; the chamber center is offset by 370\mm\ from
the interaction point (IP). } \label{\secname fig:sidedc}
\end{figure*}

The DCH is relatively small in diameter, but almost 3\m\ long,
with 40 layers of small hexagonal cells providing up to 40 spatial
and ionization loss measurements for charged particles with
transverse momentum greater than 180\mevc.  Longitudinal position
information is obtained by placing the wires in 24 of the 40
layers at small angles with respect to the $z$-axis. By choosing
low-mass aluminum field wires and a helium-based gas mixture, the
multiple scattering inside the DCH is held to a minimum, less than
0.2\%\Xrad\ of material.  The properties of the chosen gas, a
80:20 mixture of helium:isobutane, are presented in
Table~\ref{tab:gasproperties}.  This mixture has a radiation
length that is five times larger than commonly used argon-based
gases.  The smaller Lorentz angle results in a rather uniform
time-distance relationship and thereby improved spatial
resolution.

\begin{table}[!htb]
\caption{Properties of helium-isobutane gas mixture at atmospheric
pressure and 20\degc. The drift velocity is given for operation
without magnetic field, while the Lorentz angle is stated for a
1.5\tesla\ magnetic field.}
\centering
\vspace{\baselineskip}
\begin{tabular}{ll}\hline
Parameter\rule[-5pt]{0pt}{17pt}  & Values \\ \hline
Mixture $\rm He:C_4 H_{10}$\rule{0pt}{12pt} & 80:20 \\
Radiation Length & 807\m \\
Primary Ions & 21.2/cm \\ Drift Velocity & 22\mum/\ns \\
Lorentz Angle &32\degrees \\
$dE/dx$ Resolution\rule[-5pt]{0pt}{0pt} & 6.9\% \\
\hline
\end{tabular}
\label{tab:gasproperties}
\end{table}

The inner cylindrical wall of the DCH is kept thin to facilitate
the matching of the SVT and DCH tracks, to improve the track
resolution for high momentum tracks, and to minimize the
background from photon conversions and interactions.  Material in
the outer wall and in the forward direction is also minimized so
as not to degrade the performance of the DIRC and the EMC. For
this reason, the HV distribution and all of the readout
electronics are mounted on the backward endplate of the chamber.
This choice also eliminates the need for a massive, heavily
shielded cable plant.

A longitudinal cross section and dimensions of the DCH are shown
in Figure~\ref{\secname fig:sidedc}. The DCH is bounded radially
by the support tube at its inner radius and the DIRC at its outer
radius.  The device is asymmetrically located with respect to the
IP.  The forward length of 1749\mm\ is chosen so that particles
emitted at polar angles of 17.2\degrees\ traverse at least half of
the layers of the chamber before exiting through the front
endplate. In the backward direction, the length of 1015\mm\ means
that particles with polar angles down to 152.6\degrees\ traverse
at least half of the layers.  This choice ensures sufficient
coverage for forward-going tracks, and thus avoids significant
degradation of the invariant mass resolution, while at the same
time maintaining a good safety margin on the electrical stability
of the chamber. The DCH extends beyond the endplate by 485\mm\ at
the backward end to accommodate the readout electronics, cables,
and an rf shield. It extends beyond the forward endplate by 68\mm\
to provide space for wire feed-throughs and an rf shield.

\subsubsection{Structural Components}

\begin{figure*}
\centering
\includegraphics[width=12cm]{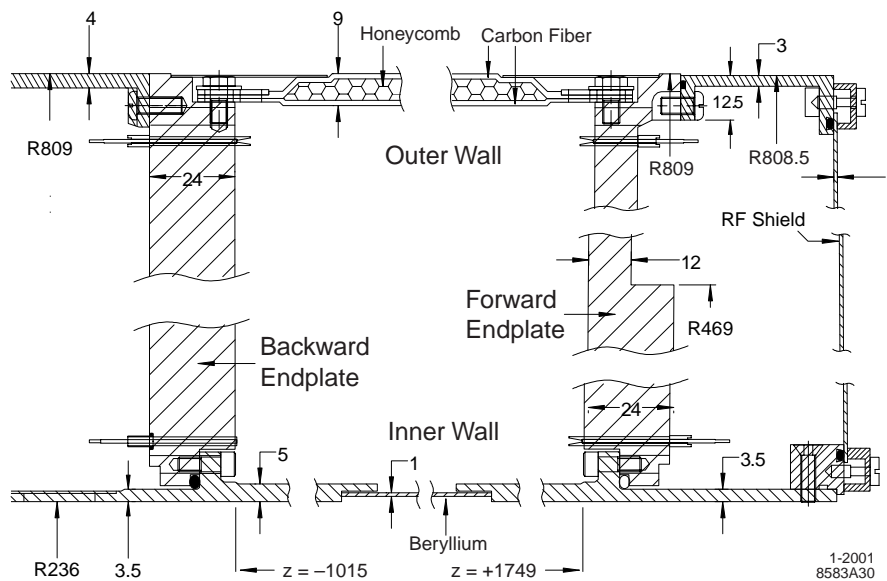}
\vspace{-2pc}
\caption{Details of the structural elements of the
DCH.  All components are made of aluminum, except for the
1\mm-thick inner beryllium wall and the 9\mm -thick outer
composite wall.} \label{\secname fig:joints}
\end{figure*}

Details of the DCH mechanical design are presented in
Figure~\ref{\secname fig:joints}.  The endplates, which carry an
axial load of 31,800\kN, are made from aluminum plates of 24\mm\
thickness.  At the forward end, this thickness is reduced to
12\mm\ beyond a radius of 46.9\cm\ to minimize the material in
front of the calorimeter endcap.  For this thickness, the
estimated safety margin on the plastic yield point for endplate
material (6061T651 aluminum) is not more than a factor of two.
The maximum total deflection of the endplates under loading is
small, about 2\mm\ or 28\% of the 7\mm\ wire elongation under
tension. During installation of the wires, this small deflection
was taken into account by over-tensioning the wires.

The inner and outer cylinder cylindrical walls are load bearing to
reduce the maximum stress and deflections of the endplates.  The
stepped forward endplate created a complication during the assembly,
because the thinner forward endplate would deflect more than the
thicker backward endplate.  The outside rim of the forward endplate
had to be pre-loaded, \ie\ displaced by 2.17\mm\ in the forward
direction, to maintain the inside and outside rims of the rear
endplate at the same longitudinal position after the load of the wires
was transfered from the stringing fixture to the outer cylinder.

Prior to installation on the inner cylinder, the two endplates
were inspected on a coordinate-measuring machine. All sense wire
holes, as well as 5\% of the field and clearing field wire holes,
were measured to determine their absolute locations. The achieved
accuracy of the hole placement was 38\mum\ for both sense and
field wires, better than the specification by more than a factor
of two. In addition, the diameters of the same sample of endplate
holes were checked with precision gauge pins. All holes passed the
diameter specification ($4.500 \pm ^{0.025} _{0.000}$~for sense
wires and $2.500\pm ^{0.025}_{0.00}$ for the field and guard
wires).

The inner cylindrical wall of the DCH, which carries 40\% of the
wire load, was made from five sections, a central 1\mm-thick
beryllium tube with two aluminum extensions which were in turn
electron-beam welded to two aluminum end flanges to form a
3\m-long cylindrical part.  The central section was made from
three 120\degrees\ segments of rolled and brazed beryllium. The
end flanges have precision surfaces onto which the endplates were
mounted. These surfaces set the angles of the two endplates with
respect to the axis and significantly constrain the concentricity
of the tube.  The inner cylinder also provides a substantial rf
shield down to the \pep2\ bunch-gap frequency of 136\khz.

The outer wall bears 60\% of axial wire load between the
endplates. To simplify its installation, this external wall was
constructed from two half-cylinders with longitudinal and
circumferential joints. The gas and electrical seals for these
joints were made up \emph{in situ}. The main structural element
consists of two 1.6\mm-thick (0.006\Xrad) carbon fiber skins
laminated to a 6\mm-thick honeycomb core. The outer shell is
capable of withstanding a differential pressure of 30\mbar\ and
temperature variations as large as $\pm 20$\degrees C, conditions
that could be encountered during shipping or installation.
Aluminum foil, 25\mum-thick on the inside surface and 100\mum\ on
the outside, are in good electrical contact with the endplates,
thereby completing the rf shield for the chamber.

The total thickness of the DCH at normal incidence is 1.08\%\Xrad, of
which the wires and gas mixture contribute 0.2\%\Xrad, and the inner
wall 0.28\%\Xrad.

\subsubsection{Wire Feed-Throughs}

A total of five different types of feed-throughs were required for the
chamber to accommodate the sense, field, and clearing field wires, as
well as two different endplate thicknesses.  The five types are
illustrated in Figure~\ref{\secname fig:FT_Assembly}.  They
incorporate crimp pins~\cite{pins} of a simple design which fasten and
precisely locate the wires.  The choice of pin material (gold-plated
copper for the signal wires and gold-plated aluminum for the field
wires) and wall thickness in the crimp region was optimized to provide
an allowable range of almost 150\mum\ in crimp size, as a primary
means for avoiding wire breakage.

Crimp pins were either press-fit into an insulator made from a single
piece of injection-molded thermoplastic reinforced with 30\% silica
glass fiber~\cite{feedthrough}, or swaged into a copper jacket for the
field wires. The plastic insulates the sense, guard, and clearing
field wires from the electrically grounded endplates, while the metal
jackets provide good ground contact for field wires $(<0.1\Omega)$ on
the backward endplate.  The outer diameter of the field and clearing
field feed-throughs was maintained at $2.000^{+0.000}_{-0.025}$\mm\
while the sense wire feed-through had a larger
($4.500^{+0.000}_{-0.025}$\mm) outer diameter and a longer body
(41.7\mm). This choice provided both thicker insulating walls and a
longer projection into the gas volume to better shield the HV from the
grounded endplate.

\begin{figure}
\centering
\includegraphics[width=6.5cm]{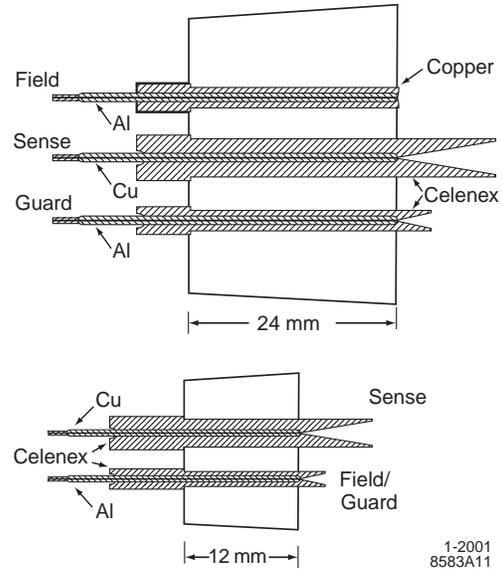}
\vspace{-2pc}
\caption{Design of the five DCH wire feed-throughs
for the 24\mm-thick endplates and the 12\mm-thick endplate.  The
copper jacketed feed-through is for grounded field wires, the
other four are for sense wires (4.5\mm\ diameter), and guard and
clearing field wires (2.5\mm\ diameter), all made from a Celenex
insulator surrounding the crimp pins.} \label{\secname
fig:FT_Assembly}
\end{figure}

\subsubsection{Assembly and Stringing}

Assembly of the chamber components and installation of the wires
was carried out in a large clean room (Class 10,000) at TRIUMF in
Vancouver. The wires were strung horizontally without the outer
cylindrical shell in place. The endplates were mounted and aligned
onto the inner cylinder which in turn was supported by a central
shaft in a mobile fixture. The endplates were mounted on the inner
cylinder at the inside rim and attached to support rings at the
outside.  These rings were connected by radial \emph{spiders} to
the central shaft of the stringing frame.

Two teams of two operators each worked in parallel as the wires
were strung from the inner radius outward. The two teams were each
assisted by an automated wire transporter~\cite{ref:robots}. A
wire was attached to a needle which was inserted through one of
the endplate hole, captured magnetically by one of the
transporters, and then transported and inserted though the
appropriate hole in the other endplate.  The wire was then
threaded through the feed-throughs, which were glued into the
endplates, and the wire was tensioned and crimped. The automated
wire transporters were largely built from industrial components,
employing commercial software and hardware.  The semi-automatic
stringing procedure ensured the correct hole selection,
accelerated the stringing rate and greatly improved the
cleanliness and quality of the stringing process. The installation
of a total of 28,768 wires was completed in less than 15 weeks.

\subsection{Drift Cells}
\label{\secname section:drift_system}

\subsubsection{Layer Arrangement}

\begin{figure}
\centering
\includegraphics[width=6.0cm]{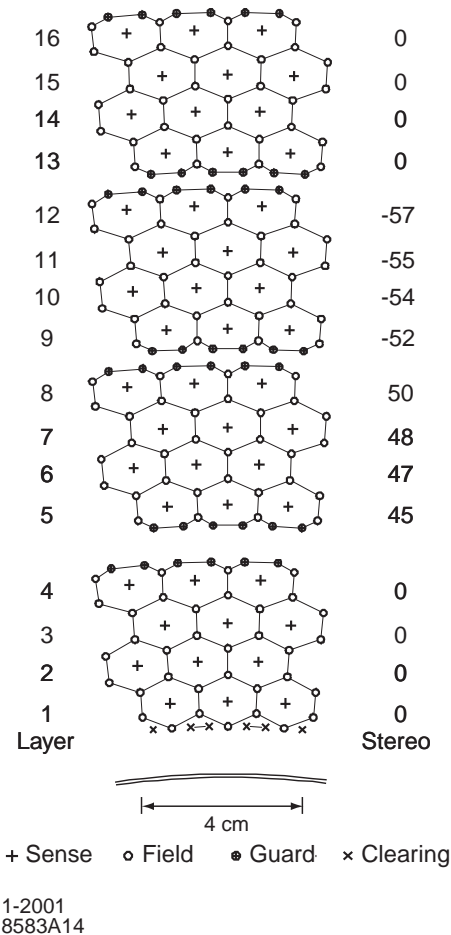}
\vspace{-2pc}

\caption{Schematic layout of drift cells for the four innermost superlayers.
Lines have been added between field wires to aid in visualization of
the cell boundaries.  The numbers on the right side give the stereo
angles (mrad) of sense wires in each layer.  The 1\mm-thick beryllium
inner wall is shown inside of the first layer.}
\label{\secname fig:BabarCells}
\end{figure}

The DCH consists of a total of 7,104 small drift cells, arranged in
40~cylindrical layers.  The layers are grouped by four into ten
superlayers, with the same wire orientation and equal numbers of cells
in each layer of a superlayer.  Sequential layers are staggered by
half a cell. This arrangement enables local segment finding and
left-right ambiguity resolution within a superlayer, even if one out
of four signals is missing. The stereo angles of the superlayers
alternate between axial (A) and stereo (U,V) pairs, in the order
AUVAUVAUVA, as shown in Figure~\ref{\secname fig:BabarCells}.  The
stereo angles vary between $\pm 45\mrad$ and $\pm 76\mrad$; they have
been chosen such that the drilling patterns are identical for the two
endplates. The hole pattern has a 16-fold azimuthal symmetry which is
well suited to the modularity of the electronic readout and trigger
system.  Table~\ref{\secname tab:cellcount} summarizes parameters for
all superlayers.

\begin{table}[!htb]
\caption{The DCH superlayer (SL) structure, specifying the number
of cells per layer, radius of the innermost sense wire layer, the
cell widths, and wire stereo angles, which vary over the four
layers in a superlayer as indicated.  The radii and widths are
specified at the mid-length of the chamber.} \label{\secname
tab:cellcount}

\vspace{\baselineskip}
\centering
\begin{tabular}{ccccc}    \hline
\rule{0pt}{12pt}   & \# of & Radius & Width & Angle \\
SL\rule[-5pt]{0pt}{0pt}& Cells & (mm) & (mm) &    (mrad) \\
\hline
\rule{0pt}{12pt}1 & 96 & 260.4 &17.0-19.4 & 0 \\ 2 & 112 & 312.4
&17.5-19.5 & 45-50
   \\ 3 & 128 & 363.4 &17.8-19.6 & -(52-57) \\ 4 & 144 & 422.7
   &18.4-20.0 & 0 \\ 5 & 176 & 476.6 &16.9-18.2 & 56-60 \\ 6 & 192 &
   526.1 & 17.2-18.3 & -(63-57) \\ 7 & 208 & 585.4 &17.7-18.8 & 0 \\ 8
   & 224 & 636.7 &17.8-18.8 & 65-69 \\ 9 & 240 & 688.0 &18.0-18.9 &
   -(72-76) \\
10\rule[-5pt]{0pt}{0pt} & 256 & 747.2 &18.3-19.2 & 0 \\
\hline
\end{tabular}
\end{table}

\subsubsection{Cell Design and Wires}
\label{\secname ssec:dchcell}

The drift cells are hexagonal in shape, 11.9\mm\ by approximately
19.0\mm\ along the radial and azimuthal directions, respectively.  The
hexagonal cell configuration is desirable because approximate circular
symmetry can be achieved over a large portion of the cell.  The choice
of aspect ratio has the benefit of decreasing the number of wires and
electronic channels, while allowing a 40-layer chamber in a confined
radial space.
Each cell consists of one sense wire surrounded by six field wires, as
shown in Figure~\ref{\secname fig:BabarCells}.  The properties of the
different types of gold-coated wires that make up the drift cells are
given in Table~\ref{\secname tab:wires}.  The sense wires are made of
tungsten-rhenium~\cite{luma}, 20\mum\ in diameter and tensioned with a
weight of 30\gm. The deflection due to gravity is 200\mum\ at
mid-length. Tungsten-rhenium has a substantially higher linear
resistivity (290\ohm/m), compared to pure tungsten (160\ohm/m), but it
is considerably stronger and has better surface quality.  While the
field wires are at ground potential, a positive high voltage is
applied to the sense wires.  An avalanche gain of approximately
$5\times 10^4$ is obtained at a typical operating voltage of
$1960$\volt\ and a 80:20 helium:isobutane gas mixture.

\begin{table}[!htb]
\caption{DCH wire specifications (all wires are gold plated).}
\label{\secname tab:wires}
\vspace{\baselineskip}
\footnotesize
\centering
\begin{tabular}{ccccc}
\hline
\rule{0pt}{12pt}     & & Diameter&Voltage& Tension \\
Type\rule[-5pt]{0pt}{0pt}    &Material &($\mu$m) & (V) &    (g) \\
\hline
\rule{0pt}{12pt}   Sense & W-Re & 20 & 1960 & 30 \\
  Field & Al & 120 & 0 & 155 \\
  Guard  & Al & 80 & 340 & 74 \\
\rule[-5pt]{0pt}{0pt}  Clearing& Al & 120 & 825 & 155 \\
\hline
\end{tabular}
\normalsize
\end{table}

The relatively low tension on the approximately 2.75\m-long sense
wires was chosen so that the aluminum field wires have matching
gravitational sag and are tensioned well below the elastic limit.
A simulation of the electrostatic forces shows that the cell
configuration has no instability problems.  At the nominal
operating voltage of 1960\volt, the wires deflect by less then
60\mum.

The field wires~\cite{field_wires} are tensioned with 155\gm\ to match
the gravitational sag of the sense wires to within 20\mum. This
tension is less than one-half the tensile yield strength of the
aluminum wire.  For cells at the inner or outer boundary of a
superlayer, two guard wires are added to improve the electrostatic
performance of the cell and to match the gain of the boundary cells to
those of the cells in the inner layers.  At the innermost boundary of
layer 1 and the outermost boundary of layer 40, two clearing wires
have been added per cell to collect charges created through photon
conversions in the material of the walls.

\begin{figure}
\centering
\includegraphics[width=6.0cm]{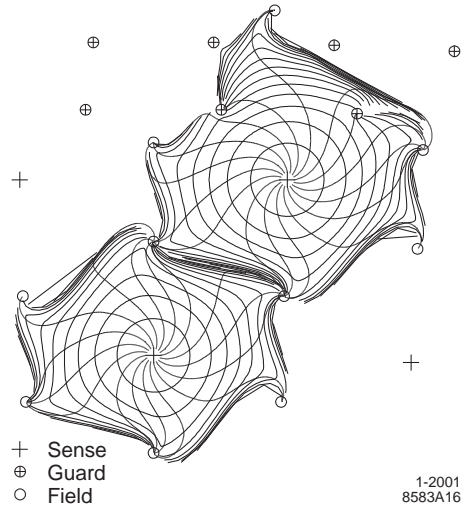}
\vspace{-2pc}

\caption{ Drift cell isochrones, \ie\ contours of equal drift
times of ions in cells of layers 3 and 4 of an axial superlayer.
The isochrones are spaced by 100\ns. They are circular near the
sense wires, but become irregular near the field wires, and extend
into the gap between superlayers.} \label{\secname fig:isochrones}
\end{figure}

\subsubsection {Drift Isochrones}

\begin{figure*}
\centering
\includegraphics[width=12cm]{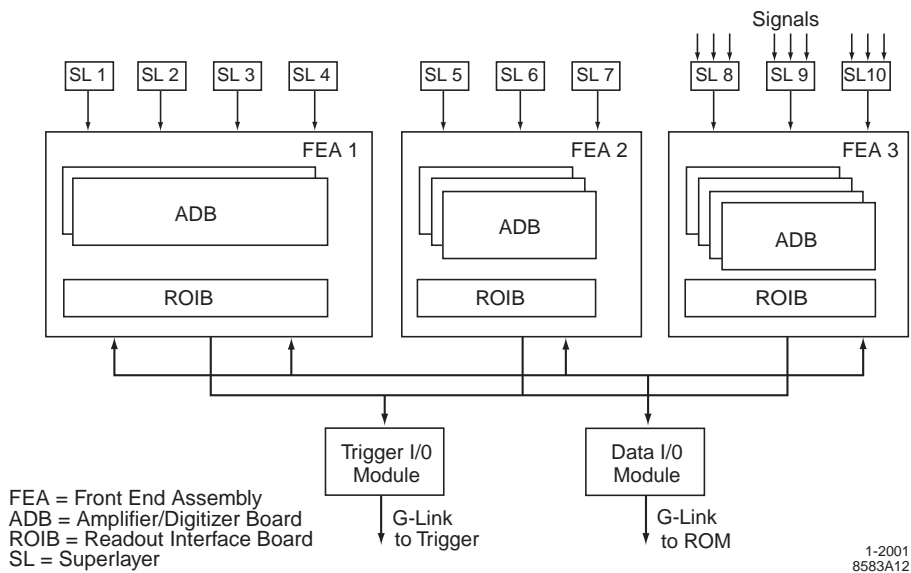}
\vspace{-2pc}

\caption{Block diagram for a 1/16$^{th}$ wedge of the DCH readout
system, showing logical organization of the three front-end
assemblies and their connections to the trigger and data I/O
modules} \label{\secname fig:pain2}
\end{figure*}

The calculated isochrones and drift paths for ions in adjacent cells
of layer 3 and 4 of an axial superlayer are presented in
Figure~\ref{\secname fig:isochrones}.  The isochrones are circular
near the sense wires, but deviate greatly from circles near the field
wires. Ions originating in the gap between superlayers are collected
by cells in the edge layers after a delay of several $\mu$s.  These
lagging ions do not affect the drift times measurements, but they
contribute to the \dedx\ measurement.

\subsubsection {Cross Talk}

A signal on one sense wire produces oppositely-charged signals on
neighboring wires due to capacitive coupling.  The cross talk is
largest between adjacent cells of adjacent layers, ranging from
$-0.5$\% at a superlayer boundary to $-2.7$\% for internal layers
within superlayers. For adjacent cells in the same layer, the cross
talk ranges from $-0.8$ to $-1.8$\%, while for cells separated by two
layers it is less than 0.5\%.

\subsection{Electronics}
\label{\secname section:electronics}

\subsubsection{Design Requirements and Overview}

The DCH electronic system is designed to provide a measurement of
the drift time and the integrated charge, as well as a single bit
to the trigger system~\cite{dct} for every wire with a signal.  In
the 80:20 helium:isobutane gas mixture, there are on average some
22 primary and 44 total ionization clusters produced per cm.  The
position of the primary ionization clusters is derived from timing
of the leading edge of the amplified signal. The design goal was
to achieve a position resolution of 140\mum, averaged over the
cells.  To reduce the time jitter in the signal arrival and at the
same time maintain a good signal-to-noise ratio, the signal
threshold was set at about 2.5 primary electrons.  For the \dedx\
measurement, a resolution of 7\% was projected for a 40-layer
chamber.

The small cell size and the difficult access through the DIRC
strong support tube require a very high density of electronics
components. As a consequence, a compact and highly modular design
was chosen.  The readout is installed in well shielded assemblies
that are plugged into the endplate and are easily removable for
maintenance.

A schematic overview of the DCH electronics is presented in
Figure~\ref{\secname fig:pain2}~\cite{ref:dch_IEEE}.  The 16-fold
azimuthal symmetry of the cell pattern is reflected in the readout
segmentation.  The DCH amplifier and digitizer electronics are
installed in electronics front-end assemblies (FEAs) that are
mounted directly onto the rear endplate. There are three FEAs in
each of the 16 sectors. These sectors are separated by brass
cooling bars that extend from the inner to the outer chamber
walls.  These bars provide mechanical support and water cooling
channel for the FEAs. The assemblies connect to the sense wires
through service boards, which route the signals and HV
distribution. A readout interface board (ROIB) in each FEA
organizes the readout of the digitized data.  Data I/O and trigger
I/O modules multiplex serial data from the FEAs to high-speed
optical fibers for transfer to the readout modules that are
located in the electronics building.

\subsubsection{Service Boards}

\begin{figure}
\centering
\includegraphics[width=6.6cm]{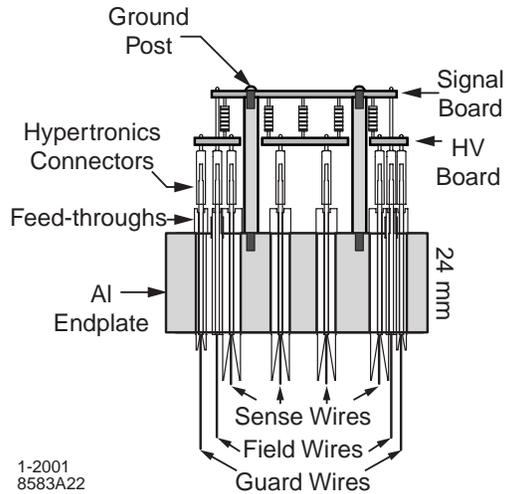}
\vspace{-2pc}

\caption{Side view of service boards showing two-tiered structure for
DCH HV distribution and signal collection.}
\label{\secname fig:svb_side_new}
\end{figure}

Service boards provide the electrostatic potentials for signal,
guard, and clearing wires, and pass signals and ground to the
front-end readout electronics. A side view of a service board is
shown in Figure~\ref{\secname fig:svb_side_new}.  The HV board
contains the HV buses and filtering, current limiting resistors,
and blocking capacitors. Jumpers connect adjacent boards.  The
stored energy is minimized by using a 220\pf\ HV blocking
capacitors.

The signals are connected via series resistors to the upper signal
board which contains the protection diodes and standard output
connectors. Mounting posts, anchored into the rear endplate, also
serve as ground connections.

\subsubsection{Front-End Assemblies}

\begin{figure}
\centering
\includegraphics[width=6.3cm]{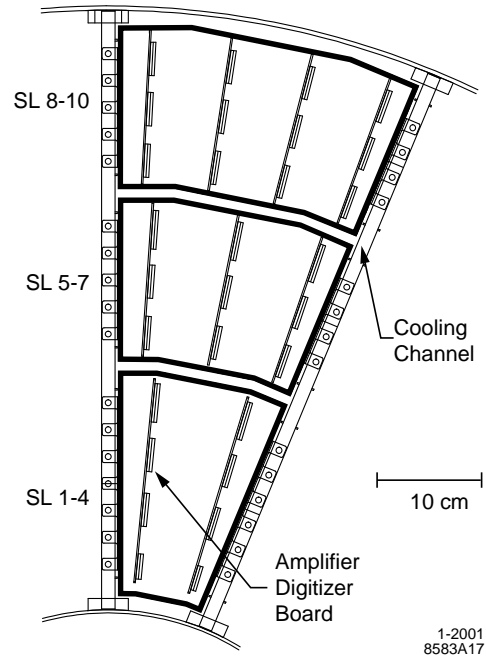}
\vspace{-2pc}
\caption{Layout of $1/16^{th}$ of the DCH rear
endplate, showing three FEA boxes between water cooled channels.}
\label{\secname fig:wdg_box1}
\end{figure}

The FEAs plug into connectors on the back side of the service boards.
These custom wedge-shaped crates are aluminum boxes that contain a
ROIB and two, three, or four amplifier/digitizer boards (ADB) for
superlayers 1--4, 5--7, and 8--10, respectively, as shown in
Figure~\ref{\secname fig:wdg_box1}.  The crates are mounted with good
thermal contact to the water cooled radial support bars. The total
heat load generated by the FEAs is 1.3\kw.

The ADBs are built from basic building blocks consisting of two
4-channel amplifier ICs~\cite{dcac} feeding a single 8-channel
digitizer custom ASIC~\cite{elephant}.  The number of channels
serviced by an ADB is 60, 48, or 45, for the inner, middle, and outer
FEA modules, respectively.

The custom amplifier IC receives the input signal from the sense
wire and produces a discriminator output signal for the drift time
measurement and a shaped analog signal for the \dedx\ measurement.
Both outputs are fully differential. The discriminator has gain
and bandwidth control, and a voltage controlled threshold. The
analog circuit has integrator and gain control.

The custom digitizer IC incorporates a 4-bit TDC for time
measurement and a 6-bit 15\mhz\ FADC to measure the total
deposited charge.  The TDC is a phase-locked digital delay linear
vernier on the sample clock of 15\mhz, which achieves a 1\ns\
precision for leading edge timing. The FADC design is based on a
resistor-divider comparator ladder that operates in bi-linear mode
to cover the full dynamic range.  The digitized output signals are
stored in a trigger latency buffer for 12.9\mus, after which a L1
\emph{Accept} initiates the transfer of a 2.2\mus\ block of data
to the readout buffer. In addition, trigger information is
supplied for every channel, based either on the presence of a TDC
hit during the sample period or FADC differential pulse height
information, should a higher discriminator level be desirable.

The ROIB interprets FCTS commands to control the flow of data and
trigger information. Data are moved to FIFOs on the ROIBs, and
then to data and trigger I/O modules via 59.5\mhz\ serial links. A
total of four such links are required per 1/16$^{th}$ wedge, one
for each of the outer two FEAs and two for the innermost of the
FEA. Each data I/O module services all FEAs one quadrant and
transmits the data to a single ROM via one optical fiber link. The
trigger stream is first multiplexed onto a total of 30 serial
lines per wedge for transmission to the trigger I/O module.
Trigger data from two wedges of FEAs are then transmitted to the
trigger system via three optical links. Thus, a total of 28
optical fibers, four for the data and 24 for the charged particle
trigger, are required to transfer the DCH data to the readout.

\subsubsection{Data Acquisition}
The data stream is received and controlled by four \babar\
standard readout modules. Drift chamber-specific feature
extraction algorithms convert the raw FADC and TDC information
into drift times, total charge, and a status word.  The time and
charge are corrected channel-by-channel for time offsets,
pedestals, and gain constants.  Based on measurements of the noise
a threshold is typically 2--3 electrons is applied to discriminate
signals.  These algorithms take about 1\mus\ per channel, and
reduce the data volume by roughly a factor of four.

\subsubsection{High Voltage System}

The HV bias lines on the chamber are daisy-chained together so that
each superlayer requires only four power supplies, except for
superlayer 1 which has eight.  The voltages are supplied to the sense,
guard, and clearing wires by a CAEN SY527 HV
mainframe~\cite{power_supply}, equipped with 24-channel plug-in
modules.  The sense wires are supplied by 44~HV channels providing up
to 40\muA\ of current each that can be monitored with a resolution of
0.1\%.

\subsection{Gas System}
\label{\secname sec:dchgas}

The gas system has been designed to provide a stable 80:20
helium:isobutane mixture at a constant over pressure of
4\mbar~\cite{dch_gas}.  The chamber volume is about 5.2\m$^3$.  Gas
mixing and recirculation is controlled by precise mass flow
controllers; the total flow is tuned to 15\liter/min, of which
2.5\liter/min are fresh gas. During normal operation, the complete DCH
gas volume is re-circulated in six hours, and one full volume of fresh
gas is added every 36 hours.  The pressure in the DCH is measured by
two independent pressure gauges, one of which is connected to a
regulator controlling the speed of the compressor.  The relative
pressure in the chamber is controlled to better than $\pm 0.05$ mbar.

Oxygen is removed from the gas mixture using a palladium catalytic
filter.  The water content is maintained at $3500\pm 200$ ppm by
passing an adjustable fraction of the gas through a water
bubbler. This relatively high level of water vapor is maintained to
prevent electrical discharge. In addition to various sensors to
monitor pressure, temperature, and flow at several points of the
system, a small wire chamber with an $^{55}\mathrm{Fe}$ source
continuously monitors gain of the gas mixture.

\subsection{Calibrations and Monitoring}

\subsection{Electronics Calibration}

The front-end electronics (FEEs) are calibrated daily to determine
the channel-by-channel correction constants and thresholds.
Calibration pulses are produced internally and input to the
preamplifier at a rate of about 160 Hz.  The calibration signals
are processed in the ROM to minimize the data transfer and fully
exploit the available processing power. The results are stored for
subsequent feature extraction.  The entire online calibration
procedure takes less than two minutes.

\subsubsection{Environmental Monitoring}
\label{\secname sec:dchenv}

The operating conditions of the DCH are monitored in realtime by a
variety of sensors and read out by the detector-wide CAN bus system.
These sensors monitor the flow rate, pressure, and gas mixture; the
voltages and currents applied to the wires in the chamber; the
voltages and currents distributed to the electronics from power
supplies and regulators; instantaneous and cumulative radiation doses;
temperature and humidity around the chamber electronics and in the
equipment racks.  Additional sensors monitor the atmosphere in and
around the detector for excess isobutane, which could pose a
flammability or explosive hazard in the event of a leak.

Many of the sensors are connected to hardware interlocks, which
ensure that the chamber is automatically put into a safe state in
response to an unsafe condition.  All of these systems have
performed reliably. In addition, automated software monitors raw
data quality, chamber occupancies and efficiencies to sense
variations in electronics performance that might indicate more
subtle operational problems.

\subsubsection{Operational Experience}

The design of the DCH specifies a voltage of 1960\volt\ on the sense
wires to achieve the desired gain and resolution.  The chamber voltage
was lowered for part of the run to 1900\volt\ out of concern for a
small region of the chamber that was damaged during the commissioning
phase by inadvertently applying 2\kv\ to the guard wires.  Wires in
this region (10.4\% of superlayer 5, and 4.2\% of superlayer 6) were
disconnected when continuous discharge was observed over extended
periods of time.

\subsection{Performance}
The DCH was first operated with full magnetic field immediately after
the installation into \babar. Cosmic ray data were recorded and
extensive studies of the basic cell performance were performed to
develop calibration algorithms for the time-to-distance and \dedx\
measurements.  These algorithms were then implemented as described
below for colliding beam data.  Calibrations are monitored
continuously to provide feedback to the operation; some time varying
parameters are updated continuously as part of OPR. For charge
particle tracking the DCH and SVT information is combined; the
performance of the combined tracking system is described in
Section~\ref{sec:trk}.

\subsubsection{Time-to-Distance Relation }

The precise relation between the measured drift time and drift
distance is determined from samples of $\ep \en$ and $\mu^+ \mu^-$
events.  For each signal, the drift distance is estimated by computing
the distance of closest approach between the track and the wire. To
avoid bias, the fit does not use the hit on the wire under
consideration.  The estimated drift distances and measured drift times
are averaged over all wires in a layer, but the data are accumulated
separately for tracks passing on the left of a sense wire and on the
right.  The time-distance relation is fit to a sixth-order Chebychev
polynomial.  An example of such a fit is shown in Figure \ref{\secname
fig:dchttodfit}.

An additional correction is made for tracks with varying entrance
angle into the drift cell.  This angle is defined relative to the
radial vector from the IP to the sense wire.  The correction is
applied as a scale factor to the drift distance and was determined
layer-by-layer from a Garfield \cite{ref:garfield} simulation. The
entrance angle correction is implemented as a tenth-order Chebychev
polynomial of the drift distance, with coefficients which are
functions of the entrance angle.

\begin{figure}
\centering
\includegraphics[width=6.5cm]{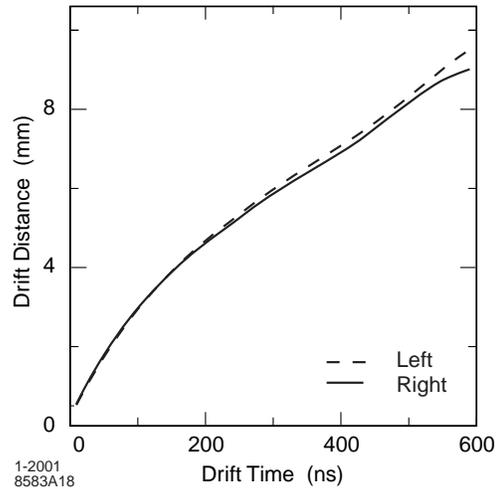}
\vspace{-2pc}
\caption{ The drift time versus distance relation
for left and right half of a cell.  These functions are obtained
from the data averaged over all cells in a single layer of the
DCH.} \label{\secname fig:dchttodfit}

\vspace{-.5\baselineskip}
\end{figure}

\begin{figure}
\centering
\includegraphics[width=6.5cm]{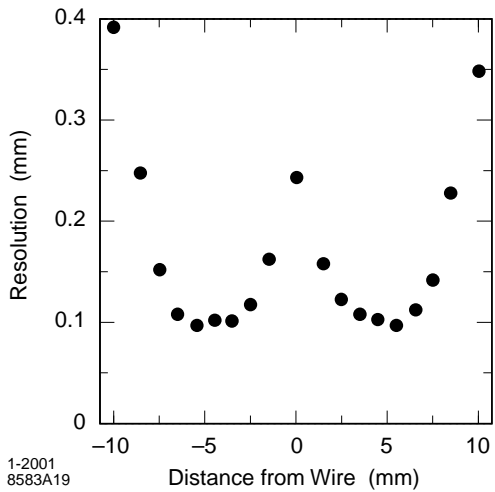}
\vspace{-2pc}
\caption{ DCH position resolution as a function of
the drift distance in layer 18, for tracks on the left and right
side of the sense wire. The data are averaged over all cells in
the layer.} \label{\secname fig:dchresolution}
\end{figure}

Figure~\ref{\secname fig:dchresolution} shows the position resolution
as a function of the drift distance, separately for the left and the
right side of the sense wire.  The resolution is taken from Gaussian
fits to the distributions of residuals obtained from unbiased track
fits.  The results are based on multi-hadron events, for data averaged
over all cells in layer 18.

\subsubsection{Charge Measurement}

\begin{figure}
\centering
\includegraphics[width=6.5cm]{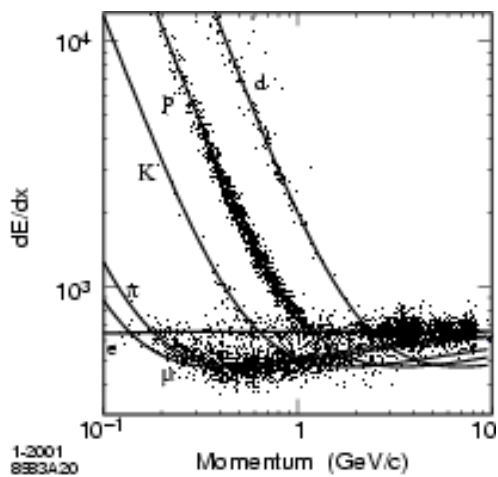}
\vspace{-2pc}
\caption{ Measurement of \dedx\ in the DCH as a
function of track momenta. The data include large samples of beam
background triggers, as evident from the high rate of protons. The
curves show the Bethe-Bloch predictions derived from selected
control samples of particles of different masses.} \label{\secname
fig:dedxvsp}
\end{figure}

The specific energy loss, \dedx, for charged particles traversing the
DCH is derived from measurement of total charge deposited in each
drift cell.  The charge collected per signal cell is measured as part
of the feature extraction algorithm in the ROM. Individual
measurements are corrected for gain variations, pedestal-subtracted
and integrated over a period of approximately 1.8\mus.

The specific energy loss per track is computed as a truncated mean
from the lowest 80\% of the individual \dedx\ measurements.
Various corrections are applied to remove sources of bias that
degrade the accuracy of the primary ionization measurement.  These
corrections include the following:
\begin{itemize}
\item
changes in gas pressure and temperature, leading to $\pm9\%$ variation
in \dedx, corrected by a single overall multiplicative constant;
\item
differences in cell geometry and charge collection ($\pm 8\%$
variation), corrected by a set of multiplicative constants for each
wire;
\item
signal saturation due to space charge build-up ($\pm 11\%$ variation),
corrected by a second-order polynomial in the dip angle, $\lambda$, of
the form $1 / \sqrt{\sin^2\lambda + const}$;
\item
non-linearities in the most probable energy loss at large dip angles
($\pm 2.5\%$ variation), corrected with a fourth-order Chebychev
polynomial as a function of $\lambda$; and
\item
variation of cell charge collection as a function entrance angle ($\pm
2.5\%$ variation), corrected using a sixth-order Chebychev polynomial
in the entrance angle.
\end{itemize}
The overall gas gain is updated continuously based on calibrations
derived as part of prompt reconstruction of the colliding beam data;
the remaining corrections are determined once for a given HV voltage
setting and gas mixture.

\begin{figure}
\centering
\includegraphics[width=6cm]{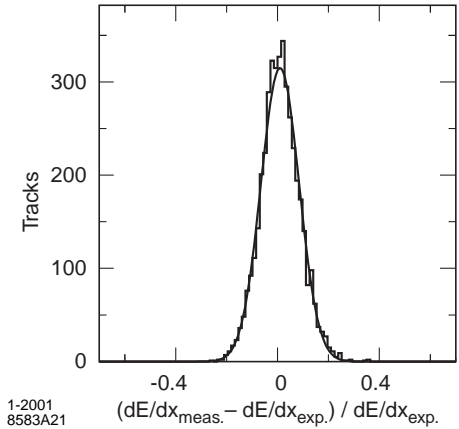}
\vspace{-2pc}
\caption{ Difference between the measured and
expected energy loss \dedx\ for \epm\ from Bhabha scattering,
measured in the DCH at an operating voltage of 1900\volt.  The
curve represents a Gaussian fit to the data with a resolution of
7.5\%.} \label{\secname fig:dedxres}
\end{figure}

Corrections applied at the single-cell level can be large compared
to the single-cell \dedx\ resolution, but have only a modest
impact on the average resolution of the ensemble of hits.  Global
corrections applied to all hits on a track are therefore the most
important for the resolution.

Figure~\ref{\secname fig:dedxvsp} shows the distribution of the
corrected \dedx\ measurements as a function of track momenta. The
superimposed Bethe-Bloch predictions for particles of different masses
have been determined from selected control samples.

The measured \dedx\ resolution for Bhabha events is shown in
Figure~\ref{\secname fig:dedxres}. The rms resolution achieved to
date is typically 7.5\%, limited by the number of samples and
Landau fluctuations. This value is close to the expected
resolution of 7\%. Further refinements and additional corrections
are being considered to improve performance.

\subsection{Conclusions}
The DCH has been performing close to design expectations from the
start of operations.  With the exception of a small number of wires
that were damaged by an unfortunate HV incident during the
commissioning phase, all cells are fully operational. The DCH
performance has proven very stable over time. The design goal for the
intrinsic position and \dedx\ resolution have been met. Backgrounds
are acceptable at present beam currents, but there is concern for
rising occupancies and data acquisition capacity at the high end of
the planned luminosity upgrades.

\renewcommand{\secname}{trk_}
\renewcommand{\sectiondir}{sec14_trk}
\section{Performance of the Charged Particle Tracking Systems}
\label{sec:trk}

Charged particle tracking has been studied with large samples of
cosmic ray muons, $\ep \en$, $\mu^+ \mu^-$, and $\tau^+ \tau^-$
events, as well as multi-hadrons.  At this time, these studies are
far from complete and the results represent the current status. In
particular, many issues related to the intrinsic alignment of the
SVT and the DCH, the variation with time of the relative alignment
of the SVT and the DCH, and movement of the beam position relative
to \babar\ remain under study.

\subsection{Track Reconstruction}

The reconstruction of charged particle tracks relies on data from both
tracking systems, the SVT and the DCH.  Charged tracks are defined by
five parameters $(d_0 ,\phi_0 ,\omega ,z_0,\tan\lambda )$ and their
associated error matrix.  These parameters are measured at the point
of closest approach to the $z$-axis; $d_0$ and $z_0$ are the
distances of this point from the origin of the coordinate system in
the $x$--$y$ plane and along the $z$-axis, respectively.  The angle
$\phi_0$ is the azimuth of the track, $\lambda$ the dip angle relative to the
transverse plane, and $\omega=1/p_t$ is its curvature. $d_0$ and
$\omega$ are signed variables; their sign depends on the charge of the
track.  The track finding and the fitting procedures make use of
Kalman filter algorithm~\cite{Kalman} that takes into account the detailed
distribution of material in the detector and the full map of the
magnetic field.

The offline charged particle track reconstruction builds on
information available from the L3 trigger and tracking algorithm.  It
begins with an improvement of the event start time $t_0$, obtained
from a fit to the parameters $d_0,\ \phi_0,$ and $t_0$ based on the
four-hit track segments in the DCH superlayers.  Next,
tracks are selected by performing helix fits to the hits found by the
L3 track finding algorithm. A search
for additional hits in the DCH that may belong on these tracks is
performed, while $t_0$ is further improved by using only hits
associated with tracks.  Two more sophisticated tracking
procedures are applied which are designed to find tracks that either
do not pass through the entire DCH or do not
originate from the IP. These algorithms primarily use
track segments that have not already been assigned to other tracks,
and thus benefit from a progressively cleaner tracking environment
with a constantly improving $t_0$.  At the end of this process, tracks
are again fit using a Kalman filter method.

The resulting tracks are then extrapolated into the SVT, and SVT track
segments are added, provided they are consistent with the expected
error in the extrapolation through the intervening material and
inhomogeneous magnetic field.  Among the possible SVT segments, those
with the smallest residuals and the largest number of SVT layers are
retained and a Kalman fit is performed to the full set of DCH and SVT
hits.

\label{sec:\secname:TrkEfficiency}

\begin{figure}
\centering
\includegraphics[width=6.5cm]{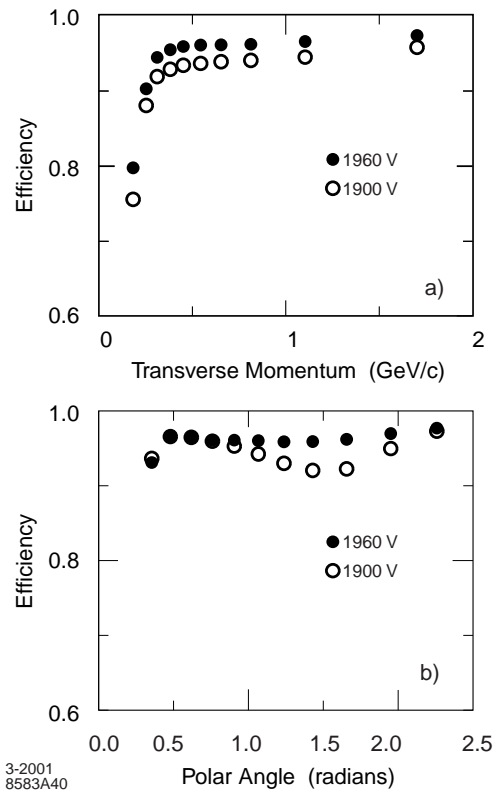}
\vspace{-1.5pc}

\caption{ The track reconstruction efficiency in the DCH at
operating voltages of 1900\volt\ and 1960\volt, as a function of
a) transverse momentum, and b) polar angle. The efficiency is
measured in multi-hadron events as the fraction of all tracks
detected in the SVT for which the DCH portion is also
reconstructed.} \label{\secname fig:efficiency}
\end{figure}

Any remaining SVT hits are then passed to two complementary
standalone track finding algorithms. The first reconstructs tracks
starting with triplets of space points (matched $\phi$ and $z$
hits) in layers 1, 3, and 5 of the SVT, and adding consistent
space points from the other layers. A minimum of four space points
are required to form a good track.  This algorithm is efficient
over a wide range of $d_0$  and $z_0$ values. The second algorithm
starts with circle trajectories from $\phi$ hits and then adds $z$
hits to form helices.  This algorithm is less sensitive to large
combinatorics and to missing $z$ information for some of the SVT
modules.

\begin{figure}
\centering
\includegraphics[width=6.2cm]{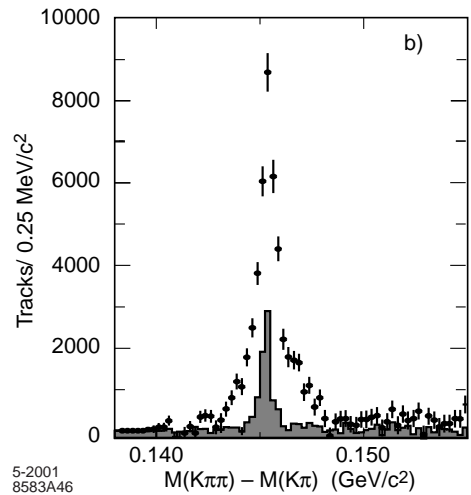}
\vspace{-1.5pc}

\caption{Reconstruction of low momentum tracks: the mass
difference, $\Delta M = M(K^- \pi^+ \pi^+) - M(K^- \pi^+)$, both
for all detected events (data points) and for events in which the
low momentum pion is reconstructed both in the SVT and DCH
(histogram). Backgrounds from combinatorics and fake tracks, as
well as non-resonant data have been subtracted.  } \label{\secname
fig:slowpi}
\end{figure}

Finally, an attempt is made to combine tracks that are only found
by one of the two tracking systems and thus recover tracks
scattered in the material of the support tube.

\subsection{Tracking Efficiency }

The efficiency for reconstructing tracks in the DCH has been
measured as a function of transverse momentum, polar and azimuthal
angles in multi-track events.  These measurements rely on specific
final states and exploit the fact that the track reconstruction
can be performed independently in the SVT and the DCH.

\begin{figure}
\centering
\includegraphics[width=6.2cm]{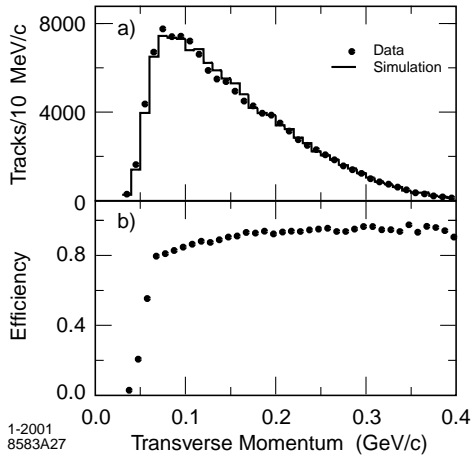}
\vspace{-2pc}

\caption{Monte Carlo studies of low momentum tracks in the SVT: a)
comparison of data (contributions from combinatoric background and
non-\BB\ events have been subtracted) with simulation of the
transverse momentum spectrum of pions from $D^{*+}\ra D^0\pi^+$ in
\BB\ events, and b) efficiency for slow pion detection derived
from simulated events.  } \label{\secname fig:mcpions}
\end{figure}

The absolute DCH tracking efficiency is determined as the ratio of
the number of reconstructed DCH tracks to the number of tracks
detected in the SVT, with the requirement that they fall within
the acceptance of the DCH. Such studies have been performed for
different samples of multi-hadron events. Figure~\ref{\secname
fig:efficiency} shows the result of one such study for the two
voltage settings. The measurement errors are dominated by the
uncertainty in the correction for fake tracks in the SVT. At the
design voltage of 1960\volt, the efficiency averages $98\pm 1$\%
per track above 200\mevc\ and polar angle $\theta >$ 500\mrad. The
data recorded at 1900\volt\ show a reduction in efficiency by
about 5\% for tracks at close to normal incidence, indicating that
the cells are not fully efficient at this voltage.

\begin{figure*}
\begin{center}
\includegraphics[width=14cm]{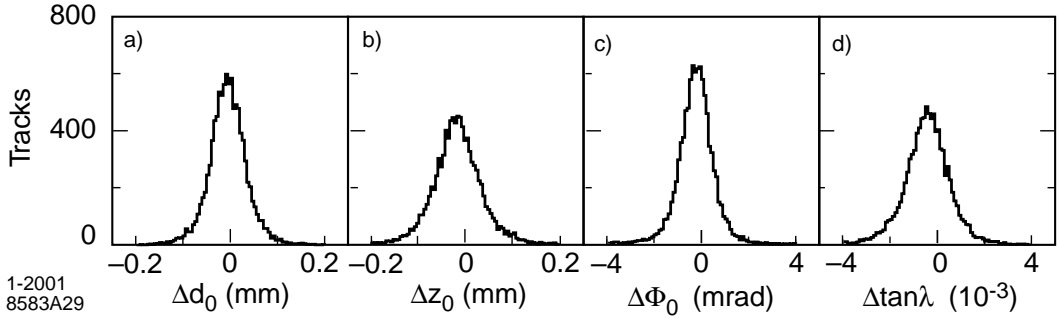}
\vspace{-2pc} \caption{ Measurements of the differences between
the fitted track parameters of the two halves of cosmic ray muons,
with transverse momenta above 3\gevc, a) $\Delta d_0$, b) $\Delta
z_0$, c) $\Delta \phi_0$, and d) $\Delta \tan\lambda$.}
\label{\secname fig:dch_resol}
\end{center}
\end{figure*}

The standalone SVT tracking algorithms have a high efficiency for
tracks with low transverse momentum. This feature is important for
the detection of $D^*$ decays. To study this efficiency, decays
$D^{*+}\ra D^0\pi^+$ are selected by reconstructing events of the
type $\bar B\ra D^{*+}X$ followed by $D^{*+}\ra D^0\pi^+ \ra K^-
\pi^+\pi^+$. For the majority of these low momentum pions the
momentum resolution is limited by multiple scattering, but the
production angle can be determined from the signals in innermost
layers of the SVT.  Figure~\ref{\secname fig:slowpi} shows the
mass difference $\Delta M = M(K^- \pi^+ \pi^+) - M(K^-\pi^+)$, for
the total sample and the subsample of events in which the slow
pion has been reconstructed in both the SVT and the DCH.  The
difference in these two distributions demonstrates the
contribution from SVT standalone tracking, both in terms of the
gain of signal events and of resolution. The gain in efficiency is
mostly at very low momenta, and the resolution is impacted by
multiple scattering and limited track length of the slow pions. To
derive an estimate of the tracking efficiency for these low
momentum tracks, a detailed Monte Carlo simulation was performed.
Specifically, the pion spectrum was derived from simulation of the
inclusive $D^*$ production in \BB\ events, and the Monte Carlo
events were selected in the same way as the data. A comparison of
the detected slow pion spectrum with the Monte Carlo prediction is
presented in Figure~\ref{\secname fig:mcpions}. Based on this very
good agreement, the detection efficiency has been derived from the
Monte Carlo simulation. The SVT significantly extends the
capability of the charged particle detection down to transverse
momenta of $\sim$50\mevc.

\subsection{Track Parameter Resolutions}

The resolution in the five track parameters is monitored in OPR using
$\ep \en$ and $\mu^+ \mu^-$ pair events. It is further investigated
offline for tracks in multi-hadron events and cosmic ray muons.

Cosmic rays that are recorded during normal data-taking offer a
simple way of studying the track parameter resolution.  The upper
and lower halves of the cosmic ray tracks traversing the DCH and
the SVT are fit as two separate tracks, and the resolution is
derived from the difference of the measured parameters for the two
track halves.  To assure that the tracks pass close to the beam
interaction point, cuts are applied on the $d_0$, $z_0$, and
$\tan\lambda$. The results of this comparison for the coordinates
of the point of closest approach and the angles are shown in
Figure~\ref{\secname fig:dch_resol} for tracks with momenta above
$p_t$ of 3\gevc. The distributions are symmetric; the non-Gaussian
tails are small.  The distributions for the differences in $z_0$
and $\tan\lambda$ show a clear offset, attributed to residual
problems with the internal alignment of the SVT. Based on the full
width at half maximum of these distributions the resolutions for
single tracks can be parametized as
\begin{displaymath}
\begin{array}{ll}
\sigma_{d_0}=23 \mu \mathrm{m} &\sigma_{\phi_0}=0.43 \mrad\\
 \sigma_{z_0}=29 \mu \mathrm{m} &\sigma_{\tan\lambda}=0.53\cdot10^{-3}.
 \end{array}
\end{displaymath}

\begin{figure}[!htb]
\centering
\includegraphics[width=6.5cm]{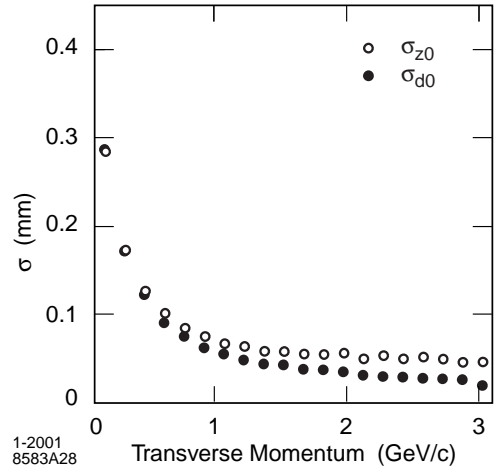}
\vspace{-2pc}

\caption{
Resolution in the parameters $d_0$ and $z_0$ for tracks in multi-hadron
events as a function of the transverse momentum.  The data are
corrected for the effects of particle decays and vertexing errors.}
\label{\secname fig:doca}
\end{figure}

The dependence of the resolution in $d_0$ and
$z_0$ on the transverse momentum $p_t$ is presented in
Figure~\ref{\secname fig:doca}.  The measurement is based on tracks in
multi-hadron events.  The resolution is determined from the width of the
distribution of the difference between the measured parameters,
$d_0$ and $z_0$, and the coordinates of the vertex reconstructed from the remaining
tracks in the event. These distributions peak at zero, but have a tail
for positive values due to the effect of particle decays.  Consequently,
only the negative part of the distributions reflects the measurement
error and is used to extract the resolution.  Event shape cuts and a cut on
the $\chi^2$ of the vertex fit are applied to reduce the
effect of weak decays on this measurement.  The contribution from the
vertex errors are removed from the measured resolutions in
quadrature.  The $d_0$ and $z_0$ resolutions so measured are about
25\mum\ and 40\mum\ respectively at $p_t = 3$\gevc. These
values agree well with expectations, and are also in reasonable agreement
with the results obtained from cosmic rays.

\begin{figure}
\centering
\includegraphics[width=6.5cm]{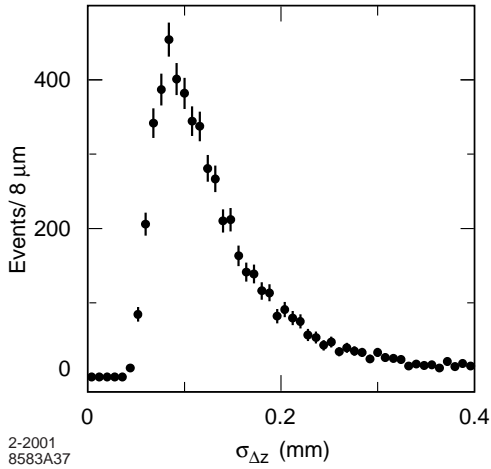}
\vspace{-2pc}

\caption{ Distribution of the error on the difference $\Delta z$
between the $B$ meson vertices for a sample of events in which one
$\Bz$ is fully reconstructed.} \label{\secname fig:vertex_error}
\end{figure}

Figure~\ref{\secname fig:vertex_error} shows the estimated error
in the measurement of the difference along the $z$-axis between
the vertices of the two neutral $B$ mesons, one of them is fully
reconstructed, the other serves as a flavor tag. The rms width of
190\mum\ is dominated by the reconstruction of the tagging $B$
vertex, the rms resolution for the fully reconstructed $B$ meson
is 70\mum. The data meet the  design expectation \cite{R_LOI}.

\begin{figure}
\begin{center}
\includegraphics[width=6.5cm]{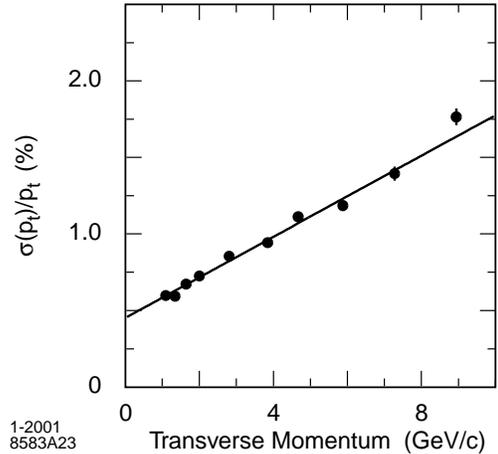}
\vspace{-2pc}

 \caption{ Resolution in the transverse momentum \pt\
determined from cosmic ray muons traversing the DCH and SVT.}
\label{\secname fig:ptresol}
\end{center}
\end{figure}

While the position and angle measurements near the IP are
dominated by the SVT measurements, the DCH contributes primarily to
the $p_t$ measurement. Figure~\ref{\secname fig:ptresol} shows the
resolution in the transverse momentum derived from cosmic muons.  The
data are well represented by a linear
function,
\begin{displaymath}
\sigma_{p_t}/p_t= (0.13\pm 0.01)\% \cdot p_t + (0.45 \pm 0.03) \%,
\end{displaymath}
where the transverse momentum $p_t$ is measured in \gevc. These
values for the resolution parameters are very close to the initial
estimates and can be reproduced by Monte Carlo simulations. More
sophisticated treatment of the DCH time-to-distance relations and
overall resolution function are presently under study.

Figure~\ref{\secname fig:psimass} shows the mass resolution for
\jpsi\ mesons reconstructed in the \mumu\ final state, averaged
over all data currently available. The reconstructed peak is
centered 0.05\% below the expected value, this difference is
attributed to the remaining inaccuracies in the SVT and DCH
alignment and in the magnetic field parameterization. The observed
mass resolution differs by 15\% for data recorded at the two DCH
HV settings, it is $13.0 \pm 0.3$~\mevcc\ and $11.4 \pm
0.3$\mevcc\ at 1900\volt\ and 1960\volt, respectively.

\begin{figure}
\begin{center}
\includegraphics[width=6.5cm]{8583A35}
\vspace{-2pc}

 \caption{ Reconstruction of the decay $\jpsi \ra
\mu^+ \mu^-$ in selected \BB\ events. } \label{\secname
fig:psimass}
\end{center}
\end{figure}

\subsection{Summary}
The two tracking devices, the SVT and DCH, have been performing close
to design expectations from the start of operations.  Studies of track
resolution at lower momenta and as a function of polar and azimuthal
angles are still under way.  Likewise, the position and angular
resolution at the entrance to the DIRC or EMC are still being studied.
Such measurements are very sensitive to internal alignment of the SVT
and relative placement of the SVT and the DCH.  A better understanding
will not only reduce the mass resolution for the reconstruction of
exclusive states, it will also be important for
improvement of the performance of the DIRC.

\renewcommand{\secname}{drc_}
\renewcommand{\sectiondir}{sec06_drc}
\section{DIRC}
\label{sec:drc}
%
%
%
%
\newcommand{\dirc}{DIRC}

\subsection{Purpose and Design Requirements}

The study of \emph{CP}-violation  requires the ability to tag the
flavor of one of the $B$ mesons via the cascade decay
$b\rightarrow c \rightarrow s$, while fully reconstructing the
second B decay. The momenta of the kaons used for flavor tagging
extend up to about 2\ GeV$/c$, with most of them below 1\ GeV$/c$.
On the other hand, pions and kaons from the rare two-body decays
$B^0\rightarrow \pi^+\pi^-$ and $ B^0 \to K^+ \pi^-$ must be
well-separated. They have momenta between 1.7 and 4.2\ GeV/$c$
with a strong momentum-polar angle correlation of the tracks
(higher momenta occur at more forward angles because of the
c.m. system boost)~\cite{R_PBook}.

The \emph{Particle Identification} (PID) system should be thin and
uniform in terms of radiation lengths (to minimize degradation of
the calorimeter energy resolution) and small in the radial
dimension to reduce the volume, and hence, the cost of the
calorimeter. Finally, for operation at high luminosity, the PID
system needs fast signal response, and should be able to tolerate
high backgrounds.

The PID system being used in \babar\ is a new kind of ring-imaging
Cherenkov detector called the \dirc\ \cite{four} (the acronym
\dirc\ stands for \emph{Detector of Internally Reflected
Cherenkov} light). It is expected to be able to provide $\pi/K$
separation of $\sim 4~\sigma$ or greater, for all tracks from
$B$-meson decays from the pion Cherenkov threshold up to 4.2\
GeV/c. PID below 700\ MeV/c relies primarily on the $dE/dx$
measurements in the DCH and SVT.

\subsection{DIRC Concept}

The \dirc\ is based on the principle that the magnitudes of angles
are maintained upon reflection from a flat surface.
Figure~\ref{\secname fig:princip} shows a  schematic of the \dirc\
geometry that illustrates the principles of light production,
transport, and imaging. The radiator material of the \dirc\ is
synthetic, fused silica in the form of long, thin bars with
rectangular cross section. These bars serve both as radiators and
as light pipes for the portion of the light trapped in the
radiator by total internal reflection. Fused, synthetic silica
(Spectrosil~\cite{six}) is chosen because of its resistance to
ionizing radiation, its long attenuation length, large index of
refraction, low chromatic dispersion within the wavelength
acceptance of the \dirc, and because it allows an excellent
optical finish on the surfaces of the bars~\cite{quartzpaper}.

In the following, the variable $\theta_c$ is used to designate the
Cherenkov angle, $\phi_c$ denotes the azimuthal angle of a
Cherenkov photon around the track direction, and $n$ represents
the mean index of refraction of fused silica ($n = 1.473$), with
the familiar relation $\cos \theta_c = 1/n\beta \ (\beta = v/c, \
v = $ velocity of the particle, $c = $ velocity of light).

For particles with $\beta \approx 1$, some photons will always lie
within the total internal reflection limit, and will be
transported to either one or both ends of the bar, depending on
the particle incident angle. To avoid instrumenting both ends of
the bar with photon detectors, a mirror is placed at the forward
end, perpendicular to the bar axis, to reflect incident photons to
the backward, instrumented  end.

\begin{figure}
\includegraphics[width=7.5cm]{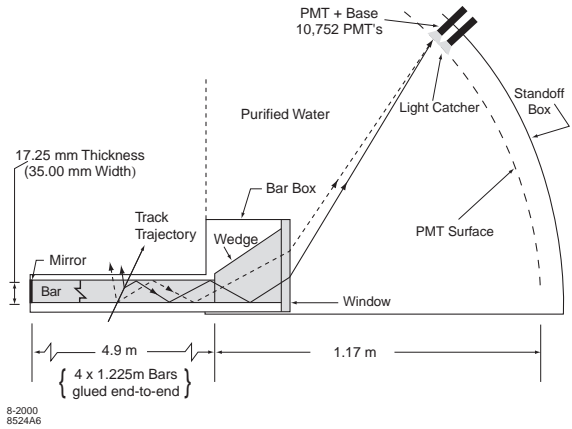}
\vspace{-2pc}

 \caption{Schematics of the \dirc\ fused silica
radiator bar and imaging region. Not shown is a 6\mrad\ angle on
the bottom surface of the wedge (see text). } \label{\secname
fig:princip}
\end{figure}

Once photons arrive at the instrumented end, most of them emerge
into a water-filled expansion region, called the \emph{standoff
box}. A fused silica \emph{wedge} at the exit of the bar reflects
photons at large angles relative to the bar axis. It thereby
reduces the size of the required detection surface and recovers
those photons that would otherwise be lost due to internal
reflection at the fused silica/water interface. The photons are
detected by an array of densely packed photomultiplier tubes
(PMTs), each surrounded by reflecting \emph{light catcher}
cones~\cite{lightcatcher} to capture light which would otherwise
miss the active area of the PMT. The PMTs are placed at a distance
of about 1.2\ m from the bar end. The expected Cherenkov light
pattern at this surface is essentially a conic section, where the
cone opening-angle is the Cherenkov production angle modified by
refraction at the exit from the fused silica window.

The \dirc\ is intrinsically a three-dimensional imaging device,
using the position and arrival time of the PMT signals.  Photons
generated in a bar are focused onto the phototube detection
surface via a ``pinhole'' defined by the exit aperture of the bar.
In order to associate the photon signals with a track traversing a
bar, the vector pointing from the center of the bar end to the
center of each PMT is taken as a measure of the photon propagation
angles $\alpha_x, \alpha_y$, and $\alpha_z$.  Since the track
position and angles are known from the tracking system, the three
$\alpha$ angles can be used to determine the two Cherenkov angles
$\theta_c$ and $\phi_c$.  In addition, the arrival time of the
signal provides an independent measurement of the propagation of
the photon, and can be related to the propagation angles $\alpha$.
This over-constraint on the angles and the signal timing are
particularly useful in dealing with ambiguities in the signal
association (see Section \ref{sec:drcreco}) and high background
rates.

\subsection{Mechanical Design\\ and Physical Description}

The \dirc\ bars are arranged in a 12-sided polygonal barrel.
Because of the beam energy asymmetry, particles are produced
preferentially forward in the detector. To minimize interference
with other detector systems in the forward region, the \dirc\
photon detector is placed at the backward end.

\begin{figure}
\centering
\includegraphics[width=6cm]{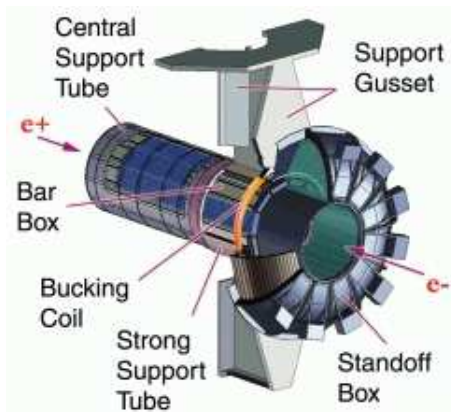}

\vspace{-2pc}
 \caption{Exploded view of the \dirc\ mechanical
support structure. The steel magnetic shield is not shown.}
\label{\secname fig:mechelmsa}
\end{figure}

\begin{figure}
\centering
\includegraphics[width=7cm]{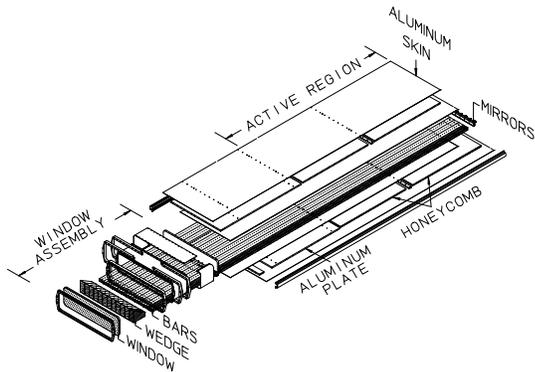}
\vspace{-2pc}
 \caption{Schematics of the \dirc\ bar box assembly.}
\label{\secname fig:mechelmsb}
\end{figure}

The principal components of the \dirc\ are shown schematically in
Figure~\ref{\secname fig:mechelmsa}. The bars are placed into 12
hermetically sealed containers, called \emph{bar boxes}, made of
very thin aluminum-hexcel panels.  Each bar box, shown in
Figure~\ref{\secname fig:mechelmsb}, contains 12 bars, for a total
of 144 bars. Within a bar box the 12 bars are optically isolated
by a $\sim 150\mu m$ air gap between neighboring bars, enforced by
custom shims made from aluminum foil.

The bars are 17 mm-thick, 35 mm-wide, and 4.9 m-long. Each bar is
assembled from four 1.225 m pieces that are glued end-to-end; this
length is the longest high-quality bar currently
obtainable~\cite{quartzpaper,eight}.

The bars are supported at 600 mm intervals by small nylon buttons
for optical isolation from the bar box. Each bar has a fused
silica wedge glued to it at the readout end. The wedge, made of
the same material as the bar, is 91 mm-long with very nearly the
same width as the bars (33 mm) and a trapezoidal profile (27
mm-high at bar end, and 79 mm at the light exit end). The bottom
of the wedge (see Figure~\ref{\secname fig:princip})  has a slight
($\sim 6$\mrad) upward slope to minimize the displacement of the
downward reflected image due to the finite bar thickness. The
twelve wedges in a bar box are glued to a common 10 mm-thick fused
silica window, that provides the interface and seal to the
purified water in the standoff box.

The mechanical support of the \dirc, shown in Figure~\ref{\secname
fig:mechelmsa}, is cantilevered from the steel of the IFR. The
\emph{Strong Support Tube} (SST) is a steel cylinder located
inside the end doors of the IFR and provides the basic support for
the entire \dirc. In turn, the SST is supported by a steel
\emph{support gusset} that fixes it to the barrel magnet steel. It
also minimizes the magnetic flux gap caused by the \dirc\ bars
extending through the flux return, and supports the axial load of
the inner magnetic plug surrounding the beam in this region.

The bar boxes are supported in the active region
by an aluminum tube, the \emph{Central Support Tube} (CST), attached to the
SST via an aluminum transition flange. The CST is a thin,
double-walled, cylindrical shell, using aircraft-type construction
with stressed aluminum skins and bulkheads having riveted or glued
joints.  The CST also provides the support for the DCH.

\begin{figure*}
\centering
\includegraphics[width=12 cm]{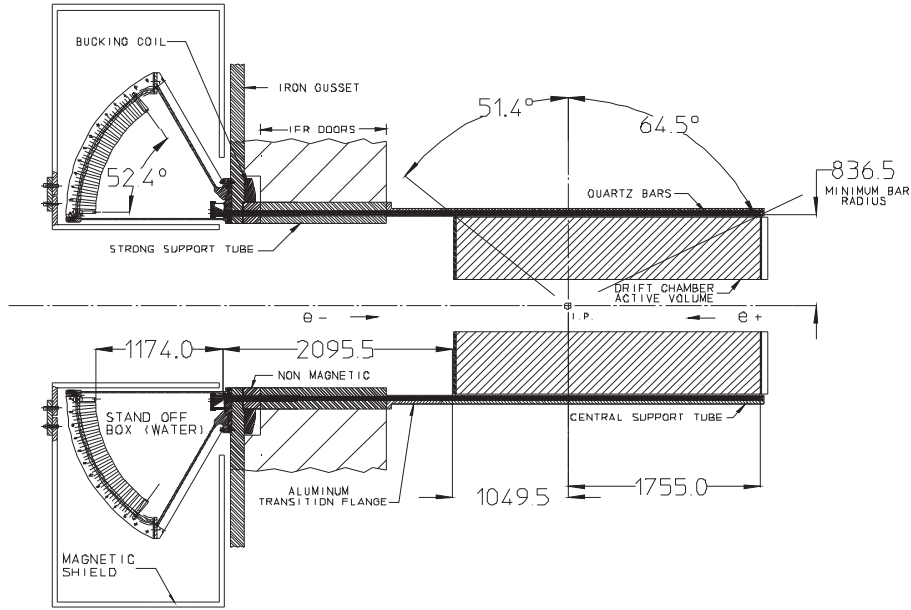}
\vspace{-2pc} \caption{Elevation view of the nominal \dirc\ system
geometry. For clarity, the end plug is not shown. All dimensions
are given in mm.} \label{\secname fig:geoview1}
\end{figure*}

The standoff box is made of stainless steel, consisting of a cone,
cylinder, and 12 sectors of PMTs. It contains about 6,000 liters
of purified water. Water is used to fill this region because it is
inexpensive and has an average index of refraction ($n \approx
1.346$) reasonably close to that of fused silica, thus minimizing
the total internal reflection at silica-water interface.
Furthermore, its chromaticity index is a close match to that of
fused silica, effectively eliminating dispersion at the
silica-water interface. The steel  gusset supports the standoff
box. A steel shield, supplemented by a \emph{bucking coil},
surrounds the standoff box to reduce the field in the PMT region
to below 1 Gauss~\cite{bfield}.

\begin{figure}
\includegraphics[width=7.5 cm]{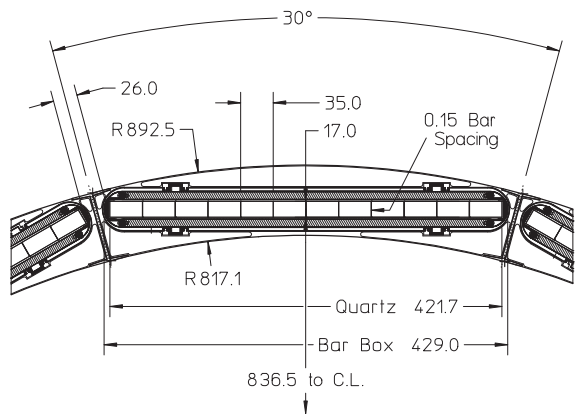}
\vspace{-2.5pc}

 \caption{Transverse section of the nominal \dirc\
bar box imbedded in the CST. All dimensions are given in mm.}
\label{\secname fig:geoview2}
\end{figure}

The PMTs at the rear of the standoff box lie on a surface that is
approximately toroidal. Each of the 12 PMT sectors contains 896
PMTs (ETL model 9125 \cite{nine,drcpmt}) with 29\ mm-diameter, in
a closely packed array inside the water volume. A double o-ring
water seal is made between the PMTs and the wall of the standoff
box. The PMTs are installed from the inside of the standoff box
and connected via a feed-through to a base mounted outside. The
hexagonal light catcher cone is mounted in front of the
photocathode of each PMT, which results in an effective active
surface area light collection fraction of about 90\%. The geometry
of the \dirc\ is shown in Figures~\ref{\secname fig:geoview1} and
\ref{\secname fig:geoview2}.

The \dirc\ occupies 80\ mm of radial space in the central detector
volume including supports and construction tolerances, with a
total of about 17\%\ radiation length thickness at normal
incidence. The radiator bars subtend a solid angle corresponding
to about 94\%\ of the azimuth and 83\%\ of the c.m. polar angle
cosine.

The distance from the end of the bar to the PMTs is $\sim 1.17$\
m, which together with the size of the bars and PMTs, gives a
geometric contribution to the single photon Cherenkov angle
resolution of $\sim 7$\mrad. This value is slightly larger than
the rms spread of the photon production (dominated by a $\sim
5.4$\mrad\ chromatic term) and transmission dispersions. The
overall single photon resolution is estimated to be about 10\mrad.

\subsubsection{Cherenkov Photon  Detection\\ Efficiency}

\begin{figure}
\includegraphics[width=7.5 cm]{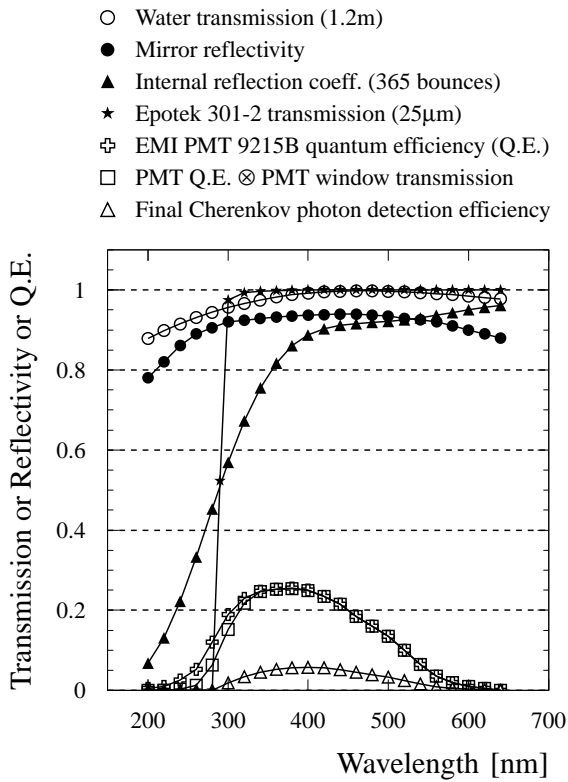}
\vspace{-2pc} \caption{Transmission, reflectivity and quantum
efficiency for various components of the \dirc\ as a function of
wavelength for a $\beta=1$ particle at normal incidence to the
center of a bar~\cite{jjv99}.} \label{\secname fig:lightbudget}
\end{figure}

Figure \ref{\secname fig:lightbudget} shows the contribution of
various optical and electronic components of the \dirc\ to the
Cherenkov photon detection efficiency  as a function of
wavelength. The data points pertain to a particle entering the
center of the bar at 90\degrees. A typical design goal for the
photon transport in the bar was that no single component should
contribute more than 10--20\% loss of detection efficiency.
Satisfying this criterion required an extremely high internal
reflection coefficient of the bar surfaces (greater than 0.9992
per bounce), so that about 80\% of the light is maintained after
multiple bounces along the bars (365 bounces in the example of
Figure~\ref{\secname fig:lightbudget}). The ultraviolet cut-off is
at $\sim 300$\ nm, determined by the epoxy
(\mbox{Epotek~301-2~\cite{epotek}}) used to glue the fused silica
bars together. The dominant contributor to the overall detection
efficiency is the quantum efficiency of the PMT. Taking into
account additional wavelength independent factors, such as the PMT
packing fraction and the geometrical efficiency for trapping
Cherenkov photons in the fused silica bars via total internal
reflection, the expected number of  photoelectrons $(N_{pe})$ is
$\sim 28$ for a $\beta=1$ particle entering normal to the surface
at the center of a bar, and increases by over a factor of two in
the forward and backward directions.

\subsubsection{DIRC Water System}
The \dirc\ water system is designed to maintain good transparency
at wavelengths as small as 300\ nm. One way to achieve this is to
use ultra-pure, de-ionized water, close to the theoretical limit
of 18\ M$\Omega$cm resistivity. In addition, the water must be
de-gassed and the entire system kept free of bacteria. Purified
water is aggressive in attacking many materials, and those in
contact with the water were selected based on known compatability
with purified water. To maintain the necessary level of water
quality, most plumbing components are made of stainless steel or
polyvinylidene fluoride.

The system contains an input line with six mechanical filters
(three 10\ $\mu$m, two 5\ $\mu$m, and one 1\ $\mu$m), a reverse
osmosis unit, de-ionization beds, a Teflon microtube de-gasser and
various pumps and valves. To prevent bacteria growth, it is
equipped with a UV lamp (254\ nm wavelength) and filters (two 1\
$\mu$m, two 0.2\ $\mu$m, and charcoal filters). Sampling ports are
provided to check the water quality and to monitor resistivity,
pH-value, temperature, and flow. A gravity feed return system
prevents overpressure. The entire standoff box water volume can be
recirculated up to four times a day.

The operating experience with the water system so far has been
very good. The water volume is exchanged every ten hours and the
resistivity of the water is typically 18\ M$\Omega$cm in the
supply line and 8--10\ M$\Omega$cm in the return line at a
temperature of about 23--26\degc. The pH-value is about 6.5 and
6.6-6.7 in the supply and return water, respectively. The water
transparency is routinely measured using lasers of three different
wavelengths. The transmission is better than 92\% per meter at
266~nm and exceeds 98\% per meter at 325\ nm and 442\ nm.

Potential leaks from the  water seals between the bar boxes and
the standoff box are detected by a water leak detection system of
20 custom water sensors in and about the bar box slots. Two
commercial ultrasonic flow sensors are used to monitor water flow
in two (normally dry) drain lines in addition to the 12 humidity
sensors on a nitrogen gas output line from each bar box (see
below). Should water be detected, a valve in a 100\ mm diameter
drain line is opened, and the entire system is drained in about 12
minutes.

All elements inside the standoff box (PMT, plastic PMT housing,
gaskets, light catchers) were tested at normal and elevated
temperatures to withstand the highly corrosive action of
ultra-pure water and to prevent its pollution. For instance,
rhodium-plated mirrors on ULTEM support had to be used for the
light catchers~\cite{lightcatcher}.

\subsubsection{DIRC Gas System}
Nitrogen gas from liquid nitrogen boil-off is used to prevent
moisture from condensing on the bars, and used also to detect
water leaks. The gas flows through each bar box at the rate of
100--200 cm$^{3}$/min, and is monitored for humidity to ensure
that the water seal around the bar box  remains tight. The gas is
filtered through a molecular sieve and three mechanical filters to
remove particulates (7 $\mu$m, 0.5 $\mu$m, and 0.01 $\mu$m). Dew
points of the gas returned from the bar boxes are about -40\degc.
Approximately one third of the input nitrogen gas leaks from the
bar boxes and keeps the bar box slots in the mechanical support
structure free of condensation.

\subsection{Electronics}
\subsubsection{DIRC PMT Electronics}
The \dirc\ PMT base contains a single printed circuit board,
equipped with surface mounted components. The operating high
voltage (HV) of the PMTs is $\sim~1.14$ kV, with a range between
0.9 and 1.3 kV. Groups of 16 tubes are selected for uniformity of
gain to allow their operation at a common HV provided from a
single distribution board.

The HV is provided by a CAEN SY-527 high voltage distribution
system. Each of the 12 sectors receives HV through 56 high voltage
channels, distributed through a single cable bundle. Each voltage
can be set between 0 and 1.6 kV.

\subsubsection{DIRC Front-End Electronics}

The DIRC front-end electronics (FEE) is designed to measure the
arrival time of each Cherenkov photon detected by the PMT
array~\cite{drcFee} to an accuracy that is limited by the
intrinsic 1.5 ns transit time spread of the PMTs. The design
contains a pipeline to deal with the L1 trigger latency of
$12\mu\,$s, and can handle random background rates of up to 200
kHz/PMT with zero dead time. In addition, the pulse height spectra
can be measured to ensure that the PMTs operate on the HV plateau.
However, because the ADC information is not needed to reconstruct
events, 64 PMTs are multiplexed onto a single ADC for monitoring
and calibration.

The \dirc\ FEE is mounted on the outside of the standoff box and
is highly integrated in order to minimize cable lengths and to
retain the required single photoelectron sensitivity. Each of the
168 \dirc\ \emph{Front-end Boards} (DFBs) processes 64 PMT inputs,
containing eight custom analog chips along with their associated
level translators, four custom-made TDC ICs, one 8-bit flash ADC
(FADC), two digitally controlled calibration signal generators,
multi-event buffers, and test hardware.

The PMT signals are amplified, and pulse-shaped by an
eight-channel analog IC~\cite{analogchip}. A digital pulse timed
with the peak of the input pulse is output by a zero-crossing
discriminator, as well as a pulse shaped by a CR-RC filter with
80\ ns peaking time, which was chosen to allow for the ADC
multiplexing. The multiplexer selects the channel to be digitized
by the FADC for calibration.

The TDC IC~\cite{tdc} is a 16-channel TDC with 0.5~ns binning,
input buffering, and selective readout of the data in time with
the trigger. To cope with the L1 maximum trigger latency of
12~$\mu{\rm s}$ and jitter of 1$\mus$, the selective readout
process extracts data in time with the trigger within a
programmable time window. The acceptance window width is
programmable between 64~ns and 2~$\mus$ and is
typically set at 600~ns. The twelve DIRC Crate Controllers (DCCs)
that form the interface to the VME front-end crates are connected
to six ROMs via 1.2~Gbit/s optical fibers.

\subsubsection{DAQ Feature Extraction}
Raw data from the DFBs are processed in the ROMs by a feature
extraction algorithm before being transmitted to the segment and
event builder. This software algorithm reduces the data volume by
roughly 50\% under typical background conditions. DFB data that
contain errors are flagged and discarded. The only data errors
seen to date have been traced to damaged DFBs that were replaced
immediately. Because the dataflow system can reliably transmit at
most 32 kBytes/crate, the feature extraction must sometimes
truncate data to limit the event size. Event data are replaced
with a per-DFB occupancy summary when a ROM's hit occupancy
exceeds 56\%, which occurs about once in $10^{4}$ events. An
appropriate flag is inserted into the feature extraction output
whenever truncation or deletion occurs. Errors, truncation, and
feature extraction performance are continuously monitored online,
and exceptions are either immediately corrected or logged for
future action.

\subsubsection{DIRC Calibration}
The \dirc\ uses two independent approaches for a calibration of
the unknown PMT time response and the delays introduced by the FEE
and the fast control system. The first is a conventional pulser
calibration. The second uses reconstructed tracks from collision
data.

The pulser calibration is performed online using a light pulser
system which generates precisely timed 1 ns duration light pulses
from twelve, blue LEDs, one per sector. The LEDs are triggered by
the global fast control calibration strobe command sent to the
DCCs. The DCC triggers an individual LED for each sector upon
receipt of calibration strobe. Pulses in adjacent sectors are
staggered by 50 ns to prevent light crosstalk between sectors. The
pulser is run at roughly 2 kHz for the time delay calibration. The
LED light is transmitted through approximately 47 m-long optical
fibers to diffusers mounted on the inner surface of the standoff
box wall opposite the PMTs. This light produces about 10\%
photoelectron occupancy nearly uniformly throughout the standoff
box.

Histograms of TDC times for each PMT are accumulated in parallel
in the ROMs, and then fit to an asymmetric peak function. About
65,000 light pulses are used to determine the mean time delay of
each of the PMTs in the standoff box to a statistical accuracy of
better than 0.1 ns. The LED pulser is also used to monitor the
phototube gains using the ADC readout. As with the TDC
calibration, histograms and fits of the ADC spectrum are
accumulated and fit in the ROM. A calibration run including both
TDC and ADC information for all PMTs requires a few minutes, and
is run once per day. Daily calibrations not only verify the time
delays, but allow the detection of hardware failures.

The data stream calibration uses reconstructed tracks from the
collision data. For calibration of the global time delay, the
observed, uncalibrated times minus the expected arrival times,
$\Delta t_\gamma$, are collected during the online prompt
reconstruction processing. To calculate individual channel
calibrations, $\Delta t_\gamma$ values for each \dirc\ channel are
accumulated until statistics equivalent to 100,000 tracks are
collected. The distribution for each channel is fit to extract the
global time offset calibration.

The data stream and online pulser calibrations of the electronic
delays, and of the PMT time response and gain yield fully
consistent results, although the data stream results in 15\%
better timing resolution than the pulser calibration. The time
delay values per channel are typically stable to an $rms$ of less
than 0.1\ ns over more than one year of daily calibrations.


\subsubsection{DIRC Environmental  Monitoring\\ System}

The \dirc\ environmental monitoring system is divided into three
parts, corresponding to three separate tasks. The first deals with
the control and monitoring of the HV system for the PMTs. The
second is devoted to monitoring low voltages related to the FEE.
The third controls a variety of other detector parameter settings.
An interlock system, based on a standard VME module (SIAM), is
provided. For the purposes of the \dirc, three dedicated VME CPUs
run the application code. The communication between the HV
mainframes and the monitoring crate is achieved by a CAENET
controller (V288). The HV monitor task controls the step sizes for
ramping the HV up or down, as well as the communication of alarm
conditions, and the values and limits for the HV and current of
each channel.

The purpose of FEE monitoring is to control and monitor parameters
related to the FEE. For each \dirc\ sector, a custom multi-purpose
board, the DCC, equipped with a micro-controller
\cite{drcDCCmicrocontroler} incorporating the appropriate
communication protocol (CANbus), is situated in the same crate as
the DFB. All monitoring and control tasks are implemented on this
card. The parameters monitored are the low voltages for the DFBs
and DCCs, the status of the optical link (Finisar), the
temperature on supply boards, and the VME crate status.

The third part of the monitoring system is based on a custom ADC
VME board (VSAM), used to monitor various type of sensors:
magnetic field sensors, an ensemble of 12 beam monitoring scalers,
16 CsI radiation monitors, the water level in the standoff box as
well as its pH-value, resistivity, and temperature.

\subsection{Operational Issues}

\begin{figure*} 
\centerline{
\includegraphics[width=7.5 cm]{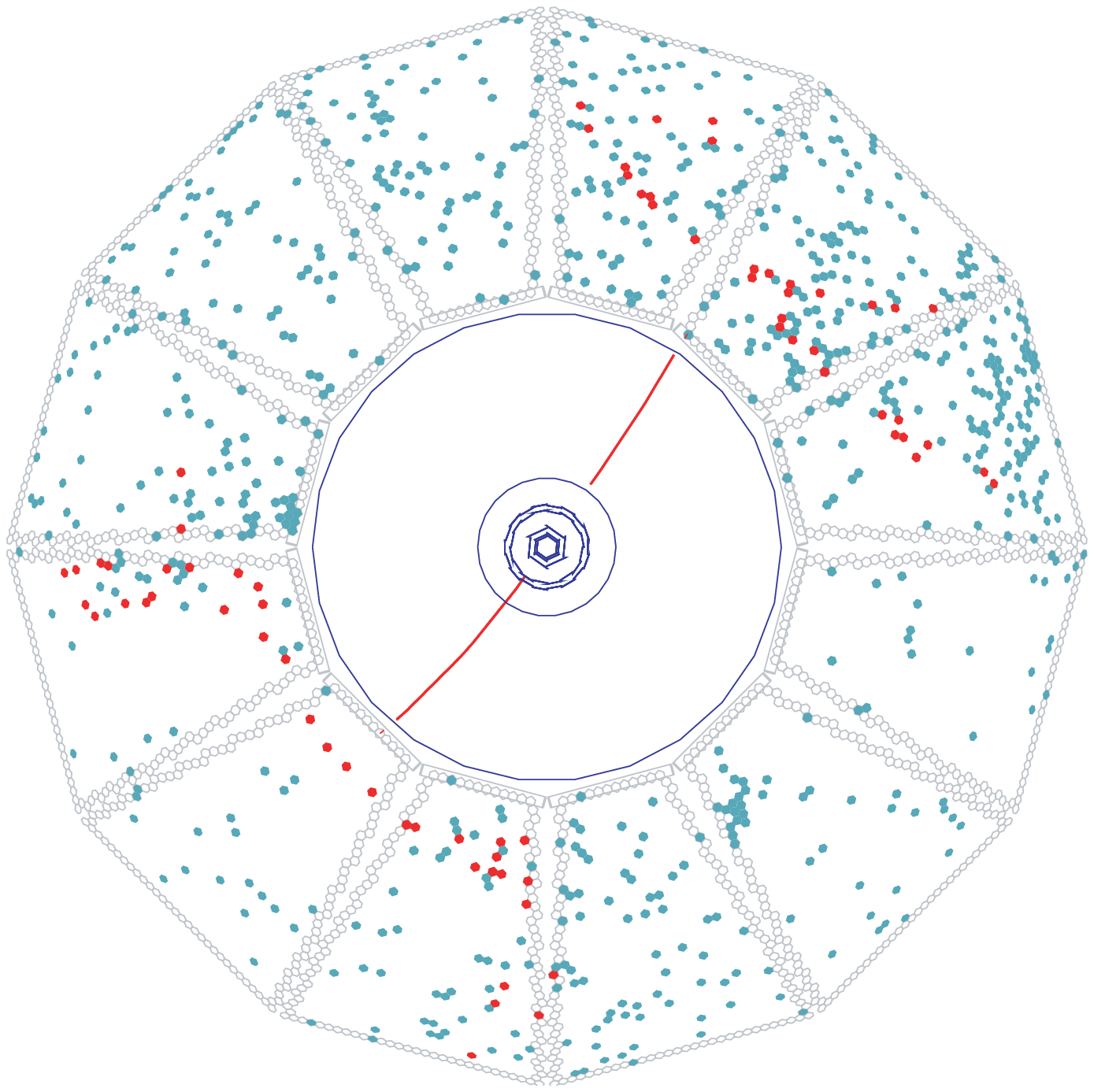}\hspace{1pc}
\includegraphics[width=7.5 cm]{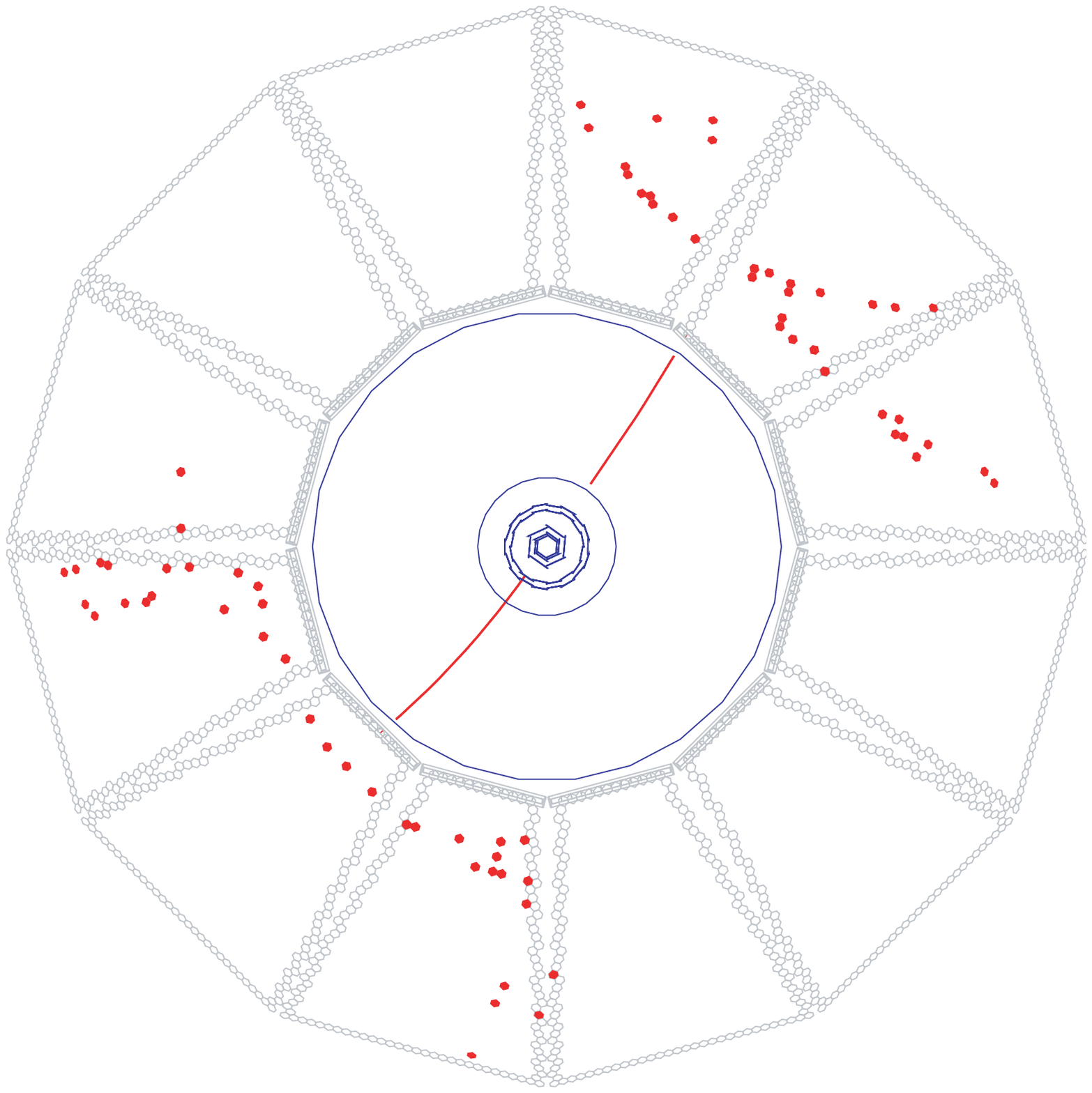}}
\vspace{-2pc}
\caption{Display of an $e^+e^-\rightarrow\mu^+\mu^-$ event
reconstructed in \babar\ with two different time cuts. On the
left, all \dirc\ PMTs with signals within the $\pm$300 ns trigger
window are shown. On the right, only those PMTs with signals
within 8 ns of the expected Cherenkov photon arrival time are
displayed.} \label{\secname fig:event}
\end{figure*}

The \dirc\ was successfully commissioned and attained performance
close to that expected from Monte Carlo simulation. The \dirc\ has
been robust and stable, and, indeed, serves also as a background
detector for \pep2\ tuning. Figure~\ref{\secname fig:event} shows
a typical di-muon event ($e^{+}e^{-} \rightarrow \mu^{+}\mu^{-}$).
In addition to the signals caused by the Cherenkov light from the
two tracks, about 500 background signals can be seen in the
readout window of $\pm$300~ns. This background is dominated by low
energy photons from the \pep2\ machine hitting the standoff box.
Some care in machine tuning is required to stay under a noise
limit of about 200 kHz/tube imposed by limited DAQ throughput.
Lead shielding has been installed around the beam line components
just outside the backward endcap, and has substantially reduced
this background.

After about two years of running, approximately 99.7\% of PMTs and
electronic channels are operating with nominal performance.

Some deterioration of the PMT front glass windows (made of B53
Borosilicate glass) that are immersed in the ultra-pure water of
the standoff box has been observed. For most of the tubes, the
observable effect is typically a slight cloudiness, but for about
50 tubes, it is much more pronounced. Extensive R\&D has
demonstrated that the effect is associated with a loss of sodium
and boron from the surface of the glass~\cite{IEEE2000}. For most
tubes, the leaching rate is a few microns per year, and is
expected to be acceptable for the full projected ten year lifetime
of the experiment. However, for the $\sim 50$ tubes, the incorrect
glass was used by the manufacturer. That glass does not contain
zinc, making it much more susceptible to rapid leaching. This
leaching may eventually lead to either a loss of performance, or
some risk of mechanical failure of the face plates for these
tubes. Direct measurements of the number of Cherenkov photons
observed in di-muon events as a function of time suggest that the
total loss of photons from all sources is less than 2\%/year.

\subsection{Data Analysis and Performance}

Figure~\ref{\secname fig:event} shows the pattern of Cherenkov
photons in a di-muon event, before and after reconstruction. The
time distribution of real Cherenkov photons from a single event is
of order $\sim 50$ ns wide, and during normal data-taking they are
accompanied by hundreds of random photons in a flat background
within the trigger acceptance window. Given a track pointing at a
particular fused silica bar and a candidate signal in a PMT within
the optical phase space of that bar, the Cherenkov angle is
determined up to a 16-fold ambiguity: top or bottom, left or
right, forward or backward, and wedge or no-wedge reflections. The
goal of the reconstruction program is to associate the correct
track with the candidate PMT signal, with the requirement that the
transit time of the photon from its creation in the bar to its
detection at the PMT be consistent with the measurement error of
$\sim 1.5$ ns.

\subsubsection{Reconstruction\label{sec:drcreco} }
An  unbinned maximum likelihood formalism is
used to incorporate all information provided by the space and time
measurements from the \dirc.

The emission angle and the arrival time of the Cherenkov photons
are reconstructed from the observed space-time coordinates of the
PMT signals, transformed into the Cherenkov coordinate system
($\theta_c$, $\phi_c$, and $\delta t$) as follows: The known
spatial position of the bar through which the track passed and the
PMTs whose signal times lie within the readout window of $\pm$300
ns from the trigger are used to calculate the three-dimensional
vector pointing from the center of the bar end to the center of
each tube. This vector is then extrapolated into the radiator bar
(using Snell's law). This procedure defines, up to the 16-fold
ambiguity described above, the angles $\theta_c$ and $\phi_c$ of a
photon.

The DIRC time measurement represents the third dimension of the
photomultiplier hit reconstruction. The timing resolution is not
competitive with the position information for Cherenkov angle
reconstruction, but timing information is used to suppress
background hits from the beam induced background and, more
importantly, exclude other tracks in the same event as the source
of the photon. Timing information is also used to resolve the
forward-backward and wedge ambiguities in the hit-to-track
association.

The relevant observable to distinguish between signal and
background photons is the difference between the measured and
expected photon arrival time, $\Delta t_\gamma$. It is calculated
for each photon using the track time-of-flight (assuming it to be
a charged pion), the measured time of the candidate signal in the
PMT and the photon propagation time within the bar and the water
filled standoff box. The time information and the requirement of
using only physically possible photon propagation paths reduce the
number of ambiguities from 16 to typically 3. Applying the time
information also substantially improves the correct matching of
photons with tracks and reduces the number of accelerator induced
background signals by approximately a factor 40, as illustrated in
Figure~\ref{\secname fig:event}.

The reconstruction routine currently provides a likelihood value
for each of the five stable particle types (e,$\mu$,$\pi$,K,p) if
the track passes through the active volume of the \dirc . These
likelihoods are calculated in an iterative process by maximizing
the likelihood value for the entire event while testing different
hypotheses for each track. If enough photons are found, a fit of
$\theta_c$ and the number of observed signal and background
photons are calculated for each track.

\subsubsection{Results}

The parameters of expected \dirc\ performance were derived from
extensive studies with a variety of prototypes, culminating with a
full-size prototype in a beam at CERN~\cite{ten}. The test beam
results were well-described by Monte Carlo simulations of the
detector. The performance of the full detector is close to
expectations, and additional offline work, particularly  on
geometrical alignment, is expected to lead to further
improvements.

\begin{figure}
\centering
\includegraphics[width=7cm]{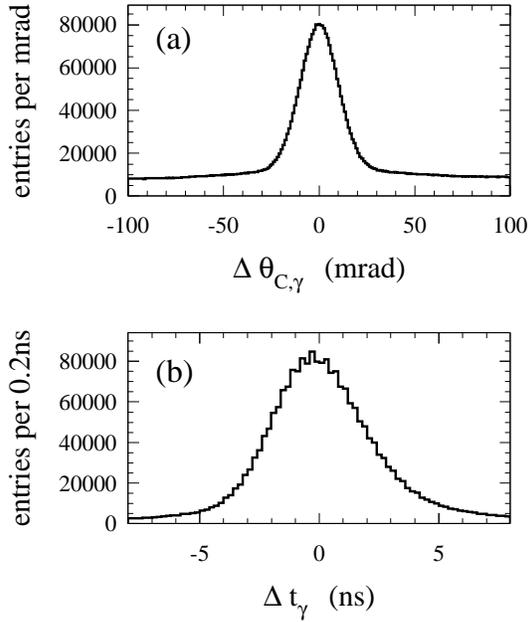}
\vspace{-2pc}

\caption{ The difference between (a)~the measured and expected
Cherenkov angle for single photons, $\Delta \theta_{c,\gamma}$,
and (b)~the measured and expected photon arrival time, for single
muons in $\mu^+\mu^-$ events.} \label{\secname fig:single}
\end{figure}

In the absence of correlated systematic errors, the resolution
($\sigma_{C,\mathrm{track}}$) on the track Cherenkov angle should
scale as
\begin{equation}
    \sigma_{C,\mathrm{track}} = {\sigma_{C,\gamma}\over \sqrt{N_{pe}}} ,
\end{equation}
where $\sigma_{C,\gamma}$ is the single photon Cherenkov angle
resolution, and $N_{pe}$ is the number of photons detected.
Figure~\ref{\secname fig:single}(a) shows the single photon
angular resolution obtained from di-muon events. There is a broad
background of less than 10\% relative height under the peak, that
originates mostly from track-associated sources, such as $\delta$
rays, and combinatorial background. The width of the peak
translates to a resolution of about 10.2\mrad, in good agreement
with the expected value. The measured time resolution (see
Figure~\ref{\secname fig:single}(b)) is 1.7\ns, close to the
intrinsic 1.5\ns\ transit time spread of the PMTs.

\begin{figure}
\includegraphics[width=7.5 cm]{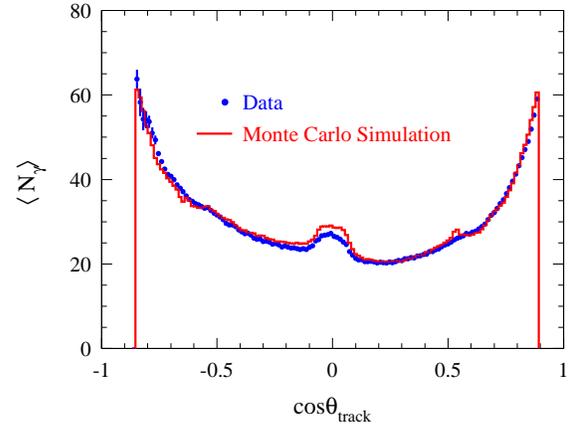}
\vspace*{-15mm} \caption{Number of detected photons versus track
polar angle for reconstructed tracks in di-muon events compared to
Monte Carlos simulation. The mean number of photons in the
simulation has been tuned to match the data.} \label{\secname
fig:nphot}
\end{figure}

The number of photoelectrons shown in Figure~\ref{\secname
fig:nphot} varies between 20 for small polar angles at the center
of the barrel and 65 at large polar angles. This variation is well
reproduced by Monte Carlo simulation, and can be understood from
the geometry of the \dirc . The number of Cherenkov photons varies
with the pathlength of the track in the radiator, it is smallest
at perpendicular incidence at the center and increases towards the
ends of the bars. In addition, the fraction of photons trapped by
total internal reflection rises with larger values of
$\vert\cos\theta_{\mathrm{track}}\vert$. This increase in the
number of photons for forward going tracks is a good match to the
increase in momentum and thus benefits the \dirc\ performance.

\begin{figure}
\includegraphics[width=7.5 cm]{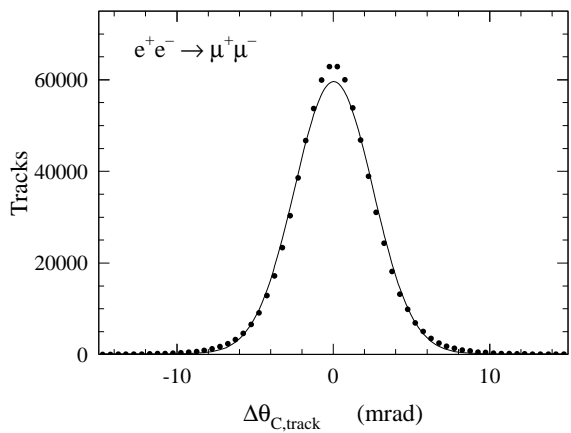}
\vspace*{-2pc}

\caption{The difference between the measured and expected
Cherenkov angle, $\Delta \theta_{c,{\mathrm track}}$, for single
muons in $\mu^+\mu^-$ events. The curve represents a Gaussian
distribution fit to the data with a width of 2.5 \mrad. }
\label{\secname fig:trackres}
\end{figure}

\begin{figure}
\includegraphics[width=7.5 cm]{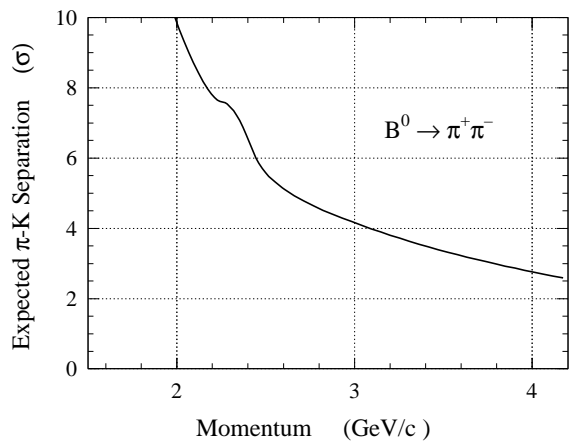}
\vspace*{-2pc}

\caption{Expected $\pi$-K separation in $B^0\rightarrow
\pi^+\pi^-$  events versus track momentum inferred from the
measured Cherenkov angle resolution and number of Cherenkov
photons per track in di-muon events.} \label{\secname fig:sep}
\end{figure}

With the present alignment, the track Cherenkov angle resolution
for di-muon events is shown in Figure~\ref{\secname fig:trackres}.
The width of the fitted Gaussian distribution is 2.5\mrad\
compared to the design goal of 2.2\mrad. From the measured single
track resolution versus momentum in di-muon events and the
difference between the expected Cherenkov angles of charged pions
and kaons, the pion-kaon separation power of the \dirc\ can be
inferred. As shown in Figure~\ref{\secname fig:sep}, the expected
separation between kaons and pions at 3\ GeV/$c$ is about
$4.2\sigma$, within 15\% of the design goal.

\begin{figure}
\includegraphics[width=.8\columnwidth]{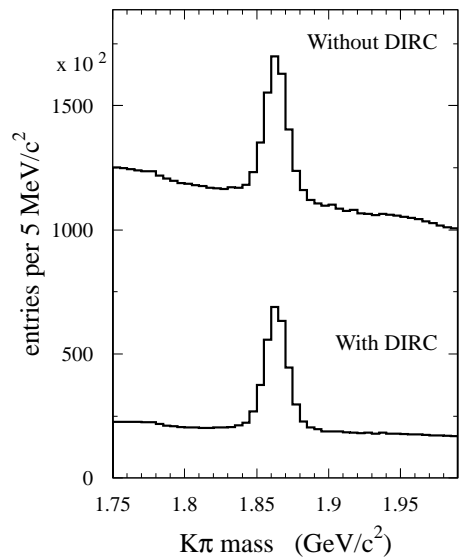}
\vspace*{-2pc}

\caption{Invariant K$\pi$ inclusive mass spectrum with and without
the use of the \dirc\ for kaon identification. The mass peak
corresponds to the decay of the $D^0$ particle.} \label{\secname
fig:kpi}
\end{figure}

Figure~\ref{\secname fig:kpi} shows an example of the use of the
\dirc\ for particle identification. The $K\pi$ invariant mass
spectra are shown with and without the use of the \dirc\ for kaon
identification. The peak corresponds to the decay of the $D^0$
particle.

The efficiency for correctly identifying a charged kaon that
traverses a radiator bar and the probability to wrongly identify a
pion as a kaon are determined using $D^0 \rightarrow K^-\pi^+$
decays selected kinematically from inclusive $D^{*}$ production
and are shown as a function of the track momentum in
Figure~\ref{\secname fig:kpi_eff} for a particular choice of
particle selection criteria. The mean kaon selection efficiency
and pion misidentification are $96.2 \pm 0.2$\% (stat.) and $2.1
\pm 0.1$\% (stat.), respectively.

\begin{figure}
\includegraphics[width=7.5 cm]{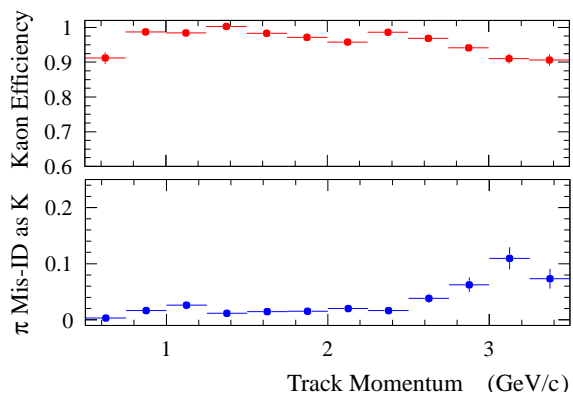}
\vspace*{-2pc}

\caption{Efficiency and misidentification probability for the
selection of charged kaons as a function of track momentum,
determined using $D^0 \rightarrow K^-\pi^+$ decays selected
kinematically from inclusive $D^{*}$ production.} \label{\secname
fig:kpi_eff}
\end{figure}

\subsection{Summary}

The \dirc\ is a novel ring-imaging Cherenkov detector that is
well-matched to the hadronic PID requirements of \babar . The
\dirc\ has been robust and stable and, two years after
installation, about 99.7\% of all PMTs and electronic channels are
operating with nominal performance. Additional shielding in the
standoff box tunnel region should reduce the sensitivity to
beam-induced backgrounds, as should faster FEE, both installed
during the winter 2000-2001 shutdown. At luminosities around
$1\times 10^{34}\cms$, the TDC IC will have to be replaced with a
faster version and deeper buffering. The design process for this
is underway.

The detector performance achieved is rather close to that
predicted by the Monte Carlo simulations. Alignment and further
code developments are expected to further improve performance.

\renewcommand{\secname}{emc_}
\renewcommand{\sectiondir}{sec07_emc}
\section{Electromagnetic Calorimeter}
\label{sec:emc}
\newcommand{\CsI}{\rm CsI(Tl)}

\subsection{Purpose and Design}

The electromagnetic calorimeter (EMC) is designed to measure
electromagnetic showers with excellent efficiency, and energy and
angular resolution over the energy range from 20\mev\ to 9\gev.
This capability allows the detection of photons from \piz\ and
$\eta$ decays as well as from electromagnetic and radiative
processes. By identifying electrons, the EMC contributes to the
flavor tagging of neutral $B$ mesons via semi-leptonic decays, to
the reconstruction of vector mesons like \jpsi, and to the study
of semi-leptonic and rare decays of $B$ and $D$ mesons, and $\tau$
leptons.  The upper bound of the energy range is set by the need
to measure QED processes, like $\epem \to \epem (\gamma)$ and
$\epem \to \gamma \gamma$, for calibration and luminosity
determination. The lower bound is set by the need for highly
efficient reconstruction of $B$-meson decays containing multiple
$\piz$s and $\eta^0$s.

\subsubsection{Requirements}

The measurement of extremely rare decays of $B$ mesons containing
$\piz$s (\eg\ $B^0 \to \piz \piz$) poses the most stringent
requirements on energy resolution, namely of order 1--2\%. Below
energies of 2\gev, the \piz\ mass resolution is dominated by the
energy resolution. At higher energies, the angular resolution
becomes dominant, and therefore it is required to be of the order
of a few mrad.

Furthermore, the EMC has to be compatible with the 1.5\tesla\
field of the solenoid and operate reliably over the anticipated
ten-year lifetime of the experiment. To achieve excellent
resolution, stable operating conditions have to be maintained.
Temperatures and the radiation exposure must be closely monitored,
and precise calibrations of the electronics and energy response
over the full dynamic range must be performed frequently.

\subsubsection{Design Considerations}

The requirements stated above lead to the choice of a hermetic,
total-absorption calorimeter, composed of a finely segmented array
of thallium-doped cesium iodide (\CsI) crystals.  The crystals are
read out with silicon photodiodes that are matched to the spectrum
of scintillation light. Recent experience at CLEO~\cite{cleo_emc}
has demonstrated the suitability of this choice for physics at the
\FourS\ resonance.

The energy resolution of a homogeneous crystal calorimeter can be
described empirically in terms of a sum of two terms added in
quadrature
\begin{equation}
\frac{\sigma_{E}}{E}= \frac{a}{^{4}\sqrt{E (\gev)}}\oplus b,
\label{caleqn::res}
\end{equation}
where $E$ and $\sigma_{E}$ refer to the energy of a photon and its
rms error, measured in \gev. The energy dependent term $a$ arises
primarily from the fluctuations in photon statistics, but it is
also impacted by electronic noise of the photon detector and
electronics. Furthermore, beam-generated background will lead to
large numbers of additional photons that add to the noise. This
term is dominant at low energies. The constant term, $b$, is
dominant at higher energies ($>1$\gev). It arises from
non-uniformity in light collection,  leakage or absorption in the
material between and in front of the crystals, and uncertainties
in the calibrations. Most of these effects can be influenced by
design choices, and they are stable with time. Others will be
impacted by changes in the operating conditions, like variations
in temperature, electronics gain, and noise, as well as by
radiation damage caused by beam-generated radiation.

The angular resolution is determined by the transverse crystal size
and the distance from the interaction point.  It can also be
empirically parameterized as a sum of an energy dependent and a
constant term,
\begin{equation}\label{caleqn::posres}
     \sigma_{\theta}=\sigma_{\phi} = \frac{c}{\sqrt{E (\gev)}} + d,
\end{equation}
where the energy $E$ is measured in \gev.  The design of the EMC
required a careful optimization of a wide range of choices,
including the crystal material and dimensions,  the choice of the
photon detector and readout electronics, and the design of a
calibration and monitoring system.  These choices were made on the
basis of extensive studies, prototyping and beam
tests~\cite{beamtest}, and Monte Carlo simulation, taking into
account limitations of space and the impact of other \babar\
detector systems.

Under ideal conditions, values for the energy resolution
parameters $a$ and $b$ close to 1--2\% could be obtained.  A
position resolution of a few mm will translate into an angular
resolution of a few mrad; corresponding parameter values are
$c\approx 3\mrad$ and $d\approx 1\mrad$.

However in practice, such performance is very difficult to achieve
in a large system with a small, but unavoidable amount of inert
material and gaps, limitations of electronics, and background in
multi-particle events, plus contributions from beam-generated
background.

Though in \CsI\ the intrinsic efficiency for the detection of photons
is close to 100\% down to a few \mev, the minimum measurable energy in
colliding beam data is expected to be about 20~\mev, a limit that is
largely determined by beam- and event-related background and the
amount of material in front of the calorimeter. Because of the
sensitivity of the \piz\ efficiency to the minimum detectable photon
energy, it is extremely important to keep the amount of material in
front of the EMC to the lowest possible level.

\subsubsection{CsI(Tl) Crystals}

Thallium-doped CsI meets the needs of \babar\ in several ways. Its
properties are listed in Table~\ref{tab:csiproperties}. The high
light yield and small Moli\`{e}re radius allow for excellent
energy and angular resolution, while the short radiation length
allows for shower containment at \babar\ energies with a
relatively compact design. Furthermore, the high light yield and
the emission spectrum permit efficient use of silicon photodiodes
which operate well in high magnetic fields. The transverse size of
the crystals is chosen to be comparable to the Moli\`{e}re radius
achieving the required angular resolution at low energies while
appropriately limiting the total number of crystals (and readout
channels).

\begin{table}
\caption{Properties of \CsI\ .}

\vspace{.5\baselineskip}
\begin{center}
\begin{tabular}{ll}\hline
Parameter\rule[-5pt]{0pt}{17pt} & Values \\ \hline
\rule{0pt}{12pt}Radiation Length & 1.85 \cm\\ Moli\`{e}re Radius &
3.8 \cm \\ Density & 4.53 $\mathrm{g/cm^{3}}$ \\ Light Yield &
50,000 ${\rm \gamma/\mev}$ \\ Light Yield Temp. Coeff.  &
0.28\%/\degc\ \\ Peak Emission $\lambda_\mathrm{max}$ & 565 \nm \\
Refractive Index ($\lambda_\mathrm{max}$) &1.80 \\ Signal Decay
Time & 680 \ns\ (64\%) \\ \rule[-5pt]{0pt}{0pt}& 3.34 \mus\
(36\%)\\ \hline
\end{tabular}
\end{center}
\label{tab:csiproperties} \vspace{-.5\baselineskip}
\end{table}

\subsection{Layout and Assembly}
\label{calsec::mechanics}

\begin{figure*}
\begin{center}
\includegraphics[width=12.5cm]{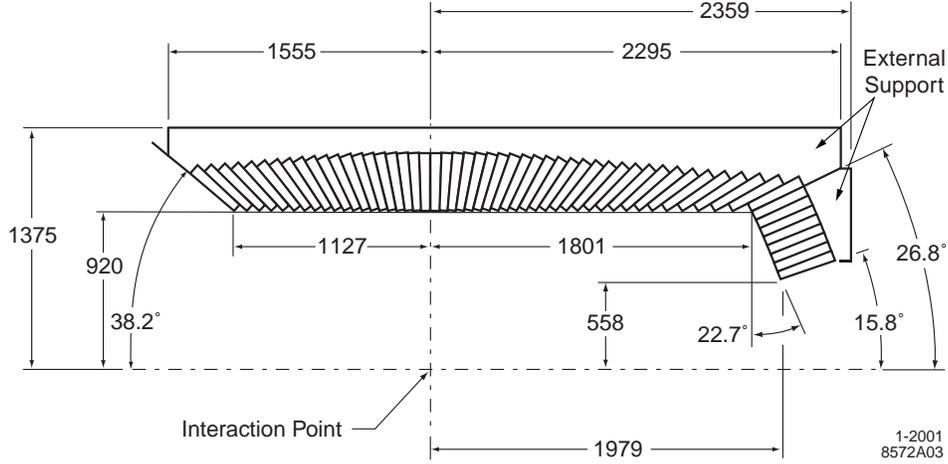}
\vspace{-2pc}
\caption{A longitudinal cross section of the EMC (only the top half is
shown) indicating the arrangement of the 56 crystal rings.  The
detector is axially symmetric around the $z$-axis. All dimensions are
given in mm.}
\label{calfig::detector}
\end{center}
\end{figure*}

\subsubsection{Overall Layout}

The EMC consists of a cylindrical barrel and a conical forward
endcap. It has full coverage in azimuth and extends in polar angle
from 15.8\degrees\ to 141.8\degrees\ corresponding to a
solid-angle coverage of 90\% in the c.m. system (see
Figure~\ref{calfig::detector} and Table~\ref{tab:csigeometry}).
The barrel contains 5,760 crystals arranged in 48 distinct rings
with 120 identical crystals each. The endcap holds 820 crystals
arranged in eight rings, adding up to a total of 6,580 crystals.
The crystals have a tapered trapezoidal cross section. The length
of the crystals increases from 29.6\cm\ in the backward to
32.4\cm\ in the forward direction to limit the effects of shower
leakage from increasingly higher energy particles.

\begin{table}[!htb]
\caption{Layout of the EMC, composed of 56 axially symmetric
rings, each consisting of CsI crystals of identical dimensions. }
\vspace{.5\baselineskip}
\begin{center}
\begin{tabular}{cccc} \hline
\rule{0pt}{12pt}$\theta$ Interval & Length & \# & Crystals \\
\rule[-5pt]{0pt}{0pt}(radians) &($X_{0}$) &Rings & /Ring \\ \hline
\multicolumn{4}{c}{\rule[-5pt]{0pt}{17pt}Barrel}\\ \hline
\rule{0pt}{12pt}$2.456-1.214$ & 16.0 & 27 & 120 \\
$1.213-0.902$ & 16.5 & 7 & 120 \\
$0.901-0.655$ & 17.0 & 7 & 120 \\
\rule[-5pt]{0pt}{0pt}$0.654-0.473$ & 17.5 & 7 & 120 \\ \hline
\multicolumn{4}{c}{\rule[-5pt]{0pt}{17pt}Endcap} \\ \hline
\rule{0pt}{12pt}$0.469-0.398$ & 17.5 & 3 & 120 \\
$0.397-0.327$ & 17.5 & 3 & 100 \\
$0.326-0.301$ & 17.5 & 1 & 80 \\
\rule[-5pt]{0pt}{0pt}$0.300-0.277$ & 16.5 & 1 & 80 \\ \hline
\end{tabular}
\end{center}
\label{tab:csigeometry}
\end{table}

To minimize the probability of pre-showering, the crystals are
supported at the outer radius, with only a thin gas seal at the
front. The barrel and outer five rings of the endcap have less than
0.3--0.6\Xrad\ of material in front of the crystal faces. The SVT
support structure and electronics, as well as the B1 dipole shadow the
inner three rings of the endcap, resulting in up to 3.0\Xrad\ for
the innermost ring. The principal purpose of the two innermost rings
is to enhance shower containment for particles close to the acceptance
limit.

\subsubsection{Crystal Fabrication and Assembly}

\begin{figure}
\begin{center}
\includegraphics[width=6.2cm]{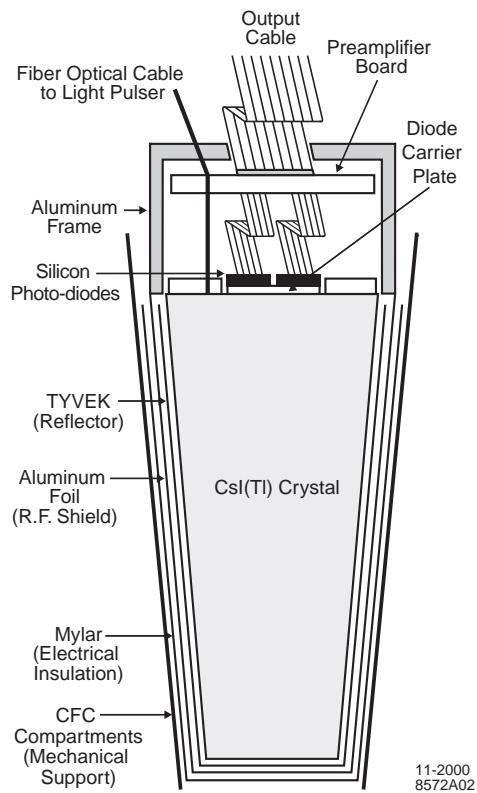}
\vspace{-2pc}
       \caption{A schematic of the wrapped \CsI\ crystal and the
       front-end readout package mounted on the rear face. Also
       indicated is the tapered, trapezoidal CFC compartment, which is
       open at the front.  This drawing is not to scale.}
\label{calfig::crystal}
\end{center}
\end{figure}

The crystals were grown in boules from a melt of CsI salt doped
with 0.1\% thallium~\cite{salt}. They were cut from the boules,
machined into tapered trapezoids (Figure~\ref{calfig::crystal}) to
a tolerance of $\pm 150$\mum, and then polished~\cite{vendor}.
The transverse dimensions of the crystals for each of the 56 rings
vary to achieve the required hermetic coverage.  The typical area
of the front face is $4.7 \times 4.7 \cma$, while the back face
area is typically $6.1 \times 6.0 \cma$.  The crystals act not
only as a total-absorption scintillating medium, but also as a
light guide to collect light at the photodiodes that are mounted
on the rear surface.  At the polished crystal surface light is
internally reflected, and a small fraction is transmitted.  The
transmitted light is recovered in part by wrapping the crystal
with two layers of diffuse white reflector~\cite{tyvek,gerd}, each
165\mum\ thick. The uniformity of light yield along the wrapped
crystal was measured by recording the signal from a highly
collimated radioactive source at 20 points along the length of the
crystal.  The light yield was required to be uniform to within
$\pm 2$\% in the front half of the crystal; the limit increased
linearly up to a maximum of $\pm 5$\% at the rear face.
Adjustments were made on individual crystals to meet these
criteria by selectively roughing or polishing the crystal surface
to reduce or increase its reflectivity.

Following these checks, the crystals were further wrapped in 25\mum\
thick aluminum foil which was electrically connected to the metal
housing of the photodiode-preamplifier assembly to provide a Faraday
shield.  The crystals were covered on the outside with a 13\mum-thick
layer of mylar to assure electrical isolation from the external
support.

\subsubsection{Photodiodes\\ and Preamplifier
Assembly}

The photon detector consists of two $2 \times 1\cma $ silicon PIN
diodes glued to a transparent 1.2\mm-thick polysterene substrate
that, in turn, is glued to the center of the rear face of the crystal by
an optical epoxy~\cite{epoxy} to maximize light
transmission~\cite{readout}. The surrounding area of the crystal face
is covered by a plastic plate coated with white reflective
paint~\cite{paint}.  The plate has two 3\mm-diameter penetrations for
the fibers of the light pulser monitoring system.

 As part of the quality control process, the 1.836\mev\ photon line
 from a $^{88}$Y radioactive source was used to measure the light
 yield of every crystal-diode assembly, employing a
 preamplifier with 2\mus\ Gaussian shaping.  The resulting signal
 distribution had a mean and rms width of 7300 and
 890~photoelectrons/MeV, respectively; none of the crystals had
 a signal of less than 4600~photoelectrons/MeV~\cite{readout,emc_footnote}.

Each of the diodes is directly connected to a low-noise
preamplifier. The entire assembly is enclosed by an aluminum
fixture as shown in Figure~\ref{calfig::crystal}.  This fixture is
electrically coupled to the aluminum foil wrapped around the
crystal and thermally coupled to the support frame to dissipate
the heat load from the preamplifiers.

Extensive aging tests were performed to ascertain that the diodes and
the preamplifiers met the ten-year lifetime requirements. In
addition, daily thermal cycles of $\pm$5\degc\ were run for many
months to assure that the diode-crystal epoxy joint could sustain
modest temperature variations.

\subsubsection{Crystal Support Structure}

The crystals are inserted into modules that are supported individually
from an external support structure.  This structure is built in three
sections, a cylinder for the barrel and two semi-circular structures
for the forward endcap. The barrel support cylinder carries the load
of the barrel modules plus the forward endcap to the magnet iron
through four flexible supports.  These supports decouple and dampen
any acceleration induced by movements of the magnet iron during a
potential earthquake.

\begin{figure*}
\begin{center}
\includegraphics[width=15.0cm]{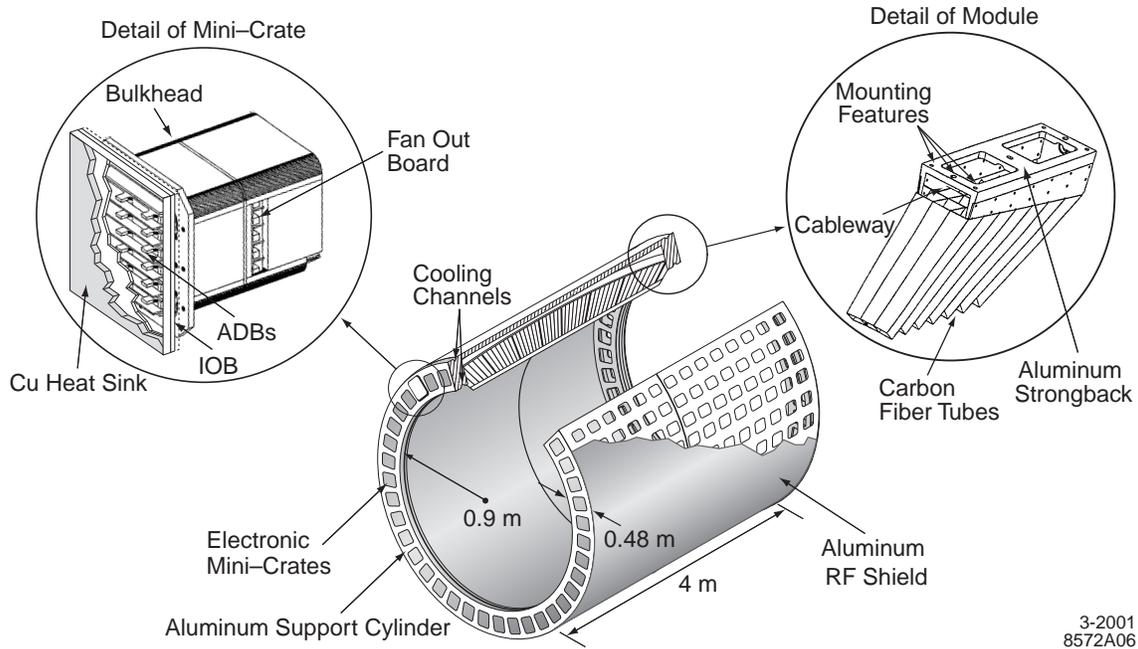}
\vspace{-2pc}
       \caption{The EMC barrel support structure, with details on the
       modules and electronics crates (not to scale).}
\label{calfig::construction}
\end{center}
\end{figure*}

The modules are built from tapered, trapezoidal compartments made from
carbon-fiber-epoxy composite (CFC) with 300\mum-thick walls
(Figure~\ref{calfig::construction}).  Each compartment loosely holds a
single wrapped and instrumented crystal and thus assures that the
forces on the crystal surfaces never exceed its own weight. Each
module is surrounded by an additional layer of 300\mum\ CFC to
provide additional strength.  The modules are bonded to an aluminum
strong-back that is mounted on the external support.  This scheme
minimizes inter-crystal materials while exerting minimal force on the
crystal surfaces; this prevents deformations and surface degradation
that could compromise performance.  By supporting the modules at the
back, the material in front of the crystals is kept to a minimum.

The barrel section is divided into $280$ separate modules, each
holding 21 crystals ($7 \times 3$ in $\theta \times \phi$).
After the insertion of the crystals, the aluminum readout frames,
which also stiffen the module, are attached with thermally-conducting
epoxy to each of the CFC compartments.  The entire 100\kg-module is
then bolted and again thermally epoxied to an aluminum strong-back.
The strong-back contains alignment features as well as channels that
couple into the cooling system. Each module was installed into the
2.5\cm-thick, 4\m-long aluminum support cylinder, and subsequently
aligned.  On each of the thick annular end-flanges this cylinder
contains access ports for digitizing electronics crates with
associated cooling channels, as well as mounting features and
alignment dowels for the forward endcap.

The endcap is constructed from 20 identical CFC modules (each with 41
crystals), individually aligned and bolted to one of two semi-circular
support structures.  The endcap is split vertically into two halves
to facilitate access to the central detector components.

The entire calorimeter is surrounded by a double Faraday shield
composed of two 1\mm-thick aluminum sheets so that the diodes and
preamplifiers are further shielded from external noise.  This cage
also serves as the environmental barrier, allowing the slightly
hygroscopic crystals to reside in a dry, temperature controlled
nitrogen atmosphere.

\subsubsection{Cooling System}

The EMC is maintained at constant, accurately monitored
temperature. Of particular concern are the stability of the
photodiode leakage current which rises exponentially with
temperature, and the large number of diode-crystal epoxy joints
that could experience stress due to differential thermal
expansion. In addition, the light yield of \CsI\ is weakly
temperature dependent.

The primary heat sources internal to the calorimeter are the
preamplifiers ($2 \times 50$\mw/crystal) and the digitizing electronics
(3\kw\ per end-flange).  In the barrel, the preamplifier heat is
removed by conduction to the module strong backs which are directly
cooled by Fluorinert (polychlorotrifluoro-ethylene)~\cite{fluorinert}.
The digitizing electronics are housed in 80 mini-crates, each in
contact with the end-flanges of the cylindrical support structure.
These crates are indirectly cooled by chilled water pumped through
channels milled into the end-flanges close to the inner and outer
radii.  A separate Fluorinert system in the endcap cools both the 20
mini-crates of digitizing electronics and the preamplifiers.

\subsection{Electronics}
\label{calsec::electronics}

\begin{figure*}
\begin{center}
\includegraphics[width=12cm]{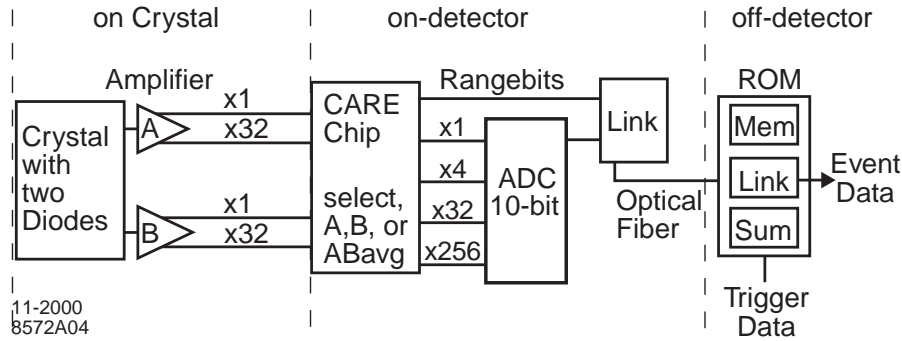}
\vspace{-2pc} \caption{Schematic diagram of the EMC readout
electronics.} \label{calfig::electronics}
\end{center}
\end{figure*}

The EMC electronics system, shown schematically in
Figure~\ref{calfig::electronics}, is required to have negligible
impact on the energy resolution of electromagnetic showers
from 20\mev\ to 9\gev, while accommodating the use of a 6.13\mev\
radioactive source for calibration.  These requirements set a limit of
less than 250\kev\ equivalent noise energy (ENE) per crystal and
define an 18-bit effective dynamic range of the digitization
scheme. For source calibrations, the least significant bit is set to
50\kev, while for colliding beam data it is set to 200\kev.  To
reach the required energy resolution at high energies, the coherent
component has to be significantly smaller than the incoherent noise
component. In addition, the impact of high rates of low energy
($<$5\mev) beam-induced photon background needs to be minimized.

\subsubsection{Photodiode Readout\\ and Preamplifiers}

The ENE is minimized by maximizing the light yield and collection,
employing a highly efficient photon detector, and a low-noise
electronic readout. The PIN silicon photodiodes~\cite{hamamatsu}
have a quantum efficiency of 85\% for the $\CsI$ scintillation
light~\cite{diodes}. At a depletion voltage of 70\volt, their
typical dark currents were measured to be 4\nA\ for an average
capacitance of 85\pf; the diodes are operated at a voltage of
50\volt. The input capacitance to the preamplifier is minimized by
connecting the diodes to the preamplifier with a very short cable.
The preamplifier is a low-noise charge-sensitive amplifier
implemented as a custom application specific integrated circuit
(ASIC)~\cite{freytag}. It shapes the signal and acts as a
band-pass filter to remove high- and low-frequency noise
components. The optimum shaping time for the $\CsI$-photodiode
readout is 2--3\mus, but a shorter time was chosen to reduce the
probability of overlap with low-energy photons from beam
background. The commensurate degradation in noise performance is
recovered by implementing a realtime digital signal-processing
algorithm following digitization.

To achieve the required operational reliability~\cite{reliability} for
the inaccessible front-end readout components, two photodiodes were
installed, each connected to a preamplifier. In addition, all
components were carefully selected and subjected to rigorous tests,
including a 72-hour burn-in of the preamplifiers at 70\degc\ to avoid
infant mortality.  The dual signals are combined in the
postamplification/digitization circuits, installed in mini-crates at
the end-flanges, a location that is accessible for maintenance.

\subsubsection{Postamplification, Digitization\\ and Readout}

The two preamplifiers on each crystal, A and B, each provide
amplification factors of 1 and 32 and thus reduce the dynamic
range of the signal that is transmitted to the mini-crates to
13-bits.  A custom auto-range encoding (CARE)
circuit~\cite{freytag} further amplifies the signal to arrive at a
total gain of 256, 32, 4 or 1 for four energy ranges, 0--50\mev,
50--400\mev, 0.4--3.2\gev, and 3.2--13.0\gev, respectively. The
appropriate range is identified by a comparator and the signal is
digitized by a 10-bit, 3.7\mhz\ ADC. Data from 24 crystals are
multiplexed onto a fiber-optic driver and sent serially at a rate
of 1.5\gbps\ across a 30\m-long optical fiber to the ROM. In the
ROM, the continuous data stream is entered into a digital
pipeline. A correction for pedestal and gain is applied to each
sample.  The pipeline is then tapped to extract the input to the
calorimeter trigger.

Upon receipt of the L1 \emph{Accept} signal, data samples within a
time window of $\pm$1\mus\ are selected for the feature
extraction.  Up to now, the calorimeter feature extraction
algorithm performs a parabolic fit to the peak of the signal
waveform to derive its energy and time. In the future, it is
planned to employ a digital filter prior to the signal fit to
further reduce noise.  For this filter algorithm, the frequency
decomposition of an average signal pulse and the typical noise
spectrum are measured for all channels and subsequently used to
derive an optimum set of weights that maximizes the
signal-to-noise ratio. These weights are then applied to
individual samples to obtain a filtered waveform.

\begin{figure}
\begin{center}
\includegraphics[width=\columnwidth]{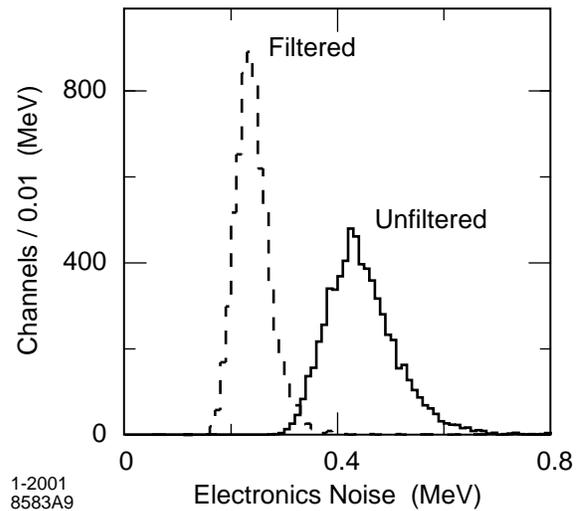}
\vspace{-1.25pc} \caption{The distribution of equivalent noise
energy (ENE) or all channels of the EMC with and without digital
filtering.  The data were recorded in the absence of beams by a
random trigger.} \label{calfig::enoise}
\end{center}
\end{figure}

The magnitude of the electronic noise is measured as the rms width
of the pedestal distribution as shown in
Figure~\ref{calfig::enoise}. The observed distribution for all
channels translates to an ENE of 230\kev\ and 440\kev\ with and
without digital filtering; this result is comparable to design
expectations. Measurements of the auto-correlation function
indicate that the coherent noise component is negligible compared
to the incoherent noise, except for regions where the
preamplifiers saturate (see below).

During data-taking, the data acquisition imposes a single-crystal
readout threshold in order to keep the data volume at an
acceptable level.  This energy threshold is currently set to
1\mev\ and during stable colliding beam conditions on average
1,000 crystals are read out (measured with 600\mA\ of \en\ and
1100\mA\ of \ep\ and a random clock trigger), corresponding to an
average occupancy of 16\%. The electronic noise accounts for about
10\%, while the remaining signals originate from beam-generated
background (see Section~\ref{sec:ir}). A typical hadronic event
contributes signals in 150 crystals.

\subsubsection{Electronics Calibration\\ and Linearity}

To measure pedestal offsets, determine the overall gain, and to
remove non-linearities the FEE are calibrated by precision charge
injection into the preamplifier input.  Initially, residual
non-linearities of up to 12\% in limited regions near each of the
range changes were observed and corrected for
offline~\cite{nonlin}. These non-linearities were traced to
oscillations on the ADC cards that have since been corrected.  The
correction resulted in markedly improved energy resolution at high
energies. Residual non-linearities (typically 2--4\%) arise
primarily from cross-talk, impacting both the electronics
calibrations and the colliding-beam data. The effect is largest at
about ~630\mev\ (950\mev) in a high (low) gain preamplifier
channel, inducing a 2\mev\ (6\mev) cross-talk signal in an
adjacent channel. The implementation of an energy dependent
correction is expected to significantly reduce this small,
remaining effect, and lead to a further improvement of the energy
resolution.

\subsubsection{Electronics Reliability}

With the exception of minor cable damage during installation
(leaving two channels inoperative), the system of 13,160 readout
channels has met its reliability requirements. After the
replacement of a batch of failing optical-fiber drivers, the
reliability of the digitizing electronics improved substantially,
averaging channel losses of less than 0.1\%.

\subsection{Energy Calibration}

The energy calibration of the EMC proceeds in two steps. First,
the measured pulse height in each crystal has to be translated to
the actual energy deposited. Second, the energy deposited in a
shower spreading over several adjacent crystals has to be related
to the energy of the incident photon or electron by correcting for
energy loss mostly due to leakage at the front and the rear, and
absorption in the material between and in front of the crystals,
as well as shower energy not associated with the cluster.

The offline pattern recognition algorithm that groups adjacent
crystals into clusters is described in detail in
Section~\ref{calsec::reco}.

\subsubsection{Individual Crystal Calibration}
\label{calsec::calcrystal}

In spite of the careful selection and tuning of the individual
crystals, their light yield varies significantly and is generally
non-uniform. It also changes with time under the impact of
beam-generated radiation.  The absorbed dose is largest at the
front of the crystal and results in increased attenuation of the
transmitted scintillation light.  The light yield must therefore
be calibrated at different energies, corresponding to different
average shower penetration, to track the effects of the radiation
damage.

The calibration of the deposited energies is performed at two energies
at opposite ends of the dynamic range, and these two measurements are
combined by a logarithmic interpolation.  A 6.13\mev\ radioactive
photon source~\cite{source} provides an absolute calibration at low
energy, while at higher energies (3--9\gev) the relation between
polar angle and energy of \epm\ from Bhabha events is
exploited~\cite{bhabha1}.

A flux of low-energy neutrons ($4\times 10^{8}/s$) is used to
irradiate Fluorinert~\cite{fluorinert} to produce photons of
6.13\mev\ via the reaction $ ^{19}F + n \to ^{16}N+\alpha$,
$^{16}N \to ^{16}O^{*} + \beta$, $^{16}O^{*} \rightarrow ^{16}O +
\gamma$. The activated $ ^{16}N $ has a half-life of 7 seconds and
thus does not cause radiation damage or long-term activation.  The
fluid is pumped at a rate of 125\liter/\s\ from the neutron
generator to a manifold of thin-walled (0.5\mm) aluminum pipes
that are mounted immediately in front of the crystals. At this
location, the typical rate of photons is 40\hz/crystal.

\begin{figure}
\centering
\includegraphics[height=7.5cm]{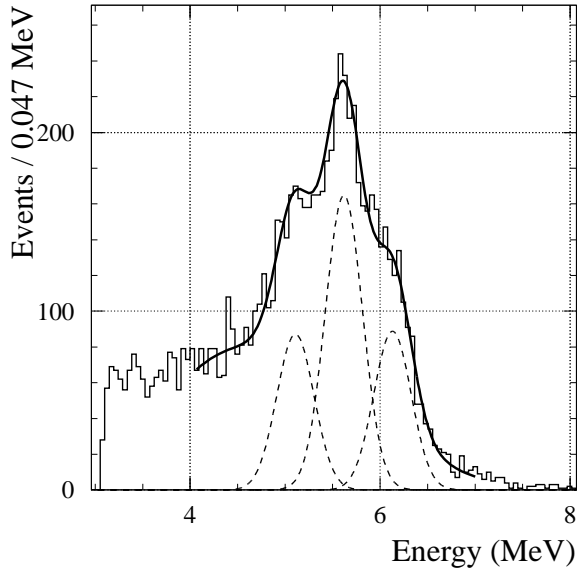}
\vspace{-2pc}
       \caption{A typical pulse-height spectrum recorded with the
       radioactive source to calibrate the single-crystal energy scale
       of the EMC. The spectrum shows the primary 6.13\mev\ peak and
       two associated escape peaks at 5.62\mev\ and 5.11\mev.
       The solid line represents a fit to the total spectrum, the dotted
       lines indicate the contributions from the three individual photon
       spectra.}
\label{calfig::source}
\end{figure}

Figure~\ref{calfig::source} shows a typical source spectrum that was
derived from the raw data by employing a digital filter algorithm.
For a 30-minute exposure, a statistical error of 0.35\% is obtained,
compared to a systematic uncertainty of less than 0.1\%. This
calibration is performed weekly.

At high energies, single crystal calibration is performed with a
pure sample of Bhabha events~\cite{bhabha1}. As a function of the
polar angle of the \epm, the deposited cluster energy is
constrained to equal the prediction of a GEANT-based Monte Carlo
simulation~\cite{geant}.  For a large number of energy clusters, a
set of simultaneous linear equations relates the measured to the
expected energy and thus permits the determination of a gain
constant for each crystal.  In a 12-hour run at a luminosity of
$3\times 10^{33}$\cms\ some 200 \epm\ per crystal can be
accumulated, leading to a statistical error of 0.35\%.  This
calibration has been performed about once per month, and will be
fully automated in the future.

\subsubsection{Cluster Energy Correction}
The correction for energy loss due to shower leakage and
absorption is performed as a function of cluster energy and polar
angle. At low energy ($\mathrm{E} < 0.8\gev$), it is derived from
$\piz$ decays~\cite{pizero}.  The true energy of the photon is
expressed as a product of the measured deposited energy and a
correction function which depends on $\ln E$ and $\cos\theta$. The
algorithm constrains the two-photon mass to the nominal $\piz$
mass and iteratively finds the coefficients of the correction
function. The typical corrections are of order $6 \pm 1\%$. The
uncertainty in the correction is due to systematic uncertainties
in the background estimation and the fitting technique.

At higher energy ($  0.8 < E < 9\gev $) the correction is
estimated from single-photon Monte Carlo simulations. A second
technique using radiative Bhabha events~\cite{radbhabha} is being
developed. The beam energy and the precise track momenta of the
$e^+$ and $e^-$, together with the direction of the radiative
photon, are used to fit the photon energy.  This fitted value is
compared to the measured photon energy to extract correction
coefficients, again as a function of $\ln E$ and $\cos \theta$.

\subsection{Monitoring}
\subsubsection{Environmental Monitoring}

The temperature is monitored by 256 thermal sensors that are
distributed over the calorimeter, and has been maintained at
$20\pm 0.5$\degc.  Dry nitrogen is circulated throughout the
detector to stabilize the relative humidity at $1\pm 0.5$\%.

\subsubsection{Light-Pulser System}

The light response of the individual crystals is measured daily
using a light-pulser system~\cite{kocian,pulser}. Spectrally
filtered light from a xenon flash lamp is transmitted through
optical fibers to the rear of each crystal. The light pulse is
similar in spectrum, rise-time and shape to the scintillation
light in the CsI(Tl) crystals. The pulses are varied in intensity
by neutral-density filters, allowing a precise measurement of the
linearity of light collection, conversion to charge,
amplification, and digitization. The intensity is monitored
pulse-to-pulse by comparison to a reference system with two
radioactive sources, $^{241}\mathrm{Am}$ and $^{148}\mathrm{Gd}$,
that are attached to a small $\CsI$ crystal that is read out by
both a photodiode and a photomultiplier tube. The system is stable
to 0.15\% over a period of one week and has proven to be very
valuable in diagnosing problems. For example, the ability to
accurately vary the light intensity led to the detection of
non-linear response in the electronics~\cite{kocian}.

\subsubsection{Radiation Monitoring and Damage}

\begin{figure}
\centering
\includegraphics[width=7.0cm]{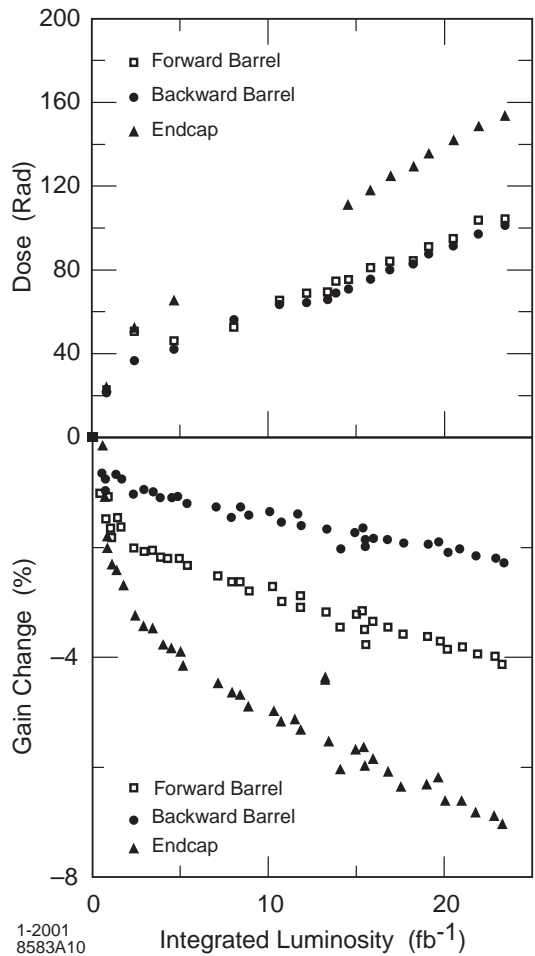}
\vspace{-1.5pc}

\caption{ Impact of beam-generated radiation on the CsI(Tl)
crystals: a) the integrated dose measured with RadFETs placed in
front of the crystals, b) the degradation in light yield measured
with the radioactive-source calibration system.}
\label{calfig::radiation}
\end{figure}

The radiation exposure is monitored by 56 and 60 realtime
integrating dosimeters (RadFETs)~\cite{radfets} placed in front of
the barrel and endcap crystals. In Figure~\ref{calfig::radiation},
the accumulated dose is compared to the observed loss in
scintillation light, separately for the endcap, the forward, and
the backward barrel.  The dose appears to follow the integrated
luminosity, approximately linearly.  The light loss is greatest in
the forward region corresponding to the area of highest integrated
radiation dose. The size of the observed light loss is close to
expectations, based on extensive irradiation tests.

\subsection{Reconstruction Algorithms}
\label{calsec::reco}

A typical electromagnetic shower spreads over many adjacent
crystals, forming a \emph{cluster} of energy deposits. Pattern
recognition algorithms have been developed to efficiently identify
these clusters and to differentiate single clusters with one
energy maximum from merged clusters with more than one local
energy maximum, referred to as a \emph{bumps}. Furthermore, the
algorithms determine whether a bump is generated by a charged or a
neutral particle.

Clusters are required to contain at least one seed crystal with an
energy above 10\mev.  Surrounding crystals are considered as part of
the cluster if their energy exceeds a threshold of 1\mev, or if
they are contiguous neighbors (including corners) of a crystal with at
least 3\mev.  The value of the single crystal threshold is set by the
data acquisition system in order to keep the data volume at an
acceptable level, given the current level of electronics noise and
beam-generated background.  It is highly desirable to reduce this
threshold since fluctuations in the effective energy loss at the edges
of a shower cause a degradation in resolution, particularly at low
energies.

Local energy maxima are identified within a cluster by requiring
that the candidate crystal have an energy, $E_\mathrm{LocalMax}$,
which exceeds the energy of each of its neighbors, and satisfy the
following condition: $ 0.5 (N-2.5) >
E_\mathrm{NMax}/E_\mathrm{LocalMax}$, where $E_\mathrm{NMax}$ is
the highest energy of any of the neighboring $N$ crystals with an
energy above 2\mev.

Clusters are divided into as many bumps as there are local maxima.
An iterative algorithm is used to determine the energy of the
bumps. Each crystal is given a weight, $w_{i}$, and the bump
energy is defined as $E_\mathrm{bump} = \sum_{i} w_{i} E_{i}$,
where the sum runs over all crystals in the cluster.  For a
cluster with a single bump, the result is $w_{i} \equiv 1$.  For a
cluster with multiple bumps, the crystal weight for each bump is
calculated as
\begin{equation}
w_{i} = E_{i} \frac{\exp(-2.5 r_{i}/r_M)}{\sum_{j} E_{j} \exp(-2.5
r_{j}/r_M)},
\end{equation}
where the index $j$ runs over all crystals in the cluster.  $r_M$
refers to the Moli\`{e}re radius, and $r_i$ is the distance of the
$i$th crystal from the centroid of the bump.  At the outset, all
weights are set to one. The process is then iterated, whereby the
centroid position used in calculating $r_{i}$ is determined from the
weights of the previous iteration, until the bump centroid position is
stable to within a tolerance of 1\mm.

The position of a bump is calculated using a center-of-gravity
method with logarithmic, rather than linear
weights~\cite{brabson,otto}, $ W_i = 4.0 + \ln
{E_i/E_\mathrm{bump}}$, where only crystals with positive weights,
\ie\ $E_i > 0.0184 \times E_\mathrm{bump}$, are used in the
calculation. This procedure emphasizes lower-energy crystals,
while utilizing only those crystals that make up the core of the
cluster. A systematic bias of the calculated polar angle
originates from the non-projectivity of the crystals. This bias is
corrected by a simple offset of $-2.6$\mrad\ for $\theta
>90$\degrees\ and $+2.6$\mrad\ for $\theta < 90$\degrees.

A bump is associated with a charged particle by projecting a track to
the inner face of the calorimeter.  The distance between the track
impact point and the bump centroid is calculated, and if it is
consistent with the angle and momentum of the track, the bump is
associated with this charged particle.  Otherwise, it is assumed to
originate from a neutral particle.

On average, 15.8 clusters are detected per hadronic event, of
which 10.2 are not associated with charged particle tracks.  At
current operating conditions, beam-induced background contributes
on average 1.4 neutral clusters with energies above 20\mev.  This
number is significantly smaller than the average number of
crystals with energies above 10\mev\ (see Section~\ref{sec:ir}).

\subsection{Performance}

\subsubsection{Energy Resolution}

\begin{figure}
\centering
\includegraphics[width=6.5cm]{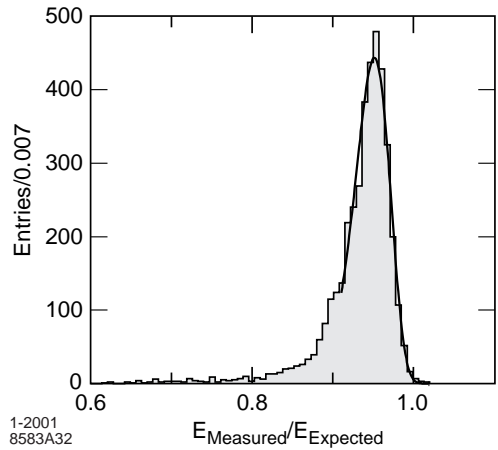}
\vspace{-2pc}
       \caption{The ratio of the EMC measured energy to the expected
       energy for electrons from Bhabha scattering of 7.5\gevc. The
       solid line indicates a fit using a logarithmic function.}
\label{calfig::bhabha}
\end{figure}

\begin{figure}
\centering
\includegraphics[width=6.5cm]{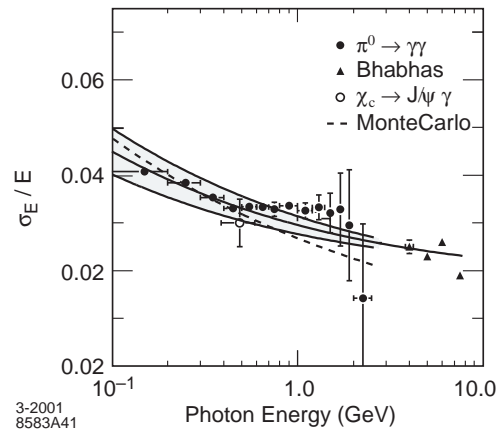}
\vspace{-2pc} \caption{The energy resolution for the ECM measured
for photons and electrons from various processes. The solid curve
is a fit to Equation~\ref{caleqn::res} and the shaded area denotes
the rms error of the fit.} \label{calfig::eres}
\end{figure}

At low energy, the energy resolution of the EMC is measured
directly with the radioactive source yielding $\sigma_{E}/E = 5.0
\pm 0.8\%$ at 6.13\mev\ (see Figure~\ref{calfig::source}).  At
high energy, the resolution is derived from Bhabha scattering,
where the energy of the detected shower can be predicted from the
polar angle of the \epm. The measured resolution is $\sigma_{E}/E
= 1.9 \pm 0.07\%$ at 7.5\gev\ (see Figure~\ref{calfig::bhabha}).
Figure~\ref{calfig::eres} shows the energy resolution extracted
from a variety of processes as a function of energy. Below 2\gev,
the mass resolution of $\piz$ and $\eta$ mesons decaying into two
photons of approximately equal energy is used to infer the EMC
energy resolution ~\cite{pizero}. The decay $\chi_{c1} \to \jpsi
\gamma$ provides a measurement at an average energy of about
500\mev, and measurements at high energy are derived from Bhabha
scattering.  A fit to the energy dependence results in
\begin{equation}
\frac{\sigma_{E}}{E}= \frac{(2.32 \pm 0.30)\%}{^{4}\sqrt{E
(\gev)}} \oplus (1.85 \pm 0.12)\%. \label{caleqn::resfit}
\end{equation}
Values of these fitted parameters are higher than the somewhat
optimistic design expectations, but they agree with detailed Monte
Carlo simulations which include the contributions from electronic
noise and beam background, as well as the impact of the material
and the energy thresholds.

\subsubsection{Angular Resolution}

\begin{figure}
\centering
\includegraphics[width=6.5cm]{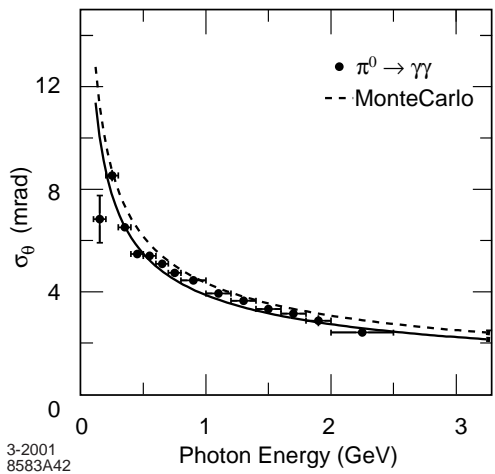}
\vspace{-2pc} \caption{The angular resolution of the EMC for
photons from \piz\  decays. The solid curve is a fit to
Equation~\ref{caleqn::posres}.} \label{calfig::pres}
\end{figure}

The measurement of the angular resolution is based on the analysis
of \piz\ and $\eta$ decays to two photons of approximately equal
energy. The result is presented in Figure~\ref{calfig::pres}.  The
resolution varies between about 12\mrad\ at low energies and
3\mrad\ at high energies.  A fit to an empirical parameterization
of the energy dependence results in
\begin{eqnarray}
\sigma_{\theta} &= &\sigma_{\phi}\nonumber\\
      &= &(\frac{3.87 \pm
0.07}{\sqrt{E (\gev)}}~+~0.00 \pm 0.04)~\mrad.
\end{eqnarray}
These fitted values are slightly better than would be expected
from detailed Monte Carlo simulations.

\subsubsection{$\piz$ Mass and Width}

\begin{figure}
\centering
\includegraphics[width=6.5cm]{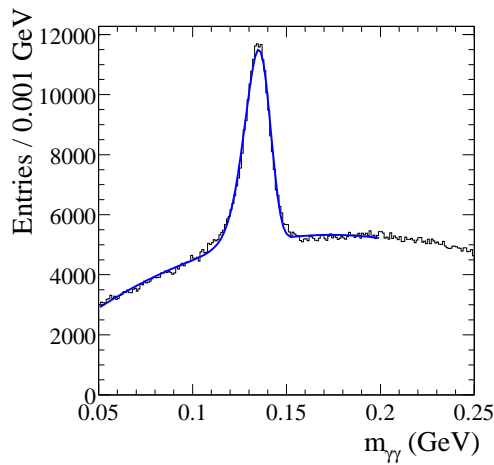}
\vspace{-2pc} \caption{Invariant mass of two photons in \BB\
events. The energies of the photons and the \piz\ are required to
exceed 30\mev\ and 300\mev, respectively. The solid line is a fit
to the data.} \label{calfig::pi0}
\end{figure}

Figure~\ref{calfig::pi0} shows the two-photon invariant mass in
\BB\ events.  The reconstructed $\piz$ mass is measured to be
135.1\mevcc\ and is stable to better than 1\% over the full photon
energy range.  The width of 6.9\mevcc\ agrees well with the
prediction obtained from detailed Monte-Carlo simulations. In
low-occupancy $\tau^{+}\tau^{-}$ events, the width is slightly
smaller, 6.5\mevcc, for \piz\ energies below 1\gev. A similar
improvement is also observed in analyses using selected isolated
photons in hadronic events.

\subsubsection{Electron Identification}

\begin{figure}
\centering
\includegraphics[width=6.5cm]{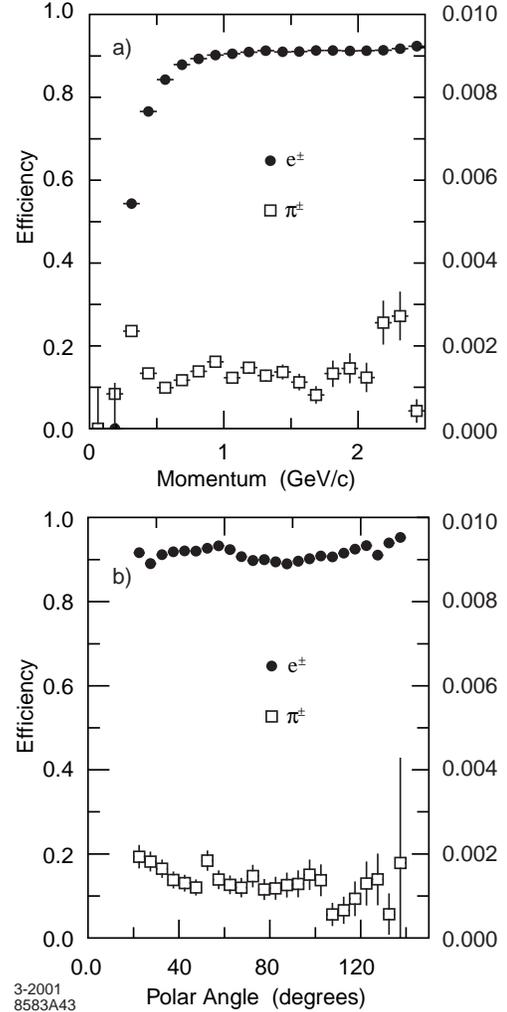}
\vspace{-2pc} \caption{The electron efficiency and pion
mis-identification probability as a function of a) the particle
momentum and b) the polar angle, measured in the laboratory
system.} \label{calfig::electronpure}
\end{figure}

Electrons are separated from charged hadrons primarily on the
basis of the shower energy, lateral shower moments, and track
momentum.  In addition, the \dedx\ energy loss in the DCH and the
DIRC Cherenkov angle are required to be consistent with an
electron.  The most important variable for the discrimination of
hadrons is the ratio of the shower energy to the track momentum
$(E/p)$.  Figure~\ref{calfig::electronpure} shows the efficiency
for electron identification and the pion mis-identification
probability as a function of momentum for two sets of selection
criteria. The electron efficiency is measured using radiative
Bhabhas and $e^+ e^- \to e^+ e^- e^+ e^- $ events. The pion
misidentification probability is measured for selected charged
pions from \KS\ decays and three-prong $\tau$ decays. A tight
(very tight) selector results in an efficiency plateau at 94.8\%
(88.1\%) in the momentum range $0.5 < p < 2\gevc$.  The pion
misidentification probability is of order 0.3\% (0.15\%) for the
tight (very tight) selection criteria.

\subsection{Summary}

The EMC is presently performing close to design
expectations. Improvements in the energy resolution are expected from
the optimization of the feature-extraction algorithms designed to
further reduce the electronics noise.  Modifications to the
electronics should allow for more precise calibrations.  The expected
noise reduction should permit a lower single-crystal readout
threshold. However, this decrease in noise might be offset by an
increase in the beam background that is expected for higher
luminosities and beam currents.

\renewcommand{\secname}{ifr_}
\renewcommand{\sectiondir}{sec09_ifr}
\section{Detector for Muons and Neutral Hadrons}
\label{sec:ifr}

\subsection {Physics Requirements and Goals}

The Instrumented Flux Return (IFR) was designed to identify muons
with high efficiency and good purity, and to detect neutral
hadrons (primarily \KL\ and neutrons) over a wide range of momenta
and angles. Muons are important for tagging the flavor of neutral
\B\ mesons via semi-leptonic decays, for the reconstruction of
vector mesons, like the \jpsi, and for the study of semi-leptonic
and rare decays involving leptons of $B$ and $D$ mesons and $\tau$
leptons. \KL\ detection allows the study of exclusive $B$ decays,
in particular \CP\ eigenstates. The IFR can also help in vetoing
charm decays and improve the reconstruction of neutrinos.

The principal requirements for IFR are large solid angle coverage,
good efficiency, and high background rejection for muons down to
momenta below 1\gevc. For neutral hadrons, high efficiency and
good angular resolution are most important.  Because this system
is very large and difficult to access, high reliability and
extensive monitoring of the detector performance and the
associated electronics plus the voltage distribution are required.

\subsection {Overview and RPC Concept}

The IFR uses the steel flux return of the magnet as a muon filter
and hadron absorber.  Single gap resistive plate chambers
(RPCs)~\cite{bib:santon} with two-coordinate readout have been
chosen as detectors.

\begin{figure*}
\centering
\includegraphics[width=15cm]{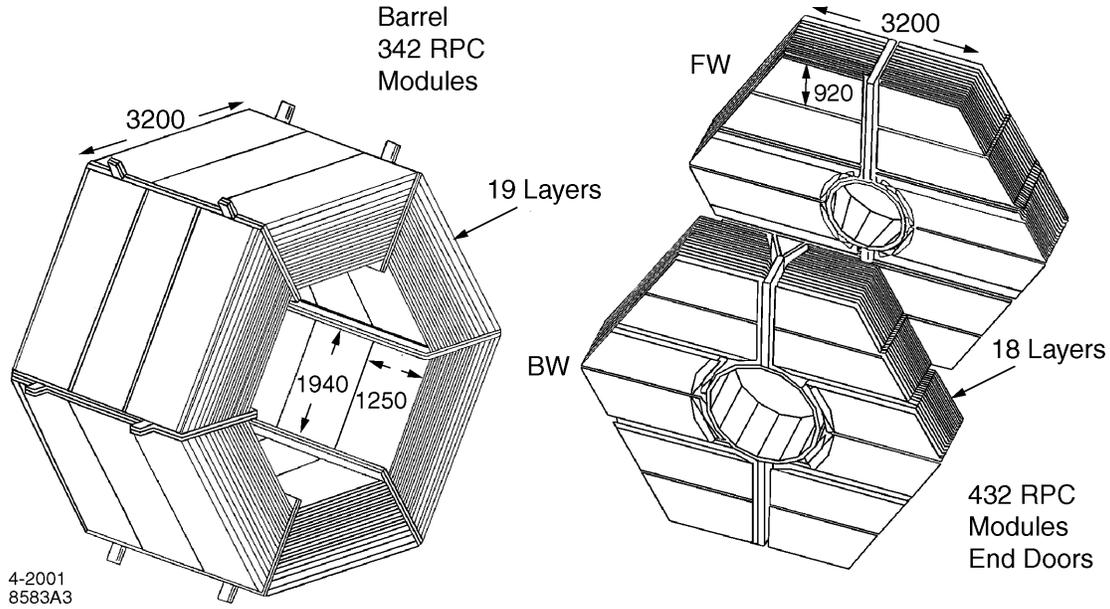}
\vspace{-2pc} \caption{Overview of the IFR: Barrel sectors and
forward (FW) and backward (BW) end doors; the shape of the RPC
modules and their dimensions are indicated.} \label{\secname
fig:ifr_det}
\end{figure*}

The RPCs are installed in the gaps of the finely segmented steel
(see Section \ref{sec:magnet}) of the barrel and the end doors of
the flux return, as illustrated in Figure \ref{\secname
fig:ifr_det}.  The steel segmentation has been chosen on the basis
of Monte Carlo studies of muon penetration and charged and neutral
hadron interactions. The steel is segmented into 18 plates,
increasing in thickness from 2\cm\ for the inner nine plates to
10\cm\ for the outermost plates. The nominal gap between the steel
plates is 3.5\cm\ in the inner layers of the barrel and 3.2\cm\
elsewhere.  There are 19 RPC layers in the barrel and 18 in the
endcaps.  In addition, two layers of cylindrical RPCs are
installed between the EMC and the magnet cryostat to detect
particles exiting the EMC.

RPCs detect streamers from ionizing particles via capacitive
readout strips. They offer several advantages: simple, low cost
construction and the possibility of covering odd shapes with
minimal dead space. Further benefits are large signals and fast
response allowing for simple and robust front-end electronics and
good time resolution, typically 1--2\ns.  The position resolution
depends on the segmentation of the readout; a value of a few \mm\
is achievable.

The construction of the planar and
cylindrical RPCs differ in detail, but they are based on the same
concept. A cross section of an RPC is shown schematically in
Figure~\ref{\secname fig:ifr_f01}.

\begin{figure}
\centering
\includegraphics[width=6.5cm]{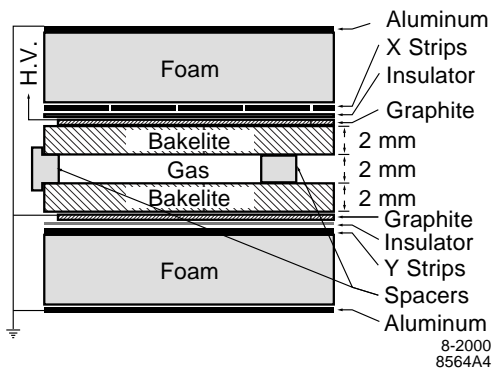}
\vspace{-1.5pc} \caption{Cross section of a planar RPC with the
schematics of the high voltage (HV) connection.} \label{\secname
fig:ifr_f01}
\end{figure}

The planar RPCs consist of two bakelite (phenolic polymer) sheets,
2\mm -thick and separated by a gap of 2\mm. The gap is enclosed at
the edge by a 7\mm\ wide frame.  The gap width is kept uniform by
polycarbonate spacers (0.8\cma) that are glued to the bakelite,
spaced at distances of about 10\cm.  The bulk resistivity of the
bakelite sheets has been especially tuned to $10^{11}$--$10^{12}$\ohm\cm.
The external surfaces are coated with graphite to achieve a
surface resistivity of $\sim 100$\kohm /square.  These two graphite
surfaces are connected to high voltage ($\sim 8$\kv) and ground, and
protected by an insulating mylar film.  The bakelite surfaces facing
the gap are treated with linseed oil.  The RPCs are operated in
limited streamer mode and the signals are read out capacitively, on
both sides of the gap, by external electrodes made of aluminum strips
on a mylar substrate.

The cylindrical RPCs have resistive electrodes made of a special
plastic composed of a conducting polymer and ABS plastic.  The gap
thickness and the spacers are identical to the planar RPCs.  No
linseed oil or any other surface treatments have been applied. The
very thin and flexible electrodes are laminated to fiberglass
boards and foam to form a rigid structure. The copper readout
strips are attached to the fiberglass boards.

\subsection {RPC Design and Construction}

The IFR detectors cover a total active area of about 2,000\m$^2$.
There are a total of 806 RPC modules, 57 in each of the six barrel
sectors, 108 in each of the four half end doors, and 32 in the two
cylindrical layers.  The size and the shape of the modules are
matched to the steel dimensions with very little dead space. More
than 25 different shapes and sizes were built.  Because the size
of a module is limited by   the maximum size of the material
available, \ie\ 320$\times$130\cma\ for the bakelite sheets, two or
three RPC modules are joined to form a gap-size chamber. The
modules of each chamber are connected to the gas system in series,
while the high voltage is supplied separately to each module.

In the barrel sectors, the gaps between the steel plates extend
375\cm\ in the $z$ direction and vary in width from 180\cm\ to
320\cm. Three modules are needed to cover the whole area of the gap,
as shown in Figure~\ref{\secname fig:ifr_det}.  Each barrel module has
32 strips running perpendicular to the beam axis to measure the $z$
coordinate and 96 strips in the orthogonal direction extending over
three modules to measure $\phi$.

Each of the four half end doors is divided into three sections by
steel spacers that are needed for mechanical strength. Each of these
sections is covered by two RPC modules that are joined to form a
larger chamber with horizontal and vertical readout strips.

The readout strips are separated from the ground aluminum plane by a
4\mm-thick foam sheet and form strip lines of 33\ohm\ impedance.
The strips are connected to the readout electronics at one end and
terminated with a 2\kohm\ resistor at the other.  Even and odd
numbered strips are connected to different front-end cards (FECs), so that a
failure of a card does not result in a total loss of signal, since a
particle crossing the gap typically generates signals in two or more
adjacent strips.

The cylindrical RPC is divided into four sections, each covering a
quarter of the circumference.  Each of these sections has four
sets of two single gap RPCs with orthogonal readout strips, the
inner with helical $u$--$v$ strips that run parallel to the
diagonals of the module, and the outer with strips parallel to
$\phi$ and $z$.  Within each section, the strips of the four sets
of RPCs in a given readout plane are connected to form long strips
extending over the whole chamber. Details of the segmentation and
dimensions can be found in Table~\ref{\secname tab:segmentation}.

\begin{table*}[!htb]
\caption{IFR Readout segmentation. The total number of channels is
close to 53,000. }
\vspace{.5\baselineskip}
\centering
\begin{tabular}{lccccccr}   
\hline\hline
\rule{0pt}{12pt}& \# of &  & \# of readout & \#
strips & strip length & strip width & total \#\\
\rule[-5pt]{0pt}{0pt} section& sectors
& coordinate & layers & layer/sect & (cm) & (mm) & channels\\ \hline
\rule{0pt}{12pt} barrel &
6 & $\phi$ & 19 & 96 & 350 & 19.7-32.8 & $\approx 11,000$\\
& & z& 19 & 96 & 190-318 & 38.5 & $\approx 11,000$\\
endcap & 4 & y &18 & 6x32 & 124-262 & 28.3 & 13,824 \\
& & x & 18 & 3x64 & 10-180&38.0 & $\approx 15,000$\\
cylinder & 4 & $\phi$ & 1 & 128 & 370 &16.0 & 512\\
& & z & 1 & 128 & 211 & 29.0 & 512\\
& & u & 1 & 128& 10-422 & 29.0 & 512\\
\rule[-5pt]{0pt}{0pt}& & v & 1 & 128 & 10-423 & 29.0 & 512\\
\hline\hline
\end{tabular}
\label{\secname tab:segmentation}
\end{table*}

Prior to shipment to SLAC, all RPC modules were tested with cosmic rays.
The single rates, dark currents, and efficiency were measured as a
function of HV.  In addition, detailed studies of the efficiency,
spatial resolution, and strip multiplicity were
performed~\cite{bib:test1,bib:test2}.

After the assembly of RPC modules into gap-size chambers, a new series
of cosmic rays tests was performed to assure stable and efficient
operation.  Before the installation of the steel flux return, the
planar chambers were inserted into the gaps.  The cylindrical chambers
were inserted after the installation of the solenoid and the EMC.

For each module, test results and conditions are retained in a
database, together with records of the critical parameters of the
components, the assembly and cabling.  In addition, operational data
are stored, such as the results of the weekly efficiency measurements
that are used in the reconstruction and simulation software.

\subsection {Power and Utilities}

Once the return flux assembly was completed, the
FECs~\cite{bib:fec} were installed and the low (LV) and high
voltage (HV), and the gas system were connected. There are
approximately 3,300 FECs, most  placed inside the steel gaps,
while the remainder was installed in custom crates mounted on the
outside of the steel.

Each FEC is individually connected to the LV power distribution.
The total power required by the entire system is about 8\kw\ at
+7.0\volt\ and 2.5\kw\ at -5.2\volt.  The LV power is supplied by
custom-built switching devices with load and line regulation of
better than 1\%.  Additional features are precision shunts to
measure output currents and TTL logic to inhibit output.

The HV power system is a custom adaptation by CAEN \cite{bib:caen}.
Each HV mainframe can hold up to ten cards, each carrying two
independent 10\kv\ outputs at 1\mA\ and 2\mA. The RPC modules are
connected via a distribution box to the HV supplies.
Each distribution box services six RPC modules and up to six
distribution boxes are daisy-chained to one HV output. Provisions are
made for monitoring the currents drawn by each module.  To reduce
noise, the RPC ground plane is decoupled from the HV power supply
ground by a 100\kohm\ resistor.

The RPCs operate with a non-flammable gas mixture, typically
56.7\% Argon, 38.8\% Freon 134a (1,1,1,2 tetrafluoroethane), and
4.5\% isobutane.  This mixture is drawn from a 760 liter tank that
is maintained at an absolute pressure of 1500--1600~Torr. The
mixing tank is filled on demand with the three component gases
under control of mass-flow meters.  Samples are extracted from the
mixing tank periodically and analyzed to verify the correct
mixture.

The mixed gas is distributed at a gauge pressure of approximately 6.5
Torr through a parallel manifold system of 12.7\mm-diameter copper
tubing.  Each chamber is connected to the manifold through several
meters of 6\mm-diameter plastic tubing (polyamide or Teflon).  The
flow to each of these is adjusted individually with a small multi-turn
metering valve.  Protection against overpressure is provided by an oil
bubbler to atmosphere in parallel with each chamber, limiting the
gauge pressure in the chamber to a maximum of about 1~Torr.  Return
flow of gas from each chamber is monitored by a second oil bubbler
which creates a back pressure of about 0.2~Torr.  The total flow
through the entire system is approximately 5\liter/minute and
corresponds on average to two gas exchanges per day.

\subsection{Electronics}

\begin{figure}
\begin{center}
\includegraphics[width=7.5cm]{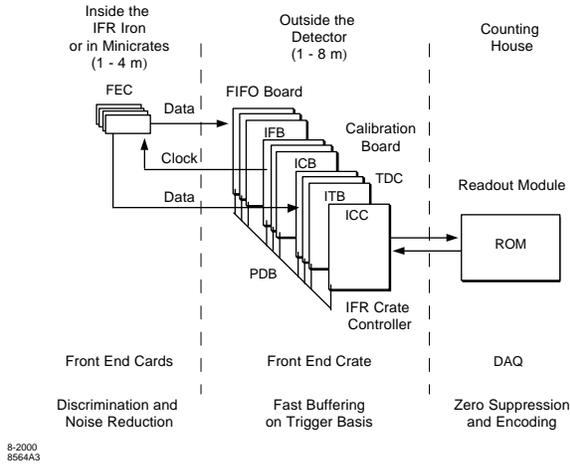}
\vspace{-2.5pc} \caption{Block diagram of the IFR electronics.}
\label{\secname fig:ifr_f05}
\end{center}
\end{figure}

A block diagram of the IFR electronics system \cite{bib:daq} is shown
in Figure~\ref{\secname fig:ifr_f05}.  It includes the FECs, the
data acquisition, and the trigger.

The FECs service 16 channels each. They shape and discriminate the
input signals and set a bit for each strip with a signal above a
fixed threshold.  The input stage operates continuously and is
connected directly to the strips which act as transmission lines.
A fast OR of all FEC input signals provides time information and
is also used for diagnostic purposes. Two types of FECs are
employed to handle inputs of different polarity for signals from
the opposite sides of the gap.  Because of the very low occupancy
there is no provision for buffering during the trigger latency
\cite{bib:fec}.

Signals from 3,300 FECs are transmitted to eight custom IFR
front-end crates that are located near the detector. Each
front-end crate houses up to 16 data handling cards, four trigger
cards and a crate controller card (ICC) that collects data from
the DAQ cards and forwards them to a ROM.  There are three kinds
of data cards: the FIFO boards (IFBs) that buffer strip hits, the
TDC boards (ITBs) that provide time information, and the
calibration boards (ICBs) that inject test pulses into the FECs.
To deliver the data and clock signals to all the boards in the
front-end crate, a custom backplane (PDB) for the standard 6U
Eurocard crate was designed using 9-layer strip line technology.
Each board is connected to the ICC via three point-to-point lines
for three single-end signals (data-in, data-out and clock), all of
the same length and impedance (50\ohm).

The IFB reads the digital hit patterns from the FECs in less than
2.2\ms, stores the data into FIFOs and transfers the FIFO contents
into one of the ROMs.  Each IFB handles 64 FECs, acting as an
acquisition master. It receives commands via the PDB, and
transmits and receives data patterns from the ROM (via G-Link and
ICC). This card operates with the system clock frequency of
59.5\mhz.

The ICB is used for front-end tests and calibrations.  A signal with
programmable amplitude and width is injected into the FEC input stage.
To provide timing calibration and to determine the correct readout
delay, the board is also used together with the TDCs.

The ICC interfaces the crate backplane with the G-Link.  The physical
interface is the Finisar transceiver, a low cost and highly reliable
data link for applications up to 1.5\gbps.

The TDC boards exploit the excellent time resolution of the RPCs.
Each board has 96 ECL differential input channels for the fast OR
signals from the FECs. Time digitization is achieved by three custom
TDCs, designed at CERN \cite{bib:cern}. Upon receipt of a L1
\emph{Accept}, data are selected and stored until readout by the ROM.
The 59.5\mhz\ clock signal is synchronized with the data and
distributed to the 16 boards.  High performance drivers provide a
reliable clock distribution with a jitter of less than 0.5\ns.

\subsection {Slow Controls\\ and Online Monitoring }

The IFR is a system with a large number of components and electronics
distributed all over the \babar\ detector. To assure safe and stable
operation, an extensive monitoring and control system was installed.
The IFR Online Detector Control (IODC) monitors the performance of
the RPCs by measuring the singles counting rate and the dark current
of every module. It also controls and monitors the operation of the
electronics, the DAQ and trigger, as well as the LV, the HV, and the
gas system.  The total number of hardware channels
is close to 2,500~\cite{bib:iodc1}.

The system has been easy to operate. HV trips are rare.  Temperature
monitoring in the steel structure and the electronics crates has
proven very useful for the diagnosis of operational problems.  The
occupancy is extremely low everywhere, except in layer 18 of the
forward end door which lacks adequate shielding from machine-generated
background.  On average, there are about 100--150 strip signals per event.

\subsection{Efficiency Measurements\\ and Performance}

The efficiency of the RPCs is evaluated  both for normal collision
data and for cosmic ray muons recorded with the IFR trigger. Every
week, cosmic ray data are recorded at different voltage settings
and the efficiency is measured chamber-by-chamber as a function of
the applied voltage. The absolute efficiency at the nominal
working voltage (typically 7.6\kv) is stored in the database for
use in the event reconstruction software.

To calculate the efficiency in a given chamber, nearby hits in a
given layer and hits in different layers are combined to form
clusters.  Two different algorithms are used. The first is based
solely on the IFR information and uses data recorded with a
dedicated IFT trigger; the second matches the IFR clusters with
the tracks reconstructed in the DCH.  Both these algorithms start
from one-dimensional IFR clusters defined as a group of adjacent
hits in one of the two readout coordinates.  The cluster position
is defined as the centroid of the strips in the cluster.  In the
first algorithm, two-dimensional clusters are formed by joining
one-dimensional clusters (of the same readout coordinate) in
different layers. In each sector, two-dimensional clusters in
different coordinates are combined into three-dimensional clusters
provided there are fewer than three layers missing in one of the
two coordinates.  The second algorithm extrapolates charged tracks
reconstructed by the DCH. IFR clusters which are less than 12\cm\
from the extrapolated track are combined to form three-dimensional
or two-dimensional clusters.  A detailed discussion of the
clustering algorithm can be found elsewhere~\cite{bib:luca}.

The residual distributions from straight line fits to two-dimensional
clusters typically have an rms width of less than 1\cm. An RPC is
considered efficient if a signal is detected at a distance of less
than 10\cm\ from the fitted straight line in either of the two
readout planes.
Following the installation and commissioning of the IFR system, all
RPC modules were tested with cosmic rays and their efficiency was
measured.  The results are presented in Figure~\ref{\secname
fig:ifr_f06}.  Of the active RPC modules, 75\% exceed an efficiency of
90\%.

\begin{figure}
\centering
\includegraphics[width=6.5cm]{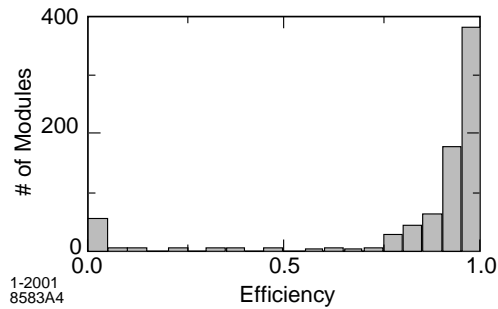}
\vspace{-2pc} \caption{Distribution of the efficiency for all
RPC modules measured with cosmic rays in June 1999. Some 50
modules were not operational at that time.} \label{\secname
fig:ifr_f06}
\end{figure}

Early tests indicated that the RPC dark current was very
temperature dependent, specifically, the current increases
14--20\% per \degc. Because the IR experimental hall does not have
temperature regulation this presents a serious problem. The FECs
that are installed in the steel gaps dissipate 3\watt\ each,
generating a total power of 3.3\kw\ in the barrel and 1.3\kw\ in
the forward end door.

During the first summer of operation, the daily average temperature in
the IR hall was 28\degc\ and the maximum hall temperature frequently
exceeded 31\degc. The temperature inside the steel rose to more than
37\degc\ and the dark currents in many modules exceeded the
capabilities of the HV system and some RPCs had to be temporarily
disconnected.

To overcome this problem, water cooling was installed on the
barrel and end door steel, removing $\approx$10\kw\ of heat and
stabilizing the temperature at 20--21\degc\ in the barrel,
22\degc\ in the backward and 24\degc\ in the forward end doors.
Figure~\ref{\secname fig:history} shows the history of temperature
in the hall and temperature and total dark current in the backward
end door. While the current closely follows the temperature
variations, the range of change is now limited to a few degrees.

\begin{figure}
\begin{center}
\includegraphics[width=7.0cm]{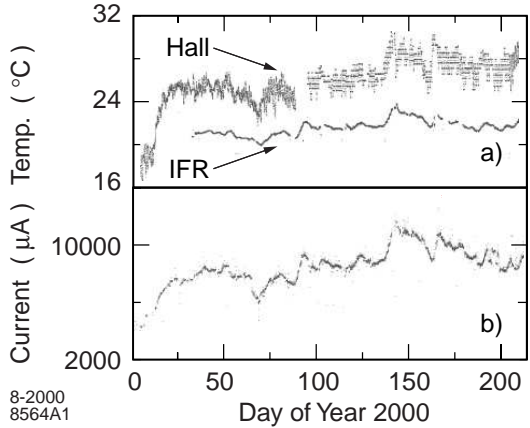}
\vspace{-2pc} \caption{History of the temperature and dark current
in the RPC modules since January 2000. a) temperature in the IR-2
hall and in the backward end door; b) total dark current in the
216 modules of the backward end door.} \label{\secname
fig:history}
\end{center}
\end{figure}

During operation at high temperatures, a large fraction of the
RPCs ($>$50\%) showed not only very high dark currents, but also
some reduction in efficiency compared to earlier
measurements~\cite{bib:zallo}. After the cooling was installed and
the RPCs were reconnected, some of them continued to deteriorate
while others remained stable, some of them ($>30$\%) at full
efficiency. (see Figure~\ref{\secname fig:ifr_f08}).  Detailed
studies revealed large regions of very low efficiency in these
modules, but no clear pattern was identified.

\begin{figure}
\centering
\includegraphics[width=6.5cm]{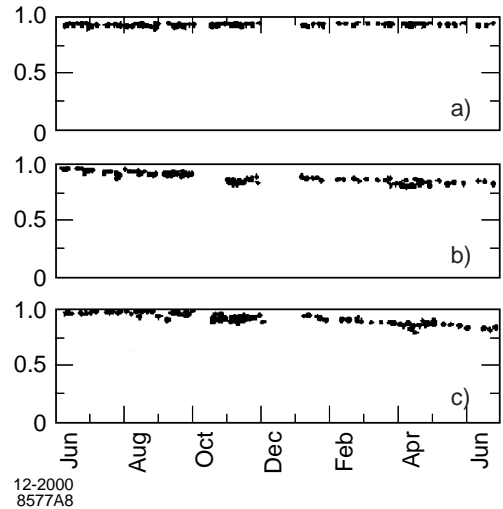}
\vspace{-2pc}
\caption{Efficiency history for 12 months starting in June 1999
for RPC modules showing different performance: a) highly efficient
and stable; b) continuous slow decrease in efficiency; c) more
recent, faster decrease in efficiency.} \label{\secname
fig:ifr_f08}
\end{figure}

The cause of the efficiency loss remains under investigation.
Several possible causes have been excluded as the primary source
of the problem, such as a change in the bakelite bulk resistivity,
loosened spacers, gas flow, or gas composition.  A number of
prototype RPCs developed similar efficiency problems after being
operated above a temperature 36\degc\ for a period of two weeks.
In some of these modules, evidence was found that the linseed oil
had failed to cure and had accumulated at various spots under the
influence of the electric field.

\subsection{ Muon Identification}

While muon identification relies almost entirely on the IFR, other
detector systems provide complementary information. Charged particles
are reconstructed in the SVT and DCH and muon candidates are required
to meet the criteria for minimum ionizing particles in the EMC.
Charged tracks that are reconstructed in the tracking systems are
extrapolated to the IFR taking into account the non-uniform magnetic
field, multiple scattering, and the average energy loss.  The projected
intersections with the RPC planes are computed and for each readout
plane all clusters detected within a predefined distance from the
predicted intersection are associated with the track.

A number of variables are defined for each IFR cluster associated
with a charged track to discriminate muons from charged hadrons:
1) the total number of interaction lengths traversed from the IP
to the last RPC layer with an associated cluster, 2) the
difference between this measured number of interaction lengths and
the number of interaction lengths predicted for a muon of the same
momentum and angle, 3) the average number and the rms of the
distribution of RPC strips per layer, 4) the $\chi^2$ for the
geometric match between the projected track and the centroids of
clusters in different RPC layers, and 5) the $\chi^2$ of a
polynomial fit to the two-dimensional IFR clusters. Selection
criteria based on these variables are applied to identify muons.

The performance of muon identification has been tested on samples
of muons from $\mu\mu e e$ and $\mu\mu\gamma$ final states and
pions from three-prong $\tau$ decays and $K_S \rightarrow
\pi^+\pi^-$ decays. The selection of these control samples is
based on kinematic variables, and not on variables used for muon
identification. As illustrated in Figure~\ref{\secname fig:muEff},
a muon detection efficiency of close to 90\% has been achieved in
the momentum range of $1.5<p<3.0$\gevc\ with a fake rate for pions
of about 6--8\%.  Decays in flight contribute about 2\% to the
pion misidentification probability.  The hadron misidentification
can be reduced by a factor of about two by tighter selection
criteria which lower the muon detection efficiency to about 80\%.

\begin{figure}
\begin{center}
\includegraphics[width=6.5cm]{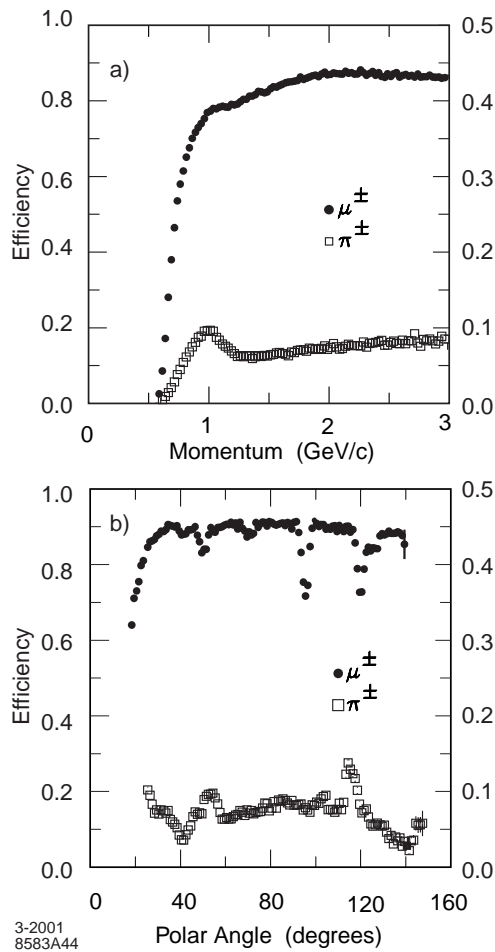}
\vspace{-2pc}
\caption{Muon efficiency (left scale) and pion misidentification
probability (right scale) as a function of a) the laboratory track
momentum, and b) the polar angle (for $1.5<p<3.0$\gevc\ momentum),
obtained with loose selection criteria.} \label{\secname
fig:muEff}
\end{center}
\end{figure}

\subsection{\KL\ and Neutral Hadron Detection}

\KL's and other neutral hadrons interact in the steel of the IFR
and can be identified as clusters that are not associated with a
charged track.  Monte Carlo simulations predict that about 64\% of
$K_L$'s above a momentum of 1\gevc\ produce a cluster in the
cylindrical RPC, or a cluster with hits in two or more planar RPC
layers.

Unassociated clusters that have an angular separation of $\leq 0.3$~rad
are combined into a composite cluster, joining clusters that originate
from showers that spread into adjacent sectors of the barrel, several
sections of the end doors and/or the cylindrical RPC.  This procedure
also combines multiple clusters from large fluctuations in the
hadronic showers. The direction of the neutral hadron is determined
from the event vertex and the centroid of the neutral cluster.  No
information on the energy of the cluster can be obtained.

Since a significant fraction of hadrons interact before reaching
the IFR, information from the EMC and the cylindrical RPCs is
combined with the IFR cluster information. Neutral showers in the
EMC are associated with the neutral hadrons detected in the IFR,
based on a match in production angles. For a good match, a
$\chi^2$ probability of $\ge 1\%$ is required.

An estimate of the angular resolution of the neutral hadron
cluster can be derived from a sample of \KL's produced in the
reaction $\epem \to \phi \gamma \to \KL \KS \gamma$. The \KL\
direction is inferred from the missing momentum computed from the
measured particles in the final state. The data in Figure
\ref{\secname fig:ifr_f13} indicate that the angular resolution of
the \KL\ derived from the IFR cluster information is of the order
of 60\mrad.  For \KL's interacting in the EMC, the resolution is
better by about a factor of two.

\begin{figure}
\centering
\includegraphics[width=6.5cm]{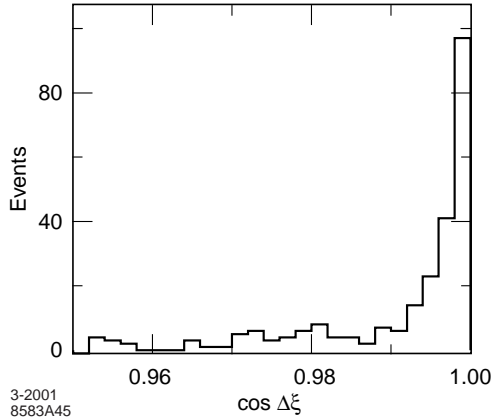}
\vspace{-2pc}
\caption{Angular difference, $\cos \Delta \xi$, between the
direction of the missing momentum and the closest neutral IFR
cluster for a sample of $\phi$ mesons produced in the reaction
$\epem \to \phi \gamma$ with $\phi \to \KL \KS$. } \label{\secname
fig:ifr_f13}
\end{figure}

\begin{figure}
\centering
\includegraphics[width=6.5cm]{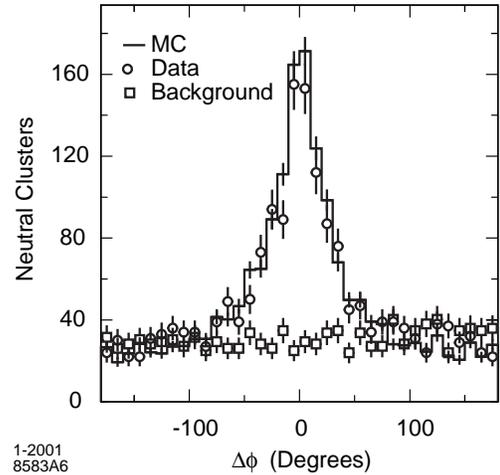}
\vspace{-2pc}

\caption{Difference between the direction of the reconstructed
neutral hadron cluster and the missing transverse momentum  in
events with a reconstructed \jpsi\ decay. The Monte Carlo
simulation is normalized to the luminosity of the data; the
background is obtained using neutral hadrons and the missing
momentum from different events.} \label{\secname fig:ifr_f14}
\end{figure}

For multi-hadron events with a reconstructed \jpsi\ decay,
Figure~\ref{\secname fig:ifr_f14} shows the angular difference,
$\Delta\phi$, between the missing momentum and the direction of the
nearest neutral hadron cluster. The observed peak demonstrates clearly
that the missing momentum can be associated with a neutral hadron,
assumed to be a \KL.  The \KL\ detection efficiency increases roughly
linearly with momentum; it varies between 20\% and 40\% in the
momentum range from 1\gevc\ to 4\gevc\ (EMC and IFR combined).

\subsection{Summary and Outlook}

The IFR is the largest RPC system built to date.  It provides
efficient muon identification and allows for the detection of
\KL's interacting in the steel and the calorimeter.  During the
first year of operation, a large fraction of the RPC modules have
suffered significant losses in efficiency.  This effect appears to
be correlated with high temperatures, but the full extent of the
problem and its cause remain under study.  Thanks to the large
number of RPC layers, this problem has not yet impacted the
overall performance severely.  But present extrapolations, even
after installation of water cooling on the steel, indicate a
severe problem for the future operation. Recently, 24 end door
modules have been replaced by new RPCs with a substantially
thinner coating of linseed oil and improved treatment of the
bakelite surfaces.  Results with these new RPCs and other tests
will need to be evaluated before decisions on future improvements
of the IFR can be made. Furthermore, it is planned to reduce the
contamination from hadron decays and punch through by increasing
the absorber thickness, \ie\ adding more steel on the outside and
replacing a few of the RPCs with absorber plates.

\renewcommand{\secname}{trg_}
\renewcommand{\sectiondir}{sec10_trg}
\section{Trigger}
\label{\secname sec:trg}

\subsection{Trigger Requirements}

The basic requirement for the trigger system is the selection of
events of interest (see Table~\ref{\secname tab:PhysRates}) with a
high, stable, and well-understood efficiency while rejecting
background events and keeping the total event rate under 120\hz.
At design luminosity, beam-induced background rates are typically
about 20\khz\ each for one or more tracks in the drift chamber
with $\pxy > 120$\mevc\ or at least one EMC cluster with
$E>100$\mev. Efficiency, diagnostic, and background studies
require prescaled samples of special event types, such as those
failing the trigger selection criteria, and random beam crossings.

The total trigger efficiency is required to exceed 99\% for all
$\BB$ events and at least 95\% for continuum events.  Less
stringent requirements apply to other event types, \eg\ $\tautau$
events should have a 90-95\% trigger efficiency, depending on the
specific $\tau^\pm$ decay channels.

The trigger system must be robust and flexible in order to function
even under extreme background situations. It must also be able to
operate in an environment with dead or noisy electronics channels.
The trigger should contribute no more than 1\% to dead time.

\begin{table}[b]
\centering

\caption{Cross sections, production and trigger rates
for the principal physics processes at 10.58\gev\ for a luminosity
of $3\times 10^{33}\cms$. The \epem\ cross section refers to
events with either the $e^+$, $e^-$, or both inside the EMC
detection volume.} \label{\secname tab:PhysRates} \vspace{.2cm}

\vspace{.5\baselineskip}

\small
\begin{tabular}{lr@{.}lr@{.}lr@{.}l}\hline
&\multicolumn{2}{c}{Cross\rule{0pt}{12pt}} &\multicolumn{2}{c}{Production}
&\multicolumn{2}{c}{Level~1}\\
Event  &\multicolumn{2}{c}{section} &\multicolumn{2}{c}{Rate}
&\multicolumn{2}{c}{Trigger}\\
type   &\multicolumn{2}{c}{(nb)}
&\multicolumn{2}{c}{(Hz)\rule[-5pt]{0pt}{0pt}}
&\multicolumn{2}{c}{Rate (Hz)}\\ \hline
\bbbar \rule{0pt}{12pt}      &1&1          &3&2        &3&2\\
other \qqbar &3&4          &10&2       &10&1\\
\epem        &\multicolumn{2}{c}{$\sim$53}
&\multicolumn{2}{c}{\hphantom{3}159}
&\multicolumn{2}{c}{\hphantom{3}156}\\
\mumu        &1&2          &3&5        &3&1\\
\tautau      &\hphantom{33}0&9
&\hphantom{3333}2&8\rule[-5pt]{0pt}{0pt}        &\hphantom{3333}2&4\\
\hline
\end{tabular}
\normalsize
\end{table}

\subsection{Trigger Overview}

The trigger is implemented as a two-level hierarchy, the Level~1
(L1) in hardware followed by the Level~3 (L3) in software. It is
designed to accommodate up to ten times the initially projected
\cite{ref:R_TDR} \pep2 background rates at design luminosity and
to degrade slowly for backgrounds above that level. Redundancy is
built into the system to measure and monitor trigger efficiencies.

During normal operation, the L1 is configured to have an output
rate of typically 1\khz.  Triggers are produced within a
fixed latency window of 11--12\mus\ after the \epem collision, and
delivered to the Fast Control and Timing System (FCTS).  Data used
to form the trigger decision are preserved with each event for
efficiency studies.

The L3 receives the output from L1, performs a second stage rate
reduction for the main physics sources, and identifies and flags
the special categories of events needed for luminosity
determination, diagnostic, and calibration purposes.  At design
luminosity, the L3 filter acceptance for physics is $\sim$90\hz,
while $\sim$30\hz\ contain the other special event categories. The
L3 algorithms comply with the same software conventions and
standards used in all other \babar\ software, thereby simplifying
its design, testing, and maintenance.

\subsection{Level~1 Trigger System}

The L1 trigger decision is based on charged tracks in the DCH
above a preset transverse momentum, showers in the EMC, and tracks
detected in the IFR.  Trigger data are processed by three
specialized hardware processors. As described below, the drift
chamber trigger (DCT) and electromagnetic calorimeter trigger
(EMT) both satisfy all trigger requirements independently with
high efficiency, and thereby provide a high degree of redundancy,
which enables the measurement of trigger efficiency. The
instrumented flux return trigger (IFT) is used for triggering
$\mumu$ and cosmic rays, mostly for diagnostic purposes.

\begin{figure}
\centering
\includegraphics[width=7.5cm]{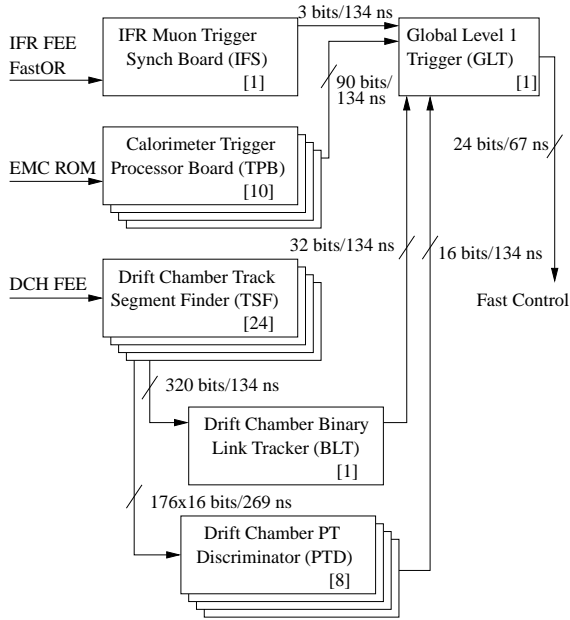}

\vspace{-2pc} \caption{Simplified L1 trigger schematic. Indicated
on the figure are the number of components (in square brackets),
and the transmission rates between components in terms of total
signal bits.} \label{\secname fig:trg-L1sys}
\end{figure}

The overall structure of the L1 trigger is illustrated in
Figure~\ref{\secname fig:trg-L1sys}.  Each of the three L1 trigger
processors generates trigger \emph{primitives}, summary data on
the position and energy of particles, that are sent to the global
trigger (GLT) every 134\ns.  The DCT and EMT primitives sent to
the GLT are \emph{$\phi$-maps}.  An individual $\phi$-map consists
of an \textit{n}-bit word representing a particular pattern of
\emph{trigger objects} as distributed in fixed-width $\phi$
regions from $0$ to $2\pi$.  A trigger object is a quantity
indicating the presence of a particle, such as a drift chamber
track or a calorimeter energy deposit.  The IFT primitive is a
three-bit pattern representing the hit topology in the IFR.  The
meaning of the various trigger primitive inputs to the GLT are
summarized in Table~\ref{\secname tab:trg-primitive}.

The GLT processes all trigger primitives to form specific triggers
and then delivers them to the FCTS.  The FCTS can optionally mask
or prescale any of these triggers.  If a valid trigger remains, a
\emph{L1 Accept} is issued to initiate event readout.  The trigger
definition logic, masks, and prescale values are all configurable
on a per run basis.

The L1 hardware is housed in five 9U VME crates.  The L1 trigger
operates in a continuous sampling mode, generating trigger
information at regular, fixed time intervals. The DCH front-end
electronics (FEEs) and the EMC \emph{untriggered personality
cards} (UPCs) send raw data to the DCT and EMT about 2\mus\ after
the \epem\ collision. The DCT and EMT event processing times are
4--5\mus, followed by another $\sim$3\mus\ of processing in the
GLT to issue a L1 trigger. The L1 trigger takes approximately
1\mus\ to propagate through the FCTS and the \emph{readout
modules} (ROMs) to initiate event readout.  These steps are all
accomplished within the 12.8\mus\ FEE buffer capacity limit.

\begin{table*}
\caption{Trigger primitives for the DCT and EMT. Most energy
thresholds are adjustable; those listed are typical values.}
\label{\secname tab:trg-primitive} \vspace{.2cm}
\centering
\begin{tabular}{lllcc}
\hline\hline
\rule[-5pt]{0pt}{17pt}& Description & Origin & No. of bits & Threshold
\\ \hline
B\rule{0pt}{12pt}    & Short track reaching DCH superlayer 5 & BLT & 16
& 120\mevc \\
A    & Long track reaching DCH superlayer 10 & BLT &16 & 180\mevc \\
A$^\prime$\rule[-5pt]{0pt}{0pt} & High \pt\ track & PTD & 16 &800\mevc \\ \hline
M\rule{0pt}{12pt}    & All-$\theta$ MIP energy & TPB & 20 &100\mev  \\
G    & All-$\theta$ intermediate energy & TPB & 20 &250\mev  \\
E    & All-$\theta$ high energy & TPB & 20 & 700\mev\\
X    & Forward endcap MIP & TPB & 20 & 100\mev  \\
Y    &\rule[-5pt]{0pt}{0pt}Backward barrel high energy & TPB & 10 & 1\gev    \\
\hline\hline
\end{tabular}
\end{table*}

The DCT, EMT and GLT each maintain a four-event buffer to hold
information resulting from the various stages of the L1 trigger.
These data are read out by the normal data acquisition system.

\subsubsection{Level~1 Drift Chamber Trigger}

The input data to the DCT consist of one bit for each of the 7104
DCH cells.  These bits convey time information derived from the
sense wire signal for that cell.  The DCT output primitives are
candidate tracks encoded in terms of three 16-bit $\phi$-maps as
listed in Table~\ref{\secname tab:trg-primitive}.

The DCT algorithms are executed in three types of
modules~\cite{ref:dct}.  First, track segments, their $\phi$ positions
and drift time estimates are found using a set of 24 Track Segment
Finder (TSF) modules~\cite{ref:tsf}.  These data are then passed to
the Binary Link Tracker (BLT) module~\cite{ref:blt}, where segments
are linked into complete tracks.  In parallel, the $\phi$ information
for segments found in axial superlayers is transmitted to eight
transverse momentum discriminator (PTD) modules~\cite{ref:ptd}, which
search for tracks above a set
\pt\ threshold.

Each of the three DCT modules (TSF, BLT, and PTD) relies heavily on
multiple FPGA's~\cite{ref:ORCA} which perform the control and
algorithmic functions.  All cabling is handled by a small (6U)
back-of-crate interface behind each main board.

\paragraph{Track Segment Finder}

\begin{figure}
\centering
\includegraphics[width=7.5cm]{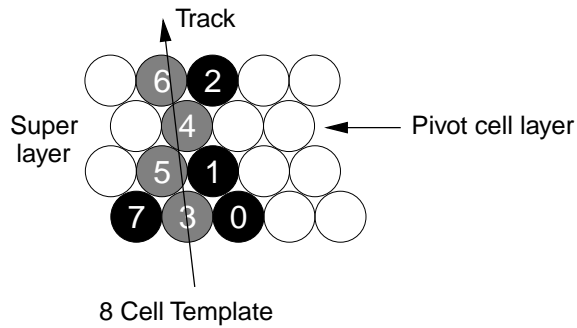}
\vspace{-2.5pc}
\caption{Track Segment Finder pivot group.}
\label{\secname fig:pivotcell}
\end{figure}

The TSF modules are responsible for finding track segments in 1776
overlapping eight-cell \emph{pivot groups}.  A pivot group is a
contiguous set of cells that span all four layers within a superlayer.
The pivot group shape is such that only reasonably straight tracks
originating from the interaction point can produce a valid segment.
Figure~\ref{\secname fig:pivotcell} shows the arrangement of cells
within a pivot group.  Cell 4 is called the \emph{pivot cell}; the TSF
algorithm is optimized to find track segments that \emph{pivot} about
this cell.

The DCH signals are sampled every 269\ns.  The passage of a single
particle through the DCH will produce ionization that drifts to the
sense wires in typically no more than four of these clock ticks.  Each
cell is associated with a two-bit counter that is incremented at every
clock tick for which a signal is present.  In this way, a short time
history of each cell is preserved.  For each clock tick, the
collection of two-bit counters for each pivot group forms a 16-bit
value used to address a look-up-table.  This look-up-table contains
two-bit weights indicating whether there is no acceptable segment, a
low-quality segment, a three-layer segment (allowing for cell
inefficiencies), or a four-layer segment.  When an acceptable segment
is found, that pivot group is examined to determine which of three
subsequent clock ticks produce the highest weight or \emph{best}
pattern.

The look-up-table also contains position and time information which,
along with a summary of cell occupancies, forms the basis of data sent
to the BLT and PTD.  The TSF algorithm uses the time-variation of the
look-up-table weights to refine both the event time and its
uncertainty, thus enabling it to output results to the BLT every
134\ns.

The position resolution as measured from the data after
calibration, is typically $\sim$600\mum\ for a four-layer segment
and $\sim$900\mum\ for a three-layer segment. For tracks
originating from the IP, the efficiency for finding TSF segments
is 97\%, and the efficiency for high-quality three-layer or
four-layer TSF segments is 94\%.

\paragraph{Binary Link Tracker}

The BLT receives segment hit information from all 24 TSF's at a
rate of 320 bits every 134\ns\ and links them into complete
tracks.  The segment hits are mapped onto the DCH geometry in
terms of 320 \emph{ supercells}, 32 sectors in $\phi$ and ten
radial superlayers.  Each bit indicates whether a segment is found
in that supercell or not. The BLT input data are combined using a
logical OR with a programmable mask pattern. The masking allows
the system to activate track segments corresponding to dead or
highly inefficient cells to prevent efficiency degradation. The
linking algorithm uses an extension of a method developed for the
CLEO-II trigger~\cite{ref:CLEO-BLT}. It starts from the innermost
superlayer, A1, and moves radially outward.

Tracks that reach the outer layer of the DCH (superlayer A10) are
classified as type~A. Tracks that reach the middle layer
(superlayer U5) are classified as type~B. An A track is found if
there is a segment in at least eight superlayers and if the
segments in two consecutive superlayers fall azimuthally within
three to five supercells of each other (depending on the
superlayer type). This allows for track curvature and dip angle
variations.    The data are compressed and output to the GLT in
the form of two 16-bit $\phi$-maps, one each for A and B tracks.

\paragraph{PT Discriminator}

The eight PTD modules receive $\phi$ information of high quality track
segments in the axial superlayers (A1, A4, A7 and A10), and determine
if the segments are consistent with a track \pt\ greater than a
configurable minimum value.  An envelope for tracks above the minimum
\pt\ is defined using the IP, and a track segment position in one of
the \emph{seed} superlayers, A7 or A10.  A high \pt\ candidate, denoted
as A$^\prime$, is identified when sufficient track segments with
accurate $\phi$ information from the other axial superlayers lie
within this envelope.

\begin{figure}
\centering
\includegraphics[width=7cm]{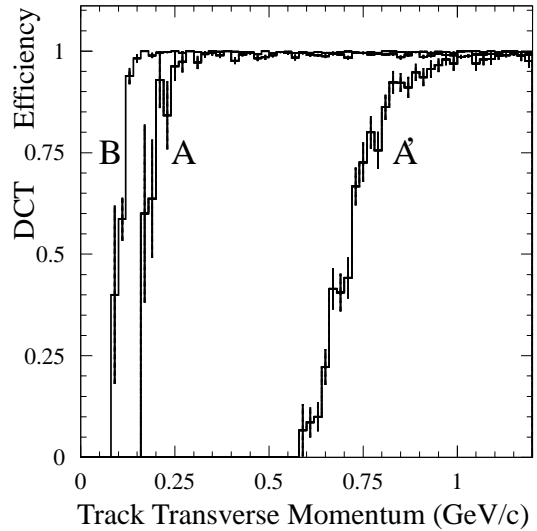}

\vspace{-2pc}
\caption{DCT track efficiency versus transverse momentum for A, B,
and A$^\prime$ tracks.  The A$^\prime$ threshold is set to 800\mevc.}
\label{\secname fig:dct-ptturnon}
\end{figure}

Each PTD module searches for seed segments in superlayers A7 and A10,
and within a 45-degree azimuthal wedge of the DCH.  This search region
spans eight supercells, and the processing for each supercell is
performed by its own processing engine on the PTD.  The principal
components in each engine are an algorithmic processor and
look-up-tables containing the limits for each individual seed
position. The contents of the look-up-tables specify the allowed track
segment positions for each of the three other axial superlayers and
consequently define the effective
\pt\ discrimination threshold.  The resulting \pt\ threshold for the PTD
A$^\prime$ tracks is shown in Figure~\ref{\secname fig:dct-ptturnon}
together with the BLT A, B track efficiency.

\subsubsection{Level~1 Calorimeter Trigger}

For trigger purposes, the EMC is divided into 280 towers, $7\times40$
($\theta\times\phi$).  Each of the barrel's 240 towers is composed of
24 crystals in a $8\times3$ ($\theta\times\phi$) array.  The endcap is
divided into 40 towers, each forming a wedge in $\phi$ containing
19--22 crystals.  For each tower, all crystal energies above a
threshold of 20\mev\ are summed and sent to the EMT every 269\ns.

The conversion of the tower data into the GLT $\phi$-maps is performed
by ten Trigger Processor Boards (TPBs). The TPBs determine energies in
the 40 $\phi$ sectors, summing over various ranges of $\theta$,
compare these energies against thresholds for each of the trigger
primitives (see Table~\ref{\secname tab:trg-primitive}), estimate the
time of energy deposition, correct for timing jitter, and then
transmit the result to the GLT.

Each TPB receives data from 28 towers, corresponding to an array of $7
\times 4$ in $\theta\times\phi$, or four \emph{$\phi$-sectors}.  Each
of the 40 $\phi$-sectors is summed independently.  To identify energy
deposits that span two adjacent $\phi$-sectors, the energy of each sector
is also made available to the summing circuit for a single adjoining
sector in such a way that all possible pairs of adjacent $\phi$-sectors
are summed.  These energy sums are compared against thresholds to form
trigger objects.  Each sum is also sent to an eight-tap finite impulse
response (FIR) digital filter which is used to estimate the energy
deposition time.  A look-up-table is used to make an energy-dependent
estimate of the timing jitter which, along with the FIR output, is used to
time the transmission of any trigger objects to the GLT.  Pairs of
$\phi$-sectors are ORed to form 20-bit $\phi$-maps for the M, G, E, and X
primitives, while for the Y primitive, groups of four are ORed to form a
10-bit $\phi$-map.  The complete algorithm is implemented in one
FPGA~\cite{ref:Xilinx} for each $\phi$-sector, with four
identical components per TPB.  Further details of the EMT system can be
found in~\cite{ref:EMT}.

\begin{figure}
\centering
\includegraphics[width=7cm]{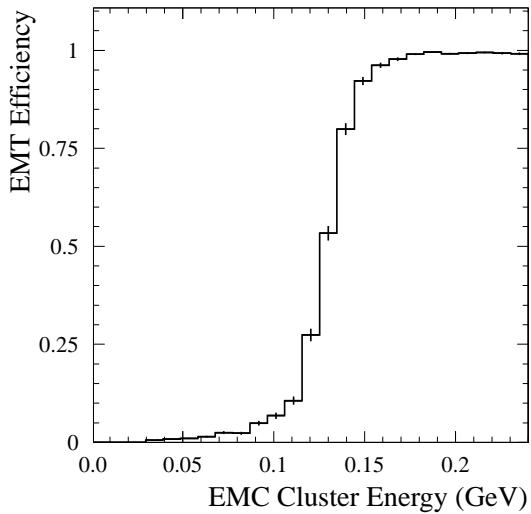}
\vspace{-2pc}
\caption{EMT M efficiency vs. EMC cluster energy for an M threshold
setting of 120\mev.}
\label{\secname fig:trg-emt1}
\end{figure}

The basic performance of the EMT can be expressed in terms of the
efficiency and timing jitter of the trigger primitives.  The
efficiency of the primitives can be measured by the number of times a
trigger bit is set for a specific energy reconstructed offline in
events from a random trigger.  Figure~\ref{\secname fig:trg-emt1}
shows this efficiency for energies near the M threshold.  The
efficiency changes from 10\% to 90\% in the range of 110 to 145\mev,
and reaches 99\% at 180\mev, close to the average energy deposition of
a minimum ionizing particle at normal incidence.

The EMT time jitter is measured by comparing the time centroid of
$\phi$-strip M hits in $\mumu$ events with the DCH track start time,
$\t_0$. The difference has an rms width of 90\ns\ with $>$99.9\% of the
matching M hits within a $\pm$500\ns\ window.

\subsubsection{Level~1 IFR Trigger}

The IFT is used for triggering on $\mumu$ and cosmic rays.  For the
purposes of the trigger, the IFR is divided into ten sectors, namely
the six barrel sextants and the four half end doors.  The inputs to
the IFT are the \emph{Fast OR} signals of all $\phi$ readout strips in
eight selected layers in each sector.

A majority logic algorithm defines trigger objects for every
sector in which at least four of the eight trigger layers have
hits within a time window of 134\ns.  The IFR trigger
synchronization module processes the trigger objects from the ten
sectors and generates the three-bit trigger word (U) encoding
seven exclusive trigger conditions, as defined in
Table~\ref{\secname tab:IFT-pattern}.  The trigger $\mathrm{U}
\geq 5$, for example, covers all $\mumu$ topologies of interest.

\begin{table}
\centering
\caption{IFR trigger pattern (U) definition, where $\mu$ refers to a
signal within a sector.}
\label{\secname tab:IFT-pattern}
\vspace{.5\baselineskip}
\begin{tabular}{cl}
\hline
\rule[-5pt]{0pt}{17pt} U &Trigger condition \\
\hline
 1\rule{0pt}{12pt} &$\geq 2  \mu$ topologies other than $\mathrm{U}=5-7$  \\
 2 &1 $\mu$ in backward endcap \\
 3 &1 $\mu$ in forward endcap \\
 4 &1 $\mu$ in barrel  \\
 5 &2 back-back $\mu$'s in barrel $+ 1$ forward $\mu$ \\
 6 &1 $\mu$ in barrel $+ 1$ forward $\mu$ \\
 7\rule[-5pt]{0pt}{0pt} &  2 back-back $\mu$'s in barrel \\
\hline
\end{tabular}
\end{table}

The efficiency of the IFT has been evaluated using cosmic rays
triggered by the DCT and crossing the detector close to the IP. For
these events, 98\% were triggered by the IFT as events with at least
one track, and 73\% as events with two tracks, inside the geometrical
region of the IFR.  Most of the IFT inefficiency is concentrated at
the boundaries between sectors.

\subsubsection{Global Trigger}

The GLT receives the eight trigger primitives in the form of
$\phi$-maps as listed in Table~\ref{\secname tab:trg-primitive} along
with information from the IFT (Table~\ref{\secname tab:IFT-pattern})
to form specific triggers that are then passed to the FCTS for the
final trigger decision.  Due to the different latencies associated
with the production of these primitives, the GLT forms a time
alignment of these input data using configurable delays.

The GLT then forms some additional combined $\phi$-maps from the
DCT and EMT data.  These maps include matched objects such as BM for B
tracks matched to an M cluster in $\phi$, back-to-back objects, B$^*$
and M$^*$, which require a pair of $\phi$ bits separated by a
configurable angle of typically $\sim120^\circ$, and an EM$^*$ object
for back-to-back EM pairs.

All 16 $\phi$-maps are then used to address individual GLT
look-up-tables which return three-bit counts of trigger objects
contained within those maps, \eg\ the number of B tracks or number
of M clusters.  To count as distinct trigger objects, the map bits
are typically required to have a separation of more than one
$\phi$ bin. The resulting 16 counts plus the IFT hit pattern are
then tested in logical operations.  The permissible operations
include: always-pass; or a comparison ($\geq$, $=$, or $<$) with a
configurable selection parameter.  A trigger line is then set as
the logical AND of these 17 operations.  This process is performed
for each of the 24 trigger lines.

The GLT derives the L1 trigger time from the centroid of the
timing distribution of the highest priority trigger, binned in the
134\ns\ interval and spanning about 1\mus. Other trigger signals
compatible with this time are retained and cached. The average
time is calculated to the nearest 67\ns\ and the 24-bit GLT output
signal is sent to the FCTS every 67\ns.  The achieved timing
resolution for hadronic events has an rms width of 52\ns; and 99\%
of the events are within 77\ns.

The GLT hardware consists of a single 9U VME module.  Most of the
logic, including diagnostic and DAQ memories, are implemented in
FPGA's~\cite{ref:ORCA}.  The look-up-table section is implemented as
an array of 16 memory chips with 8\mbytes\ of configuration data.

\subsection{Level~1 Trigger Performance\\ and Operational Experience}

The L1 trigger configuration consists of DCT-only, EMT-only, mixed
and prescaled triggers, aimed not only for maximum efficiency and
background suppression, but also for the convenience of trigger efficiency
determination.

Although most triggers target a specific physics source, they often also
select other processes.  For example, two-track triggers are not only
efficient for Bhabha, $\mumu$, and $\tautau$ events, but are also
useful for selecting jet-like hadronic events and some rare $B$ decays.

The efficiencies and rates of selected L1 triggers for various
physics processes are listed in Table~\ref{\secname tab:trg-L1eff}.
Although triggering on generic \BB\ events is relatively easy, it is
essential to ensure high efficiencies for the important rare
low-multiplicity $B$ decays.  For this reason, efficiencies for
$\Bz\to\piz\piz$ and $\Bub\to\taum\overline{\nu}$ are also listed in
Table~\ref{\secname tab:trg-L1eff}.

\begin{table*}
\caption{Level~1 Trigger efficiencies (\%) and rates (Hz) at a luminosity of
$2.2\times 10^{33}$\cms\ for selected triggers applied to various
physics processes. The symbols refer to the counts for each object.}
\label{\secname tab:trg-L1eff}
\vspace{.5\baselineskip}
\centering
\begin{tabular}{lrrrrrrrrr}
\hline\hline
\rule[-5pt]{0pt}{17pt} Level~1 Trigger  & $\epsilon_{\BB}$
            & $\epsilon_{B\to\piz\piz}$ & $\epsilon_{B\to\tau\overline{\nu}}$
            & $\epsilon_{\ccbar}$ & $\epsilon_{uds}$
            & $\epsilon_{ee}$  & $\epsilon_{\mu\mu}$ & $\epsilon_{\tau\tau}$
            & Rate \\
\hline
\rule{0pt}{12pt}
A$\geq$3 \& B$^*\geq$1              & 97.1 & 66.4 & 81.8 & 88.9 & 81.1
                                    & ---  & ---  & 17.7 & 180  \\
A$\geq$1 \& B$^*\geq$1 \& A$^\prime\geq$1 & 95.0 & 63.0 & 83.2 & 89.2 & 85.2
                                    & 98.6 & 99.1 & 79.9 & 410  \\
Combined DCT (ORed)                 & 99.1 & 79.7 & 92.2 & 95.3 & 90.6
                                    & 98.9 & 99.1 & 80.6 & 560  \\
\hline
\rule{0pt}{12pt}
M$\geq$3 \& M$^*\geq$1              & 99.7 & 98.6 & 93.7 & 98.5 & 94.7
                                    & ---  & ---  & 53.7 & 160  \\
EM$^*\geq$1                         & 71.4 & 94.9 & 55.5 & 77.1 & 79.5
                                    & 97.8 & ---  & 65.8 & 150  \\
\rule[-5pt]{0pt}{0pt}
Combined EMT (ORed)                 & 99.8 & 99.2 & 95.5 & 98.8 & 95.6
                                    & 99.2 & ---  & 77.6 & 340  \\
\hline
\rule{0pt}{12pt}
B$\geq$3 \& A$\geq$2 \& M$\geq$2    & 99.4 & 81.2 & 90.3 & 94.8 & 87.8
                                    & ---  & ---  & 19.7 & 170  \\
M$^*\geq$1 \& A$\geq$1 \& A$^\prime\geq$1 & 95.1 & 68.8 & 83.7 & 90.1 & 87.0
                                    & 97.8 & 95.9 & 78.2 & 250  \\
E$\geq$1 \& B$\geq$2 \& A$\geq$1    & 72.1 & 92.4 & 60.2 & 77.7 & 79.2
                                    & 99.3 & ---  & 72.8 & 140  \\
M$^*\geq$1 \& U$\geq$5 ($\mu$-pair) & ---  & ---  & ---  & ---  &---
                                    & ---  & 60.3 & ---  &  70  \\
\hline
\rule[-5pt]{0pt}{17pt}
Combined Level~1 triggers           &$>$99.9& 99.8 & 99.7 & 99.9 & 98.2
                                    &$>$99.9& 99.6 & 94.5 & 970  \\
\hline\hline
\end{tabular}
\end{table*}

The efficiencies listed for the hadronic events are absolute and
include acceptance losses based on Monte Carlo simulation, and
local inefficiency effects.  The efficiencies for $\tau$-pair
events are for \emph{fiducial} events, \ie\ events with two or
more tracks with $\pt>120$\mevc\ and polar angle $\theta$ to reach
at least DCH superlayer U5. The Bhabha and $\mu$-pair efficiencies
are determined from the data, for events with two high momentum
particles, which are back-to-back in the c.m. system, and within
the EMC fiducial volume.  The data in Table~\ref{\secname
tab:trg-L1eff} demonstrate that the DCT and the combined EMT/IFT
provide fully efficient, independent triggers for most physics
processes, although independent triggers for \mumu\ and \tautau\
are not individually fully efficient.  The efficiencies predicted
by the Monte Carlo simulation are generally in good agreement with
data when tested using events passing typical analysis selections
and based on orthogonal triggers. Prescaled triggers with a very
open acceptance of physics events, such as (B$\geq$2 \& A$\geq$1)
or (M$\geq$2) are also used to measure the trigger efficiencies.

The trigger rates listed in Table~\ref{\secname tab:trg-L1eff} are for
a typical run with HER (LER) currents at 650\mA\ (1350\mA) and a
luminosity of $2.2\times 10^{33}$\cms. These rates are stable to
within 20\% for the same \pep2\ configuration, but they are impacted
by changes in vacuum conditions, beam currents, and orbits.  There are
occasional background spikes which can double the L1 rate. However,
due to the 2\khz\ capability of the data acquisition, these spikes do
not induce significant dead time.

\begin{figure}
\centering
\includegraphics[width=7cm]{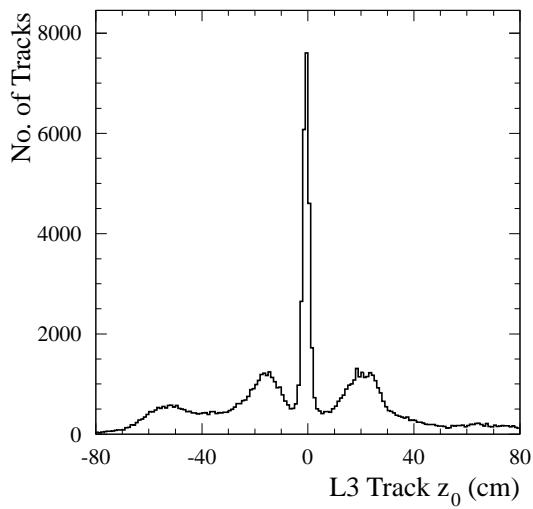}
\vspace{-2pc}

\caption{Single track $z_0$ for all L1 tracks, reconstructed by L3.}
\label{\secname fig:trg-l3trkz0}
\end{figure}

For a typical L1 rate of 1\khz, Bhabha and annihilation physics
events contribute $\sim$130\hz. There are also 100\hz\ of cosmic
ray and 20\hz\ of random beam crossing triggers.  The remaining
triggers are due to lost particles interacting with the beam pipe
or other components.  The distribution of single track $z_0$
values as reconstructed by L3 for all L1 triggers is shown in
Figure~\ref{\secname fig:trg-l3trkz0}.  The most prominent peaks
at $z=\pm 20$\cm\ correspond to a flange of the beam pipe. The
peak at $z_0=-55$\cm\ corresponds to a step in the synchrotron
mask.

The L1 trigger hardware operation has been very stable. For the first
one and half years of operation, there have been only four hardware
failures in the L1 system, mainly auxiliary or communication
boards. Occasional adjustments to the EMT tower mask were used to
temporarily suppress noisy channels in the EMC electronics.

\subsection{Level~3 Trigger System}

The L3 trigger software comprises event reconstruction and
classification, a set of event selection filters, and monitoring.
This software runs on the online computer farm.  The filters have
access to the complete event data for making their decision, including
the output of the L1 trigger processors and FCTS trigger scalers.  L3
operates by refining and augmenting the selection methods used in L1.
For example, better DCH tracking (vertex resolution) and EMC
clustering filters allow for greater rejection of beam backgrounds and
Bhabha events.

The L3 system runs within the Online Event Processing (OEP) framework
(see Section~\ref{sec:online}).  OEP delivers events to L3, then
prescales and logs those which pass the L3 selection criteria.

To provide optimum flexibility under different running conditions, L3
is designed according to a general logic model that can be configured
to support an unlimited variety of event selection mechanisms. This
provides for a number of different, independent classification tests,
called \emph{ scripts}, that are executed independently, together with a
mechanism for combining these tests into the final set of
classification decisions.

The L3 trigger has three phases.  In the first phase, events are
classified by defining L3 input lines, which are based on a logical OR
of any number of the 32 FCTS output lines.  Any number of L3 input
lines may be defined.

The second phase comprises a number of scripts.  Each script
executes if its single L3 input line is true and subsequently produces
a single pass--fail output flag.  Internally, a script may execute one
or both of the DCH or EMC algorithms, followed by one or more filters.
The algorithms construct quantities of interest, while the filters
determine whether or not those quantities satisfy the specific
selection criteria.

In the final phase, the L3 output lines are formed.  Each output line
is defined as the logical OR of selected script flags.  L3 can treat
script flags as vetoes, thereby rejecting, for example, carefully
selected Bhabha events which might otherwise satisfy the selection
criteria.

L3 utilizes the standard event data analysis framework and depends
crucially on several of its aspects.  Any code in the form of
\emph{modules} can be included and configured at run time.  A sequence of
these software modules compose a script.  The same instance of a
module may be included in multiple scripts yet it is executed only
once, thus avoiding significant additional CPU overhead.

\subsubsection{Level~3 Drift Chamber\\ Tracking Algorithm}
Many events which pass L1 but must be rejected by L3 are
beam-induced charged particle background that are produced in
material close to the IP.  L1 does not currently have sufficient
tracking resolution to identify these background tracks. The
DCH-based algorithm, L3Dch, performs fast pattern recognition
(track finding) and track fitting, which determines the five helix
track parameters for tracks with \pt\ above 250\mevc.  To speed up
the process of pattern recognition, L3Dch starts with the track
segments from the TSF system and improves the resolution by making
use of the actual DCH information.

For those TSF segments that have a simple solution to the left-right
ambiguity, a track $t_0$ is determined.  The $t_0$ values for each
segment in an event are binned and the mean produced from the values
in the most populated bin is used as the estimated event $t_0$.  All
events which pass L1 typically have enough segments to form a $t_0$\
estimate. The measured rms resolution on this estimate is 1.8\ns\ for
Bhabha events and 3.8\ns\ for hadronic events.

The pattern recognition for L3Dch is done with a look-up-table.  For
this track table, the DCH is divided into 120 $\phi$-sectors,
corresponding to the number of cells in the innermost layers. The
track table is populated with the hit patterns of Monte Carlo
generated tracks with a \pt\ above 250\mevc\ and originating within
2\cm\ of the IP in the $x$--$y$ plane, and within 10\cm\ in $z$.  The
pattern recognition algorithm searches the table entries looking for
matches to segments found by the TSFs.  The matched set of segments
for a given track is then passed to the track fitting algorithm. The
track table allows for up to two missing DCH TSF segments per track.

The track fitting algorithm is provided with both the track segments
found in pattern recognition and the individual hits within those
segments.  From this information the five helix parameters are fitted.
The fit is then iterated, adding segments close to the initially
fitted track, and dropping hits with large residuals.  The final fit
does not demand that the track originate from the IP.

\begin{figure}
\includegraphics[width=6.5cm]{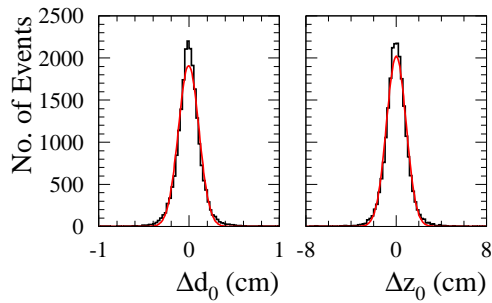}
\vspace{-2pc}
\caption{Transverse and longitudinal miss distances between
the two tracks in Bhabha events.}
\label{\secname fig:L3-2prong-miss}
\end{figure}

The two-track miss distances for Bhabha events are plotted in
Figure~\ref{\secname fig:L3-2prong-miss}. The resolutions for
individual tracks are 0.80\mm\ and 6.1\mm\ for $d_0$ and $z_0$,
respectively.  Similarly, the $1/\pxy$ difference between the two
tracks in $\mu$-pair events yields a $\pxy$ resolution of
$\delta\pxy/\pxy\sim 0.019\cdot\pxy$, with $\pt$ in \gevc.

\subsubsection{Level~3 Calorimeter\\ Clustering Algorithm}

The all-neutral trigger for L3 is based on information from the EMC.
In addition, calorimeter information is a vital complement to the DCH
data for the identification of Bhabha events.

The L3 EMC-based trigger, L3Emc, identifies energy clusters with a
sensitivity sufficient for finding minimum ionizing particles. EMC
data are processed in two steps: first, lists of crystals with
significant energy deposits are formed; and second, clusters are
identified.  The EMC typically sends data for $\sim$1400 crystals
(of 6580 total). The majority of these are caused by electronics
noise and beam-induced background.  For each crystal, these data
include the peak energy and time of the crystal waveform.  To
filter out noise, L3Emc rejects individual crystal signals below
an energy threshold of 20\mev\ or which lie outside a 1.3\mus\
time window around the event time.  For the remaining crystals,
raw energies and times are converted into physical units and added
to the L3Emc crystal list. Clusters are formed using an optimized
look-up-table technique requiring only a single pass over the
crystal list.  Clusters with a total energy above 100\mev\ are
retained, and the energy weighted centroid and average time, the
number of crystals, and a lateral moment describing the shower
shape for particle identification are calculated.

\subsubsection{Level~3 Filters}

Based on the L3 tracks and clusters, a variety of filters perform
event classification and background reduction.  The logging decision
is primarily made by two orthogonal filters, one based exclusively on
DCH data and the other based only on EMC data.

The drift chamber filters select events with one \emph{tight} (high
\pt) track or two \emph{loose} tracks originating from the IP,
respectively.  To account for the fact that the IP is not exactly at
the origin, track selection is based on its $x$--$y$ closest approach
distance to the IP, $d_0^{IP}$, and $z_0^{IP}$, the corresponding $z$
coordinate for that point.  The IP position is a fixed location close
to the average beam position over many months.  The high \pt\ track is
required to have a transverse momentum of
$\pt > 600\mevc$ and to satisfy a vertex condition defined as
$|d_0^{IP}|< 1.0$\cm, and $|z_0^{IP}-z_{IP}|< 7.0$\cm.  Two tracks are
accepted with \pt$>$250\mevc\ and a somewhat looser vertex condition
defined as $|d_0^{IP}|< 1.5$\cm, $|z_0^{IP}-z_{IP}|< 10.0$\cm.

Two calorimeter cluster filters select events with either high energy
deposits or high cluster multiplicity.  Each filter also requires a
high effective mass calculated from the cluster energy sums and the
energy weighted centroid positions of all clusters in the event
assuming massless particles.  The first filter requires at least two
clusters of $E_{CM}>350$\mev\ (c.m. system energy) and event mass
greater than 1.5\gev; the second filter requires at least four
clusters, and an event mass greater than 1.5\gev.

At current luminosities, the output of both the DCH and EMC filters is
dominated by Bhabha events, which need to be rejected. This is
accomplished by a Bhabha veto filter that selects one-prong (with only a
positron in the backward part of the detector) and two-prong events
(with both $\ep$ and $\en$ detected).  Stringent criteria on EMC
energy deposits are imposed, relying on the track momenta and on
$E/p$.  The two-prong veto requires either colinearity between the
tracks in the c.m. system or an acolinearity that is consistent with
initial state radiation (ISR).

For purposes of calibration and offline luminosity measurements,
Bhabha, radiative Bhabha, $\gamma\gamma$ final state, and cosmic ray
events are flagged.  The output rate of flagged Bhabha events is
adjusted to generate an approximately flat distribution of events in
polar angle. Radiative Bhabha events are identified by selecting
two-prong events with missing energy and requiring an EMC cluster in
the direction of the missing momentum.  Events with two high energy
clusters, back-to-back in the c.m. system select the
$\ep\en\ra\gamma\gamma$ process.  The cosmic ray selection is
DCH-based and requires two back-to-back tracks in the laboratory frame
with nearly equal impact parameters and curvature.  A significant
background from ISR Bhabha events faking this topology is removed
using the same kinematic constraints used in the two-prong veto.

The online luminosity monitoring and energy scale monitoring are
performed in L3. A track-based lepton-pair selection with a well known
efficiency monitors the luminosity.  Hadronic filters for selection of
continuum and \BB-enriched samples monitor the energy scale.  The
latter two categories are distinguished by an event shape selection
using a ratio of Fox-Wolfram moments~\cite{ref:fox-wolfram}.  The
ratio of the \BB-enriched sample to the luminosity is a sensitive
measure of relative position on the \FourS\ peak and thereby monitors
the beam energies.

\subsection{Level~3 Performance\\ and Operational Experience}

The L3 trigger efficiency for Monte Carlo simulated events are
tabulated in Table~\ref{\secname tab:trg-L3eff} for events passing
Level 1.
High efficiencies are independently achieved for the DCH and EMC based
filters applied to simulated hadronic events.  The comparison between
data and Monte Carlo L3 trigger pass fractions for the various filters
also show good agreement when requiring tracking, and EMC based
hadronic event selections in turn.

\begin{figure*}
\centering
\fbox{
\includegraphics[width=0.9\textwidth]{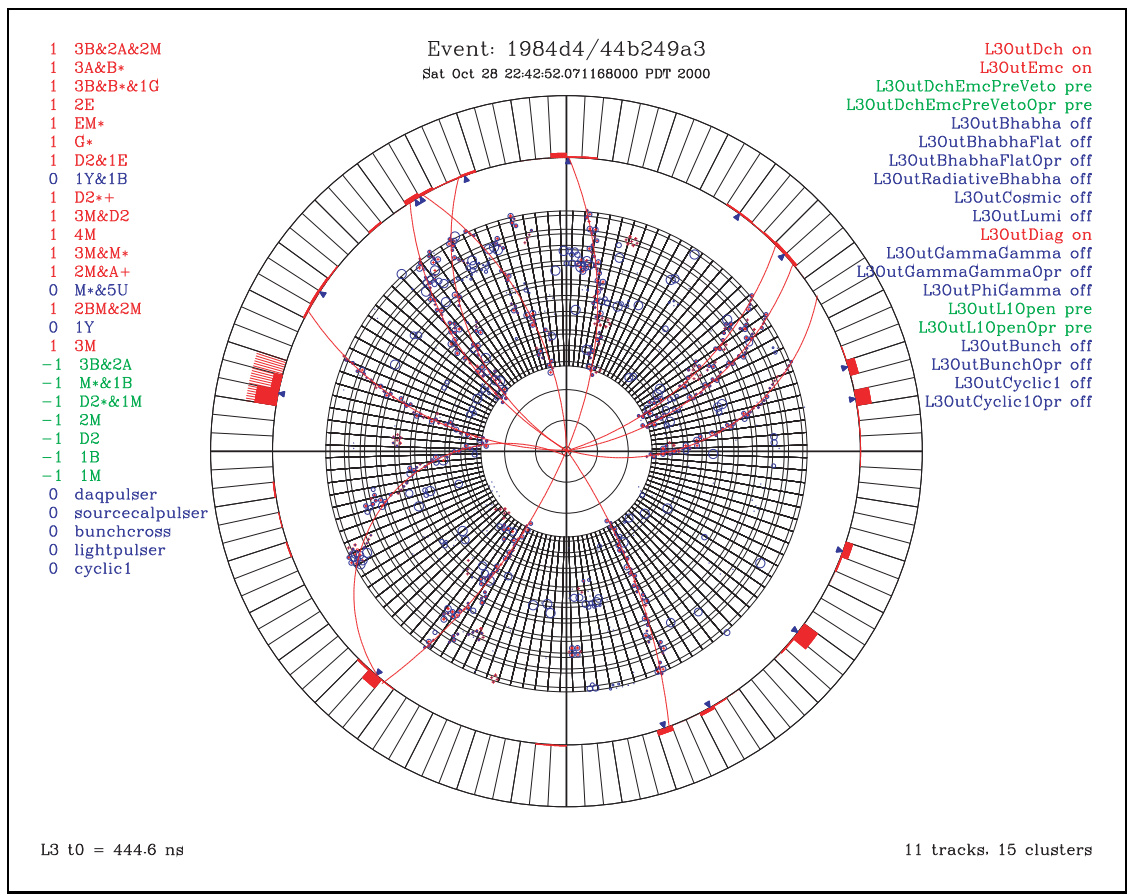}
}
\caption{A Level~3 event display. The small circles and small crosses in the
DCH volume are DCH hits and TSF segment hit wires respectively. The
filled EMC crystals represent energy deposit ($\mbox{full crystal
depth}=2\gev$) from Level~3 EMC clusters while the small triangles just
inside the EMC indicate the location of the cluster centroid.}
\label{\secname fig:trg-L3-evdisp}
\end{figure*}

An example of the event display used for online trigger monitoring is
shown in Figure~\ref{\secname fig:trg-L3-evdisp}.  L3 reconstructed
tracks and EMC clusters are shown together with the L1 and L3 trigger
line states for the event.  The left column lists the L1 trigger lines
and their states: on (1); off (0); or on but ignored due to prescale
factor (-1).  The right column shows the same information for the L3
trigger lines.

\begin{table*}
\centering
\caption{L3 trigger efficiency (\%) for various physics processes,
derived from Monte Carlo simulation.}
\label{\secname tab:trg-L3eff}
\vspace{.5\baselineskip}
\begin{tabular}{lrrrrrr}
\hline\hline
\rule[-5pt]{0pt}{17pt}
L3 Trigger                  & $\epsilon_{\BB}$
& $\epsilon_{B\to\piz\piz}$ & $\epsilon_{B\to\tau\nu}$
& $\epsilon_{\ccbar}$       & $\epsilon_{uds}$ & $\epsilon_{\tau\tau}$ \\
\hline
\rule{0pt}{12pt}
1 track filter           & 89.9 & 69.9 & 86.5 & 89.2 & 88.2 & 94.1 \\
2 track filter           & 98.9 & 84.1 & 94.5 & 96.1 & 93.2 & 87.6 \\
\rule[-5pt]{0pt}{0pt}
Combined DCH filters     & 99.4 & 89.1 & 96.6 & 97.1 & 95.4 & 95.5 \\
\hline
\rule{0pt}{12pt}
2 cluster filter         & 25.8 & 91.2 & 14.5 & 39.2 & 48.7 & 34.3 \\
4 cluster filter         & 93.5 & 95.2 & 62.3 & 87.4 & 85.5 & 37.8 \\
\rule[-5pt]{0pt}{0pt}
Combined EMC filters     & 93.5 & 95.7 & 62.3 & 87.4 & 85.6 & 46.3 \\
\hline
\rule[-5pt]{0pt}{17pt}
Combined DCH+EMC filters&$>$99.9& 99.3 & 98.1 & 99.0 & 97.6 & 97.3 \\
\hline
\rule[-5pt]{0pt}{17pt}
Combined L1+L3          &$>$99.9& 99.1 & 97.8 & 98.9 & 95.8 & 92.0 \\
\hline\hline
\end{tabular}
\end{table*}

For a typical run on the $\FourS$ peak with an average
luminosity of 2.6$\times 10^{33}$\cms, the L3 event composition is
tabulated in Table~\ref{\secname tab:L3-composition}. The desired
physics events contribute 13\% of the total output while the
calibration and diagnostic samples comprise 40\%.

\begin{table}
\centering
\caption{Composition of the L3 output at a luminosity of 2.6$\times 10^{33}\cms$.}
\label{\secname tab:L3-composition}
\vspace{.5\baselineskip}

\begin{tabular}{lr}
\hline\hline
Event type\rule[-5pt]{0pt}{17pt} &Rate (Hz)\\
\hline
Hadrons, $\tau\tau$, and $\mu\mu$\rule{0pt}{12pt} &16\\
Other QED, 2-photon events       &13\\
Unidentified Bhabha backgrounds  &18\\
Beam-induced backgrounds\rule[-5pt]{0pt}{0pt}         &26\\
\hline
Total physics accept \rule[-5pt]{0pt}{17pt}            &\textbf{73}\\
\hline\hline
Calibration Bhabhas ($\epem$)\rule{0pt}{12pt}    &30\\
$\gamma\gamma$, Radiative Bhabhas ($\epem\gamma$) &10\\
Random triggers and cosmic rays  &2\\
L1,L3 pass through diagnostics\rule[-5pt]{0pt}{0pt}   &7\\
\hline
Total calibration/diagnostics\rule[-5pt]{0pt}{17pt}    &\textbf{49}\\
\hline\hline
\end{tabular}
\end{table}

The L3 executable currently takes an average processing time of
8.5\ms\ per event per farm computer.  A Level 1 input rate of
2700\hz\ saturates the Level 3 processors, well above the 2\khz\
design requirement.  At this input rate the L3 process consumes
$\sim$72\% of the CPU time, the rest is spent in OEP, including the
network event builder and in the operating system kernel.

\subsection{Summary and Outlook}

Both the L1 and L3 trigger systems have met their original design
goals at a luminosity of $3\times 10^{33}$\cms.  The triggering
efficiencies for \BB\ events generally meet the 99\% design goal for
both L1 and L3. The orthogonal triggers based on DCH-only and EMC-only
information have successfully delivered stable and measurable overall
trigger efficiency. The current system also provides a solid
foundation for an upgrade path to luminosities of $10^{34}$\cms\ or
more.

Short-term L1 trigger improvements will primarily come from
further background rejection, afforded by algorithm refinements
and upgrades of the DCT. This is essential for reducing the load
on the DAQ and L3. The new PTD algorithm will effectively narrow
the track $d_0$ acceptance window, while a new BLT algorithm will
narrow the track $z_0$ acceptance.

For the longer term future, a major DCT upgrade is planned.  By adding
the stereo layer information, a $z_0$ resolution of 4\cm\ is expected,
allowing for an efficient rejection of beam-induced background beyond
$z=\pm20\ \cm$.

Future improvements for L3 will also emphasize background rejection.
Improvements in the L3 IP track filter are expected to further reduce
beam-induced background to about one third of current levels.  The
physics filter algorithms will be tuned and improved, primarily for
rejecting Bhabha, QED, and two-photon events.  Improvements in the L3
tracking algorithms are expected to lower the $\pt$ thresholds below
250\mevc.  A moderate CPU upgrade for the L3 online farm will be
sufficient to keep up with luminosities of $\sim10^{34}$\cms.

\renewcommand{\secname}{online_}
\renewcommand{\sectiondir}{sec12_online}
\section{The Online Computing System}
\label{sec:online}

\subsection{Overview}
\label{sec:online-overview}

The \babar\ online computing system comprises the data acquisition
chain from the common FEE, through the embedded processors in the data
acquisition system and the L3 trigger, to the logging of event data.
It also includes those components required for detector and data
acquisition control and monitoring, immediate data quality monitoring,
and online calibration.

\subsubsection{Design Requirements}

The data acquisition chain was designed to meet the following
basic performance requirements. It must support a L1 trigger
accept rate of up to 2\khz, with an average event size of
$\sim$32\kbytes\ and a maximum output (L3 trigger accept) rate of
120\hz.  While performing these functions it should not contribute
more than a time-averaged 3\% to deadtime during normal data
acquisition.

The online system is also required to be capable of performing
data acquisition simultaneously on independent \emph{
partitions}---sets of detector system components---to support
calibrations and diagnostics.

Normal detector operation, data acquisition and routine
calibrations are performed efficiently and under the control of a
simple user interface with facilities for detecting, diagnosing,
and recovering from common error conditions.

Following standard practice, the event data acquired by the system
are subjected to monitoring.  Such monitoring is configurable by
experts and designed to detect anomalies in the detector systems
which, if present, are reported to operators for rapid assessment
and, if necessary, corrective action.

Environmental conditions of the detector, such as the state of low
and high voltage power, the purity of gas supplies, and the
operating conditions of the accelerator, such as beam luminosity
and currents, are measured and recorded in a fashion that permits
the association with the event data logged.  Conditions relevant
to data quality are monitored for consistency with specified
standards.  Operators are alerted if these are not met.
Data-taking is inhibited or otherwise flagged if conditions are
incompatible with maintaining the quality of the data.

Operational configurations, calibration results, active software
version numbers, and routine messages and error messages are also
recorded. During data analysis or problem diagnosis, these data
help in reconstructing the detailed operating conditions.

\subsubsection{System Components}

The online computing system is designed as a set of subsystems
using elements of a common software infrastructure running on
a dedicated collection of hardware.

The major subsystems are:

\begin{itemize}
  \item Online Dataflow (ODF)---responsible
        for communication with and control of the detector systems'
        front-end electronics, and the acquisition and building of event
        data from them;
  \item Online Event Processing (OEP)---responsible for processing of
        complete events, including L3 (software) triggering, data
        quality monitoring, and the final stages of calibrations;
  \item Logging Manager (LM)---responsible for receiving selected events sent
        from OEP and writing them to disk files for use as input to the
        \emph{Online Prompt Reconstruction} processing;
  \item Online Detector Control (ODC)---responsible for the control and monitoring of
        environmental conditions of the detector systems;
  \item Online Run Control (ORC)---ties together all the other components, and is
        responsible for sequencing their operations, interlocking them as
        appropriate, and providing a \emph{graphical user interface}
        (GUI) for operator control.
\end{itemize}

Each of these components, as well as a selection of the common tools
which tie them together are described below.

The entire system is coded primarily in the C++ language, with some
use of Java for graphical user interfaces.  Object-oriented analysis
and design techniques have been used throughout.  This has been an
important feature, enhancing development speed, maintainability, and
extensibility.

\subsubsection{Hardware Infrastructure}

\begin{figure*}
\centering
\includegraphics[width=12cm]{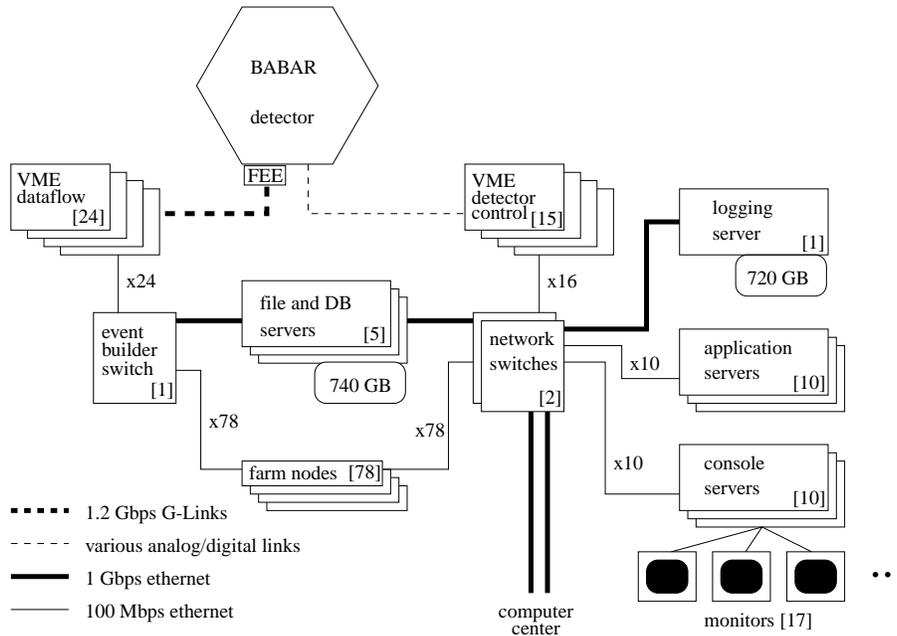}

\vspace{-2pc} \caption{Physical infrastructure of the \babar\
online system, including VME crates, computers, and networking
equipment.} \label{\secname fig:phys-overview}
\end{figure*}

The hardware infrastructure for the online system is shown schematically
in Figure~\ref{\secname fig:phys-overview}.

The data from the FEEs of the various detector systems are routed
via optical fiber links to a set of 157 custom VME \emph{Readout
Modules} (ROMs).  These ROMs are grouped by detector system and
housed in 23 data acquisition VME crates that are controlled by
the ODF software. One ROM in each crate aggregates the data and
forwards them for event building to 32 commercial Unix
workstations~\cite{ref:ov-farm} which are part of the online farm.
Other farm machines perform data monitoring and calibrations.  The
crates and farm machines communicate via full-duplex 100\mbsps\
Ethernet, linked by a network switch---the \emph{event builder
switch}~\cite{ref:ov-eb-switch}.  The ROMs are supported by a
\emph{boot server} providing core and system-specific code and
configuration information~\cite{ref:ov-boot-logging}.

The thirty-two online farm machines host the OEP and L3 trigger
software. The events accepted by the trigger are logged via TCP/IP
to a \emph{logging server}~\cite{ref:ov-boot-logging} and written
to a disk buffer for later reconstruction and archival storage.
Various data quality monitoring processes run on farm machines not
used for data acquisition.

Several additional file servers hold the online databases and
production software releases.  A further set of \emph{application
servers} host the central functions of the various online
subsystems. Operator displays are supported by a group of ten
\emph{console servers}~\cite{ref:ov-servers}.

An additional set of 15 VME crates, each with an embedded processor,
contain the data acquisition hardware for the detector control
subsystem.

All VME crates, the online farm, and all the application and
console servers are connected via a switched 100\mbsps\ Ethernet
network distinct from that used for event building, with 1\gbsps\
fiber Ethernet used for the file servers and inter-switch links.

\subsubsection{User interaction}

Operator control of the online system is achieved primarily through a
custom Motif GUI for run control and an extensive hierarchy of
displays for detector control, including control panels, strip charts
and an alarm handler.  An electronic logbook is made available through
a Web browser interface.  These and other GUIs are organized across
seventeen displays for the use of the experiment's operators.  This
operator environment provides for basic control of data acquisition,
the overall state of the detector, and certain calibration tasks.

Each detector system has developed a set of specialized calibration
and diagnostic applications using the tools provided in the online
system.  A subset of these calibrations has been specified to be run
once per day, during a ten-minute scheduled beam-off period.  The run
control logic, combined with the capability for creating partitions,
allows calibrations for all detector systems to be run in parallel and
provides the operator with basic feedback on the success or failure of
each.

\subsection{Online Dataflow}
\label{sec:online-odf}

ODF handles data acquisition and processing from the detector
systems' FEE through the delivery of complete events to the online
farm~\cite{ref:odf}.  The ODF subsystem receives the L1 trigger
outputs, filters and distributes them to the FEE, reads back the
resulting data and assembles them into events.  It provides
interfaces for control of data acquisition, processing and
calibration of detector system data, and FEE configuration.
Multiple independent \emph{partitions} of the detector may be
operated simultaneously.

Event data acquisition proceeds from a trigger decision formed in
the Fast Control and Timing System (FCTS)~\cite{ref:fcts} based on
inputs from the L1 trigger.  \emph{L1 Accepts} are distributed, in
the full detector configuration, to the 133 ROMs connected via
optical fibers to the detector system FEE.  These ROMs read and
process the data from the FEE.  One to ten such ROMs from a single
detector system are located in each of the  data acquisition VME
crates.  ODF builds complete events from these ROMs, first
collecting the data in each crate into an additional dedicated
ROM, and then collecting the data from the 23 of these, across the
event builder network switch, into the online farm.

The operation of the system is controlled by ODF software running on
one of the application servers, under the direction of run control.  A
single ROM in the VME crate containing the central FCTS hardware
supports the software interface to ODF.  The distribution of ROMs by
detector system is shown in Table~\ref{\secname tab:odf-hardware}.
The numbers of ROMs is shown as a sum of those connected directly to
the detector FEE, and those used for event building.

\begin{table}
\begin{center}
\caption{VME crates and ROMs used by ODF}

\vspace{.5\baselineskip}
\begin{tabular}{lcc}
\hline
Detector\rule{0pt}{12pt} &VME &Readout\\
System\rule[-5pt]{0pt}{0pt} &Crates &Modules\\ \hline
SVT\rule{0pt}{12pt}  &5 &14+5\\
DCH  &2 &4+2\\
DIRC &2 &6+2\\
EMC &10 &100+10\\
IFR  &1 &4+1 \\
EMT  &1 &1+1 \\
DCT  &1 &3+1 \\
GLT  &1 &1+1 \\
FCTS\rule[-5pt]{0pt}{0pt} &1 &1\\
\hline
Total\rule[-5pt]{0pt}{17pt}Total &24 &157\\
\hline
\end{tabular}
\label{\secname tab:odf-hardware}
\end{center}
\end{table}

All of the ROM CPUs boot via NFS over the event building network from
the boot server described above.  About 1.5\mbytes\ of core ODF code
plus another $\sim$4\mbytes\ of detector-specific code are loaded into
each ROM.  This, along with the booting process, takes about
40~seconds.

The ODF software allows all the components of this heterogeneous
system to be represented in a uniform object-oriented application
framework.  These components are organized into five \emph{levels}
which map closely onto the physical structure.

\begin{figure*}
\centering
\includegraphics[width=14cm]{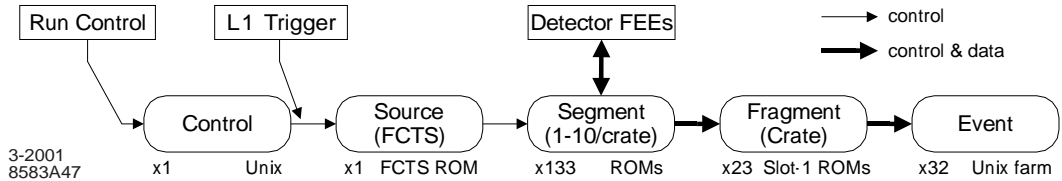}
\vspace{-2pc} \caption{Schematic of the ODF \emph{levels}, their
mapping onto physical components, and the flow of control signals
and data between them.} \label{\secname fig:odf-levels}
\end{figure*}

For each component at each level, its operation is abstracted as a
finite state machine.  The complete set of these machines is kept
coherent by passing messages and data regarding state transitions
along the chain of levels.  The basic flow of control and data is
shown in Figure~\ref{\secname fig:odf-levels}.  The mapping of levels
to components is as follows:

\emph{Control} \quad The Unix-based process controlling the
operation of each partition and the source of all state
transitions except for \emph{L1 Accept}.  It transmits state
transition messages over the network to the source level, waiting
for acknowledgement of their successful processing by all levels.

\emph{Source}\quad The FCTS hardware and the software running in
the ROM located in the FCTS VME crate.  For each partition, its
source level receives control level transitions and L1 trigger
outputs and distributes them via the FCTS hardware to all ROMs in
the VME crates included in the partition.  L1 triggers are modeled
in the subsystem as an additional, idempotent state transition,
\emph{L1 Accept,} and are treated uniformly with the others
wherever possible.

\emph{Segment}\quad The ROMs connected to the detector FEE, with
their ODF and detector system-specific software.  Each segment
level ROM receives state transition messages from the source level
and runs appropriate core and detector system-specific tasks in
response. These tasks include the acquisition of raw data from the
FEE in response to \emph{L1 Accepts}, and \emph{feature
extraction}. Output data resulting from this processing is
attached to the transition messages, which are then forwarded over
the VME backplane to the fragment level ROM in each crate.

\emph{Fragment}\quad The per-crate event builder ROMs and
software. The single fragment level ROM in each crate aggregates
the messages from the crate's segment level ROMs---the first stage
of event building---and forwards the combined message to one of
the event level Unix nodes.

\emph{Event}\quad The processes on the online farm nodes receiving
complete events and handing them over to OEP for filtering and
logging.  The ODF event level code aggregates messages, with their
attached data, from all the crates in a partition---the second and
final stage of event building.  The resulting data may be further
processed by user code in the event level, but are normally just
passed on to OEP.  The control level is notified of the completion
of processing of all transitions other than \emph{L1 Accept}.
Both the fragment and event level event builders use a data-driven
``push'' model, with a back pressure mechanism to signal when they
are unable to accept more data.

Test stands of varying complexity are supported.  The simplest
possible consists of a single Unix machine which runs both control and
event level code, with two FCTS modules and a single ROM, running
source, segment, and fragment level code, in one VME crate.
Configuration is detected at run-time, so the same code that runs in
the full system can also run in test stand systems.

\subsubsection {Control and Source Levels}

The control level sends state transition messages for a partition
over the network, using the User Datagram Protocol,
UDP~\cite{ref:udpip}, to the source level in the single ROM inside
the FCTS crate.  In the source level, the transition message is
sent over VME to an FCTS module which forwards it as a 104-bit
59.5\mhz\ serial word to all VME crates in the relevant partition.
This serial word contains a 56-bit event time stamp (counting at
59.5\mhz), a 32-bit transition-specific word and additional
control bits.  \emph{L1 Accept} transitions and calibration
sequences, however, originate in the source level and the same
mechanism is used to transmit them through the system.

The FCTS hardware receives the 24 L1 trigger output lines and eight
additional external trigger lines.  The FCTS crate is a 9U VME crate,
with a custom P3 backplane on which all the trigger lines are bussed.
For each partition, an FCTS module receives these lines.  It is
configurable with a bit mask specifying the trigger lines enabled for
its partition, and an optional prescale factor for each line.  A
trigger decision is formed for the partition by taking the logical OR
of the enabled prescaled lines.  Twelve of these modules are installed
in the full system, thus setting its maximum number of partitions.  A
detector system can belong to only one partition at a time.

The FCTS crate receives two timing signals from the accelerator: a
476\mhz\ clock tied to the RF structure of \pep2\ and a 136\khz\
fiducial that counts at the beam revolution frequency.  The former is
divided by eight to create a 59.5\mhz\ system clock.  The fiducial is
used to start timing counters and to check the synchronization of the
clocks.

There are two types of deadtime in the ODF subsystem.  The first
arises from the minimum 2.7\mus\ spacing between \emph{L1~Accept}
transitions.  This restriction simplifies the logic design of the
FEE readout, because each signal in the silicon tracker and drift
chamber is thus associated with only one \emph{L1~Accept}.  The
FCTS hardware enforces this minimum separation between
transitions, introducing an irreducible, yet minimal dead time of
0.54\% at 2\khz.

The second type of deadtime arises when all FEE buffers are full
and thus unable to accept another event.  In a time required to be
less than the inter-command spacing, each VME crate in a partition
may send back a \textsc{full} signal indicating that it is no
longer able to process further \emph{L1~Accept} transitions.  The
FCTS hardware detects these signals and disables triggering until
the FEE are once again prepared to accept data.

An actual \emph{L1 Accept} signal is only generated from a
partition's trigger decision when neither form of dead time is
asserted.

\subsubsection {Segment and Fragment Levels}

The segment and fragment levels reside in the 23 detector system VME
crates.  These are standard 9U crates with a custom P3 backplane.

The 104-bit serial transition messages that leave the source level are
received by a FCTS module in each VME crate in a partition.  This
module in turn forwards these messages to the ROMs in the crate over
the custom backplane, along with the 59.5\mhz\ system clock.

A ROM consists of four components (see Figure~\ref{\secname
fig:rom}), a commercial single-board computer
(SBC)~\cite{ref:mvme2306} and three custom boards. The custom
boards include: a \emph{controller card} for receiving FCTS
commands and supporting FEE reads and writes; a \emph{personality
card} that transmits commands to and receives data from the FEE;
and a PCI mezzanine card with a 33\mhz\ Intel i960 I/O processor.
The SBCs run the VxWorks~\cite{ref:vxworks} operating system with
custom code written in C++ and assembly language.

\begin{figure}
\centering
\includegraphics[width=7.5cm]{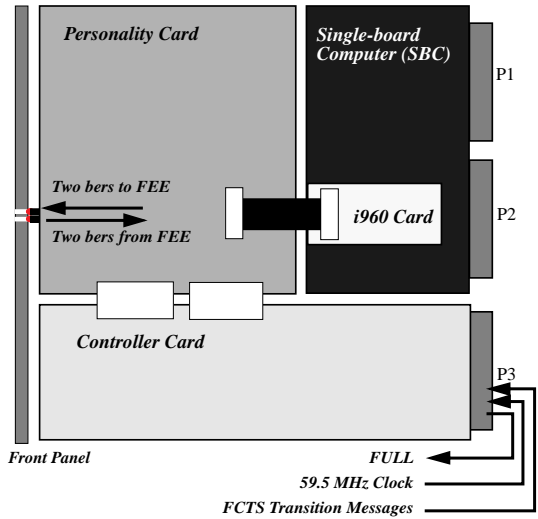}
\vspace{-2pc}
\caption{A ROM with a triggered personality card
(TPC)} \label{\secname fig:rom}
\end{figure}

There are two styles of personality cards in the system: triggered
(TPC) and untriggered (UPC).  UPCs are used only in the EMC
system.  UPCs accept data continuously from the FEE into a buffer
pipeline, at a rate of 3.7\mhz.  From these samples EMC trigger
information is derived and sent over a dedicated serial link to
the trigger hardware, providing it with a continuous data stream.
An \emph{L1~Accept} causes up to 256 samples of the raw data
stream to be saved to an intermediate memory on the UPC.

A TPC (used in all other systems) reads out FEE data only when an
\emph{L1~Accept} signal is received, again saving it into an
intermediate memory.  Each detector reads out data in a time
window around the trigger signal, large enough to allow for
trigger jitter and detector time resolution.  For instance, this
window is about 500\ns\ wide for the SVT.  The actual event time
within this window is estimated in the L3 trigger software and
then refined offline in the course of full event reconstruction.

FEE commands are sent and data received by the personality cards over
uni-directional 1.2\gbsps\ serial optical fiber
links~\cite{ref:odf-glink}.  All FEEs provide zero suppression in
hardware except in the EMC and IFR.  Data are transferred from the
personality card to the SBC memory using the i960 as a direct memory
access (DMA) engine.  This DMA runs at nearly the ideal 133\mbps\ rate
of the PCI bus.

The FEEs for various systems are able to buffer data for three to
five \emph{L1 Accept} transitions.  The ROM keeps track of the
buffer occupancy and sends, when necessary, a \textsc{full} signal
as previously described.  The \textsc{full} condition is removed
when event reading by the ROM frees sufficient buffer space.  This
mechanism handles back pressure from any stage of the data
acquisition through to data logging by OEP.

The ODF application framework provides uniform software entry points
for the insertion of user code at each level of the system.  This
capability is used primarily at the segment level, for FEE
configuration and feature extraction.  Table~\ref{\secname
tbl:evtsize} presents typical data contributions from each detector
system and the trigger.

\begin{table}
\caption{Typical event sizes from detector systems}

\vspace{.5\baselineskip}

\begin{tabular}{lcr@{.}lr@{.}l} \hline
\rule{0pt}{12pt}&Hit Size &\multicolumn{2}{c}{Total Size}
&\multicolumn{2}{c}{Overhead}\\
Detector\rule[-5pt]{0pt}{0pt} &\multicolumn{1}{c}{(bytes)} &\multicolumn{2}{c}{(kB)}
&\multicolumn{2}{c}{(kB)}\\ \hline
SVT\rule{0pt}{12pt}  &\hphantom{1}2 &4&9 &0&4\\
DCH &10  &4&8 &0&2\\
DIRC &\hphantom{1}4  &3&1 &0&3\\
EMC  &\hphantom{1}4  &9&1 &3&0\\
IFR  &\hphantom{1}8  &1&2 &0&2\\
EMT  &--- &1&2 &$<\,$0&1\\
DCT  &--- &2&7 &0&1\\
\rule[-5pt]{0pt}{0pt}GLT  &--- &0&9 &$<\,$0&1\\ \hline
\rule[-5pt]{0pt}{17pt}Total& &\hphantom{33}27&9 &\hphantom{333}4&2\\
\hline
\end{tabular}
\label{\secname tbl:evtsize}
\end{table}

Data from the segment level ROMs in a crate are gathered by the
fragment level ROM using a \emph{chained} sequence of DMA operations.
The maximum throughput of the fragment level event builder is about
31\mbsps.

In calibrations, ODF may be operated in a mode in which \emph{L1
Accept} data are not transferred out of the segment level ROMs.
This allows for calibration data accumulation at high rates inside
the ROMs, not limited by the throughput of the event builders or
any downstream software.  Completed calibration results are
computed, read out, and written to a database.

\subsubsection {Event Level}

For each \emph{L1 Accept} transition passing through the ODF
subsystem, all fragment ROM data are sent to one of the farm
machines.  The destination is chosen by a deterministic
calculation based on the \emph{L1 Accept}'s 56-bit time stamp,
available from the FCTS in each ROM. This technique produces a
uniform quasi-random distribution and introduces no detectable
inefficiency.  Events sent to a farm machine still busy with a
previous event are held in a buffer to await processing.

All fragment data for an event are sent over the switched 100\mbsps\
Ethernet event building network to the selected farm machine.  The
connectionless UDP was chosen as the data transport
protocol~\cite{ref:neteb}, allowing a flow control mechanism to be
tailored specifically to this application.  Dropped packets are
minimized by the network's purely point-to-point, full duplex switched
architecture, and by careful tuning of the buffering in the network
switch and other parameters.  The rare instance of packet loss is
detected by the event builder and the resulting incomplete event
is flagged.

The event level provides the standard software entry points for user
code.  During normal operation, these are used only to transfer events
via shared memory to the OEP subsystem for L3 triggering, monitoring,
and logging.

\subsubsection{System Monitoring}

It is critical that the clocks of the FEEs stay synchronized with
the rest of the system.  Each FEE module maintains a time counter
which is compared to the time stamp of each \emph{L1 Accept} in
order to ensure that the system remains synchronized.  If it
becomes unsynchronized, a special \emph{synch} command can be sent
through the FCTS, causing all systems to reset their clocks.

To ensure that the data from the correct event is retrieved from
the FEEs, a five-bit counter is incremented and sent from the FCTS
to the FEEs with each \emph{L1 Accept}.  These bits are stored in
the FEEs along with the data and are compared on read-back.  If
they disagree, a special \emph{clear-readout} command is sent
which resynchronizes ROM buffer pointers with FEE buffer pointers.

All transitions, including \emph{L1 Accept}, are logged in a
4\kbytes-deep by 20~byte-wide FIFO as they pass through the FCTS
crate.  The transition type, the event time stamp, a bit list of
the trigger lines contributing to the decision, and the current
\textsc{full} bit list from all VME crates are recorded in this
FIFO.  There are also scalers which record delivered and accepted
luminosity, deadtime due to the 2.7~$\mu$s minimum inter-command
spacing, deadtime caused by VME crates being \textsc{full} and
triggers on each line.  These FIFOs and scalers are read out by
the FCTS ROM, which then transmits the data to monitoring programs
that calculate quantities such as luminosity, deadtime and trigger
rates.  The UDP \emph{multicast} protocol~\cite{ref:udpmulti} is
used to allow efficient simultaneous transmission of data to
multiple clients.

To provide diagnostics, a system which multicasts additional
performance information on demand from each CPU, typically at 1\hz, is
used.  This information is currently received by a single client on
one of the Unix application servers and archived.  It can be retrieved
subsequently to investigate any unusual behaviour observed in the system.

\subsection{Online Event Processing}

The online event processing (OEP) subsystem provides a framework for
the processing of complete events delivered from the ODF event
builder~\cite{ref:OEP}.  The L3 software trigger operates in this
framework, along with event-based data quality monitoring and the
final stages of online calibrations.  Figure~\ref{\secname
fig:oep-flow} shows the basic flow of data in the OEP subsystem.

The OEP subsystem serves as an adapter between the ODF event builder
interface and the application framework originally developed for the
offline computing system.  Raw data delivered from the ODF subsystem
are put into an object-oriented form and made available through the
standard event data analysis interface.

The use of this technique permits the L3 trigger and most of the
data quality monitoring software to be written and debugged within
the offline environment.  This software is decomposed into small,
reusable units---\emph{modules}, pluggable software components in
the framework---many of which are shared among multiple
applications.

The OEP interfaces allow user applications to append new data blocks
to the original raw data from ODF.  The results of L3 event
analysis are stored in this manner so that the trigger decision and
the tracks and calorimeter clusters on which it is based may be used
in later processing, such as reconstruction and trigger performance
studies.

Histograms and other monitoring data are accumulated across the farm.
A \emph{distributed histograming package} (DHP)~\cite{ref:dhp} was
developed to provide networked clients with a single view of
histograms and time history data.  This data is summed across all
nodes via CORBA-based communication
protocols~\cite{ref:corba,ref:acetao}.

\begin{figure*}
\includegraphics[width=14.5cm]{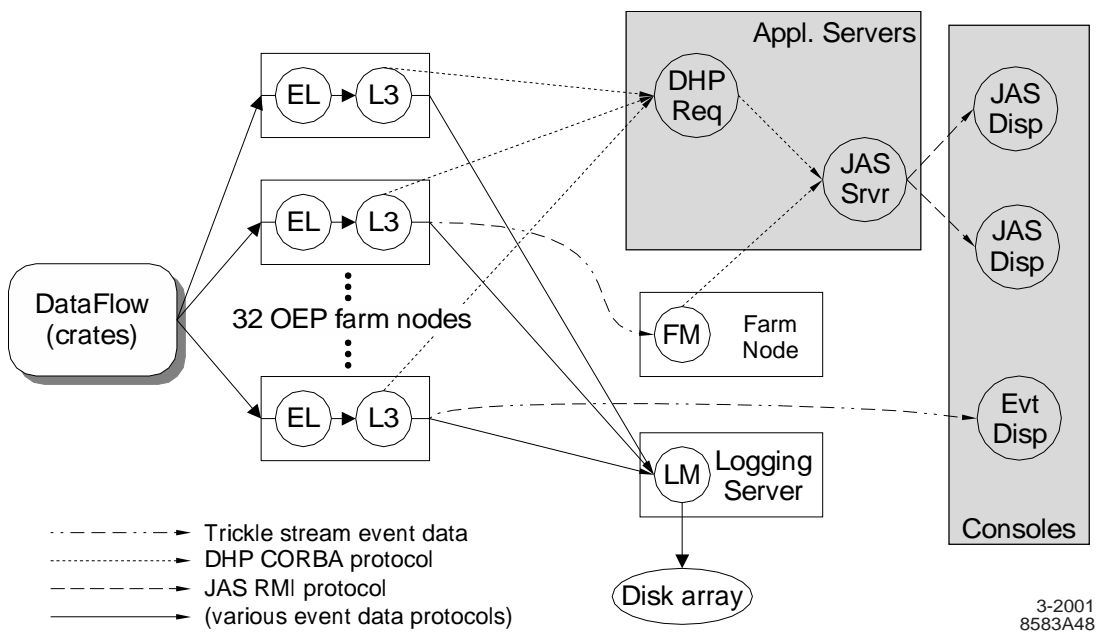}
\vspace{-2pc}
\caption{Flow of data in the OEP subsystem: ODF event level (EL) and
L3 trigger processes on each OEP node; the Logging Manager (LM)
on the logging server; the DHP ``requestor'' process that combines
histograms from all 32 L3 processes; one instance of a Fast
Monitoring (FM) process with DHP histograms; the Java server that
makes DHP histograms available to JAS clients; two such clients, and
one event display for the L3 trigger.  OEP-specific data
transport protocols are identified.}
\label{\secname fig:oep-flow}
\end{figure*}

The fast monitoring system provides automated comparisons of
monitoring data against defined references.  Statistical comparisons
of live histograms, or the results of fits to reference histograms,
analytic spectra, or nominal values of fit parameters may be performed
at configurable time intervals.  Comparison failures, tagged with
configurable severity levels based on the confidence levels of the
comparisons, are displayed to operators and logged in the common
occurrence database, described below.

The \emph{Java Analysis Studio} (JAS) package~\cite{ref:jas}
previously developed at SLAC was enhanced with the ability to serve as
a DHP client.  It is used for viewing of monitoring data.  This
feature was implemented by devising a Java server that adapts the DHP
protocol to the native JAS data protocol.

In addition to the primary triggering and monitoring functions carried
out on 32 online farm machines, OEP provides a ``trickle stream''
protocol that allows networked clients to subscribe to a sampling of
the event data.  This scheme provides support for event displays and
additional detailed data quality monitoring (``Fast Monitoring'').  A
dedicated operator console supports the JAS-based data quality
monitoring system.  This console is used to display histograms from
Fast Monitoring and the L3 trigger processes, along with any
error conditions detected by the automatic histogram analysis
facility.

\subsection{Data logging}

Events selected by the L3 trigger algorithms in OEP are retained
for subsequent full reconstruction.  The events are sent from the
32 OEP nodes via TCP/IP to a single multithreaded process, the
\emph{Logging Manager} (LM), running on the logging server.  The
LM writes these data to RAID storage arrays in a format specific
to OEP.  Data from all 32 nodes are combined into a single file
for each data-taking \emph{run} (typically two to three hours of
data acquisition, resulting in files of about 15--20\gbytes\ in
size).

Completed data files are copied to the SLAC High Performance Storage
System (HPSS)~\cite{ref:ibmhpss} system for archiving to tape.  Within
eight hours of data acquisition these files are retrieved from HPSS
for event reconstruction.  The data files are also retrievable for
other tasks such as detector system hardware diagnostics and offline
tests of the L3 trigger algorithms.

\subsection{Detector Control}

\subsubsection{Design Principles}

The Experimental Physics and Industrial Control System, EPICS
\cite{ref:EPICS}, was selected to provide the basis for the ODC
subsystem. This provides direct connection to the electrical
signals of the power supplies and other hardware, with sufficient
monitoring and control to allow commissioning, fault diagnosis,
and testing. A summary of monitoring and control points is
presented in Table~\ref{\secname tab:odc-monitor}.

\begin{table*}
\begin{center}
\caption{ODC Distribution by system of approximately 12,000
recorded monitor channels} \vspace{.2cm}
\begin{tabular}{lrrrrrrr}
\hline \rule[-5pt]{0pt}{17pt}  & SVT  & DCH  & DIRC & EMC & IFR &
Central & \\ \hline\hline
Radiation Dose\rule{0pt}{12pt} &12 &8&22 &116 &---   &--- & \\
Data Rate &---   &---   & 12 &---& 1612   &   20 &\\
Temperature  &  208 &   87 & 36 &  506&  146 &  100 &\\
Humidity      4 &    5 &   12 & 7 &--- &--- &2 & \\
Magnetic Field      ---   &---   &    8 &--- &--- &---  & 50 &\\
Position   &   30 &   22 &---   &--- &--- &---   & \\
Gas System  &    4 &  115 &   12 &    1 & 32 &--- & \\
Fluid System  &   20 &    2 &   18 &   18 & 20 &   90 &\\
Liquid Source  &---   &---   &---   &   12 &---   &    3 &\\
HV System  & 2080 & 1299 &  672 &   24 & 1574 &---   & \\
LV System (non-VME) &---   &   62 & 1080 & 442 &  94 &--- &\\
VME Crates      &    5 &    2 &   12 &   12 &  8 & 1500 & \\
CAN Micro Controller    &   48 &   40 &    3 &   47 &45 &---   & \\
Finisar Monitor \rule[-5pt]{0pt}{0pt}&   28 &   12 &   12 &---
&---   &---   & \\ \hline
System Totals\rule[-5pt]{0pt}{17pt} &
2439 & 1654 & 1899 & 1185 & 3531 & 1765 & \\ \hline\hline
\end{tabular}
\label{\secname tab:odc-monitor}
\end{center}
\end{table*}

Beyond the writing of custom drivers, only minor additions or changes
were required to EPICS.  EPICS and the additional \babar-specific
software are written in the C language.

Detector-wide standard hardware was adopted to ease development
and maintenance.  The standard ODC crate is a 6U VME chassis
containing a single board computer~\cite{ref:mvme177} serving as
an EPICS input/output controller (IOC).  Fifteen such crates are
used in the experiment.  EPICS is fully distributed. For example,
each IOC supplies its own naming service, notify-by-exception
semantics, and processing.  The IOCs boot from a dedicated server.

Analog data are either digitized by modules within the crates or, more
commonly, on digitizer boards located directly on the detector.  In
the latter case, the CANbus standard~\cite{ref:CANbus} is used for the
transport of signals to and from the detector.  A custom ``general
monitoring board'' (GMB)~\cite{ref:GMB} was developed to interface
CANbus to the on-detector electronics.  The GMB contains a
microcontroller, an ADC, multiplexors, and operational amplifiers.  It
can digitize up to 32 signals.

\subsubsection{User Interface}

The operator view of this part of the control system is via screens
controlled by the EPICS \emph{display manager} (DM).  Dedicated
control and display panels were developed using DM for each of the
detector systems, using common color rules to show the status of
devices.  A top-level panel for ODC summarizes the status of all
systems and provides access to specialized panels.

The EPICS alarm handler with some \babar-specific modifications is
used to provide operators with audible and color-coded alarms and
warnings in a hierarchical view of all the systems and components.
Conditions directly relevant to personnel or detector safety are
further enforced by hardware interlocks, the status of which are
themselves displayed in a set of uniform EPICS screens, in the
alarm handler, and on an alarm annunciator panel.

\subsubsection{Interfaces to Other\\ \babar\  Software}

A custom C++ layer above EPICS consisting of \emph{Component
Proxies} and \emph{Archivers} provides for device-oriented state
management and archival data collection.  This is ODC's interface to
the rest of the online system.

The 27 component proxies (CPs), running on a Unix application server,
each define a logical component representing some aspect of a detector
system or the experiment's central support systems, aggregated from
the $\lsim10^5$ individual EPICS records.

The CPs present a simple finite state machine model as their
interface to Run Control. The most important actions available are
\emph{Configure}, on which the CP accesses the configuration
database, retrieving set points for its component's channels, and
\emph{Begin Run}, which puts the CP into the \emph{Running} state,
in which setpoint changes are prohibited and readbacks are
required to match settings.  While in the \emph{Running} state,
the CP maintains a \emph{Runnable} flag which reflects that
requirement and allows Run Control to ensure that data acquisition
is performed only under satisfactory conditions.

The CP's other principal function is to provide an interface for the
rest of the online system to the \emph{ambient data} collected by ODC
on the state of the detector hardware and its environment.  It is the
task of the \emph{archiver} processes, each paired with a CP, to
collect the ambient data, aggregate them and write out histories
approximately every hour to the \emph{ambient database}.  These
recorded data are associated with times so that they may be correlated
with the time stamps of the event data.  Data from the archiver
processes or from the database may be viewed with a custom graphical
browser.

Ambient data typically vary only within a narrow noise range or
dead-band.  The storage of unnecessary data is avoided by recording
only those monitored quantities which move outside of a per-channel
dead-band range or across an alarm threshold.

\subsubsection{Integration With the Accelerator}

Close integration between the \babar\ detector and the \pep2\
accelerator is essential for safe and efficient data collection.  Data
from the accelerator control system are transferred via EPICS channel
access to \babar\ for display and storage, managed by a dedicated CP.
In turn, background signals from the detector are made available to
PEP-II to aid in injection and tuning, minimizing backgrounds, and
optimizing integrated luminosity.  An important component of this
communication is the ``injection request'' handshake.  When the \pep2\
operator requests a significant change in the beam conditions, such as
injection, the request can only procede following confirmation from
\babar.  This procedure complements the safety interlocks based on
radiation dose monitors.

\subsubsection{Operational Experience}

The ODC subsystem has been operational since the initial cosmic-ray
commissioning of the detector and the beginning of data-taking with
colliding beams. The core EPICS infrastructure has proven to be very
robust. The large size of the subsystem, with its 15 IOCs and $\lsim10^5$
records, produces heavy but manageable traffic on the
experiment's network.  The rate of data into the ambient database
averages 4.6\mbytes/hr or 110\mbytes/day.

\subsection{Run Control}

The ORC subsystem is implemented as an application of SMI++, a
toolkit for designing distributed control systems~\cite{ref:SMI}.
Using this software, the \babar\ experiment is modeled as a
collection of objects behaving as finite state machines.  These
objects represent both real entities, such as the ODF subsystem or
the drift chamber high voltage controller, and abstract subsystems
such as the ``calibrator,'' a supervisor for the coordination of
online components during detector calibration.  The behavior of
the objects are described in a specialized language (SML) which is
interpreted by a generic logic engine to implement the control
system.

The SML descriptions of the objects which make up the experiment
simply specify their own states and transitions as well as the
connections between the states of different objects.  Objects perform
actions on state transitions, which may include explicitly commanding
transitions in other objects; objects may also be programmed to
monitor and automatically respond to changes of state in other
objects.  Anticipated error conditions in components of the online
system are reflected in their state models, allowing many errors to be
handled automatically by the system.  To reduce complexity, logically
related objects are grouped together into a hierarchy of cooperating
domains.

The system is highly automated; user input is generally required
only to initialize the system, start and stop runs, and handle
unusual error conditions.  The user communicates with ORC via a
configurable Motif-based GUI included in SMI++.

The states and behavior of ORC objects representing external
systems are provided by a special class of intermediate software
processes called \emph{proxies}.  A proxy monitors its system,
provides an abstraction of it to ORC, and receives state
transition commands.  These commands are interpreted and applied
to the underlying hardware or software components, implementing
the transitions' actions.  The control level of an ODF partition
is an example of such a proxy.

Communication between the various proxies and the ORC engines is
provided by DIM~\cite{ref:DIM}, a fault tolerant ``publish
and subscribe'' communications package based on TCP/IP sockets,
allowing ORC to be distributed transparently over a network.

Essential to the operation of the online system is the notion of the
\emph{Runnable} status of its various ODC and data
acquisition components, indicating that they are in a state suitable
for production-quality data-taking.  The ORC logic interlocks
data-taking to the logical AND of all components' Runnable
status.  Whenever this condition is not satisfied, data-taking may not
start and any existing run will be paused with an alert sent to the
operator.

\subsection{Common Software Infrastructure}

\subsubsection{Databases}

Five major databases are used by the online system:

1. \emph{Configuration Database:} This database, implemented using the
commercial object-oriented database management system Objectivity
\cite{ref:objectivity}, allows the creation of hierarchical
associations of system-specific configuration data with a single
numeric \emph{configuration key}.  This key is distributed to all
online components, which can then use it to retrieve from the database
all the configuration information they require.  Convenient
mnemonics are associated with the keys for currently relevant
configurations, and may be selected for use via the ORC
GUI~\cite{ref:AmbConfDB}.

2. \emph{Conditions Database:} The Conditions Database is used to
record calibration and alignment constants, and the configuration keys
in force during data-taking runs.  It has the additional feature that
the data for a given time interval may be updated as they are refined
in the course of improved understanding of the apparatus
~\cite{ref:CondDB}.

The Configuration and Conditions Databases are both made available for
reconstruction and physics analysis.

3. \emph{Ambient Database:} The Ambient Database is used principally by
the ODC subsystem to record detector parameters and
environmental data at the time they are measured~\cite{ref:AmbConfDB}.

Both the Ambient and Conditions databases, are implemented using
Objectivity, and are based on the notion of time histories of various
data associated with the experiment.  The history for each item is
divided into intervals over which a specific value is consistent.

4. \emph{Occurrence (Error) Log:} Informational and error messages
generated in the online system are routed through the CMLOG system
\cite{ref:cdev-CMLOG} to a central database, from which they
are available for operators' realtime viewing or historical browsing,
using a graphical tool, as well as for subscription by online
components which may require notification of certain occurrences.

5. \emph{Electronic Logbook:} An Oracle-based~\cite{ref:Oracle}
logbook is used to maintain the history of the data-taking, organized
by runs.  It contains information on beam parameters---instantaneous
and integrated luminosity, currents, and energies---as well as
records of data acquisition parameters such as trigger rates, data
volumes, and dead times, and the detector configuration used for a run.
The logbook also contains text comments and graphics added by the
operations staff.

A number of other databases are used in the online system for various
tasks such as indexing logged data files, the repair history of online
hardware and spares, and software problem reports.

\subsubsection{Software Release Control\\ and Configuration Management}

All of the online software is maintained in the common \babar\ code
repository, based on the freely available Concurrent Versions
System software, CVS~\cite{ref:CVS}.

The online's Unix and VxWorks applications are built and maintained
with an extension of the standard \babar\ software release
tools~\cite{ref:SRT}.  At the start of every data-taking run, the
identities of the current production software release and any
installed patches are recorded; thus it is possible at a later date to
reconstruct the versions of online software used to acquire data.

\subsection{Summary and Outlook}
\label{sec:online-operation}

The online system has exceeded its data acquisition performance
goals. It is capable of acquiring colliding beam events, with an
average size of 28\kbytes, at a $\sim 2500\hz$ L1 trigger rate and
reducing this rate in L3 to the required $\sim 120\hz$ limit. This
provides comfortable margins, since under normal beam conditions
the L1 trigger rate is 800--1000\hz.

The system is capable of logging data at much higher rates; the
nominal 120\hz\ figure represents a compromise between data volume
and its consequential load on downstream processing and archival
storage, and trigger efficiency for low multiplicity final states.

During normal data-taking, the online system routinely achieves an
efficiency of over 98\%, taking both data acquisition livetime and
the system's overall reliability into account.

There are several hardware options for enhancing ODF capacity.
Currently most ROMs receive more than one fiber from the FEE.
These fibers could be distributed over more ROMs to add processing
power. There are also commercial upgrade paths for the ROMs' SBC
boards available.  Crates can be split (up to a maximum total of
32) to create more VME event building bandwidth, as well as more
fragment level CPU power and network bandwidth. Gigabit Ethernet
connections could also be installed to improve the network event
builder's bandwidth.

Various software upgrade options are being investigated, including
optimizing the VxWorks network drivers and grouping sets of events
together in order to reduce the impact of per-event overhead.

Current background projections indicate that fragment level CPU,
segment level memory bus bandwidth, and network event building
bandwidth are the most likely bottlenecks for future running.

Increases in the L1 trigger rate or in the background occupancy
and complexity of events are expected to necessitate additional
capacity for OEP, principally for L3 triggering.  The online farm
machines could be replaced with faster models.  More machines
could be added, at the expense of increases in coherent loading on
various servers and of additional management complexity.

No significant capacity upgrades to the data logging subsystem or to
ODC are foreseen at this time.

\renewcommand{\secname}{summary_}
\renewcommand{\sectiondir}{sec13_summary}
\section{Conclusions}
\label{sec:summary}

During the first year of operation, the \babar\ detector has
performed close to expectation with a high degree of reliability.
In parallel, the \pep2\ storage rings gradually increased its
performance and towards the end of the first year of data-taking
routinely delivered close to design luminosity.  In fact, the best
performance surpassed the design goals, both in terms of
instantaneous as well as integrated luminosity per day and month.
Of the total luminosity of 23.5\,\invfb\ delivered by \pep2\
during the first ten months of the year 2000, \babar\ logged more
than 92\%. The data are fully processed with a delay of only a few
hours.  They are of very high quality and have been extensively
used for physics analysis.

A large variety of improvements to the event reconstruction and
detailed simulation are presently being pursued.  They include
improvements in many aspect of the calibration and reconstruction
procedures and software, for example, the calibration and noise
suppression in the EMC,  and the development of techniques for
precision alignment of the SVT and DCH. The latter effort will not
only benefit the overall efficiency and precision of the track
reconstruction, it will also improve the matching of tracks with
signals in the DIRC and EMC.  Detailed studies and the full
integration of all available information pertinent to the
identification  of charged and neutral particles are expected to
result in better understanding and improved performance of various
techniques.

Beyond routine maintenance, minor upgrades and a few replacements
of faulty components are currently planned.  They include the
replacement of SVT modules that are expected to fail in the next
few years due to radiation damage, plus a few others that cannot
be correctly read out due to broken connections.  A large fraction
of the RPCs are showing gradually increasing losses in efficiency
and plans are being developed for the replacement of the RPC
modules over the next few years. Furthermore, 20-25\,\cm\ of
absorber will be added to the flux return to reduce the hadron
misidentification rates.

With the expected increase in luminosity, machine-induced
backgrounds will rise.  Measures are being prepared to reduce the
sources and the impact of such backgrounds on \babar. Apart from
the addition of shielding against shower debris, upgrades to the
DCH power supply system and to the DIRC electronics are presently
under way. Most important are upgrades to the trigger, both at
levels L1 and L3.  Specifically, the DCH stereo layer information
will be added to allow for a more efficient suppression of
background tracks from outside the luminous region of the beam.
The L3 processing will be refined so as to reject both backgrounds
and high rate QED processes with higher efficiency.  In addition,
data acquisition and processing capacity will be expanded to meet
the demands of higher luminosity.

In summary, the \babar\ detector is performing very well under
current conditions and is well suited to record data at significantly
higher than design luminosity.

\section*{Acknowledgements}
The authors are grateful for the tremendous support they have
received from their home institutions and supporting staff over
the past six years.  They also would like to commend their PEP-II
colleagues for their extraordinary achievement in reaching the
design luminosity and high reliability in a remarkably short time.
The collaborating institutions wish to thank SLAC for its support
and kind hospitality.

This work has been supported by the US Department of Energy and
the National Science Foundation, the Natural Sciences and
Engineering Research Council (Canada), the Institute of High
Energy Physics (P.R.~China), le Commisariat \`{a} l'Energie
Atomique and Institut National de Physique Nucl\'{e}aire et de
Physique des Particules (France), Bundesministerium f\"{u}r
Bildung und Forschung (Germany), Istituto Nazionale di Fisica
Nucleare (Italy), the Research Council of Norway, the Ministry of
Science and Technology of the Russian Federation, and the Particle
Physics and Astronomy Research Council (United Kingdom).  In
addition, individuals have received support from the Swiss
National Foundation, the A.P.~Sloan Foundation, the Research
Corporation, and the Alexander von~Humboldt Foundation.



\begin{thebibliography}{}
\bibitem{LHe}
Solenoid cool-down and cryogenic Helium is supplied by a modified
Linde TCF-200 liquefier/refrigerator.
\bibitem{airforce}
The solenoid was shipped from Genova on a C5-B transport plane of the
US Air Force.
\bibitem{emc_footnote}
The calibration procedure employed in this measurement introduces a
dependency of the light yield on the shaping time of the
preamplifier. When connected to the actual front-end electronics in
the \babar\ detector, the signal is reduced by a factor 1.29.
\bibitem{ref:ov-servers}
These file and database servers (presently five) are primarily
additional Sun Ultra~220Rs, each with about 200\gbytes\ of RAID disk.  The
ten application servers are a mix of Sun Ultra~60s, with dual 360\mhz\
UltraSPARC-II CPUs and 1\gbytes\ of RAM, and Ultra~5s; the console servers
are all Ultra~5s.
\bibitem{ref:mvme2306}
Motorola model MVME2306 boards, each with a 300\mhz\ PowerPC 604 CPU,
32\mbytes\ of RAM, 5\mbytes\ of non-volatile flash memory, two PCI mezzanine
card slots (only one of which is currently used), a 100\mbsps\ Ethernet
interface, and a Tundra Universe~II VME interface, Motorola Inc.,
Tempe, AZ, USA.
\end{thebibliography}
\end{document}